\documentclass[%
 aps,
 prb,
 amsmath,amssymb,
 showpacs,
 reprint,%
]{revtex4-1}

\usepackage{graphicx}
\usepackage{dcolumn}
\usepackage{bm}
\usepackage{verbatim}
\usepackage{multirow}
\usepackage{array}
\usepackage{makecell}
\usepackage{rotating}
\usepackage[table]{xcolor}

\begin{document}

\title{Stable pair liquid phase in fermionic systems}

\author{Pavel Kornilovitch}
 \email{pavel.kornilovich@gmail.com}
 \affiliation{Department of Physics, Oregon State University, Corvallis, Oregon 97331 USA} 

\date{\today}  

\begin{abstract}

We predict the existence of a pair-liquid phase in lattice fermion systems with {\em finite-range} attractive interactions. This exotic state competes on one side with a normal Fermi liquid of unpaired fermions and on the other side with a phase-separated state where all fermions are coupled into macroscopic clusters. We show that such a phase is absent in bosonic systems and therefore is protected by the exclusion principle. In contrast with zero-range attractive systems where clustering of more than two fermions is directly prohibited, here quantum statistics acts dynamically and in a more subtle way. Since a many-fermion wave function must have nodes, the cluster formation threshold is larger than the pair formation threshold. By directly solving a four-body Schr\"odinger equation on one- and two-dimensional lattices, we map the boundaries of pair stability. The pair liquid phase should be observable in cold atom systems. We also discuss implications for the preformed-pair mechanism of high-temperature superconductivity.        
         
\end{abstract}



\maketitle

\section{\label{4UV2d:sec:one}
Quantum pair liquids
}

The most famous example of quantum pair fluids is superconductivity. Cooper pairs, however, are not individual particles and are only stable as a collective entity. The question of whether superconductivity can be explained by particle-like {\em preformed} pairs has been debated for decades: starting even before BCS,~\cite{Ogg1946,Schafroth1954a,Schafroth1954b,Schafroth1957} then after BCS,~\cite{Eagles1969,Bogoliubov1970,Vinetskii1983,Alexandrov1981} and most intensely after the discovery of high-temperature superconductivity.~\cite{Micnas1990,Alexandrov1994,Perali2002,Chen2005,Alexandrov2013} In this physical picture, the pseudogap state of underdoped cuprates is a manifestation of the formation of real-space pairs, and Uemura scaling of the critical temperature~\cite{Uemura1989,Uemura1991} is explained by Bose-Einstein condensation of strongly anisotropic pairs.~\cite{Alexandrov1999} Recently, normal-state hole pairs were observed in the shot noise in copper oxide junctions.~\cite{Zhou2019} This, together with the growing doubts that high-temperature superconductivity can be explained by purely repulsive mechanisms,~\cite{Alexandrov2011,Qin2020,Sherman2021} lends support to the old idea~\cite{Micnas1990} that an intermediary subsystem such as phonons,~\cite{Alexandrov1981,Alexandrov2013,Varelogiannis1996} spin fluctuations,~\cite{Scalapino2012} or polarizable orbitals~\cite{Yacoby2021} is necessary to provide direct {\em weakly-retarded} attraction between the carriers. Many properties of such systems can be studied without reference to a specific pairing mechanism, but rather by postulating a phenomenological attractive potential and then analyzing the consequences. The simplest attractive Hubbard model, while providing useful insights into pseudogap physics and the nature of BCS-BEC crossover, disregards Coulomb repulsion, and can hardly be considered a microscopic model of the cuprates or other oxide superconductors. More realistic are extended attractive Hubbard models, to be called here ``$UV$ models,'' that combine a Hubbard repulsion $U$ with a {\em finite-range} attraction $V$. Such models have been studied for a long time~\cite{Micnas1990,Lin1991,Petukhov1992,Kagan1994,Alexandrov1993,Kornilovitch1995,daVeiga2002,Kornilovitch2004,Bak2007} but recently received renewed attention.~\cite{Kornilovitch2015,Nayak2018,Boudjada2020,Jiang2021,Zhang2021,Qu2021,Singh2021} They are also supported by first-principle quantum chemistry calculations.~\cite{Zhang1991,Catlow1998}    

\begin{figure*}[t]
\includegraphics[width=0.95\textwidth]{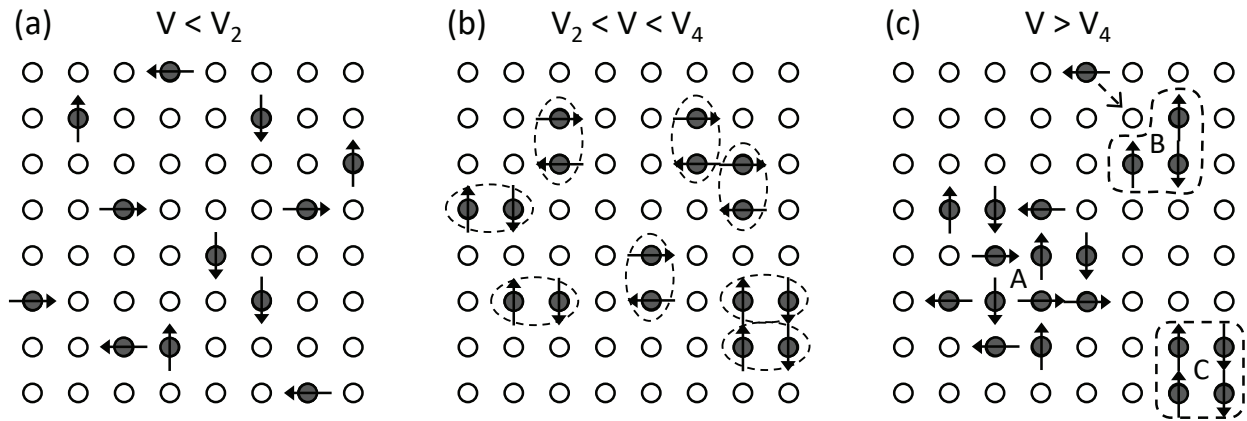}
\caption{Evolution of a Fermi-$UV$ model with increasing intersite attraction $V$. (a) $V$ is not strong enough to bind particles into pairs. Effective interaction between fermions is repulsive. The system is a Fermi liquid and may become a BCS superconductor at low temperatures. (b) $V$ is strong enough to form pairs but not strong enough to form quads. Effective interaction between {\em pairs} remains repulsive. The system is a pair liquid and may become a BEC superconductor at low temperatures. (c) $V$ is strong enough to form quads C and larger clusters A. The effective interaction between pairs becomes attractive. The system phase-separates and becomes an insulator. Quad threshold $V_4$ can be determined by solving a four-body lattice problem.}
\label{4UV2d:fig:one}
\end{figure*}

What has been largely missing from these discussions is analysis of the system's stability against phase separation. Once the pairs are formed, what prevents them from aggregating into a macroscopic cluster? In the attractive Fermi-Hubbard model, the formation of trions and larger complexes is prevented by the exclusion principle. This is not the case for $UV$ potentials. Due to a finite range of attraction, an infinite cluster will always form for sufficiently large $V$'s. To have many individual pairs, $V$ needs to be large enough to bind two carriers into a mobile pair and at the same time weak enough not to cause macroscopic phase separation. If the pairing and phase separation thresholds are different, then a {\em pair liquid} may be the system's ground state. If the two thresholds coincide, then a pair liquid state is impossible: as soon as the pairs form, the system phase-separates. The situation is illustrated in Fig.~\ref{4UV2d:fig:one}. 

We believe these considerations are of the utmost importance for pseudogap physics in particular and high-temperature superconductivity (HTSC) in general. The correct phenomenological model of cuprate superconductors must include a long-range attractive tail. If such an assumption leads to the impossibility of a pair liquid even at a model level, it would invalidate the entire real-space pairing mechanism. However, if the pair liquid state is possible, not only would it lend support to the mechanism, but it would also explain why HTSC is rare: the system's microscopic parameters must fall within a narrow region of pair stability between the two thresholds.

A different angle on these issues is provided by cold-atom technology.~\cite{Bloch2008,Blume2012,Sowinski2019,Mistakidis2022} Whereas in solid crystals the $UV$ model is an approximation to real inter-particle potentials, in optical lattices it can be precisely engineered and studied in pure form. The onsite interaction is controlled via Feshbach resonances and can be made either repulsive~\cite{Joerdens2008} or attractive.~\cite{Strohmaier2007} The intersite interaction can be controlled either by exciting dressed Rydberg atoms to large quantum numbers~\cite{Hague2012} or via proper alignment of dipolar quantum gases.~\cite{Lahaye2009} Precise manipulation of a few particles in optical traps has been demonstrated.~\cite{Serwane2011,Zurn2012,Zurn2013} Pseudogap behavior~\cite{Gaebler2010} and a pseudogap--Bose-gas evolution~\cite{Perali2011} have both been observed. Local pairing can now be measured directly using gas microscopy.~\cite{Mitra2018,Hartke2022} Thus the question of pair liquid stability can be answered experimentally.    

Rigorous analysis of phase separation with nonlocal attraction is difficult. Emin~\cite{Emin1994} argued for the existence of a stable liquid of {\em large} bipolarons. In Refs.~[\onlinecite{Alexandrov2002a},\onlinecite{Alexandrov2002b}], phase boundaries of a bipolaronic superconductor were determined by counting kinetic and lattice energy contributions following a displaced oscillator (Lang-Firsov) transformation of the bare unscreened electron-ion model. It was found that the pair (bipolaron) liquid was stable within a narrow but finite range of the polaron binding energy. Finite stability intervals were reported in a ``hole-rich'' phase of the {\it t-J} model,~\cite{Emery1990,Dagotto1993} which is equivalent to the $UV$ model in the $U \rightarrow \infty$ limit. More recently, Chakraborty, Tezuka, and Min reported~\cite{Chakraborty2014} a finite stability regions of bipolarons in the one-dimensional {\em extended} Holstein model.~\cite{Alexandrov1999b} The present author analyzed the formation of fermionic {\em trions} in one-dimensional,~\cite{Kornilovitch2013} two-dimensional,~\cite{Kornilovitch2014} and three-dimensional~\cite{Kornilovitch2020} $UV$ models. The analysis was based on direct solution of a three-body Schr\"odinger equation, which was a generalization of the approximation-free methods of Mattis~\cite{Mattis1986} and Rudin~\cite{Rudin1986} to finite-range potentials. The obtained results were numerically exact. It was found that in 1D~\cite{Kornilovitch2013} the region of pair liquid stability is finite but narrow and it shrinks to zero in the limit of large Hubbard repulsion. In contrast, in 2D~\cite{Kornilovitch2014} and 3D~\cite{Kornilovitch2020} the region of stability remains finite (of order $t$) even in the $U \rightarrow \infty$ limit.    

In this paper, we extend the integral equation method of Refs.~[\onlinecite{Kornilovitch2013,Kornilovitch2014,Kornilovitch2020,Mattis1986,Rudin1986}] to {\em four} fermions. The motivation for such an extension lies in the topology of the two-dimensional square lattice. When two pairs attempt to form a four-particle cluster (a quad), they form {\em two} new attractive bonds. This contrasts with the formation of a trion, when only one attractive bond is formed. Additionally, the kinetic energy lost in quad formation is less than in trion formation since a pair is heavier than a single particle. These two effects suggest that a quad threshold $V_4$ might be smaller than a trion threshold $V_3$, which would shrink the pair liquid domain of stability. The above picture is valid in the large-$V$ limit when pairs are tightly bound and confined to nearest neighbors. At threshold, however, pairs are loosely bound and spread over many lattice sites. The argument based on counting the number of attractive bonds may not hold at the threshold. Thus, an exact solution is required to settle the issue.   

Solving a four-body problem on large lattices is difficult because of the large number of basis states involved. Mattis' integral equation method utilizes conservation of total momentum and the finite radius of interaction to eliminate two internal variables. As a result, the four-variable Schr\"odinger equation reduces to a set of coupled two-variable integral equations, which is then solved by direct discretization. With this approach, we are able to study lattices as large as $10 \times 10$ and then perform a meaningful extrapolation to obtain thresholds for an infinite system. By restricting the symmetry of a four-body wave function from the start, we are able to study bosons and fermions separately, and in the case of fermions, separate states with different total spins $S = 0$, 1 and 2. As a biproduct, we also provide solutions of a four-body problem on the 1D chain, which is simpler numerically. We are not aware of any prior studies that utilized a similar approach.

\section{\label{4UV2d:sec:two}
The model
}

We consider a tight-binding $UV$ model with isotropic hopping $t$ and isotropic nearest-neighbor attraction on the two-dimensional (2D) square lattice 
\begin{eqnarray}
H  & = & - t \sum_{{\bf m}, {\bf b}, \sigma} 
             c^{\dagger}_{{\bf m} \sigma} c_{{\bf m} + {\bf b}, \sigma}      
\nonumber \\    
    &  & + \: \frac{U}{2} \sum_{\bf m} \hat{n}_{\bf m} \left( \hat{n}_{\bf m} - 1 \right) 
         - \frac{V}{2} \sum_{{\bf m}, {\bf b}} \hat{n}_{\bf m} \hat{n}_{{\bf m} + {\bf b}} \: .     
\label{4UV2d:eq:one}    
\end{eqnarray}
Here, $c^{\dagger}$ and $c$ are spin-$\frac{1}{2}$ fermion operators, index {\bf m} runs over lattice sites, index ${\bf b} = \pm {\bf x}, \pm {\bf y}$ enumerates the four nearest neighbors within the $xy$ plane, $\sigma = \pm \frac{1}{2}$ is the $z$-axis spin projection, and $\hat{n}_{\bf m} = \sum_{\sigma} c^{\dagger}_{{\bf m} \sigma} c_{{\bf m} \sigma}$ is the total fermion number operator on site ${\bf m}$. As discussed in the Introduction, the model is designed to describe a normal-state pseudogap and pair superconductivity with mediated attraction. Equation~(\ref{4UV2d:eq:one}) may be considered a phenomenological model of hole-doped copper-oxygen planes in the underdoped regime. With this connection in mind, we focus on the case $U > 0$, $V > 0$, although Hamiltonian~(\ref{4UV2d:eq:one}) is well defined for any values of $U$ and $V$.  

The {\em two-body} sector of Eq.~(\ref{4UV2d:eq:one}) admits an exact solution and was studied by several authors.~\cite{Micnas1990,Lin1991,Petukhov1992,Kagan1994,Kornilovitch1995,daVeiga2002,Kornilovitch2004} There are four pair states: one {\em extended} $s$-symmetric spin singlet, two $p$-symmetric spin triplets, and one $d$-symmetric spin singlet. The $s$-state is always the lowest, and it forms when 
\begin{equation}
V > V_2 = \frac{ 2 U t }{ U + 8 t } \: .
\label{4UV2d:eq:two}    
\end{equation}
One should mention that threshold $V_2$ corresponds to a stationary pair with zero total momentum. Pairs with nonzero momenta form at smaller $V$.~\cite{Kornilovitch2004,Wortis1963} This interesting effect may be related to variation of a gap function along the Fermi surface observed in some cuprates.~\cite{Damascelli2003} This topic is outside the scope of the present work and will be addressed elsewhere.  

The {\em three-body} sector of Eq.~(\ref{4UV2d:eq:one}) was studied in Ref.~[\onlinecite{Kornilovitch2014}] using the integral equation method. It was found that trion-formation threshold $V_3$ was larger than $V_2$ by about $2t$. That was explained as being due to the requirement of a three-fermion wave function to have at least one node, which increased the kinetic energy of internal motion. Since the trion is a compact object, the energy increase is of order $t$. A similar increase in $V$ is required to overcome the excess in kinetic energy, which explains why $V_{3} - V_{2} \sim t$. In determining quad threshold $V_{4}$, quad stability needs to be tested not only against decaying in two pairs $[4] \leftrightarrow [22]$, but also against decaying in a trion plus one free particle, $[4] \leftrightarrow [31]$. The results of Ref.~[\onlinecite{Kornilovitch2014}] are incorporated in the phase boundaries presented below. 

In the context of cold atoms, lattice sites ${\bf m}$ are assumed to be painted by Gaussian beams on a single optical pancake.~\cite{Henderson2009} Hopping $t$ originates from the overlap of atomic wave functions localized in neighboring wells,~\cite{Bloch2008} $U$ is controlled by Feshbach resonances, and $V$ by the excitation number of dressed Rydberg atoms. (We assume that the attraction between dressed Rydberg atoms falls off as the sixth power of separation.~\cite{Hague2012} That still results in a second-neighbor attraction of about $V/8$. We assume that the influence of second and more distant neighbors is not qualitative and can be accounted for by a redefinition of $V$. In other words, the nearest-neighbor $V$ serves as a pseudo-potential for real cold-atom systems much like $V$ approximating a complex long-range attractive tail in the solid-state case.) Thus, cold-atom technology offers great flexibility in fine-tuning model parameters as well as lattice geometry. Out of all the theoretical possibilities, in this work we will consider only one geometric extension of Eq.~(\ref{4UV2d:eq:one}), namely its one-dimensional version. The four-body problem in 1D is numerically simpler than in 2D and provides a convenient point of comparison. It also highlights the important role of orbital motion in stabilizing pair liquids in 2D. In the 1D version of Eq.~(\ref{4UV2d:eq:one}), the nearest-neighbor vector assumes only two values, ${\bf b} = \pm {\bf x}$. A 1D analog of Eq.~(\ref{4UV2d:eq:two}) reads $V_{2}({\rm 1D}) = 2Ut/(U + 4t)$.~\cite{Kornilovitch2004}     

The cold-atom technology also enables us to extend the model, Eq.~(\ref{4UV2d:eq:one}), to bosons. Indeed, bosonic atoms can be trapped by lasers as well as fermionic ones, and all the experimental knobs listed above apply equally to cold Bose systems. Hamiltonian~(\ref{4UV2d:eq:one}) remains valid for (spin zero) bosons as long as spin index $\sigma$ is omitted from summations and $c, c^{\dagger}$ are understood as Bose operators. The Bose version of Eq.~(\ref{4UV2d:eq:one}) will be referred to hereafter as the Bose-$UV$ model. Regarding pair stability, comparison between Bose and Fermi cases is critical as the wave function node argument does not apply to bosons. Indeed, we will see that Bose-$UV$ pair liquids are {\em unstable} in both 1D and 2D. Since cold-atom technology can be used to study both cases, it should be possible to demonstrate the Fermi-Bose distinction experimentally.

\section{\label{4UV2d:sec:three}
The method
}

An attempt to solve a four-body Schr\"odinger equation in real space leads to very large matrices. For example, on a $10 \times 10$ lattice, the number of states is of order $100^4 = 10^8$. Although the matrices are sparse, diagonalizing them directly is impractical. A method due to Mattis~\cite{Mattis1986} formulates the problem in momentum space and takes advantage of two reduction steps. The first reduction utilizes conservation of total momentum ${\bf P}$ and effectively reduces the number of variables by 1. The second reduction originates from a finite range of the interaction potential. The Schr\"odinger equation for a four-particle wave function $\Psi({\bf q}_1,{\bf q}_2,{\bf q}_3,{\bf q}_4)$ has the form
\begin{widetext}
\begin{eqnarray}
& & [ E - \varepsilon({\bf q}_1) - \varepsilon({\bf q}_2) - \varepsilon({\bf q}_3) 
    - \varepsilon({\bf P} - {\bf q}_1 - {\bf q}_2 - {\bf q}_3) ] \, 
\Psi( {\bf q}_1, {\bf q}_2, {\bf q}_3, {\bf P} - {\bf q}_1 - {\bf q}_2 - {\bf q}_3 ) = 
\nonumber \\
& & \hspace{0.6cm}   U \sum_{\bf k}  
 \Psi( {\bf k}, {\bf q}_1 + {\bf q}_2 - {\bf k}, {\bf q}_3, {\bf P} - {\bf q}_1 - {\bf q}_2 - {\bf q}_3 ) 
 + {\rm permutations \; of} \Psi \; {\rm arguments} 
\nonumber \\
& & \hspace{0.3cm} - V \sum_{\bf k} [ \cos{( k_x - q_{1x} )} + \cos{( k_y - q_{1y} )} ] \, 
 \Psi( {\bf k}, {\bf q}_1 + {\bf q}_2 - {\bf k}, {\bf q}_3, {\bf P} - {\bf q}_1 - {\bf q}_2 - {\bf q}_3 ) 
 + {\rm permutations \; of} \Psi \; {\rm arguments} \, , \makebox[0.3cm]{} 
\label{4UV2d:eq:three}    
\end{eqnarray}
where $\varepsilon({\bf q}) = -2t ( \cos{q_x} + \cos{q_y})$ is the one-particle dispersion and $E$ is the total energy. The right-hand-side of Eq.~(\ref{4UV2d:eq:three}) splits into a linear combination of a {\em finite} number of integrals
\begin{equation}
J( {\bf q}_1 + {\bf q}_2 , {\bf q}_3 ) = \sum_{\bf k} f({\bf k}) \, 
\Psi( {\bf k}, {\bf q}_1 + {\bf q}_2 - {\bf k}, {\bf q}_3, {\bf P} - {\bf q}_1 - {\bf q}_2 - {\bf q}_3 ) \: ,
\label{4UV2d:eq:four}    
\end{equation}
\end{widetext}
in which $f({\bf k})$'s are simple kernels: $f = 1$, $f = \cos{k_x}$, $f = \sin{k_x}$, $f = \cos{k_y}$, or $f = \sin{k_y}$. Note that $J({\bf Q}_{12},{\bf q}_3)$ is a function of only {\em two} variables: combined momentum ${\bf Q}_{12} = {\bf q}_1 + {\bf q_2}$ and one-particle momentum ${\bf q}_3$. From Eq.~(\ref{4UV2d:eq:three}), the four-particle wave function can be expressed as a linear combination of {\em two-variable} auxiliary functions $J$. Substituting the wave function back into the definitions, Eq.~(\ref{4UV2d:eq:four}), one arrives at a system of coupled integral equations for $J$'s. Thus, the original four-variable Schr\"odinger equation has been reduced to a system of two-variable integral equations. The latter system can be discretized into one large matrix equation, and energy $E$ can be determined from the condition that the matrix has an eigenvalue $\lambda = 1$. The method is fully described in Appendixes~\ref{4UV2d:sec:threetwo}-\ref{4UV2d:sec:threeseven}, and final equations are given in the Supplemental Material.~\cite{SupplMat2022}

Reduction of the number of variables by 2 is a common property of few-body lattice problems. Fundamentally, this is the reason why many two-body lattice problems can be solved analytically: the reduction procedure leads to a system of {\em linear} rather than integral equations (two minus two equals zero). Often, the system's coefficients can be evaluated analytically, which leads to an exact solution.~\cite{Petukhov1992,Alexandrov1993,Kornilovitch1995,daVeiga2002,Kornilovitch2004,Bak2007,Kornilovitch2015} A {\em three-body} lattice problem can be reduced to one-variable integral equations, which was utilized by Mattis,~\cite{Mattis1986} Rudin,~\cite{Rudin1986} and more recently by the present author.~\cite{Kornilovitch2013,Kornilovitch2014,Kornilovitch2020} In this paper, we take the next logical step and show how a four-particle problem in a model of practical interest can be analyzed using this method. Motivation for such an extension was explained in the Introduction. One should add that solving a set of two-variable equations numerically is much more demanding than one-variable equations. The situation is still manageable in low dimensions. Two 1D variables have the same number of degrees of freedom as one 2D variable. Computational complexity of a four-particle problem on a 1D chain is of the same order as that of a three-particle problem on a 2D lattice, which has already been done.~\cite{Kornilovitch2014} However, four particles in 2D take the complexity to the next level, and at this point they seem to exhaust the capabilities of desktop computers.

\section{\label{4UV2d:sec:four}
1D Results   
}

\begin{figure}[b]
\includegraphics[width=0.48\textwidth]{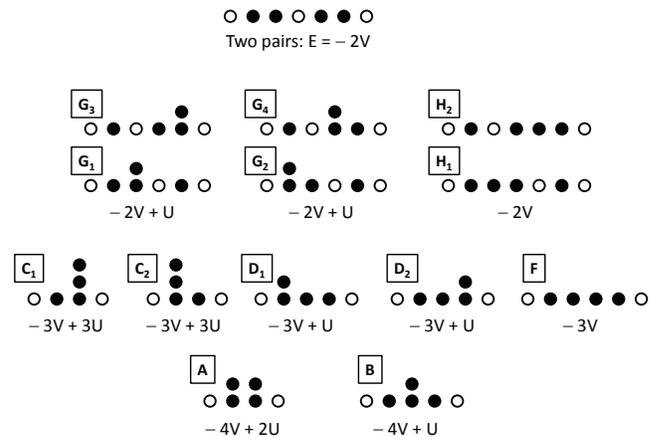}
\caption{Four-boson configurations in the 1D Bose-$UV$ model in the large-$V$ limit. Energy scaling is indicated under the diagrams.}
\label{4UV2d:fig:two}
\end{figure}

\subsection{\label{4UV2d:sec:fourone}
1D Bose-$UV$ model   
}

\begin{figure}[t]
\includegraphics[width=0.48\textwidth]{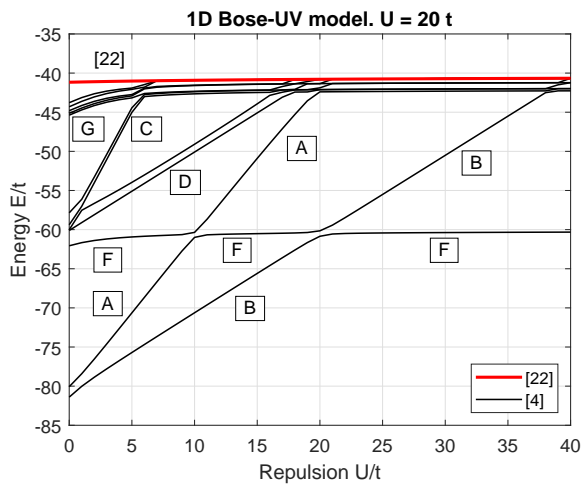}
\caption{Twelve lowest quad energies vs $U$ in the 1D Bose-$UV$ model for $V = 20 t$, $P = 0$, and chain length $N = 16$. The energies were computed with step $\triangle U = 1.0 t$. Notice how states disappear into the $[22]$ continuum with increasing $U$. The state with energy nearly independent of $U$ corresponds to configuration {\bf F} with four bosons occupying four nearest-neighbor sites. It becomes a ground state in the $U \rightarrow \infty$ limit.}
\label{4UV2d:fig:three}
\end{figure}

We begin exposition of results with an illustration of bosonic energy levels in the strong-coupling limit $V \gg t$. In this case, kinetic energy is small and the states are accurately represented by the real-space configurations shown in Fig.~\ref{4UV2d:fig:two}. The seven quad states split into two groups: two states {\bf A} and {\bf B} with a leading energy term $-4V$, and five states {\bf C}$_{1,2}$, {\bf D}$_{1,2}$, and {\bf F} with a leading energy term $-3V$. In addition, there are six $[31]$ states {\bf G}$_{1-4}$ and {\bf H}$_{1,2}$, in which a three-boson cluster (trion) coexists with a ``detached'' fourth particle. The leading energy term for these states is $-2V$, which is of the same order as for two separate boson pairs. (Since the detached particle can move freely along the lattice, energy dispersion of the $[31]$ states is of order $t$ rather than $t^2/V$.) Since not all energies have the same $U$ dependence, the states are easily identifiable when energy is plotted as a function of $U$, as done in Fig.~\ref{4UV2d:fig:three} for $V = 20 t$. Notice how energies of {\bf A} and {\bf B} are systematically pushed up by increasing $U$. By $U > V$, state {\bf F} with no double occupancy becomes the system's ground state.

\begin{figure}[t]
\includegraphics[width=0.48\textwidth]{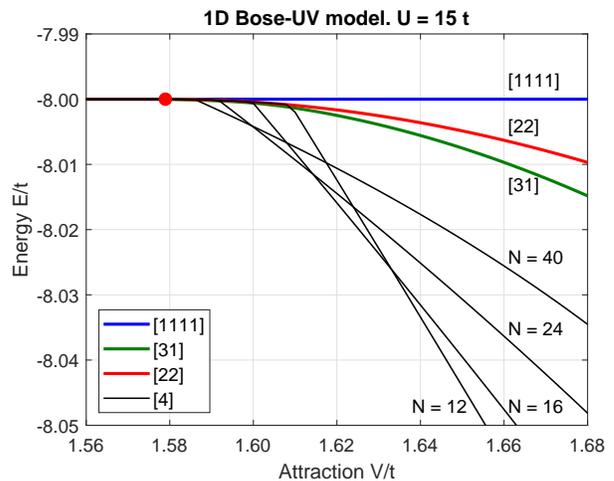}
\vspace{0.2cm}
\caption{Energy of the lowest bosonic quad state in the 1D Bose-$UV$ model as a function of $V$ and linear chain length $N$. $U = 15 t$ and $P = 0$. Quad energies were computed with step $\triangle V = 0.002 t$. The blue line is the lowest energy of four free particles, $E_{1111} = -8t$. The red line $[22]$ is the lowest energy of two bound pairs.~\cite{Kornilovitch2004} The red circle marks the pair threshold for $U = 15 t$: $V_2(15) = 2 \times 15 \, t/(15 + 4) = 1.578947 \, t$. The green line $[31]$ is the lowest energy of a trion plus one free particle. Trion energies~\cite{Kornilovitch2013} were computed for chain lengths as large as $N = 256$ and extrapolated to $N = \infty$. The $[31]$ line terminates at the red circle indicating that in this system trion threshold $V_3$ equals $V_2$. Notice how the quad threshold (intersection of black and red lines) also approaches $V_2$ as $N \rightarrow \infty$.}
\label{4UV2d:fig:four}
\end{figure}

The focus of the present work is stability of pair liquid phases. Accordingly, our main interest is $V$ values near the pair-binding threshold, Eq.~(\ref{4UV2d:eq:two}). At weak couplings, only the lowest quad state remains, and its energy becomes visibly $N$-dependent. ($N$ is the chain length, see Appendixes.) This is because the wave function of a weakly bound quad extends over many lattice sites and is affected by boundary conditions. An example of a $V$ and $N$-dependent ground-state energy is shown in Fig.~\ref{4UV2d:fig:four}.   

To obtain quad-forming threshold $V_{4}$, quad energy $E_{4}$ should be compared with the lowest energy of two bound pairs $[22]$~\cite{Kornilovitch2004} and of a $[31]$ complex~\cite{Kornilovitch2013} {\em at the same} total momentum $P$. Computing the latter energies is relatively easy at $P = 0$ since all subcomplexes have zero momenta in the ground state. The situation is more complicated at $P \neq 0$. For two pairs, one can show that the total momentum is split equally between the pairs, and $\min\{ E_{22}(P) \} = 2E_{2}(P/2)$. For the $[31]$ complex, momentum split between a trion and a free boson is unknown a priori and must be determined numerically using tabulated trion energies.~\cite{Kornilovitch2013} Fortunately, stability of pair fluids is typically considered in the ground state. In the following, we only present $P = 0$ phase diagrams where the above complications are absent. 

To determine $V_{4}$ for a given lattice size $N$, it is not necessary to compute energy curves like Fig.~\ref{4UV2d:fig:four} and then extrapolate $E_{4}(V)$ until it crosses with $E_{22}(V)$. Rather, the solution process can be organized as a direct binary search for $V_{4}$ with the response function being the presence or absence of a quad state with $E_{4} < E_{22}$. The computational cost of this method only slightly exceeds that of finding {\em one} $E_{4}$ with the same numerical accuracy.

Threshold values $V_{4}$ for decay channel $[ 4 ] \rightarrow [ 2 2 ]$ are plotted in Fig.~\ref{4UV2d:fig:five}. The entire table $V_{4}(U,N)$ is given in Supplemental Material,~\cite{SupplMat2022} Sec.~\ref{4UV2d:sec:sm:j}. One can see that $V_{4}(N)$ is smooth and has a well-defined $N \rightarrow \infty$ limit. We used cubic extrapolation to estimate $V_{4}(\infty)$. We also calculated $V_{4}$ for decay channel $[ 4 ] \rightarrow [ 3 1 ]$ in a similar way. However, as can be observed in Fig.~\ref{4UV2d:fig:four}, the energy difference $E_{22} - E_{31}$ becomes vanishingly small near the threshold. The threshold values are essentially the same as for the $[ 4 ] \rightarrow [ 2 2 ]$ channel, and therefore are not presented here.   

Finally, we plot $V_{4}(U,\infty)$ together with pair threshold $V_2(U)$ in Fig.~\ref{4UV2d:fig:six}. Within our numerical accuracy, $V_{4}$'s land exactly on top of the analytical solution $V_{2}(U) = 2Ut/( U + 4t )$.~\cite{Kornilovitch2004} Physically, this means that as soon as bosonic pairs form, bosonic quads form as well. In an earlier work, we found that bosonic trions form at the same threshold, too.~\cite{Kornilovitch2013} 

Taken together, these results lead to our first major conclusion: {\em A liquid of boson pairs is unstable in the 1D Bose-$UV$ model. As soon as pairs form, the entire sys\-tem phase-separates.}       

We close this section by noting that the numerical match between $V_{2}$, $V_{3}$ and $V_{4}$ seems to be too precise to be coincidental. We therefore conjecture that the above conclusion might be proven analytically. However, we are unaware of any theorems relating pair formation with instability of a many-boson system.

\begin{figure}[t]
\includegraphics[width=0.48\textwidth]{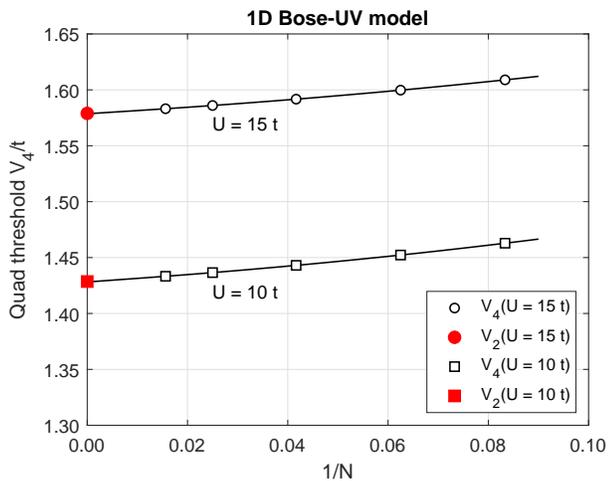}
\caption{Open symbols are quad thresholds $V_4$ of the 1D Bose-$UV$ model for two values of $U$ and several chain lengths, $N$, determined from the condition $E_{4}(V) = E_{22}(V)$. The largest $N = 64$. Numerical values are given in Supplemental Material,~\cite{SupplMat2022} Sec.~\ref{4UV2d:sec:sm:j}. The lines are cubic extrapolations to $N = \infty$. Solid symbols are pair thresholds $V_2 = 2Ut/(U + 4t)$. Notice how extrapolated $V_{4}$'s match $V_2$'s.}
\label{4UV2d:fig:five}
\end{figure}

\subsection{\label{4UV2d:sec:fourtwo}
1D Fermi-$UV$ model
}

\begin{figure}[b]
\includegraphics[width=0.48\textwidth]{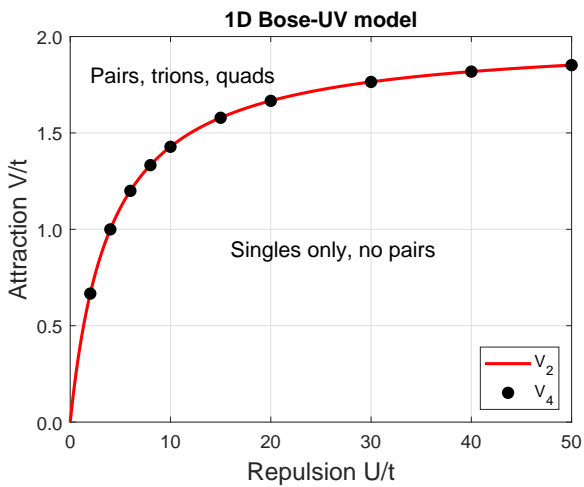}
\caption{Phase diagram of four bosons in the 1D Bose-$UV$ model. The red line is the pair threshold: $V_2 = 2Ut/(U + 4t)$. The circles are quad thresholds $V_{4}$ extrapolated to $N = \infty$.}
\label{4UV2d:fig:six}
\end{figure}

\begin{figure}[t]
\includegraphics[width=0.48\textwidth]{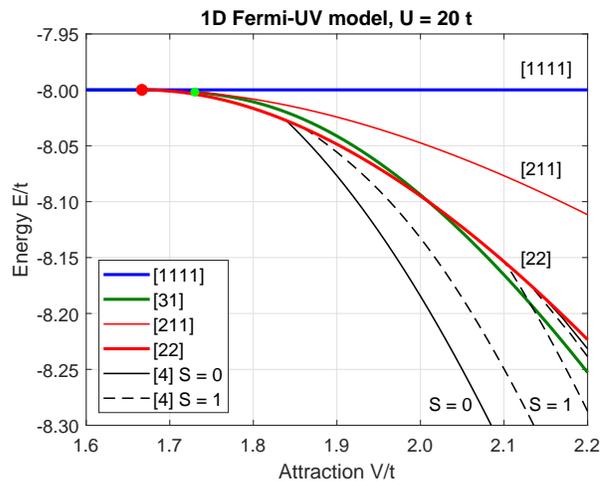}
\caption{Fermionic states at $U = 20t$ and $P = 0$ in the 1D Fermi-$UV$ model. The blue line is four free fermions, $E_{1111} = -8t$. The thin red line is one singlet pair plus two free fermions, $E_{211} = E_{2} - 4t$. The thick red line is two singlet pairs, $E_{22} = 2E_{2}$. Both $E_{211}(V)$ and $E_{22}(V)$ terminate at the pair threshold, $V_{2} = 2 \times 20t/(20 + 4) = (5/3)t$ marked by the red circle. The green line is the lowest $S = 1/2$ trion plus one free fermion, $E_{31}$. The trion energies~\cite{Kornilovitch2013} were computed for chain lengths up to $N = 256$ and extrapolated to $N \rightarrow \infty$. $E_{31}(V)$ terminates at the trion threshold $V_3 = 1.73 t$, marked by the green circle, where $E_{31} = E_{211}$. Finally, the black lines are quad energies, $E_{4}(V)$, computed for $N = 32$. Solid lines correspond to total spin $S = 0$ and dashed lines to $S = 1$. There are no $S = 2$ quad states visible within the figure window. All energies are computed with step $\triangle V = 0.01 \, t$.}
\label{4UV2d:fig:seven}
\end{figure}

Because of the spin degree of freedom, there are many more fermionic than bosonic states for the same $U$ and $V$. It is not the goal of this work to comprehensively analyze the fermion spectrum. Instead, we focus on weak coupling near threshold. Figure~\ref{4UV2d:fig:seven} shows various states with small binding energy for $U = 20t$ and $P = 0$. Notice how trion energy, $E_{31}$, crosses two-pair energy, $E_{22}$, near $V = 2.01 t$. To determine quad thresholds, it is important to know the preferred decay channel, $[ 4 ] \rightarrow [ 3 1 ]$ or $[ 4 ] \rightarrow [ 2 2 ]$. Whereas $E_{22}$ is known analytically,~\cite{Kornilovitch2004} $E_{31}$ needs to be computed numerically~\cite{Kornilovitch2013} as accurately as possible. For the present study, $E_{31}$ were computed for linear chains as long as $N = 256$ and then extrapolated to $N = \infty$. In the strong-coupling limit, and at large $U$ like $U = 20t$, the $[ 3 1 ]$ energy scales as $E_{31}(U, V \gg t) = -2V - 2t + {\cal O}(t^2/V)$ and the $[ 2 2 ]$ energy as $E_{22}(U, V \gg t) = -2V + {\cal O}(t^2/V)$. For sufficiently large $V$, $E_{31} < E_{22}$ thanks to the free fermion contribution, $E_{1} = -2t$. However, at small $V$, the relation between $E_{31}$ and $E_{22}$ is not known a priori. Exact numerical solution reveals that at least for the parameters of Fig.~\ref{4UV2d:fig:seven}, $E_{22} < E_{31}$. Thus, the quad threshold should be determined by comparing $E_{4}$ and $E_{22}$.

The technical procedure of calculating $V_{4}$ is the same as in the boson case. The ground-state energy, which has total spin $S = 0$, is computed for progressively increasing chain lengths $N$. For example, Fig.~\ref{4UV2d:fig:seven} shows quad energy for $N = 32$. For each $N$, $V_{4}$ is obtained from the condition $E_{4}(V) = E_{22}(V)$. Then $V_{4}$ versus $1/N$ is plotted as in Fig.~\ref{4UV2d:fig:five} and extrapolated to $N = \infty$. The largest chain length reached in this process was $N = 64$. Numerical values of $V_{4}(U,N)$ are given in the Supplemental Material,~\cite{SupplMat2022} Sec.~\ref{4UV2d:sec:sm:k}.     

\begin{figure}[t]
\includegraphics[width=0.48\textwidth]{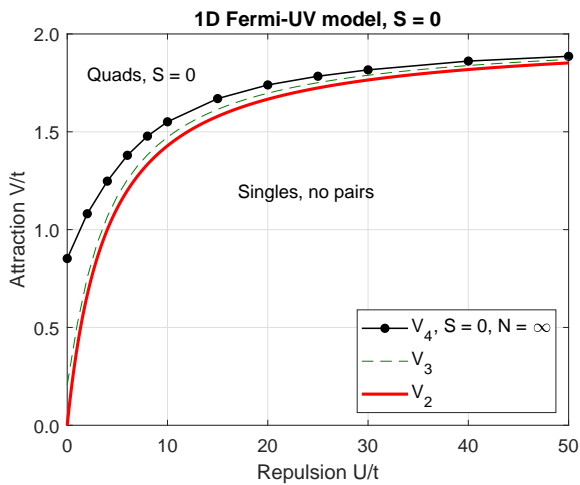}
\caption{Phase diagram of four fermions in the 1D Fermi-$UV$ model. The red line is the singlet pair threshold known analytically,~\cite{Kornilovitch2004} $V_{2} = 2Ut/(U + 4t)$. The black circles are $S = 0$ quad thresholds obtained by numerical extrapolation of $V_{4}(U,N)$ to $N = \infty$. The black line is a guide to the eye. The green dashed line is the trion threshold in a {\em three}-fermion system.~\cite{Kornilovitch2013}}
\label{4UV2d:fig:eight}
\end{figure}

The resulting phase diagram is shown in Fig.~\ref{4UV2d:fig:eight}. Its most important feature is the separation between pair and quad thresholds. The ground state {\em between} the red and black lines is two spin-singlet fermion pairs. The ground state above the black line is an $S = 0$ quad cluster. For reference, we also show the $S = 1/2$ trion threshold $V_3$ in a {\em three}-fermion system.~\cite{Kornilovitch2013} Figure~\ref{4UV2d:fig:eight} might suggest that the ground state between the green and black lines is a combination of an $S = 1/2$ trion and a free fermion. However, our numerical analysis unambiguously shows that $E_{22}$ is always lower than $E_{31}$, see for example, Fig.~\ref{4UV2d:fig:seven}. We have verified that the same conclusion holds for all $U$ within the interval studied here, $0 \leq U \leq 50t$. Thus, the region between the black and red lines is where a ``liquid'' of two fermion pairs is stable against phase separation.

An interesting feature of the phase diagram is that the gap between $V_{2}$ and $V_{4}$ shrinks with increasing $U$. This is specific to one dimension. In 1D, a large on-site repulsion imposes wave function nodes to the same effect as antisymmetry. As a result, all symmetry sectors behave similarly in the $U \rightarrow \infty$ limit, differences between bosons and fermions disappear, and the pair stability region shrinks to zero. The situation is different in 2D, as we will see in the next section.   

Figure~\ref{4UV2d:fig:eight} is qualitatively similar to the phase diagram of bipolarons in the 1D Hubbard-extended-Holstein model reported in Ref.~[\onlinecite{Chakraborty2014}]. Our conclusion from this section is as follows: {\em There is a narrow but finite region of pair stability in the 1D Fermi-$UV$ model. The region is $\approx t$ wide at $U = 0$ but shrinks to zero as $U \rightarrow \infty$.}

\section{\label{4UV2d:sec:five}
2D Results   
}

The 2D case is harder computationally than 1D, as can be appreciated by comparing matrix sizes listed in Table~\ref{4UV2d:tab:one} and Table~\ref{4UV2d:tab:two} in the Appendixes. Generally, method validation calculations were done on $4 \times 4$ and $6 \times 6$ lattices, and energy calculations were done on $6 \times 6$ and $8 \times 8$ lattices. The use of rotational symmetries in the ground state described in Appendix~\ref{4UV2d:sec:threefive} allowed us to reach the $10 \times 10$ lattice. In the latter case, we computed only quad thresholds $V_{4}(U)$, which required much less effort than full energy landscape $E_{4}(U,V)$.

\begin{figure}[t]
\includegraphics[width=0.48\textwidth]{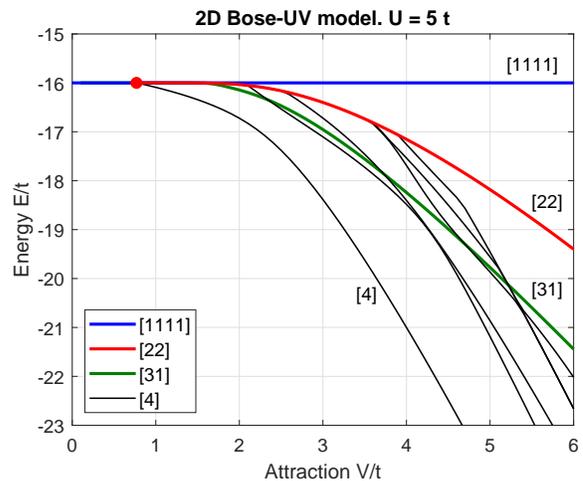}
\caption{Six lowest quad energies in the 2D Bose-$UV$ model for $U = 5t$, ${\bf P} = (0,0)$ and $N_x = N_y = 8$. The energies are computed with step $\triangle V = 0.1 \, t$. The red circle marks the pair threshold, $V_2 = 2 \times 5t/(5 + 8) = 0.769 t$. Trion energies were computed~\cite{SupplMat2014} for several lattice sizes up to $32 \times 32$ and then extrapolated to $N_x = \infty$.}
\label{4UV2d:fig:nine}
\end{figure}

\subsection{\label{4UV2d:sec:fiveone}
2D Bose-$UV$ model   
}

There are many more bosonic states in 2D than in 1D. For example, there are 22 configurations that scale as $E_{4} \propto -3V$ in the $(U,V) \rightarrow \infty$ limit (``Tetris shapes''), as compared with only one such state in 1D (state {\bf F} in Fig.~\ref{4UV2d:fig:two}). Additionally, there is one new state that scales as $E_{4} \propto -4V$, when four bosons occupy a $2 \times 2$ plaquette. Although the situation is less intuitive at weak coupling, we expect this ``plaquette'' state to always remain the ground state because of its four nearest-neighbor attractive bonds. Figure~\ref{4UV2d:fig:nine} shows the six lowest bosonic quad states for $U = 5 t$ and $N_x = 8$. Although the top five states decay into $[ 31 ]$ (the green line) as $V$ is reduced, we extend the lines to the $[ 22 ]$ continuum for visual purposes. The pairs and trions disappear below the pair threshold $V_{2} = 0.769 t$ marked by the red circle. At $V > V_{2}$, the pairs' binding energy is exponentially small as expected in 2D. The lowest quad state terminates very near $V_2$, suggesting $V_{4} \approx V_{2}$. Numerical scaling confirms that indeed $V_{4} = V_{2}$ in the $N_{x} = N_{y} = \infty$ limit. 

\begin{figure}[t]
\includegraphics[width=0.48\textwidth]{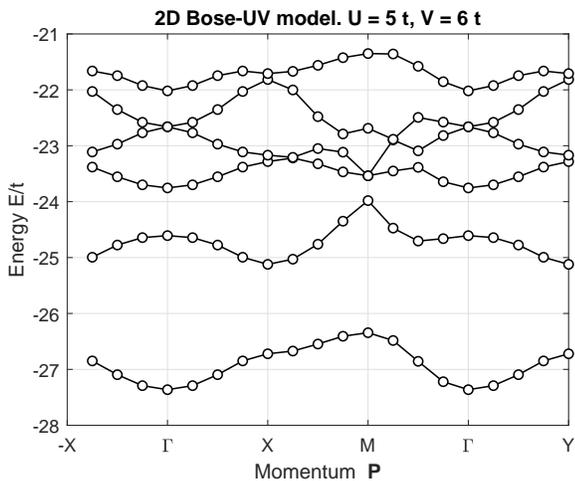}
\caption{Six lowest quad bands $E_{4}({\bf P})$ in the 2D square Bose-$UV$ model. $U = 5t$, $V = 6t$, $N_x = N_y = 8$. The circles represent calculated energies. The lines are guides to the eye. Notice a degeneracy at the $\Gamma$-point.}
\label{4UV2d:fig:ten}
\end{figure}

An example of bosonic quad dispersion is given in Fig.~\ref{4UV2d:fig:ten}. Six lowest bands were computed on the $8 \times 8$ lattice. The dispersion looks ``normal'' in the sense that it resembles dispersion of an electron in a complex potential. This confirms that quads behave as single particles. It is an additional validation of the entire method.

\begin{figure}[t]
\includegraphics[width=0.48\textwidth]{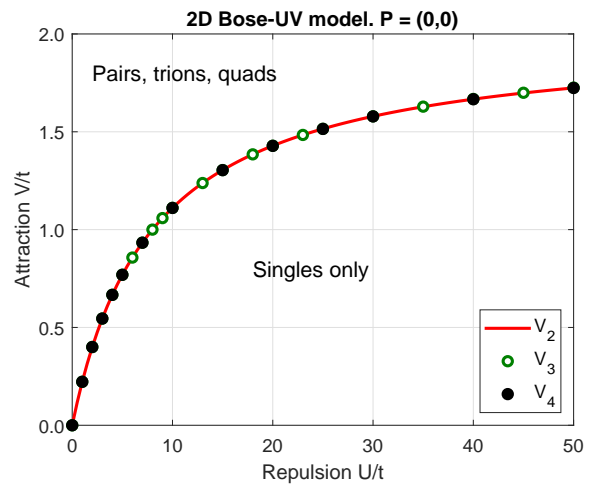}
\caption{Phase diagram of four bosons in the 2D square Bose-$UV$ model. The red line is pair threshold, Eq.~(\ref{4UV2d:eq:two}), known analytically. The open circles are trion thresholds, $V_3$, extrapolated to $N_x = \infty$. The filled circles are quad thresholds, $V_{4}$, extrapolated to $N_x = \infty$.} 
\label{4UV2d:fig:eleven}
\end{figure}

The process of determining phase boundaries between quads, pairs, and trions is the same as in 1D. There are two possible decay channels, $[ 4 ] \rightarrow [ 22 ]$ and $[ 4 ] \rightarrow [ 31 ]$. The former threshold was determined from the condition $E_{4}(V) = E_{22}(V)$ where $E_{4}$ is the energy of the ground ``plaquette'' state. Threshold values $V_{4}(U,N_x)$ for $N_x = 6$, 8, and 10 are tabulated in the Supplemental Material,~\cite{SupplMat2022} Sec.~\ref{4UV2d:sec:sm:l}. Next, data were fitted to a parabola, $V_{4}(N) = V_{4\infty} + c N^{-2}_{x}$, to estimate the quad threshold in an infinite square lattice. Those $V_{4\infty}$'s are shown in Fig.~\ref{4UV2d:fig:eleven} as black circles. To determine the second threshold, we first computed bosonic {\em trion} energies using the trion integral equations derived in an earlier work.~\cite{SupplMat2014} Then, we computed trion thresholds $V_{3}$ from the condition $E_{3}(V) = E_{21}(V)$ and extrapolated to $N_x = \infty$ using the same parabolic fit method. The $V_{3}$ values are also given in the Supplemental Material,~\cite{SupplMat2022} Sec.~\ref{4UV2d:sec:sm:l}. From that analysis, we established that $V_{3} = V_{2}$ within our numerical accuracy, i.e., bosonic trions form simultaneously with pairs. Although $E_{31} < E_{22}$ for $V > V_{2}$, $E_{31} \rightarrow E_{22}$ from below as $V \rightarrow V_{2}$ from above. As a result, the two conditions, $E_{4}(V) = E_{31}(V)$ and $E_{4}(V) = E_{22}(V)$, produce the same threshold $V_{4}$ in the $N_x = \infty$ limit, although the two are slightly different at any finite $N_x$.

The resulting phase diagram is displayed in Fig.~\ref{4UV2d:fig:eleven}. Like in 1D, there is only {\em one} boundary line, described by Eq.~(\ref{4UV2d:eq:two}). There is no pairing below the line: the ground state consists of four interacting but individual bosons. However, above the line there are all sorts of clustering: a pair forms in a two-boson system, a trion forms in a three-boson system, and a quad forms in a four-boson system. Based on these results, we expect a many-boson system to form a macroscopic cluster and phase-separate at the same threshold. One should also note the qualitative similarity between the Bose-$UV$ phase diagrams in 1D, Fig.~\ref{4UV2d:fig:six}, and 2D, Fig.~\ref{4UV2d:fig:eleven}. Quantitatively, the 2D phase boundary is slightly flatter than its 1D counterpart. 

Our conclusion about the 2D Bose-$UV$ model is the same as in 1D: {\em A liquid of boson pairs is unstable against forming larger clusters. As soon as pairs form, the entire system phase-separates.}

\subsection{\label{4UV2d:sec:fivetwo}
2D Fermi-$UV$ model   
}

We now proceed to the technically most challenging but physically most interesting case: four spin-$\frac{1}{2}$ fermions on the 2D square lattice. There is great proliferation of states compared with the 2D Bose case. There are dozens of quad states within each $S$ sector. Additionally, quads mix with $[ 31 ]$ states at elevated energies. Since $E_{31}$ are sensitive to boundary conditions, the high-energy part of the spectrum depends on lattice size. For example, the $6 \times 6$ and $8 \times 8$ energy ladders look quite different at high energies. Untangling this complicated web and making sense of the entire energy landscape amounts to a separate investigation that is not attempted here. Suffice it to say that we conducted rigorous validation checks described in Appendix~\ref{4UV2d:sec:threeseven} to make sure energies and wave functions produced by the integral equation method are numerically accurate.

\begin{figure}[t]
\includegraphics[width=0.48\textwidth]{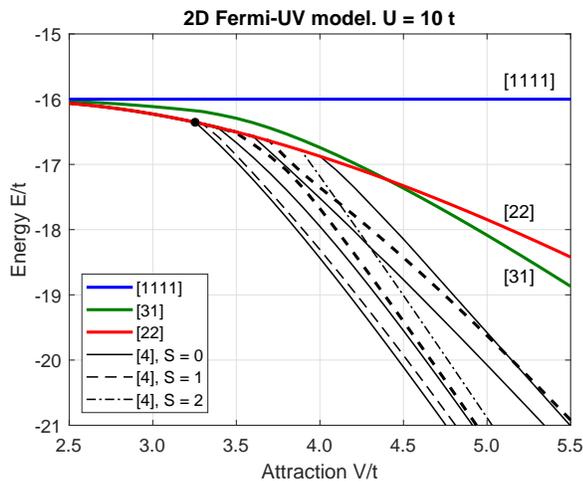}
\caption{The ten lowest quad states in the 2D Fermi-$UV$ model. $U = 10 \, t$, ${\bf P} = (0,0)$, $N_x = N_y = 6$. All three spin sectors, $S = 0$, 1, and 2, are present. Two $S = 1$ states are double-degenerate. They are shown by bold dashed lines. The energies are computed with step $\triangle V = 0.1 \, t$. The lowest quad threshold is marked by the black circle. Trion energies were computed~\cite{Kornilovitch2014,SupplMat2014} for several lattice sizes up to $32 \times 32$ and then extrapolated to $N_x = \infty$. The pair threshold, $V_{2} = 2 \times 10 t/( 10 + 8 ) = 1.111 \, t$, is outside the $V$-axis limits.} 
\label{4UV2d:fig:twelve}
\end{figure}

Before presenting our main result, it is useful to recall the motivation for this work. In an earlier investigation,~\cite{Kornilovitch2014} we established that in the 2D Fermi-$UV$ model, pair formation and trion formation are separated by a stability interval of about $\triangle V \approx 2t$. This can be interpreted as effective repulsion between a spin-singlet fermionic pair and a free fermion. An even better predictor for pair liquid stability is the effective interaction between {\em two pairs}, which amounts to a four-body problem. The origin of the repulsion is antisymmetry of many-fermion wave functions. Being qualitative in nature, we expect this repulsion mechanism to remain effective for four, five, six fermions, and so on, and eventually in a macroscopic system. However, quantitatively, the pair stability region may shrink as density increases. The transition from three to four fermions is particularly interesting. A fourth fermion, when trying to bind with a trion, is forming {\em two} new attractive bonds rather than one; see process B in Fig.~\ref{4UV2d:fig:one}(c). Although it must create an additional wave-function node at the same time, dynamic energy gain may exceed kinematic energy loss. In other words, the quad may be more stable than the trion, and the quad threshold may be lower than the trion one. The real question is how much lower? If insignificantly, in relative terms, it would leave the wide stability region essentially unaffected. In the opposite case, the stability region could be significantly reduced, which would leave the question of what happens in the thermodynamic limit open. Determining the quad threshold and its position relative to the trion threshold is the main goal of this work. 

Strictly speaking, when determining $S = 1$ and $S = 2$ quad thresholds, conservation of total spin needs to be respected. That is, the lowest quad energy of the $S = 1$ sector should be compared with the lowest energy of one spin-singlet pair plus one spin-triplet pair. Likewise, the lowest quad energy of the $S = 2$ sector should be compared with the lowest energy of two spin-triplet pairs. However, we assume that in any real system there is enough perturbation to flip the spin if that lowers overall energy. Therefore, in the following, all spin sectors will be compared with the lowest energy of two singlet pairs.      

Figure~\ref{4UV2d:fig:twelve} shows an example of $V$-dependent quad energies. Notice how the ground state $E_{4}(V)$ intersects $E_{22}(V)$ at a $V_{4}$ substantially larger than pair threshold $V_{2}$.

\begin{figure}[t]
\includegraphics[width=0.48\textwidth]{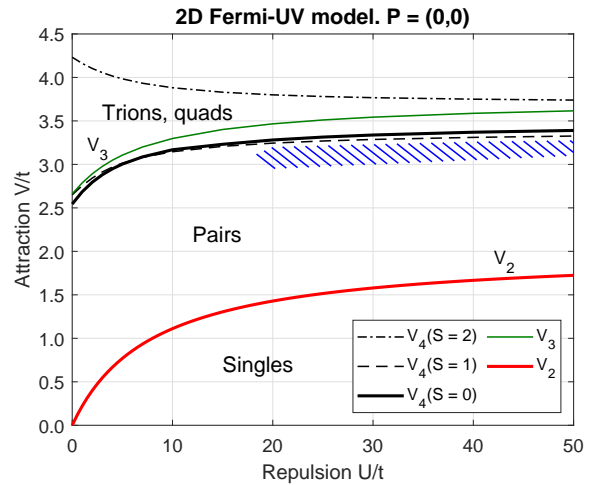}
\caption{Phase diagram of four fermions in the 2D square Fermi-$UV$ model. Pair liquid is stable in the space between $V_{2}$ and the lowest $V_{4}$. The hatched area approximately corresponds to high-temperature superconductivity.} 
\label{4UV2d:fig:thirteen}
\end{figure}

Quad thresholds for $S = 0$, 1 and 2 were computed for $6 \times 6$, $8 \times 8$, and $10 \times 10$ lattices and extrapolated to $N_x = \infty$. Values are listed in the Supplemental Material,~\cite{SupplMat2022} Sec.~\ref{4UV2d:sec:sm:m}. The resulting phase diagram is shown in Fig.~\ref{4UV2d:fig:thirteen}. Together with the quad thresholds, the pair threshold $V_2$ [Eq.~(\ref{4UV2d:fig:two})] and the trion threshold $V_3$~\cite{Kornilovitch2014} are also shown. One interesting feature of the phase diagram is the $U$-dependence of the $S = 2$ threshold. Although $E_{4}(S = 2)$ itself does not depend on $U$, $E_{22}$ does, so the threshold that follows from the two energies being equal is also $U$-dependent. As a result, $V_{4}(S = 2)$ decreases with $U$: at larger $U$ singlet pairs become less stable which expands the stability domain of $S = 2$ quads. Another interesting feature of Fig.~\ref{4UV2d:fig:thirteen} is the closeness of the $S = 0$ and $1$ thresholds. This is a consequence of the lowest $S = 0$ and $1$ states being close in energy; cf. Fig.~\ref{4UV2d:fig:twelve}. In fact, our numerical data suggest a crossover. At $U < 6.4t$, the $S = 0$ quad is lower. As a result, it forms at a slightly smaller $V_{4}$ than the $S = 1$ quad. At $U > 6.4t$, the order is reversed. 

The most important feature of the phase diagram is the relative positions of the quad, trion and pair thresholds. We first note that $V_{4} < V_{3}$. Fermionic quads are formed at a weaker attraction and are more stable than trions, as expected on qualitative grounds. At the same time, $V_{4}$ remains well-separated from the pair threshold $V_{2}$. The addition of a fourth fermion shrinks the region of pair stability, but not by much in relative terms. Another important property of Fig.~\ref{4UV2d:fig:thirteen} is that the gap $V_{4} - V_{2}$ does not tend to zero as $U \rightarrow \infty$. This is a manifestation of a fundamental difference between 1D and 2D. As noted in Sec.~\ref{4UV2d:sec:fourtwo}, in 1D a large $U$ forces the wave function to go to zero whenever two fermions occupy the same site. It is the same condition as that imposed by antisymmetry. In other words, fermions and bosons behave similarly in the $U = \infty$ limit. That is why in 1D all thresholds converge to the same limit at large $U$; see Fig.~\ref{4UV2d:fig:eight}. In contrast, in 2D an infinite on-site repulsion is {\em not} equivalent to full antisymmetry. An infinite $U$ does not prevent two fermions from being exchanged by rotating around their center of mass. The many-body wave function must possess additional nodes along angular coordinates that cannot be provided dynamically by a large $U$. These extra nodes elevate kinetic energy by an amount of order $t$. To overcome such an increase and still form a quad, $V$ needs to be raised by a value $\sim t$, too. This explains why $V_{4} - V_{2}$ remains of order $t$ for all $U$.      

The $U \rightarrow \infty$ limit of Fig.~\ref{4UV2d:fig:thirteen} is consistent with the phase-separation thresholds computed on $4 \times 4$ {\em t-J} clusters.~\cite{Emery1990,Dagotto1993} 

Here is our conclusion for the 2D square Fermi-$UV$ model: {\em Two fermion pairs are stable against quad formation within a finite $V$ interval $\approx 2t$ wide, for all values of $U$. This can be interpreted as an effective repulsion between pairs. The repulsion is kinematic in nature and originates from antisymmetry of the four-fermion wave function.}

\subsection{\label{4UV2d:sec:fivethree}
Relevance to HTSC   
}

We now apply the obtained results to high-tempera\-ture superconductivity, specifically to the superconductivity of copper oxides. Unlike the preceding discussions, this one is going to be more speculative. First, we note that any realistic model of the cuprates must include an interlayer hopping $t_{\perp}$. However, even a small $t_{\perp} \sim 0.01 \, t$ profoundly changes the phase diagram of Fig.~\ref{4UV2d:fig:thirteen} by nearly doubling the pairing threshold $V_2$. This is because in a pure 2D system, the spin-singlet fermion pair is the ground state. It is stabilized by long-range logarithmic fluctuations that lower $V_2$. Finite $t_{\perp}$ suppresses the fluctuations and raises $V_2$. Specific curves $V_2(t_{\perp})$ can be found in Refs.~[\onlinecite{Kornilovitch2015},\onlinecite{Kornilovitch2020}]. In contrast, a fermionic trion is not the lowest state of the Schr\"odinger equation (the lowest is forbidden by the exclusion principle) and is not affected by $t_{\perp}$ as much. $V_3$ changes with $t_{\perp}$ only linearly.~\cite{Kornilovitch2020} As a result, the pair stability region shrinks to about $\triangle V \sim ( 0.5 - 1.0 ) \, t$, depending on $U$ and $t_{\perp}$. Now, the present paper deals with four fermions in the pure 2D case, $t_{\perp} = 0$. The four-fermion problem in 3D has not been solved yet. Therefore, we cannot rigorously prove that quad threshold $V_4$ is also weakly affected by $t_{\perp}$. However, this is a reasonable conjecture given the closeness of $V_3$ and $V_4$ shown in Fig.~\ref{4UV2d:fig:thirteen} and the fact that a fermionic quad is not the lowest state of the Schr\"odinger equation either.  

Additionally, analysis of the pair's effective mass and radius showed~\cite{Kornilovitch2015} that the {\em close-packed} BEC temperature increases monotonically with $V$ until the system phase-separates.~\cite{Kornilovitch2020} In the preformed pair scenario, the close-packing of pairs is identified with optimal doping. Thus, the region of truly {\em high} critical temperatures is limited to a much narrower interval $\triangle V = ( 0.1 - 0.3 ) \, t$ just below $V_4$. This region is schematically shown in Fig.~\ref{4UV2d:fig:thirteen} as the hatched area.         

It is nontrivial to assign real-life values to the parameters of Fig.~\ref{4UV2d:fig:thirteen}, mainly because the $UV$ model with nearest-neighbor hopping is an effective model while any real copper oxide is a complex multi-orbital material with a multitude of interactions and hoppings. Only rough estimates can be made. According to Fig.~\ref{4UV2d:fig:thirteen}, the attraction must satisfy $V < 3.5 \, t$, otherwise the system phase-separates. Then, according to Fig.~3 of Ref.~[\onlinecite{Kornilovitch2015}], the highest close-packed $T_c \approx 0.05 \, t$ is reached at $t_{\perp} \approx 0.05 \, t$. Taking a typical $T_c \sim 50$ K, one obtains $t \sim 1000$ K $ = 0.1$ eV. That corresponds to a bandwidth $W \sim 1$ eV, which is consistent with spectroscopic evidence.~\cite{Damascelli2003} The Hubbard parameter on copper ions is $U_{\rm Cu} = 4.9$ eV~\cite{Ramadhan2022} whereas that on oxygen ions it is expected to be less. Hence, in relative units, $U/t \sim 20 - 50$, as shown in Fig.~\ref{4UV2d:fig:thirteen}.      

To summarize this section, the region of high-tempera\-ture superconductivity is much narrower than the region of pair liquid stability once interlayer hopping is taken into account. This topic will be addressed more quantitatively elsewhere.

\section{\label{4UV2d:sec:six}
Summary and discussion    
}

In this paper, we investigated the formation criteria of four-particle clusters in quantum-mechanical systems with zero-range repulsion and finite-range attraction. We directly solved a four-body Schr\"odinger equation by reducing it to a system of coupled integral equations of two variables. By comparing quad energy with the energy of two bound pairs, we determined quad-forming thresholds and mapped parameter space where pairs are stable against further clustering.    

One clear conclusion that follows from this study is the critical role of quantum statistics. Bosons and fermions behave very differently. In the case of bosons, the quad threshold exactly matches the pair threshold, $V_{4}(U) = V_{2}(U)$, within our numerical accuracy. As soon as attraction is strong enough to form Bose pairs, a Bose quad also forms. This is true in both 1D and 2D. One can state that {\em effective interaction between two Bose pairs is attractive}. The numerical match between $V_{4}$ and $V_{2}$ is so good that it suggests a mathematical theorem. We are unaware, however, of any analytical results relating $V_{4}$ and $V_{2}$ (and $V_{3}$ for that matter) in Bose systems. Extrapolating these results to many-boson systems, we conjecture that effective interactions between larger clusters --- between two quads, between two octets, and so on --- is also attractive, although we cannot prove this rigorously. This implies that the entire system phase-separates and forms one macroscopic cluster at $V = V_{2}$. {\em A liquid of Bose pairs is unstable and does not exist.}  

The situation is different for fermions. Although {\em two} fermions have a symmetric ground state and in that sense are equivalent to two bosons, a four-fermion wave function must have nodes, which elevates the quad energy. As a result, there is a {\em finite} separation between $V_{4}$ and $V_{2}$. When $V$ falls between $V_{2}$ and $V_{4}$, pairs are formed but quads are not. Effective interaction between pairs is repulsive. {\em A liquid of Fermi pairs is stable within a finite interval of $V$.} We conjecture that the stability region remains finite even in a microscopic system, at least when fermion density is small. Indeed, at a small pair density, pair-pair collisions are unaffected by the presence of other pairs. Since the pair-pair interaction is repulsive in a four-fermion system, it should remain repulsive in the presence of other pairs because they are far away. This is not a proof but a qualitative supporting argument. 

The second conclusion concerns the role of lattice dimensionality. As explained in Sec.~\ref{4UV2d:sec:fourtwo} and Sec.~\ref{4UV2d:sec:fivetwo}, in 1D a finite $U$ effectively mimics the wave-function antisymmetry, whereas in 2D it does not. In 1D, $V_{4}$ is close to $V_{2}$ and the difference shrinks to zero in the $U = \infty$ limit. In 2D, the separation remains of order $t$ for all $U$. {\em 2D is much more favorable for the existence of a stable pair liquid than 1D.}     
    
We now discuss the implications of the obtained results for real physical systems. First, we acknowledge that we have studied here a special type of potential, the $UV$ model, that neglects long-range interaction tails. However, we consider the $UV$ potential to be a pseudo-potential that captures the essential physics of relevant systems, namely short-range repulsion plus finite-range attraction. The arguments given above are qualitative in nature and expected to be applicable to other potentials from the same class. In particular, we expect the pair stability region to remain finite when the repulsive core has a finite range or when the attractive part has a long tail. In all those cases, a pair would have a symmetrical wave function, while a quad would have nodes. That creates an energy gap between the quad and two pairs, which translates into a finite region of stability. 

The paper's conclusions are also robust against modifications of the kinetic energy term. For instance, the addition of non-zero next-nearest-neighbor hopping, $t^{\prime}$, does not affect the applicability of the symmetry argument. The pair liquid will remain stable, but the domain boundaries will change.    

Many of these effects can be demonstrated in Dressed Rydberg atoms (DRA) systems. DRAs provide a convenient realization of $UV$ models, with $U$ tunable via Feshbach resonances,~\cite{Joerdens2008,Strohmaier2007} and $V$ tunable via laser intensity.~\cite{Hague2012} Additionally, optical lattices allow for direct control of lattice geometry and hopping amplitude $t$ via the positions of Gaussian beams piercing the main optical pancake.~\cite{Henderson2009} The presence or absence of bound pairs can be monitored by gas microscopy.~\cite{Mitra2018} Using these DRA techniques, it should be possible to demonstrate the existence of a stable pair phase in the case of fermions, and its non-existence in the case of bosons. By changing lattice anisotropy, it should also be possible to smoothly convert a 2D square Fermi-$UV$ system where the pair liquid is stable into a 1D Fermi-$UV$ system where it is unstable, while $U$ and $V$ remain constant. Finally, by scanning $U$, $V$, and $t$, an entire phase diagram can be mapped and compared with theory. A detailed description of DRA systems suitable for such experiments, as well as specifics of external drivers and observation methods, are outside the scope of the present study and will be addressed elsewhere. 

The results developed here also lend further support to the preformed pair mechanism of high-temperature superconductivity (HTSC).~\cite{Micnas1990,Alexandrov1994,Chen2005,Alexandrov2013} We have rigorously shown that a fermionic system can exist as a collection of interacting but distinct pairs in a finite domain of model parameters. We also expect a pair liquid to remain stable even in the presence of nonzero interlayer hopping, although the domain of stability will shrink significantly. That makes it possible for some real systems, for example HTSC cuprates, to fall within that domain even though the real inter-hole potential differs from the pure $UV$ form. The fact that the region of high critical temperatures is narrow helps explain the experimental fact that HTSC is a rare phenomenon among crystalline solids. Additionally, being close to the quad threshold on average makes a system prone to crossing over the threshold locally. That may explain the proliferation of charge-density waves, nematic orders, and other types of charge instabilities observed in HTSC. We close by noting that a natural extension of the present work would be an analysis of a four-fermion system in 3D, including anisotropic lattices.~\cite{Kornilovitch2015,Kornilovitch2020} This problem is computationally much more demanding and is left for the future.

\begin{acknowledgments}

The author wishes to thank James Hague and Viktor Kabanov for useful discussions on the subject of this paper.         

\end{acknowledgments}

%
%

\begin{appendix}

\begin{widetext}

\section{\label{4UV2d:sec:threetwo}
Four distinguishable particles  
}

In this Appendix, we derive a working set of equations for four {\em distinguishable} particles in the $UV$ model, Eq.~(\ref{4UV2d:eq:one}). We choose distinguishable particles to illustrate the method because this case admits the most transparent derivation. The resulting equations will permit solutions of different permutation symmetries, including bosonic and fermionic ones. To be able to make reasonable physical conclusions, permutation symmetry must be restricted from the start. How to do that is explained in Appendixes~\ref{4UV2d:sec:threethree} and \ref{4UV2d:sec:threefour}. The fully unsymmetrized solution derived in this section will serve as a useful self-consistency check of the entire procedure.    

Coordinate wave function $\Psi( {\bf m}_1 , {\bf m}_2 , {\bf m}_3 , {\bf m}_4 )$, when acted upon by the Hamiltonian, Eq.~(\ref{4UV2d:eq:one}), satisfies the following difference equation 
\begin{eqnarray}
& & -t \sum_{\bf b} \left[ 
\Psi( {\bf m}_1 + {\bf b} , {\bf m}_2 , {\bf m}_3 , {\bf m}_4 ) + 
\Psi( {\bf m}_1 , {\bf m}_2 + {\bf b} , {\bf m}_3 , {\bf m}_4 ) + 
\Psi( {\bf m}_1 , {\bf m}_2 , {\bf m}_3 + {\bf b} , {\bf m}_4 ) + 
\Psi( {\bf m}_1 , {\bf m}_2 , {\bf m}_3 , {\bf m}_4 + {\bf b} ) \right] + 
\nonumber \\
& & \hspace{1.0cm} 
+ \: U \left( \delta_{{\bf m}_1 {\bf m}_2} + \delta_{{\bf m}_1 {\bf m}_3} + 
              \delta_{{\bf m}_1 {\bf m}_4} + \delta_{{\bf m}_2 {\bf m}_3} + 
              \delta_{{\bf m}_2 {\bf m}_4} + \delta_{{\bf m}_3 {\bf m}_4} \right) 
\, \Psi( {\bf m}_1 , {\bf m}_2 , {\bf m}_3 , {\bf m}_4 ) - 
\nonumber \\
& & \hspace{1.0cm} 
- \: V \sum_{\bf b}
       \left( \delta_{{\bf m}_1 , {\bf m}_2 + {\bf b}} + \delta_{{\bf m}_1 , {\bf m}_3 + {\bf b}} + 
              \delta_{{\bf m}_1 , {\bf m}_4 + {\bf b}} + \delta_{{\bf m}_2 , {\bf m}_3 + {\bf b}} + 
              \delta_{{\bf m}_2 , {\bf m}_4 + {\bf b}} + \delta_{{\bf m}_3 , {\bf m}_4 + {\bf b}} \right) 
\, \Psi( {\bf m}_1 , {\bf m}_2 , {\bf m}_3 , {\bf m}_4 ) =                             
\nonumber \\
& & \hspace{1.0cm}   
= E \, \Psi( {\bf m}_1 , {\bf m}_2 , {\bf m}_3 , {\bf m}_4 )  \: ,
\label{4UV2d:eq:five}    
\end{eqnarray}
where $E$ is total energy. We define a momentum-space wave function via Fourier transformation
\begin{equation}
\Psi( {\bf m}_1 , {\bf m}_2 , {\bf m}_3 , {\bf m}_4 ) = 
\frac{1}{N^2} \sum_{{\bf q}_1 {\bf q}_2 {\bf q}_3 {\bf q}_4}
e^{ i \, ( {\bf q}_1 {\bf m}_1 + {\bf q}_2 {\bf m}_2 + {\bf q}_3 {\bf m}_3 + {\bf q}_4 {\bf m}_4 ) }  
\Psi( {\bf q}_1 , {\bf q}_2 , {\bf q}_3 , {\bf q}_4 ) \: ,
\label{4UV2d:eq:six}    
\end{equation}
where $N$ is the number of lattice sites (unit cells). Substitution of Eq.~(\ref{4UV2d:eq:six}) into Eq.~(\ref{4UV2d:eq:five}) yields 
\begin{eqnarray}
& & \left[ E - \varepsilon({\bf q}_1) - \varepsilon({\bf q}_2) 
             - \varepsilon({\bf q}_3) - \varepsilon({\bf q}_4) \right] \, 
    \Psi( {\bf q}_1 , {\bf q}_2 , {\bf q}_3 , {\bf q}_4 ) =           
\nonumber \\
& & \hspace{1.0cm} 
U \frac{1}{N} \sum_{\bf k} \left\{ 
 	    \Psi( {\bf k} , {\bf q}_1 + {\bf q}_2 - {\bf k} , {\bf q}_3 , {\bf q}_4 ) +
 	    \Psi( {\bf k} , {\bf q}_2 , {\bf q}_1 + {\bf q}_3 - {\bf k} , {\bf q}_4 ) +
 	    \Psi( {\bf k} , {\bf q}_2 , {\bf q}_3 , {\bf q}_1 + {\bf q}_4 - {\bf k} ) + \right.          
\nonumber \\
& & \hspace{2.4cm} \left. 
        \Psi( {\bf q}_1 , {\bf k} , {\bf q}_2 + {\bf q}_3 - {\bf k} , {\bf q}_4 ) +
        \Psi( {\bf q}_1 , {\bf k} , {\bf q}_3 , {\bf q}_2 + {\bf q}_4 - {\bf k} ) +
        \Psi( {\bf q}_1 , {\bf q}_2 , {\bf k} , {\bf q}_3 + {\bf q}_4 - {\bf k} )  \right\}
\nonumber \\
& & \hspace{1.0cm} 
- \: V \frac{1}{N} \sum_{\bf k} \sum_{\bf b} \left\{ 
\cos{[({\bf k} - {\bf q}_1 ){\bf b}]} \, \Psi( {\bf k} , {\bf q}_1 + {\bf q}_2 - {\bf k} , {\bf q}_3 , {\bf q}_4 ) +
\cos{[({\bf k} - {\bf q}_1 ){\bf b}]} \, \Psi( {\bf k} , {\bf q}_2 , {\bf q}_1 + {\bf q}_3 - {\bf k} , {\bf q}_4 ) +  
\right.  
\nonumber \\
& & \hspace{3.4cm} 
\cos{[({\bf k} - {\bf q}_1 ){\bf b}]} \, \Psi( {\bf k} , {\bf q}_2 , {\bf q}_3 , {\bf q}_1 + {\bf q}_4 - {\bf k} ) +
\cos{[({\bf k} - {\bf q}_2 ){\bf b}]} \, \Psi( {\bf q}_1 , {\bf k} , {\bf q}_2 + {\bf q}_3 - {\bf k} , {\bf q}_4 ) +  
\nonumber \\
& & \hspace{3.4cm} \left.
\cos{[({\bf k} - {\bf q}_2 ){\bf b}]} \, \Psi( {\bf q}_1 , {\bf k} , {\bf q}_3 , {\bf q}_2 + {\bf q}_4 - {\bf k} ) +
\cos{[({\bf k} - {\bf q}_3 ){\bf b}]} \, \Psi( {\bf q}_1 , {\bf q}_2 , {\bf k} , {\bf q}_3 + {\bf q}_4 - {\bf k} )
\right\} .
\label{4UV2d:eq:seven}    
\end{eqnarray}
Here all six pairwise interactions are written down explicitly. The Schr\"odinger equation, Eq.~(\ref{4UV2d:eq:seven}), is valid for the $UV$ model in any dimensionality. We now perform a reduction to two-variable equations in 1D. First, summation over nearest-neighbor vectors ${\bf b} = \pm {\bf x}$ yields a factor of 2. After changing to scalar notation $k$, $q$, Eq.~(\ref{4UV2d:eq:seven}) becomes  
\begin{eqnarray}
& & \left[ E - \varepsilon(q_1) - \varepsilon(q_2) 
             - \varepsilon(q_3) - \varepsilon(q_4) \right] \, 
    \Psi( q_1 , q_2 , q_3 , q_4 ) =          
\nonumber \\
& & \hspace{1.0cm} 
U \frac{1}{N} \sum_{k} \left\{ 
 	    \Psi( k , q_1 + q_2 - k , q_3 , q_4 ) +
 	    \Psi( k , q_2 , q_1 + q_3 - k , q_4 ) +
 	    \Psi( k , q_2 , q_3 , q_1 + q_4 - k ) + \right.          
\nonumber \\
& & \hspace{2.4cm} \left. 
        \Psi( q_1 , k , q_2 + q_3 - k , q_4 ) +
        \Psi( q_1 , k , q_3 , q_2 + q_4 - k ) +
        \Psi( q_1 , q_2 , k , q_3 + q_4 - k )  \right\}
\nonumber \\
& & \hspace{1.0cm} 
- \: 2V \frac{1}{N} \sum_{k} \left\{ 
\cos{( k - q_1 )} \, \Psi( k , q_1 + q_2 - k , q_3 , q_4 ) +
\cos{( k - q_1 )} \, \Psi( k , q_2 , q_1 + q_3 - k , q_4 ) +  
\right.  
\nonumber \\
& & \hspace{3.1cm} 
\cos{( k - q_1 )} \, \Psi( k , q_2 , q_3 , q_1 + q_4 - k ) +
\cos{( k - q_2 )} \, \Psi( q_1 , k , q_2 + q_3 - k , q_4 ) +  
\nonumber \\
& & \hspace{3.1cm} \left.
\cos{( k - q_2 )} \, \Psi( q_1 , k , q_3 , q_2 + q_4 - k ) +
\cos{( k - q_3 )} \, \Psi( q_1 , q_2 , k , q_3 + q_4 - k )
\right\} .
\label{4UV2d:eq:eight}    
\end{eqnarray}
Second, expand the cosines in the $V$ term. It generates a number of integrals like
\begin{equation}
\frac{1}{N} \sum_{k} \cos{(k)} \Psi( k , q_1 + q_2 - k , q_3 , q_4 ) \: , 
\label{4UV2d:eq:nine}    
\end{equation}
with various permutations of $\Psi$ arguments. The $V$ term also contains similar integrals with $\sin{(k)}$ instead of $\cos{(k)}$. The $U$ term contains terms with $1$ instead of $\cos{(k)}$. Next, we notice that due to conservation of total momentum, Eq.~(\ref{4UV2d:eq:nine}) is a function of $q_3$ and $q_4$ only. Defining 
\begin{equation}
P = q_1 + q_2 + q_3 + q_4 \: , 
\label{4UV2d:eq:ten}    
\end{equation}
Eq.~(\ref{4UV2d:eq:nine}) can be rewritten as 
\begin{equation}
C_{12}(q_3,q_4) \equiv \frac{1}{N} \sum_{k} \cos{(k)} \Psi( k , P - q_3 - q_4 - k , q_3 , q_4 ) \: ,  
\label{4UV2d:eq:eleven}    
\end{equation}
which defines a new function $C_{12}(q_3,q_4)$. The subscript $\{12\}$ indicates that this function describes interaction between the first and second particles, as determined by the order of $\Psi$ arguments. $q_3$ and $q_4$ are the actual arguments of $C_{12}$ for this particular instance. In total, the right-hand-side of Eq.~(\ref{4UV2d:eq:eight}) generates 18 auxiliary functions, 
\begin{eqnarray}
A_{12}(q_1,q_2) \, , \: C_{12}(q_1,q_2) \, , \: S_{12}(q_1,q_2) & = & 
\frac{1}{N} \sum_{k} f(k) \Psi( k , P - q_1 - q_2 - k , q_1 , q_2 ) \: ,  
\label{4UV2d:eq:twelve}     \\
A_{13}(q_1,q_2) \, , \: C_{13}(q_1,q_2) \, , \: S_{13}(q_1,q_2) & = & 
\frac{1}{N} \sum_{k} f(k) \Psi( k , q_1 , P - q_1 - q_2 - k , q_2 ) \: ,  
\label{4UV2d:eq:thirteen}   \\  
A_{14}(q_1,q_2) \, , \: C_{14}(q_1,q_2) \, , \: S_{14}(q_1,q_2) & = & 
\frac{1}{N} \sum_{k} f(k) \Psi( k , q_1 , q_2 , P - q_1 - q_2 - k ) \: ,  
\label{4UV2d:eq:fourteen}   \\
A_{23}(q_1,q_2) \, , \: C_{23}(q_1,q_2) \, , \: S_{23}(q_1,q_2) & = & 
\frac{1}{N} \sum_{k} f(k) \Psi( q_1 , k , P - q_1 - q_2 - k , q_2 ) \: ,  
\label{4UV2d:eq:fifteen}    \\
A_{24}(q_1,q_2) \, , \: C_{24}(q_1,q_2) \, , \: S_{24}(q_1,q_2) & = & 
\frac{1}{N} \sum_{k} f(k) \Psi( q_1 , k , q_2 , P - q_1 - q_2 - k ) \: ,  
\label{4UV2d:eq:sixteen}    \\
A_{34}(q_1,q_2) \, , \: C_{34}(q_1,q_2) \, , \: S_{34}(q_1,q_2) & = & 
\frac{1}{N} \sum_{k} f(k) \Psi( q_1 , q_2 , k , P - q_1 - q_2 - k ) \: .  
\label{4UV2d:eq:seventeen}
\end{eqnarray}
In each line above, $A_{i \! j}(q_1,q_2)$ corresponds to $f(k) = 1$, $C_{i \! j}(q_1,q_2)$ to $f(k) = \cos{(k)}$, and $S_{i \! j}(q_1,q_2)$ to $f(k) = \sin{(k)}$. Note that since we did not assume any permutation symmetries of $\Psi$ (the particles are distinguishable), all 18 functions are unique. With the definitions, Eqs.~(\ref{4UV2d:eq:twelve})-(\ref{4UV2d:eq:seventeen}), at hand, the full wave function is expressed from Eq.~(\ref{4UV2d:eq:eight}) as follows:
\begin{eqnarray}
\Psi( q_1 , q_2 , q_3 , q_4 ) & = &           
U \: \frac{ A_{12}( q_3 , q_4 ) + A_{13}( q_2 , q_4 ) + A_{14}( q_2 , q_3 ) + 
            A_{23}( q_1 , q_4 ) + A_{24}( q_1 , q_3 ) + A_{34}( q_1 , q_2 ) }
{ E - \varepsilon(q_1) - \varepsilon(q_2) - \varepsilon(q_3) - \varepsilon(q_4) } 
\nonumber \\
& & - \: 2V \: \frac{ \cos{q_1} \, C_{12}( q_3 , q_4 ) + \sin{q_1} \, S_{12}( q_3 , q_4 ) +
                      \cos{q_1} \, C_{13}( q_2 , q_4 ) + \sin{q_1} \, S_{13}( q_2 , q_4 ) }
{ E - \varepsilon(q_1) - \varepsilon(q_2) - \varepsilon(q_3) - \varepsilon(q_4) }  
\nonumber \\
& & - \: 2V \: \frac{ \cos{q_1} \, C_{14}( q_2 , q_3 ) + \sin{q_1} \, S_{14}( q_2 , q_3 ) +
                      \cos{q_2} \, C_{23}( q_1 , q_4 ) + \sin{q_2} \, S_{23}( q_1 , q_4 ) }
{ E - \varepsilon(q_1) - \varepsilon(q_2) - \varepsilon(q_3) - \varepsilon(q_4) }    
\nonumber \\
& & - \: 2V \: \frac{ \cos{q_2} \, C_{24}( q_1 , q_3 ) + \sin{q_2} \, S_{24}( q_1 , q_3 ) +
                      \cos{q_3} \, C_{34}( q_1 , q_2 ) + \sin{q_3} \, S_{34}( q_1 , q_2 ) }
{ E - \varepsilon(q_1) - \varepsilon(q_2) - \varepsilon(q_3) - \varepsilon(q_4) } \: .  
\label{4UV2d:eq:eighteen}    
\end{eqnarray}
Finally, a substitution of $\Psi$ into the definitions, Eqs.~(\ref{4UV2d:eq:twelve})-(\ref{4UV2d:eq:seventeen}), leads to 18 coupled integral equations. In full form, they can be found in the Supplemental Material,~\cite{SupplMat2022} Sec.~\ref{4UV2d:sec:sm:a}. Here we present the equation for $S_{23}(q_1,q_2)$ as an example: 
\begin{eqnarray}
S_{23}( q_1 , q_2 ) & = &           
      \frac{U}{N} \sum_{k} 
\frac{ \sin{k} \, A_{12}( P - q_1 - q_2 - k , q_2 ) + \sin{k} \, A_{13}( k , q_2 ) +  
       \sin{k} \, A_{14}( k , P - q_1 - q_2 - k ) }
{ E - \varepsilon(q_1) - \varepsilon(k) - \varepsilon( P - q_1 - q_2 - k ) - \varepsilon(q_2) } 
\nonumber \\ 
& & + \frac{U}{N} \sum_{k}  
\frac{ \sin{k} \, A_{23}( q_1 , q_2 ) + \sin{k} \, A_{24}( q_1 , P - q_1 - q_2 - k ) + 
       \sin{k} \, A_{34}( q_1 , k ) }
{ E - \varepsilon(q_1) - \varepsilon(k) - \varepsilon( P - q_1 - q_2 - k ) - \varepsilon(q_2) } 
\nonumber \\
& & - \frac{2V}{N} \sum_{k}  
\frac{ \sin{k}\cos{q_1} \, C_{12}( P - q_1 - q_2 - k , q_2 ) + 
       \sin{k}\sin{q_1} \, S_{12}( P - q_1 - q_2 - k , q_2 ) }
{ E - \varepsilon(q_1) - \varepsilon(k) - \varepsilon( P - q_1 - q_2 - k ) - \varepsilon(q_2) } 
\nonumber \\
& & - \frac{2V}{N} \sum_{k}  
\frac{ \sin{k}\cos{q_1} \, C_{13}( k , q_2 ) + 
       \sin{k}\sin{q_1} \, S_{13}( k , q_2 ) }
{ E - \varepsilon(q_1) - \varepsilon(k) - \varepsilon( P - q_1 - q_2 - k ) - \varepsilon(q_2) } 
\nonumber \\
& & - \frac{2V}{N} \sum_{k}  
\frac{ \sin{k}\cos{q_1} \, C_{14}( k , P - q_1 - q_2 - k ) + 
       \sin{k}\sin{q_1} \, S_{14}( k , P - q_1 - q_2 - k ) }
{ E - \varepsilon(q_1) - \varepsilon(k) - \varepsilon( P - q_1 - q_2 - k ) - \varepsilon(q_2) } 
\nonumber \\
& & - \frac{2V}{N} \sum_{k}  
\frac{ \sin{k}\cos{k} \, C_{23}( q_1 , q_2 ) + 
       \sin{k}\sin{k} \, S_{23}( q_1 , q_2 ) }
{ E - \varepsilon(q_1) - \varepsilon(k) - \varepsilon( P - q_1 - q_2 - k ) - \varepsilon(q_2) } 
\nonumber \\
& & - \frac{2V}{N} \sum_{k}  
\frac{ \sin{k}\cos{k} \, C_{24}( q_1 , P - q_1 - q_2 - k ) + 
       \sin{k}\sin{k} \, S_{24}( q_1 , P - q_1 - q_2 - k ) }
{ E - \varepsilon(q_1) - \varepsilon(k) - \varepsilon( P - q_1 - q_2 - k ) - \varepsilon(q_2) } 
\nonumber \\
& & - \frac{2V}{N} \sum_{k}  
\frac{ \sin{k}\cos{( P - q_1 - q_2 - k )} \, C_{34}( q_1 , k ) + 
       \sin{k}\sin{( P - q_1 - q_2 - k )} \, S_{34}( q_1 , k ) }
{ E - \varepsilon(q_1) - \varepsilon(k) - \varepsilon( P - q_1 - q_2 - k ) - \varepsilon(q_2) } \: .
\label{4UV2d:eq:nineteen}    
\end{eqnarray}

The equations are solved by discretizing integrals into finite sums, converting the system into a single matrix and then finding $E$ for fixed $U$, $V$, and $P$ via eigenvalue search. Eighteen equations sound like a lot but it is a constant number that does not change with the size of the lattice studied. For example, for a linear chain of 50 sites, the Brillouin zone is approximated by 50 points. Each function $A$, $C$, $S$ consists of $50^2 = 2500$ components and the entire matrix has a linear size of $M = 45000$. Such matrices can still be handled by desktop workstations and do not require supercomputers. At the same time, $N = 50$ is a long enough chain to enable a meaningful finite-size scaling analysis. 

The situation is more difficult in 2D. The base Eq.~(\ref{4UV2d:eq:seven}) is the same but there are four vectors ${\bf b}$. A two-dimensional version of Eq.~(\ref{4UV2d:eq:eight}) contains factors $\cos{(k_x - q_x)}$ and $\cos{(k_y - q_y)}$. Analogs of Eqs.~(\ref{4UV2d:eq:twelve})-(\ref{4UV2d:eq:seventeen}) comprise five functions each, corresponding to $f(k) = 1$, $\cos{(k_x)}$, $\sin{(k_x)}$, $\cos{(k_y)}$, and $\sin{(k_y)}$. The total number of auxiliary functions increases to 30. More importantly, ${\bf q}_1$ and ${\bf q}_2$ are now two-dimensional variables. For a square lattice of linear size $N_x$, the matrix size is $M = 30 \times N^4_x$. A matrix with $M = 45000$ now corresponds to a lattice of only $N_x = 6$. In 3D, the reduction procedure generates 42 unique functions, and the matrix size scales as $M = 42 \times N^6_x$.         

Permutation and rotation symmetries of $\Psi$ reduce the number of equations in a final system and increase the lattice size accessible with this method.

\section{\label{4UV2d:sec:threethree}
Permutation symmetries. Four $s = 0$ bosons 
}

The derivation presented in Appendix~\ref{4UV2d:sec:threetwo} does not take into account permutation symmetries of the four-body wave function $\Psi$. The resulting system composed of equations like Eq.~(\ref{4UV2d:eq:nineteen}) produces solutions of all possible symmetries including bosonic and fermionic states. To obtain physically meaningful results, the symmetry needs to be restricted in the base Schr\"odinger equation, Eq.~(\ref{4UV2d:eq:five}). As an added benefit, the number of integral equations in the final system is reduced. In this section, we derive equations for $s = 0$ bosons, i.e., for a fully symmetrical $\Psi$.    

A full permutation symmetry can be imposed explicitly by redefining the Fourier transformation:
\begin{eqnarray}
& & \Psi_{s}( {\bf m}_1 , {\bf m}_2 , {\bf m}_3 , {\bf m}_4 ) = 
\frac{1}{N^2} \sum_{{\bf q}_1 {\bf q}_2 {\bf q}_3 {\bf q}_4} \left\{ 
e^{ i \, ( {\bf q}_1 {\bf m}_1 + {\bf q}_2 {\bf m}_2 + {\bf q}_3 {\bf m}_3 + {\bf q}_4 {\bf m}_4 ) } + \right. 
\nonumber \\
& & \hspace{1.0cm} \left.
e^{ i \, ( {\bf q}_1 {\bf m}_1 + {\bf q}_2 {\bf m}_2 + {\bf q}_3 {\bf m}_4 + {\bf q}_4 {\bf m}_3 ) } + 
\: 22 \; {\rm more} \; {\rm permutations} \; {\rm of} \; 
[ {\bf m}_1 , {\bf m}_2 , {\bf m}_3 , {\bf m}_4 ]
\right\} \Psi_{s}( {\bf q}_1 , {\bf q}_2 , {\bf q}_3 , {\bf q}_4 ) \: .
\label{4UV2d:eq:twenty}    
\end{eqnarray}
The subscript $s$ indicates that $\Psi_{s}$ is fully symmetrical with respect to permutations of all its arguments in both representations, ${\bf m}$ and ${\bf q}$. Substitution of Eq.~(\ref{4UV2d:eq:twenty}) to Eq.~(\ref{4UV2d:eq:five}) yields, after transformations
\begin{eqnarray}
& & \left[ E - \varepsilon({\bf q}_1) - \varepsilon({\bf q}_2) 
             - \varepsilon({\bf q}_3) - \varepsilon({\bf q}_4) \right] \, 
    \Psi_{s}( {\bf q}_1 , {\bf q}_2 , {\bf q}_3 , {\bf q}_4 ) =           
\nonumber \\
& & \hspace{1.0cm} 
U \frac{1}{N} \sum_{\bf k} \left\{ 
 	    \Psi_{s}( {\bf k} , {\bf q}_3 , {\bf q}_4 , {\bf q}_1 + {\bf q}_2 - {\bf k} ) +
 	    \Psi_{s}( {\bf k} , {\bf q}_2 , {\bf q}_4 , {\bf q}_1 + {\bf q}_3 - {\bf k} ) +
 	    \Psi_{s}( {\bf k} , {\bf q}_2 , {\bf q}_3 , {\bf q}_1 + {\bf q}_4 - {\bf k} ) + \right.          
\nonumber \\
& & \hspace{2.4cm} \left. 
        \Psi_{s}( {\bf k} , {\bf q}_1 , {\bf q}_4 , {\bf q}_2 + {\bf q}_3 - {\bf k} ) +
        \Psi_{s}( {\bf k} , {\bf q}_1 , {\bf q}_3 , {\bf q}_2 + {\bf q}_4 - {\bf k} ) +
        \Psi_{s}( {\bf k} , {\bf q}_1 , {\bf q}_2 , {\bf q}_3 + {\bf q}_4 - {\bf k} )  \right\}
\nonumber \\
& & \hspace{1.0cm} 
- \: V \frac{1}{2N} \sum_{\bf k} \sum_{\bf b} \left\{ 
\left( \cos{[({\bf k} - {\bf q}_1 ){\bf b}]} + \cos{[({\bf k} - {\bf q}_2 ){\bf b}]} \right)  
\Psi_{s}( {\bf k} , {\bf q}_3 , {\bf q}_4 , {\bf q}_1 + {\bf q}_2 - {\bf k} ) +
\right.
\nonumber \\
& & \hspace{3.6cm} 
\left( \cos{[({\bf k} - {\bf q}_1 ){\bf b}]} + \cos{[({\bf k} - {\bf q}_3 ){\bf b}]} \right)  
\Psi_{s}( {\bf k} , {\bf q}_2 , {\bf q}_4 , {\bf q}_1 + {\bf q}_3 - {\bf k} ) +
\nonumber \\
& & \hspace{3.6cm} 
\left( \cos{[({\bf k} - {\bf q}_1 ){\bf b}]} + \cos{[({\bf k} - {\bf q}_4 ){\bf b}]} \right)  
\Psi_{s}( {\bf k} , {\bf q}_2 , {\bf q}_3 , {\bf q}_1 + {\bf q}_4 - {\bf k} ) +
\nonumber \\
& & \hspace{3.6cm} 
\left( \cos{[({\bf k} - {\bf q}_2 ){\bf b}]} + \cos{[({\bf k} - {\bf q}_3 ){\bf b}]} \right)  
\Psi_{s}( {\bf k} , {\bf q}_1 , {\bf q}_4 , {\bf q}_2 + {\bf q}_3 - {\bf k} ) +
\nonumber \\
& & \hspace{3.6cm} 
\left( \cos{[({\bf k} - {\bf q}_2 ){\bf b}]} + \cos{[({\bf k} - {\bf q}_4 ){\bf b}]} \right)  
\Psi_{s}( {\bf k} , {\bf q}_1 , {\bf q}_3 , {\bf q}_2 + {\bf q}_4 - {\bf k} ) +
\nonumber \\
& & \hspace{3.55cm} \left.
\left( \cos{[({\bf k} - {\bf q}_3 ){\bf b}]} + \cos{[({\bf k} - {\bf q}_4 ){\bf b}]} \right)  
\Psi_{s}( {\bf k} , {\bf q}_1 , {\bf q}_2 , {\bf q}_3 + {\bf q}_4 - {\bf k} ) \right\} .
\label{4UV2d:eq:twentyone}    
\end{eqnarray}
Equation~(\ref{4UV2d:eq:twentyone}) differs from its unsymmetrized counterpart, Eq.~(\ref{4UV2d:eq:seven}), in two aspects. First, all $\Psi_{s}$'s have been brought to the same argument order, $\Psi_{s}({\bf k},{\bf q},{\bf q},{\bf P}-{\bf q}-{\bf q}-{\bf k})$. This is allowed due to the permutation symmetry. Second, the interaction form factor in the $V$ term is symmetric with respect to the two particles interacting in each channel. For example, in the third $V$ term, particles 1 and 4 interact, and the potential is symmetric with respect to permuting ${\bf q}_1$ and ${\bf q}_4$. Note that a global factor of $\frac{1}{2}$ in front of the $V$ term ensures proper normalization.         

The form of Eq.~(\ref{4UV2d:eq:twentyone}) explains why only a few auxiliary functions are needed in the fully symmetrized case. In the distinguishable case, many functions had to be introduced because of various argument arrangements in $\Psi$; see Eqs.~(\ref{4UV2d:eq:twelve})-(\ref{4UV2d:eq:seventeen}). Here, all argument orders are equivalent, which reduces the number of needed functions by a factor of 6. At the same time, the complexity of each equation increases because the interaction form factor now consists of many more parts resulting from symmetrization.

\subsection{\label{4UV2d:sec:threethreeone}
Attractive Bose-Hubbard model
}

Before proceeding further, we digress to derive a four-body cluster equation in the attractive (not extended) Bose-Hubbard model. Although not the main object of the present work, this model is popular in the science of cold atoms, and such an equation can be derived here with little extra effort. To this end, we explicitly set $U = - \vert U \vert$, $V = 0$ in Eq.~(\ref{4UV2d:eq:twentyone}), and we define {\em one} auxiliary function 
\begin{equation}
A( {\bf q}_1 , {\bf q}_2 ) = \frac{1}{N} \sum_{\bf k} 
\Psi( {\bf k} , {\bf q}_1 , {\bf q}_2 , {\bf P} - {\bf q}_1 - {\bf q}_2 - {\bf k} ) \: .  
\label{4UV2d:eq:twentytwo}    
\end{equation}
Note that $A( {\bf q}_1 , {\bf q}_2 ) = A( {\bf q}_2 , {\bf q}_1 )$. The wave function follows from Eq.~(\ref{4UV2d:eq:twentyone}),
\begin{equation}
\Psi_{s}( {\bf q}_1 , {\bf q}_2 , {\bf q}_3 , {\bf q}_4 ) = - \vert U \vert \: 
\frac{ A( {\bf q}_3 , {\bf q}_4 ) + A( {\bf q}_2 , {\bf q}_4 ) + A( {\bf q}_2 , {\bf q}_3 ) + 
       A( {\bf q}_1 , {\bf q}_4 ) + A( {\bf q}_1 , {\bf q}_3 ) + A( {\bf q}_1 , {\bf q}_2 ) }
{ E - \varepsilon({\bf q}_1) - \varepsilon({\bf q}_2) - \varepsilon({\bf q}_3) - \varepsilon({\bf q}_4) } \: .  
\label{4UV2d:eq:twentythree}    
\end{equation}
Substituting Eq.~(\ref{4UV2d:eq:twentythree}) in Eq.~(\ref{4UV2d:eq:twentytwo}), one obtains
\begin{eqnarray}
A( {\bf q}_1 , {\bf q}_2 ) & = & - \frac{\vert U \vert}{N} \sum_{\bf k}      \left\{ 
\frac{ A( {\bf q}_2 , {\bf P} - {\bf q}_1 - {\bf q}_2 - {\bf k} ) + 
       A( {\bf q}_1 , {\bf P} - {\bf q}_1 - {\bf q}_2 - {\bf k} ) + 
       A( {\bf q}_1 , {\bf q}_2 )  }                                   
{ E - \varepsilon({\bf k}) - \varepsilon({\bf q}_1) - \varepsilon({\bf q}_2) 
    - \varepsilon( {\bf P} - {\bf q}_1 - {\bf q}_2 - {\bf k} ) }             \right.
\nonumber \\               &   &      \hspace{1.2cm}  +                      \left.
\frac{ A( {\bf k}   , {\bf P} - {\bf q}_1 - {\bf q}_2 - {\bf k} ) +
       A( {\bf k}   , {\bf q}_2 )                                 +  
       A( {\bf k}   , {\bf q}_1 )  }
{ E - \varepsilon({\bf k}) - \varepsilon({\bf q}_1) - \varepsilon({\bf q}_2) 
    - \varepsilon( {\bf P} - {\bf q}_1 - {\bf q}_2 - {\bf k} ) }             \right\} .  
\label{4UV2d:eq:twentyfour}    
\end{eqnarray}
This equation is valid in all lattice dimensions, and for all Bravais lattices for that matter. Equation~(\ref{4UV2d:eq:twentyfour}) is the natural extension of Mattis' equation~\cite{Mattis1986} to four identical bosons.

\subsection{\label{4UV2d:sec:threethreetwo}
Bose-$UV$ model in 1D. Linear chain 
}

We now return to the full Bose-$UV$ model, Eq.~(\ref{4UV2d:eq:twentyone}). In 1D, summation over ${\bf b}$ yields a global factor of 2. After expansion of the cosines, only three kinds of auxiliary integrals emerge:
\begin{eqnarray}
A(q_1,q_2) & = & 
\frac{1}{N} \sum_{k} \hspace{1.0cm} \Psi( k , q_1 , q_2 , P - q_1 - q_2 - k ) \: ,  
\label{4UV2d:eq:twentyfive}     \\
C(q_1,q_2) & = & 
\frac{1}{N} \sum_{k} \cos{(k)} \Psi( k , q_1 , q_2 , P - q_1 - q_2 - k ) \: ,  
\label{4UV2d:eq:twentysix}   \\  
S(q_1,q_2) & = & 
\frac{1}{N} \sum_{k} \sin{(k)} \Psi( k , q_1 , q_2 , P - q_1 - q_2 - k ) \: .  
\label{4UV2d:eq:twentyseven}   
\end{eqnarray}
All three functions are symmetrical with respect to permutation $q_1 \leftrightarrow q_2$. Next, we express $\Psi$ via $A$, $C$, and $S$ from Eq.~(\ref{4UV2d:eq:twentyone}) and substitute back in the definitions, Eqs.~(\ref{4UV2d:eq:twentyfive})-(\ref{4UV2d:eq:twentyseven}). This results in three coupled integral equations. In full form, they are given in the Supplemental Material,~\cite{SupplMat2022} Sec.~\ref{4UV2d:sec:sm:b}. The equation for $A$ reads      
\begin{eqnarray}
A( q_1 , q_2 ) & = &           
      \frac{U}{N} \sum_{k} 
\frac{ f(k) \, A( q_2 , P - q_1 - q_2 - k ) + f(k) \, A( q_1 , P - q_1 - q_2 - k ) + 
       f(k) \, A( q_1 , q_2 ) }
{ E - \varepsilon(k) - \varepsilon(q_1) - \varepsilon(q_2) - \varepsilon( P - q_1 - q_2 - k ) } 
\nonumber \\ 
& & + \frac{U}{N} \sum_{k}  
\frac{ f(k) \, A( k , P - q_1 - q_2 - k ) + f(k) \, A( k , q_2 ) + f(k) \, A( k , q_1 ) }
{ E - \varepsilon(k) - \varepsilon(q_1) - \varepsilon(q_2) - \varepsilon( P - q_1 - q_2 - k ) } 
\nonumber \\
& & - \frac{V}{N} \sum_{k}  
\frac{ f(k) [ \cos{k} + \cos{q_1} ] \, C( q_2 , P - q_1 - q_2 - k ) + 
       f(k) [ \sin{k} + \sin{q_1} ] \, S( q_2 , P - q_1 - q_2 - k ) }
{ E - \varepsilon(k) - \varepsilon(q_1) - \varepsilon(q_2) - \varepsilon( P - q_1 - q_2 - k ) } 
\nonumber \\
& & - \frac{V}{N} \sum_{k}  
\frac{ f(k) [ \cos{k} + \cos{q_2} ] \, C( q_1 , P - q_1 - q_2 - k ) + 
       f(k) [ \sin{k} + \sin{q_2} ] \, S( q_1 , P - q_1 - q_2 - k ) }
{ E - \varepsilon(k) - \varepsilon(q_1) - \varepsilon(q_2) - \varepsilon( P - q_1 - q_2 - k ) } 
\nonumber \\
& & - \frac{V}{N} \sum_{k}  
\frac{ f(k) [ \cos{k} + \cos{( P - q_1 - q_2 - k )} ] \, C( q_1 , q_2 ) + 
       f(k) [ \sin{k} + \sin{( P - q_1 - q_2 - k )} ] \, S( q_1 , q_2 ) }
{ E - \varepsilon(k) - \varepsilon(q_1) - \varepsilon(q_2) - \varepsilon( P - q_1 - q_2 - k ) } 
\nonumber \\
& & - \frac{V}{N} \sum_{k}  
\frac{ f(k) [ \cos{q_1} + \cos{q_2} ] \, C( k , P - q_1 - q_2 - k ) + 
       f(k) [ \sin{q_1} + \sin{q_2} ] \, S( k , P - q_1 - q_2 - k ) }
{ E - \varepsilon(k) - \varepsilon(q_1) - \varepsilon(q_2) - \varepsilon( P - q_1 - q_2 - k ) } 
\nonumber \\
& & - \frac{V}{N} \sum_{k}  
\frac{ f(k) [ \cos{q_1} + \cos{( P - q_1 - q_2 - k )} ] \, C( k , q_2 ) + 
       f(k) [ \sin{q_1} + \sin{( P - q_1 - q_2 - k )} ] \, S( k , q_2 ) }
{ E - \varepsilon(k) - \varepsilon(q_1) - \varepsilon(q_2) - \varepsilon( P - q_1 - q_2 - k ) } 
\nonumber \\
& & - \frac{V}{N} \sum_{k}  
\frac{ f(k) [ \cos{q_2} + \cos{( P - q_1 - q_2 - k )} ] \, C( k , q_1 ) + 
       f(k) [ \sin{q_2} + \sin{( P - q_1 - q_2 - k )} ] \, S( k , q_1 ) }
{ E - \varepsilon(k) - \varepsilon(q_1) - \varepsilon(q_2) - \varepsilon( P - q_1 - q_2 - k ) } \: ,
\hspace{0.5cm}
\label{4UV2d:eq:twentyeight}       
\end{eqnarray}
with $f(k) \equiv 1$. The equations for $C$ and $S$ are also described by Eq.~(\ref{4UV2d:eq:twentyeight}) but with $f(k) = \cos{(k)}$ and $f(k) = \sin{(k)}$, respectively.

\subsection{\label{4UV2d:sec:threethreethree}
Bose-$UV$ model in 2D. Square lattice 
}

To derive a working system of equations for the square lattice, we come back to Eq.~(\ref{4UV2d:eq:twentyone}) and sum over ${\bf b} = \pm {\bf x}, \pm {\bf y}$. Expansion of the cosines leads to five auxiliary functions: $A$, $C_{x}$, $S_{x}$, $C_{y}$ and $S_{y}$. $A( {\bf q}_1 , {\bf q}_2 )$ is defined similarly to Eq.~(\ref{4UV2d:eq:twentyfive}), except all the variables are two-dimensional vectors, and the integration is over a two-dimensional Brillouin zone. $C_{x}( {\bf q}_1 , {\bf q}_2 )$ and $C_{y}( {\bf q}_1 , {\bf q}_2 )$ are defined according to Eq.~(\ref{4UV2d:eq:twentysix}) but with $\cos{(k_x)}$ and $\cos{(k_y)}$ in place of $\cos(k)$, respectively. Likewise, $S_{x}( {\bf q}_1 , {\bf q}_2 )$ and $S_{y}( {\bf q}_1 , {\bf q}_2 )$ are defined according to Eq.~(\ref{4UV2d:eq:twentyseven}), but with $\sin{(k_x)}$ and $\sin{(k_y)}$ in place of $\sin(k)$. Then the wave function is expressed via $A$, $C_{x,y}$, $S_{x,y}$ and substituted back into the definitions. The resulting system consists of five coupled integral equations which are given in Supplemental Material,~\cite{SupplMat2022} section~\ref{4UV2d:sec:sm:c}.

A 3D cubic Bose-$UV$ model could be analyzed similarly. Seven auxiliary functions would be required, and the resulting system would consist of seven coupled equations. However, 3D four-body systems exceed the computational capabilities of desktop workstations and are not studied here.

\section{\label{4UV2d:sec:threefour}
Permutation symmetries. Four $s = \frac{1}{2}$ fermions 
}

The fermionic case is more complicated because four spin-$\frac{1}{2}$ fermions can form states with total spin values $S = 0, 1$, and 2. Separation of fermionic states by $S$ and separation between bosons and fermions requires symmetrization of $\Psi$ in accordance with Young's tables. Unfortunately, proper symmetrization leads to auxiliary functions losing some easy-to-check permutation properties that are useful for validating numerical results. Therefore, we adopt here {\em partial} implementation of Young's tables, which produces verifiable solutions but still enables separation of states by $S$.  

The process begins with $S = 2$. The coordinate wave function is fully antisymmetric with respect to all four arguments, which leads to unambiguous results. Technically, treatment is similar to the bosonic case but with alternating signs in the Fourier transformation. Next, one moves to $S = 1$. In this case, we antisymmetrize $\Psi$ with respect to three arguments but leave the symmetries involving the fourth argument unrestricted. As a result, states with $S = 1$ {\em and} $2$ are produced. However, since the $S = 2$ states have already been determined, they can be separated out. Finally, to access states with $S = 0$, we antisymmetrize $\Psi$ with respect to two {\em pairs} of arguments, for example, ${\bf q}_1 \leftrightarrow {\bf q}_2$ and ${\bf q}_3 \leftrightarrow {\bf q}_4$. The symmetries involving permutations of arguments from different pairs, such as ${\bf q}_2 \leftrightarrow {\bf q}_4$, remain unrestricted. The process finds solutions with all three $S = 0$, 1, and 2. However, since the $S = 2$ and $1$ sectors have been mapped in the previous steps, the $S = 0$ states can be identified. 

In what follows, we provide further details of the procedure. The overall flow parallels that of the indistinguishable and bosonic cases. Once the symmetry of $\Psi$ is defined via an appropriate Fourier transformation, the Schr\"odinger equation is converted in momentum space and a certain number of auxiliary functions is introduced. Then the wave function is expressed via the auxiliary functions and substituted back into their definitions, producing a final set of equations. Complete derivations can be found in the Supplemental Material.~\cite{SupplMat2022}

\subsection{\label{4UV2d:sec:threefourone}
Four $s = \frac{1}{2}$ fermions. $S = 2$ 
}

Full antisymmetry of $\Psi$ is imposed by utilizing a fully antisymmetric basis in Fourier transformation 
\begin{equation}
\Psi_{2}( {\bf m}_1 , {\bf m}_2 , {\bf m}_3 , {\bf m}_4 ) = 
\frac{1}{N^2} \sum_{{\bf q}_1 {\bf q}_2 {\bf q}_3 {\bf q}_4} \left\{ 
\sum_{[ {\bf m}_1 , {\bf m}_2 , {\bf m}_3 , {\bf m}_4 ]} (-1)^{C[{\bf m}]} 
e^{ i \, ( {\bf q}_1 {\bf m}_1 + {\bf q}_2 {\bf m}_2 + {\bf q}_3 {\bf m}_3 + {\bf q}_4 {\bf m}_4 ) } 
\right\} \Psi_{2}( {\bf q}_1 , {\bf q}_2 , {\bf q}_3 , {\bf q}_4 ) \: .
\label{4UV2d:eq:twentynine}    
\end{equation}
Here the inner sum is over 24 permutations of site indices ${\bf m}_i$, and $(-1)^{C[{\bf m}]} = +1$ if the permutation is even and $-1$ otherwise. The subscript 2 indicates that we are dealing with the fermionic sector $S = 2$. Substituting Eq.~(\ref{4UV2d:eq:twentynine}) in the base Schr\"odinger equation, Eq.~(\ref{4UV2d:eq:five}), one obtains 
\begin{eqnarray}
& & \left[ E - \varepsilon({\bf q}_1) - \varepsilon({\bf q}_2) 
             - \varepsilon({\bf q}_3) - \varepsilon({\bf q}_4) \right] \, 
    \Psi_{2}( {\bf q}_1 , {\bf q}_2 , {\bf q}_3 , {\bf q}_4 ) =           
\nonumber \\
& & \hspace{1.0cm} 
- \: V \frac{1}{2N} \sum_{\bf k} \sum_{\bf b} \left\{ 
\left( \cos{[({\bf k} - {\bf q}_1 ){\bf b}]} - \cos{[({\bf k} - {\bf q}_2 ){\bf b}]} \right)  
\Psi_{2}( {\bf k} , {\bf q}_1 + {\bf q}_2 - {\bf k} , {\bf q}_3 , {\bf q}_4 ) +
\right.
\nonumber \\
& & \hspace{3.6cm} 
\left( \cos{[({\bf k} - {\bf q}_1 ){\bf b}]} - \cos{[({\bf k} - {\bf q}_3 ){\bf b}]} \right)  
\Psi_{2}( {\bf k} , {\bf q}_2 , {\bf q}_1 + {\bf q}_3 - {\bf k} , {\bf q}_4 ) +
\nonumber \\
& & \hspace{3.6cm} 
\left( \cos{[({\bf k} - {\bf q}_1 ){\bf b}]} - \cos{[({\bf k} - {\bf q}_4 ){\bf b}]} \right)  
\Psi_{2}( {\bf k} , {\bf q}_2 , {\bf q}_3 , {\bf q}_1 + {\bf q}_4 - {\bf k} ) +
\nonumber \\
& & \hspace{3.6cm} 
\left( \cos{[({\bf k} - {\bf q}_2 ){\bf b}]} - \cos{[({\bf k} - {\bf q}_3 ){\bf b}]} \right)  
\Psi_{2}( {\bf q}_1 , {\bf k} , {\bf q}_2 + {\bf q}_3 - {\bf k} , {\bf q}_4 ) +
\nonumber \\
& & \hspace{3.6cm} 
\left( \cos{[({\bf k} - {\bf q}_2 ){\bf b}]} - \cos{[({\bf k} - {\bf q}_4 ){\bf b}]} \right)  
\Psi_{2}( {\bf q}_1 , {\bf k} , {\bf q}_3 , {\bf q}_2 + {\bf q}_4 - {\bf k} ) +
\nonumber \\
& & \hspace{3.55cm} \left.
\left( \cos{[({\bf k} - {\bf q}_3 ){\bf b}]} - \cos{[({\bf k} - {\bf q}_4 ){\bf b}]} \right)  
\Psi_{2}( {\bf q}_1 , {\bf q}_2 , {\bf k} , {\bf q}_3 + {\bf q}_4 - {\bf k} ) \right\} .
\label{4UV2d:eq:thirty}  
\end{eqnarray}
It is instructive to compare Eq.~(\ref{4UV2d:eq:thirty}) with its fully symmetric counterpart, Eq.~(\ref{4UV2d:eq:twentyone}). First, the $U$ term has disappeared. This is expected since fully antisymmetric wave functions should not feel a contact interaction. Second, the form factors in Eq.~(\ref{4UV2d:eq:thirty}) are now antisymmetrized with respect to permutations of interacting particles. Note that for now we have left the argument order in $\Psi_{2}$ unchanged to keep the connection between the potential and particle order clearer. By utilizing full antisymmetry, all $\Psi_{2}$ can be brought to a standard form $\Psi({\bf k},{\bf q},{\bf q},{\bf P}-{\bf q}-{\bf q}-{\bf k})$, which generates additional factors of $(-1)$. 

The rest of the derivation is relegated to the Supplemental Material.~\cite{SupplMat2022} Suffice it to say here that two auxiliary functions are needed in 1D (Sec.~\ref{4UV2d:sec:sm:d}), and four functions are needed in 2D (Sec.~\ref{4UV2d:sec:sm:e}). In both dimensionalities, the number of functions is one less than the respective bosonic numbers because the $U$ term is absent.

\subsection{\label{4UV2d:sec:threefourtwo}
Four $s = \frac{1}{2}$ fermions. $S = 1$ 
}

To access $S = 1$ fermionic states, we construct a wave function $\Psi_{1}$ that is antisymmetric with respect to permutations of three particles 1, 2, and 3:
\begin{eqnarray}
& & \Psi_{1}( {\bf m}_1 , {\bf m}_2 , {\bf m}_3 ; {\bf m}_4 ) = 
\frac{1}{N^2} \sum_{{\bf q}_1 {\bf q}_2 {\bf q}_3 {\bf q}_4} \left\{ 
e^{ i \, ( {\bf q}_1 {\bf m}_1 + {\bf q}_2 {\bf m}_2 + {\bf q}_3 {\bf m}_3 ) } - 
e^{ i \, ( {\bf q}_1 {\bf m}_1 + {\bf q}_2 {\bf m}_3 + {\bf q}_3 {\bf m}_2 ) } +
e^{ i \, ( {\bf q}_1 {\bf m}_3 + {\bf q}_2 {\bf m}_1 + {\bf q}_3 {\bf m}_2 ) } -
\right. 
\nonumber \\
& & \hspace{1.0cm} \left.
e^{ i \, ( {\bf q}_1 {\bf m}_2 + {\bf q}_2 {\bf m}_1 + {\bf q}_3 {\bf m}_3 ) } + 
e^{ i \, ( {\bf q}_1 {\bf m}_2 + {\bf q}_2 {\bf m}_3 + {\bf q}_3 {\bf m}_1 ) } - 
e^{ i \, ( {\bf q}_1 {\bf m}_3 + {\bf q}_2 {\bf m}_2 + {\bf q}_3 {\bf m}_1 ) } 
\right\} e^{i \, ( {\bf q}_4 {\bf m}_4 ) } \Psi_{1}( {\bf q}_1 , {\bf q}_2 , {\bf q}_3 ; {\bf q}_4 ) \: .
\label{4UV2d:eq:thirtyone}    
\end{eqnarray}
The semicolon notation indicates that the arguments are now split into two groups. $\Psi_{1}$ is antisymmetric with respect to permutations of the first group ${\bf m}_1$, ${\bf m}_2$, ${\bf m}_3$. At the same time, no symmetry is imposed for permuting ${\bf m}_4$ and any argument from the first group. Substitution of Eq.~(\ref{4UV2d:eq:thirtyone}) in Eq.~(\ref{4UV2d:eq:five}) yields a corresponding Schr\"odinger equation,            
\begin{eqnarray}
& & \left[ E - \varepsilon({\bf q}_1) - \varepsilon({\bf q}_2) 
             - \varepsilon({\bf q}_3) - \varepsilon({\bf q}_4) \right] \, 
    \Psi_{1}( {\bf q}_1 , {\bf q}_2 , {\bf q}_3 ; {\bf q}_4 ) =           
\nonumber \\
& & \hspace{1.0cm} 
U \frac{1}{N} \sum_{\bf k} \left\{ 
 	    \Psi_{1}( {\bf k}   , {\bf q}_2 , {\bf q}_3 ; {\bf q}_1 + {\bf q}_4 - {\bf k} ) +
 	    \Psi_{1}( {\bf q}_1 , {\bf k}   , {\bf q}_3 ; {\bf q}_2 + {\bf q}_4 - {\bf k} ) +
 	    \Psi_{1}( {\bf q}_1 , {\bf q}_2 , {\bf k}   ; {\bf q}_3 + {\bf q}_4 - {\bf k} )   \right\}          
\nonumber \\
& & \hspace{1.0cm} 
- \: V \frac{1}{2N} \sum_{\bf k} \sum_{\bf b} \left\{ 
\left( \cos{[({\bf k} - {\bf q}_1 ){\bf b}]} - \cos{[({\bf k} - {\bf q}_2 ){\bf b}]} \right)  
\Psi_{1}( {\bf k} , {\bf q}_1 + {\bf q}_2 - {\bf k} , {\bf q}_3 ; {\bf q}_4 ) +
\right.
\nonumber \\
& & \hspace{3.6cm} 
\left( \cos{[({\bf k} - {\bf q}_1 ){\bf b}]} - \cos{[({\bf k} - {\bf q}_3 ){\bf b}]} \right)  
\Psi_{1}( {\bf k} , {\bf q}_2 , {\bf q}_1 + {\bf q}_3 - {\bf k} ; {\bf q}_4 ) +
\nonumber \\
& & \hspace{3.55cm} \left.
\left( \cos{[({\bf k} - {\bf q}_2 ){\bf b}]} - \cos{[({\bf k} - {\bf q}_3 ){\bf b}]} \right)  
\Psi_{1}( {\bf q}_1 , {\bf k} , {\bf q}_2 + {\bf q}_3 - {\bf k} ; {\bf q}_4 ) \right\} .
\nonumber \\
& & \hspace{1.0cm} 
- \: V \frac{1}{N} \sum_{\bf k} \sum_{\bf b} \left\{ 
\cos{[({\bf k} - {\bf q}_1 ){\bf b}]}   
\Psi_{1}( {\bf k} , {\bf q}_2 , {\bf q}_3 ; {\bf q}_1 + {\bf q}_4 - {\bf k} ) +
\right.
\nonumber \\
& & \hspace{3.4cm} 
\cos{[({\bf k} - {\bf q}_2 ){\bf b}]} 
\Psi_{1}( {\bf q}_1 , {\bf k} , {\bf q}_3 ; {\bf q}_2 + {\bf q}_4 - {\bf k} ) +
\nonumber \\
& & \hspace{3.36cm} \left.
\cos{[({\bf k} - {\bf q}_3 ){\bf b}]}   
\Psi_{1}( {\bf q}_1 , {\bf q}_2 , {\bf k} ; {\bf q}_3 + {\bf q}_4 - {\bf k} ) \right\} .
\label{4UV2d:eq:thirtytwo}    
\end{eqnarray}
One observes that in the $U$ term only interactions between ${\bf q}_4$ and the first group remain. Interactions within the first group have been eliminated by antisymmetry. The $V$ interactions are also split into two groups. The first three terms describe interactions within the first group and are analogous to Eq.~(\ref{4UV2d:eq:thirty}). The last three $V$ terms in Eq.~(\ref{4UV2d:eq:thirtytwo}) describe unsymmetrized interactions between ${\bf q}_4$ and the first group, and they are analogous to Eq.~(\ref{4UV2d:eq:seven}). It is possible to show that the additional assumption of full antisymmetry between the last and the first three arguments of $\Psi$ reduces Eq.~(\ref{4UV2d:eq:thirtytwo}) back to the $S = 2$ version, Eq.~(\ref{4UV2d:eq:thirty}).  

The rest of the derivation can be found in the Supplemental Material.~\cite{SupplMat2022} The 1D case requires five auxiliary functions (Sec.~\ref{4UV2d:sec:sm:f}) and the 2D case nine auxiliary functions (Sec.~\ref{4UV2d:sec:sm:g}).

\subsection{\label{4UV2d:sec:threefourthree}
Four $s = \frac{1}{2}$ fermions. $S = 0$ 
}

We finally discuss the $S = 0$ fermionic sector, which often contains the system's ground state. To access $S = 0$, we construct a wave function that is antisymmetric with respect to permutations within two groups of coordinates $\{ 12 \}$ and $\{ 34 \}$,  
\begin{eqnarray}
\Psi_{0}( {\bf m}_1 , {\bf m}_2 ; {\bf m}_3 , {\bf m}_4 ) & = & 
\frac{1}{N^2} \sum_{{\bf q}_1 {\bf q}_2 {\bf q}_3 {\bf q}_4} \left\{ 
e^{ i \, ( {\bf q}_1 {\bf m}_1 + {\bf q}_2 {\bf m}_2 ) } - 
e^{ i \, ( {\bf q}_1 {\bf m}_2 + {\bf q}_2 {\bf m}_1 ) } 
\right\} \times
\nonumber \\
& & \hspace{2.0cm} \times \left\{
e^{ i \, ( {\bf q}_3 {\bf m}_3 + {\bf q}_4 {\bf m}_4 ) } - 
e^{ i \, ( {\bf q}_3 {\bf m}_4 + {\bf q}_4 {\bf m}_3 ) }  
\right\} \Psi_{0}( {\bf q}_1 , {\bf q}_2 ; {\bf q}_3 , {\bf q}_4 ) \: .
\label{4UV2d:eq:thirtythree}    
\end{eqnarray}
As a consequence, $\Psi_{0}( {\bf q}_2 , {\bf q}_1 ; {\bf q}_3 , {\bf q}_4 ) = - \Psi_{0}( {\bf q}_1 , {\bf q}_2 ; {\bf q}_3 , {\bf q}_4 )$ and $\Psi_{0}( {\bf q}_1 , {\bf q}_2 ; {\bf q}_4 , {\bf q}_3 ) = - \Psi_{0}( {\bf q}_1 , {\bf q}_2 ; {\bf q}_3 , {\bf q}_4 )$, but no definite symmetries are associated with permutations between the two groups. Substitution of Eq.~(\ref{4UV2d:eq:thirtythree}) in Eq.~(\ref{4UV2d:eq:five}) leads to
\begin{eqnarray}
& & \left[ E - \varepsilon({\bf q}_1) - \varepsilon({\bf q}_2) 
             - \varepsilon({\bf q}_3) - \varepsilon({\bf q}_4) \right] \, 
    \Psi_{0}( {\bf q}_1 , {\bf q}_2 ; {\bf q}_3 , {\bf q}_4 ) =           
\nonumber \\
& & \hspace{1.0cm} 
U \frac{1}{N} \sum_{\bf k} \left\{ 
 	    \Psi_{0}( {\bf k} , {\bf q}_2 ; {\bf q}_1 + {\bf q}_3 - {\bf k} , {\bf q}_4 ) 
 	  - \Psi_{0}( {\bf k} , {\bf q}_2 ; {\bf q}_1 + {\bf q}_4 - {\bf k} , {\bf q}_3 ) \right.
\nonumber \\
& & \hspace{3.0cm} \left.  	    
 	  - \Psi_{0}( {\bf k} , {\bf q}_1 ; {\bf q}_2 + {\bf q}_3 - {\bf k} , {\bf q}_4 )
 	  + \Psi_{0}( {\bf k} , {\bf q}_1 ; {\bf q}_2 + {\bf q}_4 - {\bf k} , {\bf q}_3 )  \right\}          
\nonumber \\
& & \hspace{1.0cm} 
- \: V \frac{1}{N} \sum_{\bf k} \sum_{\bf b} \left\{ \frac{1}{2}
\left[ \cos{({\bf k} - {\bf q}_1 ){\bf b}} - \cos{({\bf k} - {\bf q}_2 ){\bf b}} \right]  
\Psi_{0}( {\bf k} , {\bf q}_1 + {\bf q}_2 - {\bf k} ; {\bf q}_3 , {\bf q}_4 )    \right.
\nonumber \\
& & \hspace{3.5cm} +
\cos{[({\bf k} - {\bf q}_1 ){\bf b}]} \,  
\Psi_{0}( {\bf k} , {\bf q}_2 ; {\bf q}_1 + {\bf q}_3 - {\bf k} , {\bf q}_4 ) -
\cos{[({\bf k} - {\bf q}_1 ){\bf b}]} \, 
\Psi_{0}( {\bf k} , {\bf q}_2 ; {\bf q}_1 + {\bf q}_4 - {\bf k} , {\bf q}_3 ) 
\nonumber \\
& & \hspace{3.5cm} -
\cos{[({\bf k} - {\bf q}_2 ){\bf b}]} \,  
\Psi_{0}( {\bf k} , {\bf q}_1 ; {\bf q}_2 + {\bf q}_3 - {\bf k} , {\bf q}_4 ) +
\cos{[({\bf k} - {\bf q}_2 ){\bf b}]} \, 
\Psi_{0}( {\bf k} , {\bf q}_1 ; {\bf q}_2 + {\bf q}_4 - {\bf k} , {\bf q}_3 ) 
\nonumber \\
& & \hspace{3.1cm} + \left. \frac{1}{2}
\left[ \cos{({\bf k} - {\bf q}_3 ){\bf b}} - \cos{({\bf k} - {\bf q}_4 ){\bf b}} \right]  
\Psi_{0}( {\bf q}_1 , {\bf q}_2  ; {\bf k} , {\bf q}_3 + {\bf q}_4 - {\bf k} )   \right\} .
\label{4UV2d:eq:thirtyfour}    
\end{eqnarray}
Reduction of this Schr\"odinger equation to two-variable integral equations is completed in the Supplemental Material.~\cite{SupplMat2022} The 1D case requires seven auxiliary functions (Sec.~\ref{4UV2d:sec:sm:h}) and the 2D case 13 auxiliary functions (Sec.~\ref{4UV2d:sec:sm:i}).

\end{widetext}

\section{\label{4UV2d:sec:threefive}
Rotational symmetries. Ground state at ${\bf P} = 0$ 
}

In 2D, the ground state possesses rotational symmetries that link the $x$ and $y$ auxiliary functions. Those relations enable further reduction of the number of resulting integral equations.   

For $s = 0$ bosons in 2D, the auxiliary functions are defined in the Supplemental Material~\cite{SupplMat2022} [Sec.~\ref{4UV2d:sec:sm:c}, Eqs.~(\ref{4UV2d:eq:sm:twentyfour})-(\ref{4UV2d:eq:sm:twentyseven})]. In the ground state at ${\bf P} = (0,0)$, they obey the following relations:
\begin{eqnarray}
C^{b}_{x}( q_{1x} , q_{1y} ; q_{2x} , q_{2y} ) & = & + C^{b}_{y}( q_{1y} , q_{1x} ; q_{2y} , q_{2x} ) \: ,
\label{4UV2d:eq:thirtyfive}   \\  
S^{b}_{x}( q_{1x} , q_{1y} ; q_{2x} , q_{2y} ) & = & + S^{b}_{y}( q_{1y} , q_{1x} ; q_{2y} , q_{2x} ) \: ,
\label{4UV2d:eq:thirtysix}    \\
C^{b}_{y}( q_{1x} , q_{1y} ; q_{2x} , q_{2y} ) & = & + C^{b}_{x}( q_{1y} , q_{1x} ; q_{2y} , q_{2x} ) \: ,
\label{4UV2d:eq:thirtyseven}  \\  
S^{b}_{y}( q_{1x} , q_{1y} ; q_{2x} , q_{2y} ) & = & + S^{b}_{x}( q_{1y} , q_{1x} ; q_{2y} , q_{2x} ) \: .
\label{4UV2d:eq:thirtyeight}
\end{eqnarray}
In the equations themselves, for instance in Eq.~(\ref{4UV2d:eq:sm:twentynine}), the $C_{x}$ and $C_{y}$ terms can therefore be combined, but with a proper permutation of arguments. The $S_{x}$ and $S_{y}$ terms are also combined. After that, the $C_{x}$ and $C_{y}$ equations become equivalent, and the $S_{x}$ and $S_{y}$ equations become equivalent, too. Thus, the total number of equations in the full system is reduced from 5 to 3.  

In the case of $s = \frac{1}{2}$ fermions, similar relations between the $x$ and $y$ auxiliary functions hold true, but the sign is negative. For example, in the $S = 2$ sector, see the Supplemental Material,~\cite{SupplMat2022} Sec.~\ref{4UV2d:sec:sm:e}, we have 
\begin{eqnarray}
C^{f}_{x}( q_{1x} , q_{1y} ; q_{2x} , q_{2y} ) & = & - C^{f}_{y}( q_{1y} , q_{1x} ; q_{2y} , q_{2x} ) \: ,
\label{4UV2d:eq:thirtynine}   \\  
S^{f}_{x}( q_{1x} , q_{1y} ; q_{2x} , q_{2y} ) & = & - S^{f}_{y}( q_{1y} , q_{1x} ; q_{2y} , q_{2x} ) \: ,
\label{4UV2d:eq:forty}        \\
C^{f}_{y}( q_{1x} , q_{1y} ; q_{2x} , q_{2y} ) & = & - C^{f}_{x}( q_{1y} , q_{1x} ; q_{2y} , q_{2x} ) \: ,
\label{4UV2d:eq:fortyone}     \\  
S^{f}_{y}( q_{1x} , q_{1y} ; q_{2x} , q_{2y} ) & = & - S^{f}_{x}( q_{1y} , q_{1x} ; q_{2y} , q_{2x} ) \: .
\label{4UV2d:eq:fortytwo}
\end{eqnarray}
The number of equations to solve is reduced from four to two. 

A similar reasoning applies to fermionic sectors $S = 1$ and $0$. Relations like Eqs.~(\ref{4UV2d:eq:thirtynine})-(\ref{4UV2d:eq:fortytwo}) are valid within each group of functions: $\{ 12 \}$, $\{ 13 \}$, $\{ 14 \}$, and $\{ 34 \}$. In all cases, the number of $C$ functions and $S$ functions is cut in half. 

The number of integral equations that constitute the final system in various cases are summarized in Tables~\ref{4UV2d:tab:one} and \ref{4UV2d:tab:two}.

\begin{table}[b]
\renewcommand{\tabcolsep}{0.1cm}
\renewcommand{\arraystretch}{1.5}
\begin{tabular}{|l|c||c|c|c|c|}
\hline\hline
\multicolumn{2}{|c||}{1D $UV$ model}              &  \multicolumn{4}{c|}{Linear matrix size, $M$}  \\ 
\hline
 Symmetry           &  $K_{\rm eq}$  &  $N = 16$  &  $N = 32$  &  $N = 48$  &  $N = 64$  \\
\hline\hline 
 Bosons             &     3          &     768    &    3072    &    6912    &   12288    \\ \hline
 Fermions, $S = 2$  &     2          &     512    &    2048    &    4608    &    8192    \\ \hline
 Fermions, $S = 1$  &     5          &    1280    &    5120    &   11520    &   20480    \\ \hline
 Fermions, $S = 0$  &     7          &    1792    &    7168    &   16128    &   28672    \\ \hline 
 Distinguishable    &    18          &    4608    &   18432    &   41472    &   73728    \\ \hline 
\hline 
\end{tabular}
\caption{
The number of coupled integral equations, $K_{\rm eq}$, and the linear size of the final matrix, $M = K_{\rm eq} N^2$, for different symmetry sectors and different chain lengths, $N$, in the 1D $UV$ model.  
} 
\label{4UV2d:tab:one}
\end{table}

\section{\label{4UV2d:sec:threesix}
Practical implementation   
}

We now describe practical ways of solving integral equations like Eq.~(\ref{4UV2d:eq:nineteen}) and converting them to useful physical information. The first step is to approximate auxiliary functions by their values at a finite set of $q$ points spanning the entire Brillouin zone. For a regular $q$ mesh, there is a 1-to-1 correspondence between the mesh size and the lattice size $N$. It is also possible to employ a nonuniform mesh. For example, one can argue that due to the polelike structure of energy denominators, it is advantageous to have a denser mesh near $q_1 = q_2 = 0$ to better resolve fast varying kernels, especially at small binding energies. However, nonuniform meshes are incompatible with the circular coordinate $P - q_1 - q_2 - k$ because a momentum difference is not guaranteed to land on a mesh point. For that reason, we employed only regular $q$ meshes in both 1D and 2D.

\begin{table}[t]
\renewcommand{\tabcolsep}{0.1cm}
\renewcommand{\arraystretch}{1.5}
\begin{tabular}{|c|l|c||c|c|c|}
\hline\hline
\multicolumn{3}{|c||}{2D $UV$ model}        &  \multicolumn{3}{c|}{Linear matrix size, $M$}  \\ 
\hline
 ${\bf P}$ & Symmetry      &  $K_{\rm eq}$  &  $N_x = 6$  &  $N_x = 8$  &  $N_x = 10$  \\
\hline\hline 
\multirow{4}{*}{Any}
      & Bosons             &      5         &     6480    &    20480    &    50000     \\ \cline{2-6}
      & Fermions, $S = 2$  &      4         &     5184    &    16384    &    40000     \\ \cline{2-6}
      & Fermions, $S = 1$  &      9         &    11664    &    36864    &    90000     \\ \cline{2-6}
      & Fermions, $S = 0$  &     13         &    16848    &    53248    &   130000     \\ 
\hline\hline
\multirow{4}{*}{$(0,0)$}
      & Bosons             &      3         &     3888    &    12288    &    30000     \\ \cline{2-6}
      & Fermions, $S = 2$  &      2         &     2592    &     8192    &    20000     \\ \cline{2-6}
      & Fermions, $S = 1$  &      5         &     6480    &    20480    &    50000     \\ \cline{2-6}
      & Fermions, $S = 0$  &      7         &     9072    &    28672    &    70000     \\ \hline 
\hline 
\end{tabular}
\caption{
The number of coupled integral equations, $K_{\rm eq}$, and the linear size of the final matrix, $M = K_{\rm eq} N^4_{x}$, for different symmetry sectors and different square sizes, $N_x = N_y$, in the 2D square $UV$ model.    
} 
\label{4UV2d:tab:two}
\end{table}

Once the functions are discretized, $k$ integrals on the right-hand sides of the equations need to be approximated by finite sums. The trapezoidal sum rule is most natural here. Again, it seems possible to improve the accuracy of integration by employing the Simpson sum rule or other integration methods. We experimented with the Simpson rule but with no obvious benefit. All results presented in the following were obtained with the trapezoidal rule.   

The next step is to arrange variables $\{ {\bf q}_1 , {\bf q}_2 \}$ into a one-dimensional array. This task is straightforward in 1D but is more challenging in 2D. Indeed, two 2D variables are in fact four variables, $\{ {\bf q}_1 , {\bf q}_2 \} = \{ q_{1x} , q_{1y} , q_{2x} , q_{2y} \}$, and arranging them within a one-dimensional array is tricky. In particular, extreme care must be taken in calculating the index of momentum difference ${\bf P} - {\bf q}_1 - {\bf q}_2 - {\bf k}$.

\begin{table*}[t]
\renewcommand{\tabcolsep}{0.1cm}
\renewcommand{\arraystretch}{1.3}
\begin{tabular}{|c|c|c|c|c|c|c|}
\hline\hline
 State \# &   \cellcolor{pink} Bosons       & \cellcolor{yellow} Fermions, $S = 2$  
     & \cellcolor{green} Fermions, $S = 1$  &   \cellcolor{cyan} Fermions, $S = 0$  
     &     $q_1 \leftrightarrow q_2$ asymm. &  Distinguish.  \\ 
\hline\hline 
 28  &                                    &                                    
     &                                    & \cellcolor{green}  $-41.31527524$  
     &                    $-41.99878977$  & \cellcolor{green}  $-42.15329156$  \\ \hline
 27  &                                    &                                    
     &                                    & \cellcolor{green}  $-41.31731132$  
     & \cellcolor{yellow} $-42.03439805$  & \cellcolor{green}  $-42.15329156$  \\ \hline
 26  &                                    &                                    
     &                                    & \cellcolor{yellow} $-41.49968626$  
     & \cellcolor{yellow} $-42.03439805$  & \cellcolor{green}  $-42.15329156$  \\ \hline
 25  &                                    &                                    
     & \cellcolor{yellow} $-40.58697476$  & \cellcolor{yellow} $-41.49968626$  
     &                    $-42.06670947$  & \cellcolor{pink}   $-42.17973884$  \\ \hline
 24  &                                    &                                    
     & \cellcolor{yellow} $-40.58697476$  & \cellcolor{cyan}   $-41.55120278$  
     &                    $-42.07546896$  & \cellcolor{yellow} $-60.10050280$  \\ \hline
 23  &                                    &                                    
     & \cellcolor{green}  $-40.85302446$  & \cellcolor{cyan}   $-41.55420017$  
     & \cellcolor{cyan}   $-42.08827011$  & \cellcolor{green}  $-60.11042100$  \\ \hline
 22  &                                    &                                    
     & \cellcolor{green}  $-40.85477063$  & \cellcolor{green}  $-41.69263456$  
     & \cellcolor{cyan}   $-42.09099284$  & \cellcolor{green}  $-60.11042100$  \\ \hline
 21  &                                    &                                    
     & \cellcolor{green}  $-41.10899062$  & \cellcolor{green}  $-41.69483277$  
     & \cellcolor{green}  $-42.09458246$  & \cellcolor{green}  $-60.11042100$  \\ \hline
 20  &                                    &                                    
     & \cellcolor{green}  $-41.11142629$  & \cellcolor{cyan}   $-41.84474082$  
     & \cellcolor{green}  $-42.09458246$  & \cellcolor{cyan}   $-60.12180786$  \\ \hline
 19  &                                    &                                    
     & \cellcolor{green}  $-41.31527607$  & \cellcolor{cyan}   $-41.84581751$  
     & \cellcolor{green}  $-42.10708511$  & \cellcolor{cyan}   $-60.12180786$  \\ \hline
 18  &                                    &                                    
     & \cellcolor{green}  $-41.31731120$  & \cellcolor{green}  $-41.86051552$  
     & \cellcolor{green}  $-42.10708511$  & \cellcolor{green}  $-60.13484457$  \\ \hline
 17  &                                    &                                    
     & \cellcolor{yellow} $-41.49968614$  & \cellcolor{green}  $-41.86443225$  
     & \cellcolor{cyan}   $-42.12383830$  & \cellcolor{green}  $-60.13484457$  \\ \hline
 16  &                                    &                                    
     & \cellcolor{yellow} $-41.49968614$  & \cellcolor{green}  $-41.96573057$  
     &                    $-42.14551244$  & \cellcolor{green}  $-60.13484457$  \\ \hline
 15  &                                    &                                    
     & \cellcolor{green}  $-41.69263447$  & \cellcolor{green}  $-41.96916283$  
     &                    $-42.14840587$  &                    $-60.14477421$  \\ \hline
 14  &                                    &                                    
     & \cellcolor{green}  $-41.69483267$  & \cellcolor{yellow} $-42.03439890$  
     & \cellcolor{green}  $-42.15329153$  &                    $-60.14477421$  \\ \hline
 13  &                                    &                                    
     & \cellcolor{green}  $-41.86051545$  & \cellcolor{yellow} $-42.03439890$  
     & \cellcolor{green}  $-42.15329153$  &                    $-60.14477421$  \\ \hline
 12  &                                    &                                    
     & \cellcolor{green}  $-41.86443218$  & \cellcolor{cyan}   $-42.08827100$  
     & \cellcolor{yellow} $-60.10050279$  & \cellcolor{green}  $-60.15836978$  \\ \hline
 11  &                                    &                                    
     & \cellcolor{green}  $-41.96573050$  & \cellcolor{cyan}   $-42.09099277$  
     & \cellcolor{green}  $-60.11042098$  & \cellcolor{green}  $-60.15836978$  \\ \hline
 10  &                                    &                                    
     & \cellcolor{green}  $-41.96916276$  & \cellcolor{green}  $-42.09458239$  
     & \cellcolor{green}  $-60.11042098$  & \cellcolor{green}  $-60.15836978$  \\ \hline
  9  &                                    &                                    
     & \cellcolor{yellow} $-42.03439884$  & \cellcolor{green}  $-42.10708505$  
     & \cellcolor{cyan}   $-60.12180784$  &                    $-60.16831659$  \\ \hline
  8  &                                    &                                    
     & \cellcolor{yellow} $-42.03439884$  & \cellcolor{cyan}   $-42.12383823$  
     & \cellcolor{green}  $-60.13484456$  &                    $-60.16831659$  \\ \hline
  7  &                                    & \cellcolor{yellow} $-40.58697485$  
     & \cellcolor{green}  $-42.09458235$  & \cellcolor{green}  $-42.15329149$  
     & \cellcolor{green}  $-60.13484456$  &                    $-60.16831659$  \\ \hline
  6  & \cellcolor{pink}   $-41.14295746$  & \cellcolor{yellow} $-40.58697485$  
     & \cellcolor{green}  $-42.10708500$  & \cellcolor{yellow} $-60.10050274$  
     &                    $-60.14477420$  & \cellcolor{cyan}   $-60.18135998$  \\ \hline
  5  & \cellcolor{pink}   $-41.14710593$  & \cellcolor{yellow} $-41.49968598$  
     & \cellcolor{green}  $-42.15329147$  & \cellcolor{green}  $-60.11042094$  
     & \cellcolor{green}  $-60.15836977$  & \cellcolor{cyan}   $-60.18135998$  \\ \hline
  4  & \cellcolor{pink}   $-41.89815782$  & \cellcolor{yellow} $-41.49968598$  
     & \cellcolor{yellow} $-60.10050272$  & \cellcolor{cyan}   $-60.12180780$  
     & \cellcolor{green}  $-60.15836977$  &                    $-60.19279643$  \\ \hline
  3  & \cellcolor{pink}   $-41.90180085$  & \cellcolor{yellow} $-42.03439876$  
     & \cellcolor{green}  $-60.11042092$  & \cellcolor{green}  $-60.13484452$  
     &                    $-60.16831658$  &                    $-60.19279643$  \\ \hline
  2  & \cellcolor{pink}   $-42.17973872$  & \cellcolor{yellow} $-42.03439876$  
     & \cellcolor{green}  $-60.13484450$  & \cellcolor{green}  $-60.15836974$  
     & \cellcolor{cyan}   $-60.18135997$  &                    $-60.19279643$  \\ \hline
  1  & \cellcolor{pink}   $-60.20276422$  & \cellcolor{yellow} $-60.10050268$  
     & \cellcolor{green}  $-60.15836973$  & \cellcolor{cyan}   $-60.18135995$  
     &                    $-60.19279642$  & \cellcolor{pink}   $-60.20276422$  \\ \hline 
\hline 
\end{tabular}
\caption{
The lowest 28 four-particle energy levels computed to $10^{-7}$ accuracy in the 1D $UV$ model for $V = 20 \, t$, $U = 100 \, t$, and $P = 0$. The lattice chain length is $N = 16$. All energies are in units of $t$. Only levels below the combined energy of two bound pairs are shown. A singlet pair's energy at these parameters is $-20.26585624$.~\cite{Kornilovitch2004} Thus, only 6 fully symmetric bosonic states exist below the two-pair threshold of $-40.53171247$. Cells are colored according to states' permutation symmetry. Notice how the $S = 0$ system (the fourth column) produces $S = 1$ states (green) and $S = 2$ states (yellow). A match between $S = 2$ energies produced by the $S = 2$, $S = 1$, and $S = 0$ systems validates the method.   
} 
\label{4UV2d:tab:three}
\end{table*}

After the $\{ {\bf q}_1 , {\bf q}_2 \}$ sequence is defined, all auxiliary arrays are concatenated vertically to form a single array $\hat{\cal O}$. Thus, the full system of integral equations is approximated by a matrix equation 
\begin{equation}
\hat{\cal O} = \hat{M} \cdot \hat{\cal O} \: ,
\label{4UV2d:eq:fortythree}
\end{equation}
where $\hat{M}$ is a square matrix whose elements explicitly depend on model parameters $U$ and $V$, total momentum ${\bf P}$, and total energy $E$. Note that despite being a linear eigenvalue in the original Schr\"odinger equation, here $E$ appears in denominators, i.e., nonlinearly. A practical way of finding $E$ for given $\{ U , V , {\bf P} \}$ consists of rewriting Eq.~(\ref{4UV2d:eq:fortythree}) as an eigenvalue problem:
\begin{equation}
\lambda \: \hat{\cal O} = \hat{M}(U, V, {\bf P}, E) \cdot \hat{\cal O} \: . 
\label{4UV2d:eq:fortyfour}
\end{equation}
Then $E$ is varied by a root-finding algorithm until $\lambda = 1$. A side benefit of this method is automatic determination of level degeneracy. For example, if an $E$ corresponds to a triple-degenerate energy level, then three $\lambda = 1$ eigenvalues will appear all at once. By counting how many eigenvalues cross the $\lambda = 1$ threshold simultaneously, the degeneracy of $E$ is easily determined. Note also that $\hat{M}$ is a dense matrix, and no sparse algorithms are applicable here.  

Lastly, one should mention that it is not necessary to compute {\em all} eigenvalues of Eq.~(\ref{4UV2d:eq:fortyfour}). The linear size of $\hat{M}$ can reach tens of thousands, see Tables~\ref{4UV2d:tab:one} and \ref{4UV2d:tab:two}, so computation of all $\lambda$'s quickly becomes prohibitively expensive. Fortunately, only the largest eigenvalues correspond to real physical states. Limiting the search to fewer than $100$ eigenvalues near $\lambda = 1$ is sufficient for most physically interesting questions. Typically, such a subset search reduces the computation time between one and two orders of magnitude.

\section{\label{4UV2d:sec:threeseven}
Validation and finite size scaling   
}

The complexity of a four-body quantum-mechanical problem calls for thorough validation of the integral equation method developed here. In this work, we use two types of self-consistency checks inherent in the method itself to make sure final energy values and other results are error-free. 

The first check is based on verification of auxiliary functions' permutation symmetries. An advantage of the eigenvalue formulation of Eq.~(\ref{4UV2d:eq:fortyfour}) is that all eigenvectors are produced simultaneously with eigenvalues ``for free''. Knowing the arrangement of $\{ {\bf q}_1 , {\bf q}_2 \}$, an eigenvector $\hat{\cal O}$ can be split and reshaped into individual functions of two arguments to verify their permutation symmetries. For example, in the case of 1D bosons, all three auxiliary functions defined in Eqs.~(\ref{4UV2d:eq:twentyfive})-(\ref{4UV2d:eq:twentyseven}) must be symmetrical: $C( q_2 , q_1 ) = + C( q_1 , q_2 )$ and so on. Any symmetry violation indicates an analytical or coding error.  

The second check is inherent in the {\em partial} implementation of Young's tables. Essentially, each energy level is computed more than once, but by different matrix equations. When solving the $S = 0$ fermionic system, $S = 1$ and $2$ states are produced in addition to $S = 0$ states. Any mismatch with the $S = 1$ or $2$ system indicates a coding error. To cross-check bosonic states, we implemented the distinguishable set of equations described in Appendix~\ref{4UV2d:sec:threetwo}, as well as one more system antisymmetric with respect to only {\em one} pair of coordinates, $\{ 12 \}$. An example of such a cross-check is shown in Table~\ref{4UV2d:tab:three}. Because of the higher computational cost of solving the distinguishable equations, such a cross-check was conducted for smaller lattice sizes. Once the code was validated, fermionic and bosonic systems were analyzed for larger $N$.       

We now discuss finite-size scaling. The computer workstation used in this work could handle eigenvalue problems, Eq.~(\ref{4UV2d:eq:fortyfour}), with linear sizes of about $M = 80000$. In 1D, $M$ grows quadratically with the lattice size as $M = K_{\rm eq} N^2$, where $K_{\rm eq}$ is the number of integral equations. Thus, we were able to analyze linear chains as large as $N = 64$; see Table~\ref{4UV2d:tab:one}. That enabled accurate extrapolation of phase boundaries to $N = \infty$, as detailed in Sec.~\ref{4UV2d:sec:four} of the main text. In 2D, $M$ grows as the fourth power of the lattice linear size, $M = K_{\rm eq} N_{x}^{4}$. (In this work, we study only square lattice segments, $N_x = N_y$.) That effectively limits the investigation to $8 \times 8$ or smaller lattices at arbitrary ${\bf P}$, and to $10 \times 10$ or smaller lattices at ${\bf P} = (0,0)$; see Table~\ref{4UV2d:tab:two}.

\end{appendix}

%
%

\clearpage

\begin{widetext}

\section{\label{4UV2d:sec:sm}
Supplemental Material
}

\subsection{\label{4UV2d:sec:sm:a}
Four distinguishable particles in 1D
}

Below is the complete system of 18 equations that result from substitution of Eq.~(\ref{4UV2d:eq:eighteen}) in Eqs.(\ref{4UV2d:eq:twelve})-(\ref{4UV2d:eq:seventeen}). 

\begin{eqnarray}
A_{12}( q_1 , q_2 ) & = &           
      \frac{U}{N} \sum_{k} 
\frac{ A_{12}( q_1 , q_2 ) + A_{13}( P - q_1 - q_2 - k , q_2 ) + A_{14}( P - q_1 - q_2 - k , q_1 ) }
{ E - \varepsilon(k) - \varepsilon( P - q_1 - q_2 - k ) - \varepsilon(q_1) - \varepsilon(q_2) } 
\nonumber \\ 
& & + \frac{U}{N} \sum_{k}  
\frac{ A_{23}( k , q_2 ) + A_{24}( k , q_1 ) +  A_{34}( k , P - q_1 - q_2 - k ) }
{ E - \varepsilon(k) - \varepsilon( P - q_1 - q_2 - k ) - \varepsilon(q_1) - \varepsilon(q_2) } 
\nonumber \\
& & - \frac{2V}{N} \sum_{k}  
\frac{ \cos{k} \, C_{12}( q_1 , q_2 ) + 
       \sin{k} \, S_{12}( q_1 , q_2 ) }
{ E - \varepsilon(k) - \varepsilon( P - q_1 - q_2 - k ) - \varepsilon(q_1) - \varepsilon(q_2) } 
\nonumber \\
& & - \frac{2V}{N} \sum_{k}  
\frac{ \cos{k} \, C_{13}( P - q_1 - q_2 - k , q_2 ) + 
       \sin{k} \, S_{13}( P - q_1 - q_2 - k , q_2 ) }
{ E - \varepsilon(k) - \varepsilon( P - q_1 - q_2 - k ) - \varepsilon(q_1) - \varepsilon(q_2) } 
\nonumber \\
& & - \frac{2V}{N} \sum_{k}  
\frac{ \cos{k} \, C_{14}( P - q_1 - q_2 - k , q_1 ) + 
       \sin{k} \, S_{14}( P - q_1 - q_2 - k , q_1 ) }
{ E - \varepsilon(k) - \varepsilon( P - q_1 - q_2 - k ) - \varepsilon(q_1) - \varepsilon(q_2) } 
\nonumber \\
& & - \frac{2V}{N} \sum_{k}  
\frac{ \cos{( P - q_1 - q_2 - k )} \, C_{23}( k , q_2 ) + 
       \sin{( P - q_1 - q_2 - k )} \, S_{23}( k , q_2 ) }
{ E - \varepsilon(k) - \varepsilon( P - q_1 - q_2 - k ) - \varepsilon(q_1) - \varepsilon(q_2) } 
\nonumber \\
& & - \frac{2V}{N} \sum_{k}  
\frac{ \cos{( P - q_1 - q_2 - k )} \, C_{24}( k , q_1 ) + 
       \sin{( P - q_1 - q_2 - k )} \, S_{24}( k , q_1 ) }
{ E - \varepsilon(k) - \varepsilon( P - q_1 - q_2 - k ) - \varepsilon(q_1) - \varepsilon(q_2) } 
\nonumber \\
& & - \frac{2V}{N} \sum_{k}  
\frac{ \cos{q_1} \, C_{34}( k , P - q_1 - q_2 - k ) + 
       \sin{q_1} \, S_{34}( k , P - q_1 - q_2 - k ) }
{ E - \varepsilon(k) - \varepsilon( P - q_1 - q_2 - k ) - \varepsilon(q_1) - \varepsilon(q_2) }  \: .
\label{4UV2d:eq:sm:one}    
\end{eqnarray}
\begin{eqnarray}
C_{12}( q_1 , q_2 ) & = &           
      \frac{U}{N} \sum_{k} 
\frac{ \cos{k} \, A_{12}( q_1 , q_2 ) + \cos{k} \, A_{13}( P - q_1 - q_2 - k , q_2 ) +  
       \cos{k} \, A_{14}( P - q_1 - q_2 - k , q_1 ) }
{ E - \varepsilon(k) - \varepsilon( P - q_1 - q_2 - k ) - \varepsilon(q_1) - \varepsilon(q_2) } 
\nonumber \\ 
& & + \frac{U}{N} \sum_{k}  
\frac{ \cos{k} \, A_{23}( k , q_2 ) + \cos{k} \, A_{24}( k , q_1 ) + 
       \cos{k} \, A_{34}( k , P - q_1 - q_2 - k ) }
{ E - \varepsilon(k) - \varepsilon( P - q_1 - q_2 - k ) - \varepsilon(q_1) - \varepsilon(q_2) } 
\nonumber \\
& & - \frac{2V}{N} \sum_{k}  
\frac{ \cos{k}\cos{k} \, C_{12}( q_1 , q_2 ) + 
       \cos{k}\sin{k} \, S_{12}( q_1 , q_2 ) }
{ E - \varepsilon(k) - \varepsilon( P - q_1 - q_2 - k ) - \varepsilon(q_1) - \varepsilon(q_2) } 
\nonumber \\
& & - \frac{2V}{N} \sum_{k}  
\frac{ \cos{k}\cos{k} \, C_{13}( P - q_1 - q_2 - k , q_2 ) + 
       \cos{k}\sin{k} \, S_{13}( P - q_1 - q_2 - k , q_2 ) }
{ E - \varepsilon(k) - \varepsilon( P - q_1 - q_2 - k ) - \varepsilon(q_1) - \varepsilon(q_2) } 
\nonumber \\
& & - \frac{2V}{N} \sum_{k}  
\frac{ \cos{k}\cos{k} \, C_{14}( P - q_1 - q_2 - k , q_1 ) + 
       \cos{k}\sin{k} \, S_{14}( P - q_1 - q_2 - k , q_1 ) }
{ E - \varepsilon(k) - \varepsilon( P - q_1 - q_2 - k ) - \varepsilon(q_1) - \varepsilon(q_2) } 
\nonumber \\
& & - \frac{2V}{N} \sum_{k}  
\frac{ \cos{k}\cos{( P - q_1 - q_2 - k )} \, C_{23}( k , q_2 ) + 
       \cos{k}\sin{( P - q_1 - q_2 - k )} \, S_{23}( k , q_2 ) }
{ E - \varepsilon(k) - \varepsilon( P - q_1 - q_2 - k ) - \varepsilon(q_1) - \varepsilon(q_2) } 
\nonumber \\
& & - \frac{2V}{N} \sum_{k}  
\frac{ \cos{k}\cos{( P - q_1 - q_2 - k )} \, C_{24}( k , q_1 ) + 
       \cos{k}\sin{( P - q_1 - q_2 - k )} \, S_{24}( k , q_1 ) }
{ E - \varepsilon(k) - \varepsilon( P - q_1 - q_2 - k ) - \varepsilon(q_1) - \varepsilon(q_2) } 
\nonumber \\
& & - \frac{2V}{N} \sum_{k}  
\frac{ \cos{k}\cos{q_1} \, C_{34}( k , P - q_1 - q_2 - k ) + 
       \cos{k}\sin{q_1} \, S_{34}( k , P - q_1 - q_2 - k ) }
{ E - \varepsilon(k) - \varepsilon( P - q_1 - q_2 - k ) - \varepsilon(q_1) - \varepsilon(q_2) }  \: .
\label{4UV2d:eq:sm:two}    
\end{eqnarray}
\begin{eqnarray}
S_{12}( q_1 , q_2 ) & = &           
      \frac{U}{N} \sum_{k} 
\frac{ \sin{k} \, A_{12}( q_1 , q_2 ) + \sin{k} \, A_{13}( P - q_1 - q_2 - k , q_2 ) +  
       \sin{k} \, A_{14}( P - q_1 - q_2 - k , q_1 ) }
{ E - \varepsilon(k) - \varepsilon( P - q_1 - q_2 - k ) - \varepsilon(q_1) - \varepsilon(q_2) } 
\nonumber \\ 
& & + \frac{U}{N} \sum_{k}  
\frac{ \sin{k} \, A_{23}( k , q_2 ) + \sin{k} \, A_{24}( k , q_1 ) + 
       \sin{k} \, A_{34}( k , P - q_1 - q_2 - k ) }
{ E - \varepsilon(k) - \varepsilon( P - q_1 - q_2 - k ) - \varepsilon(q_1) - \varepsilon(q_2) } 
\nonumber \\
& & - \frac{2V}{N} \sum_{k}  
\frac{ \sin{k}\cos{k} \, C_{12}( q_1 , q_2 ) + 
       \sin{k}\sin{k} \, S_{12}( q_1 , q_2 ) }
{ E - \varepsilon(k) - \varepsilon( P - q_1 - q_2 - k ) - \varepsilon(q_1) - \varepsilon(q_2) } 
\nonumber \\
& & - \frac{2V}{N} \sum_{k}  
\frac{ \sin{k}\cos{k} \, C_{13}( P - q_1 - q_2 - k , q_2 ) + 
       \sin{k}\sin{k} \, S_{13}( P - q_1 - q_2 - k , q_2 ) }
{ E - \varepsilon(k) - \varepsilon( P - q_1 - q_2 - k ) - \varepsilon(q_1) - \varepsilon(q_2) } 
\nonumber \\
& & - \frac{2V}{N} \sum_{k}  
\frac{ \sin{k}\cos{k} \, C_{14}( P - q_1 - q_2 - k , q_1 ) + 
       \sin{k}\sin{k} \, S_{14}( P - q_1 - q_2 - k , q_1 ) }
{ E - \varepsilon(k) - \varepsilon( P - q_1 - q_2 - k ) - \varepsilon(q_1) - \varepsilon(q_2) } 
\nonumber \\
& & - \frac{2V}{N} \sum_{k}  
\frac{ \sin{k}\cos{( P - q_1 - q_2 - k )} \, C_{23}( k , q_2 ) + 
       \sin{k}\sin{( P - q_1 - q_2 - k )} \, S_{23}( k , q_2 ) }
{ E - \varepsilon(k) - \varepsilon( P - q_1 - q_2 - k ) - \varepsilon(q_1) - \varepsilon(q_2) } 
\nonumber \\
& & - \frac{2V}{N} \sum_{k}  
\frac{ \sin{k}\cos{( P - q_1 - q_2 - k )} \, C_{24}( k , q_1 ) + 
       \sin{k}\sin{( P - q_1 - q_2 - k )} \, S_{24}( k , q_1 ) }
{ E - \varepsilon(k) - \varepsilon( P - q_1 - q_2 - k ) - \varepsilon(q_1) - \varepsilon(q_2) } 
\nonumber \\
& & - \frac{2V}{N} \sum_{k}  
\frac{ \sin{k}\cos{q_1} \, C_{34}( k , P - q_1 - q_2 - k ) + 
       \sin{k}\sin{q_1} \, S_{34}( k , P - q_1 - q_2 - k ) }
{ E - \varepsilon(k) - \varepsilon( P - q_1 - q_2 - k ) - \varepsilon(q_1) - \varepsilon(q_2) }  \: .
\label{4UV2d:eq:sm:three}    
\end{eqnarray}
\begin{eqnarray}
A_{13}( q_1 , q_2 ) & = &           
      \frac{U}{N} \sum_{k} 
\frac{ A_{12}( P - q_1 - q_2 - k , q_2 ) + A_{13}( q_1 , q_2 ) + A_{14}( q_1 , P - q_1 - q_2 - k ) }
{ E - \varepsilon(k) - \varepsilon(q_1) - \varepsilon( P - q_1 - q_2 - k ) - \varepsilon(q_2) } 
\nonumber \\ 
& & + \frac{U}{N} \sum_{k}  
\frac{ A_{23}( k , q_2 ) + A_{24}( k , P - q_1 - q_2 - k ) + A_{34}( k , q_1 ) }
{ E - \varepsilon(k) - \varepsilon(q_1) - \varepsilon( P - q_1 - q_2 - k ) - \varepsilon(q_2) } 
\nonumber \\
& & - \frac{2V}{N} \sum_{k}  
\frac{ \cos{k} \, C_{12}( P - q_1 - q_2 - k , q_2 ) + 
       \sin{k} \, S_{12}( P - q_1 - q_2 - k , q_2 ) }
{ E - \varepsilon(k) - \varepsilon(q_1) - \varepsilon( P - q_1 - q_2 - k ) - \varepsilon(q_2) } 
\nonumber \\
& & - \frac{2V}{N} \sum_{k}  
\frac{ \cos{k} \, C_{13}( q_1 , q_2 ) + 
       \sin{k} \, S_{13}( q_1 , q_2 ) }
{ E - \varepsilon(k) - \varepsilon(q_1) - \varepsilon( P - q_1 - q_2 - k ) - \varepsilon(q_2) } 
\nonumber \\
& & - \frac{2V}{N} \sum_{k}  
\frac{ \cos{k} \, C_{14}( q_1 , P - q_1 - q_2 - k ) + 
       \sin{k} \, S_{14}( q_1 , P - q_1 - q_2 - k ) }
{ E - \varepsilon(k) - \varepsilon(q_1) - \varepsilon( P - q_1 - q_2 - k ) - \varepsilon(q_2) } 
\nonumber \\
& & - \frac{2V}{N} \sum_{k}  
\frac{ \cos{q_1} \, C_{23}( k , q_2 ) + 
       \sin{q_1} \, S_{23}( k , q_2 ) }
{ E - \varepsilon(k) - \varepsilon(q_1) - \varepsilon( P - q_1 - q_2 - k ) - \varepsilon(q_2) } 
\nonumber \\
& & - \frac{2V}{N} \sum_{k}  
\frac{ \cos{q_1} \, C_{24}( k , P - q_1 - q_2 - k ) + 
       \sin{q_1} \, S_{24}( k , P - q_1 - q_2 - k ) }
{ E - \varepsilon(k) - \varepsilon(q_1) - \varepsilon( P - q_1 - q_2 - k ) - \varepsilon(q_2) } 
\nonumber \\
& & - \frac{2V}{N} \sum_{k}  
\frac{ \cos{( P - q_1 - q_2 - k )} \, C_{34}( k , q_1 ) + 
       \sin{( P - q_1 - q_2 - k )} \, S_{34}( k , q_1 ) }
{ E - \varepsilon(k) - \varepsilon(q_1) - \varepsilon( P - q_1 - q_2 - k ) - \varepsilon(q_2) }   \: .
\label{4UV2d:eq:sm:four}    
\end{eqnarray}
\begin{eqnarray}
C_{13}( q_1 , q_2 ) & = &           
      \frac{U}{N} \sum_{k} 
\frac{ \cos{k} \, A_{12}( P - q_1 - q_2 - k , q_2 ) + \cos{k} \, A_{13}( q_1 , q_2 ) +  
       \cos{k} \, A_{14}( q_1 , P - q_1 - q_2 - k ) }
{ E - \varepsilon(k) - \varepsilon(q_1) - \varepsilon( P - q_1 - q_2 - k ) - \varepsilon(q_2) } 
\nonumber \\ 
& & + \frac{U}{N} \sum_{k}  
\frac{ \cos{k} \, A_{23}( k , q_2 ) + \cos{k} \, A_{24}( k , P - q_1 - q_2 - k ) + 
       \cos{k} \, A_{34}( k , q_1 ) }
{ E - \varepsilon(k) - \varepsilon(q_1) - \varepsilon( P - q_1 - q_2 - k ) - \varepsilon(q_2) } 
\nonumber \\
& & - \frac{2V}{N} \sum_{k}  
\frac{ \cos{k}\cos{k} \, C_{12}( P - q_1 - q_2 - k , q_2 ) + 
       \cos{k}\sin{k} \, S_{12}( P - q_1 - q_2 - k , q_2 ) }
{ E - \varepsilon(k) - \varepsilon(q_1) - \varepsilon( P - q_1 - q_2 - k ) - \varepsilon(q_2) } 
\nonumber \\
& & - \frac{2V}{N} \sum_{k}  
\frac{ \cos{k}\cos{k} \, C_{13}( q_1 , q_2 ) + 
       \cos{k}\sin{k} \, S_{13}( q_1 , q_2 ) }
{ E - \varepsilon(k) - \varepsilon(q_1) - \varepsilon( P - q_1 - q_2 - k ) - \varepsilon(q_2) } 
\nonumber \\
& & - \frac{2V}{N} \sum_{k}  
\frac{ \cos{k}\cos{k} \, C_{14}( q_1 , P - q_1 - q_2 - k ) + 
       \cos{k}\sin{k} \, S_{14}( q_1 , P - q_1 - q_2 - k ) }
{ E - \varepsilon(k) - \varepsilon(q_1) - \varepsilon( P - q_1 - q_2 - k ) - \varepsilon(q_2) } 
\nonumber \\
& & - \frac{2V}{N} \sum_{k}  
\frac{ \cos{k}\cos{q_1} \, C_{23}( k , q_2 ) + 
       \cos{k}\sin{q_1} \, S_{23}( k , q_2 ) }
{ E - \varepsilon(k) - \varepsilon(q_1) - \varepsilon( P - q_1 - q_2 - k ) - \varepsilon(q_2) } 
\nonumber \\
& & - \frac{2V}{N} \sum_{k}  
\frac{ \cos{k}\cos{q_1} \, C_{24}( k , P - q_1 - q_2 - k ) + 
       \cos{k}\sin{q_1} \, S_{24}( k , P - q_1 - q_2 - k ) }
{ E - \varepsilon(k) - \varepsilon(q_1) - \varepsilon( P - q_1 - q_2 - k ) - \varepsilon(q_2) } 
\nonumber \\
& & - \frac{2V}{N} \sum_{k}  
\frac{ \cos{k}\cos{( P - q_1 - q_2 - k )} \, C_{34}( k , q_1 ) + 
       \cos{k}\sin{( P - q_1 - q_2 - k )} \, S_{34}( k , q_1 ) }
{ E - \varepsilon(k) - \varepsilon(q_1) - \varepsilon( P - q_1 - q_2 - k ) - \varepsilon(q_2) }   \: .
\label{4UV2d:eq:sm:five}    
\end{eqnarray}
\begin{eqnarray}
S_{13}( q_1 , q_2 ) & = &           
      \frac{U}{N} \sum_{k} 
\frac{ \sin{k} \, A_{12}( P - q_1 - q_2 - k , q_2 ) + \sin{k} \, A_{13}( q_1 , q_2 ) +  
       \sin{k} \, A_{14}( q_1 , P - q_1 - q_2 - k ) }
{ E - \varepsilon(k) - \varepsilon(q_1) - \varepsilon( P - q_1 - q_2 - k ) - \varepsilon(q_2) } 
\nonumber \\ 
& & + \frac{U}{N} \sum_{k}  
\frac{ \sin{k} \, A_{23}( k , q_2 ) + \sin{k} \, A_{24}( k , P - q_1 - q_2 - k ) + 
       \sin{k} \, A_{34}( k , q_1 ) }
{ E - \varepsilon(k) - \varepsilon(q_1) - \varepsilon( P - q_1 - q_2 - k ) - \varepsilon(q_2) } 
\nonumber \\
& & - \frac{2V}{N} \sum_{k}  
\frac{ \sin{k}\cos{k} \, C_{12}( P - q_1 - q_2 - k , q_2 ) + 
       \sin{k}\sin{k} \, S_{12}( P - q_1 - q_2 - k , q_2 ) }
{ E - \varepsilon(k) - \varepsilon(q_1) - \varepsilon( P - q_1 - q_2 - k ) - \varepsilon(q_2) } 
\nonumber \\
& & - \frac{2V}{N} \sum_{k}  
\frac{ \sin{k}\cos{k} \, C_{13}( q_1 , q_2 ) + 
       \sin{k}\sin{k} \, S_{13}( q_1 , q_2 ) }
{ E - \varepsilon(k) - \varepsilon(q_1) - \varepsilon( P - q_1 - q_2 - k ) - \varepsilon(q_2) } 
\nonumber \\
& & - \frac{2V}{N} \sum_{k}  
\frac{ \sin{k}\cos{k} \, C_{14}( q_1 , P - q_1 - q_2 - k ) + 
       \sin{k}\sin{k} \, S_{14}( q_1 , P - q_1 - q_2 - k ) }
{ E - \varepsilon(k) - \varepsilon(q_1) - \varepsilon( P - q_1 - q_2 - k ) - \varepsilon(q_2) } 
\nonumber \\
& & - \frac{2V}{N} \sum_{k}  
\frac{ \sin{k}\cos{q_1} \, C_{23}( k , q_2 ) + 
       \sin{k}\sin{q_1} \, S_{23}( k , q_2 ) }
{ E - \varepsilon(k) - \varepsilon(q_1) - \varepsilon( P - q_1 - q_2 - k ) - \varepsilon(q_2) } 
\nonumber \\
& & - \frac{2V}{N} \sum_{k}  
\frac{ \sin{k}\cos{q_1} \, C_{24}( k , P - q_1 - q_2 - k ) + 
       \sin{k}\sin{q_1} \, S_{24}( k , P - q_1 - q_2 - k ) }
{ E - \varepsilon(k) - \varepsilon(q_1) - \varepsilon( P - q_1 - q_2 - k ) - \varepsilon(q_2) } 
\nonumber \\
& & - \frac{2V}{N} \sum_{k}  
\frac{ \sin{k}\cos{( P - q_1 - q_2 - k )} \, C_{34}( k , q_1 ) + 
       \sin{k}\sin{( P - q_1 - q_2 - k )} \, S_{34}( k , q_1 ) }
{ E - \varepsilon(k) - \varepsilon(q_1) - \varepsilon( P - q_1 - q_2 - k ) - \varepsilon(q_2) }   \: .
\label{4UV2d:eq:sm:six}    
\end{eqnarray}
\begin{eqnarray}
A_{14}( q_1 , q_2 ) & = &           
      \frac{U}{N} \sum_{k} 
\frac{ A_{12}( q_2 , P - q_1 - q_2 - k  ) + A_{13}( q_1 , P - q_1 - q_2 - k ) + A_{14}( q_1 , q_2 ) }
{ E - \varepsilon(k) - \varepsilon(q_1) - \varepsilon(q_2) - \varepsilon( P - q_1 - q_2 - k ) } 
\nonumber \\ 
& & + \frac{U}{N} \sum_{k}  
\frac{ A_{23}( k , P - q_1 - q_2 - k ) + A_{24}( k , q_2 ) + A_{34}( k , q_1 ) }
{ E - \varepsilon(k) - \varepsilon(q_1) - \varepsilon(q_2) - \varepsilon( P - q_1 - q_2 - k ) }
\nonumber \\
& & - \frac{2V}{N} \sum_{k}  
\frac{ \cos{k} \, C_{12}( q_2 , P - q_1 - q_2 - k  ) + 
       \sin{k} \, S_{12}( q_2 , P - q_1 - q_2 - k  ) }
{ E - \varepsilon(k) - \varepsilon(q_1) - \varepsilon(q_2) - \varepsilon( P - q_1 - q_2 - k ) }
\nonumber \\
& & - \frac{2V}{N} \sum_{k}  
\frac{ \cos{k} \, C_{13}( q_1 , P - q_1 - q_2 - k ) + 
       \sin{k} \, S_{13}( q_1 , P - q_1 - q_2 - k ) }
{ E - \varepsilon(k) - \varepsilon(q_1) - \varepsilon(q_2) - \varepsilon( P - q_1 - q_2 - k ) }
\nonumber \\
& & - \frac{2V}{N} \sum_{k}  
\frac{ \cos{k} \, C_{14}( q_1 , q_2 ) + 
       \sin{k} \, S_{14}( q_1 , q_2 ) }
{ E - \varepsilon(k) - \varepsilon(q_1) - \varepsilon(q_2) - \varepsilon( P - q_1 - q_2 - k ) }
\nonumber \\
& & - \frac{2V}{N} \sum_{k}  
\frac{ \cos{q_1} \, C_{23}( k , P - q_1 - q_2 - k ) + 
       \sin{q_1} \, S_{23}( k , P - q_1 - q_2 - k ) }
{ E - \varepsilon(k) - \varepsilon(q_1) - \varepsilon(q_2) - \varepsilon( P - q_1 - q_2 - k ) }
\nonumber \\
& & - \frac{2V}{N} \sum_{k}  
\frac{ \cos{q_1} \, C_{24}( k , q_2 ) + 
       \sin{q_1} \, S_{24}( k , q_2 ) }
{ E - \varepsilon(k) - \varepsilon(q_1) - \varepsilon(q_2) - \varepsilon( P - q_1 - q_2 - k ) }
\nonumber \\
& & - \frac{2V}{N} \sum_{k}  
\frac{ \cos{q_2} \, C_{34}( k , q_1 ) + 
       \sin{q_2} \, S_{34}( k , q_1 ) }
{ E - \varepsilon(k) - \varepsilon(q_1) - \varepsilon(q_2) - \varepsilon( P - q_1 - q_2 - k ) } \: .
\label{4UV2d:eq:sm:seven}    
\end{eqnarray}
\begin{eqnarray}
C_{14}( q_1 , q_2 ) & = &           
      \frac{U}{N} \sum_{k} 
\frac{ \cos{k} \, A_{12}( q_2 , P - q_1 - q_2 - k  ) + \cos{k} \, A_{13}( q_1 , P - q_1 - q_2 - k ) +  
       \cos{k} \, A_{14}( q_1 , q_2 ) }
{ E - \varepsilon(k) - \varepsilon(q_1) - \varepsilon(q_2) - \varepsilon( P - q_1 - q_2 - k ) } 
\nonumber \\ 
& & + \frac{U}{N} \sum_{k}  
\frac{ \cos{k} \, A_{23}( k , P - q_1 - q_2 - k ) + \cos{k} \, A_{24}( k , q_2 ) + 
       \cos{k} \, A_{34}( k , q_1 ) }
{ E - \varepsilon(k) - \varepsilon(q_1) - \varepsilon(q_2) - \varepsilon( P - q_1 - q_2 - k ) }
\nonumber \\
& & - \frac{2V}{N} \sum_{k}  
\frac{ \cos{k}\cos{k} \, C_{12}( q_2 , P - q_1 - q_2 - k  ) + 
       \cos{k}\sin{k} \, S_{12}( q_2 , P - q_1 - q_2 - k  ) }
{ E - \varepsilon(k) - \varepsilon(q_1) - \varepsilon(q_2) - \varepsilon( P - q_1 - q_2 - k ) }
\nonumber \\
& & - \frac{2V}{N} \sum_{k}  
\frac{ \cos{k}\cos{k} \, C_{13}( q_1 , P - q_1 - q_2 - k ) + 
       \cos{k}\sin{k} \, S_{13}( q_1 , P - q_1 - q_2 - k ) }
{ E - \varepsilon(k) - \varepsilon(q_1) - \varepsilon(q_2) - \varepsilon( P - q_1 - q_2 - k ) }
\nonumber \\
& & - \frac{2V}{N} \sum_{k}  
\frac{ \cos{k}\cos{k} \, C_{14}( q_1 , q_2 ) + 
       \cos{k}\sin{k} \, S_{14}( q_1 , q_2 ) }
{ E - \varepsilon(k) - \varepsilon(q_1) - \varepsilon(q_2) - \varepsilon( P - q_1 - q_2 - k ) }
\nonumber \\
& & - \frac{2V}{N} \sum_{k}  
\frac{ \cos{k}\cos{q_1} \, C_{23}( k , P - q_1 - q_2 - k ) + 
       \cos{k}\sin{q_1} \, S_{23}( k , P - q_1 - q_2 - k ) }
{ E - \varepsilon(k) - \varepsilon(q_1) - \varepsilon(q_2) - \varepsilon( P - q_1 - q_2 - k ) }
\nonumber \\
& & - \frac{2V}{N} \sum_{k}  
\frac{ \cos{k}\cos{q_1} \, C_{24}( k , q_2 ) + 
       \cos{k}\sin{q_1} \, S_{24}( k , q_2 ) }
{ E - \varepsilon(k) - \varepsilon(q_1) - \varepsilon(q_2) - \varepsilon( P - q_1 - q_2 - k ) }
\nonumber \\
& & - \frac{2V}{N} \sum_{k}  
\frac{ \cos{k}\cos{q_2} \, C_{34}( k , q_1 ) + 
       \cos{k}\sin{q_2} \, S_{34}( k , q_1 ) }
{ E - \varepsilon(k) - \varepsilon(q_1) - \varepsilon(q_2) - \varepsilon( P - q_1 - q_2 - k ) } \: .
\label{4UV2d:eq:sm:eight}    
\end{eqnarray}
\begin{eqnarray}
S_{14}( q_1 , q_2 ) & = &           
      \frac{U}{N} \sum_{k} 
\frac{ \sin{k} \, A_{12}( q_2 , P - q_1 - q_2 - k  ) + \sin{k} \, A_{13}( q_1 , P - q_1 - q_2 - k ) +  
       \sin{k} \, A_{14}( q_1 , q_2 ) }
{ E - \varepsilon(k) - \varepsilon(q_1) - \varepsilon(q_2) - \varepsilon( P - q_1 - q_2 - k ) } 
\nonumber \\ 
& & + \frac{U}{N} \sum_{k}  
\frac{ \sin{k} \, A_{23}( k , P - q_1 - q_2 - k ) + \sin{k} \, A_{24}( k , q_2 ) + 
       \sin{k} \, A_{34}( k , q_1 ) }
{ E - \varepsilon(k) - \varepsilon(q_1) - \varepsilon(q_2) - \varepsilon( P - q_1 - q_2 - k ) }
\nonumber \\
& & - \frac{2V}{N} \sum_{k}  
\frac{ \sin{k}\cos{k} \, C_{12}( q_2 , P - q_1 - q_2 - k  ) + 
       \sin{k}\sin{k} \, S_{12}( q_2 , P - q_1 - q_2 - k  ) }
{ E - \varepsilon(k) - \varepsilon(q_1) - \varepsilon(q_2) - \varepsilon( P - q_1 - q_2 - k ) }
\nonumber \\
& & - \frac{2V}{N} \sum_{k}  
\frac{ \sin{k}\cos{k} \, C_{13}( q_1 , P - q_1 - q_2 - k ) + 
       \sin{k}\sin{k} \, S_{13}( q_1 , P - q_1 - q_2 - k ) }
{ E - \varepsilon(k) - \varepsilon(q_1) - \varepsilon(q_2) - \varepsilon( P - q_1 - q_2 - k ) }
\nonumber \\
& & - \frac{2V}{N} \sum_{k}  
\frac{ \sin{k}\cos{k} \, C_{14}( q_1 , q_2 ) + 
       \sin{k}\sin{k} \, S_{14}( q_1 , q_2 ) }
{ E - \varepsilon(k) - \varepsilon(q_1) - \varepsilon(q_2) - \varepsilon( P - q_1 - q_2 - k ) }
\nonumber \\
& & - \frac{2V}{N} \sum_{k}  
\frac{ \sin{k}\cos{q_1} \, C_{23}( k , P - q_1 - q_2 - k ) + 
       \sin{k}\sin{q_1} \, S_{23}( k , P - q_1 - q_2 - k ) }
{ E - \varepsilon(k) - \varepsilon(q_1) - \varepsilon(q_2) - \varepsilon( P - q_1 - q_2 - k ) }
\nonumber \\
& & - \frac{2V}{N} \sum_{k}  
\frac{ \sin{k}\cos{q_1} \, C_{24}( k , q_2 ) + 
       \sin{k}\sin{q_1} \, S_{24}( k , q_2 ) }
{ E - \varepsilon(k) - \varepsilon(q_1) - \varepsilon(q_2) - \varepsilon( P - q_1 - q_2 - k ) }
\nonumber \\
& & - \frac{2V}{N} \sum_{k}  
\frac{ \sin{k}\cos{q_2} \, C_{34}( k , q_1 ) + 
       \sin{k}\sin{q_2} \, S_{34}( k , q_1 ) }
{ E - \varepsilon(k) - \varepsilon(q_1) - \varepsilon(q_2) - \varepsilon( P - q_1 - q_2 - k ) } \: .
\label{4UV2d:eq:sm:nine}    
\end{eqnarray}
\begin{eqnarray}
A_{23}( q_1 , q_2 ) & = &           
      \frac{U}{N} \sum_{k} 
\frac{ A_{12}( P - q_1 - q_2 - k , q_2 ) + A_{13}( k , q_2 ) + A_{14}( k , P - q_1 - q_2 - k ) }
{ E - \varepsilon(q_1) - \varepsilon(k) - \varepsilon( P - q_1 - q_2 - k ) - \varepsilon(q_2) } 
\nonumber \\ 
& & + \frac{U}{N} \sum_{k}  
\frac{ A_{23}( q_1 , q_2 ) + A_{24}( q_1 , P - q_1 - q_2 - k ) + A_{34}( q_1 , k ) }
{ E - \varepsilon(q_1) - \varepsilon(k) - \varepsilon( P - q_1 - q_2 - k ) - \varepsilon(q_2) } 
\nonumber \\
& & - \frac{2V}{N} \sum_{k}  
\frac{ \cos{q_1} \, C_{12}( P - q_1 - q_2 - k , q_2 ) + 
       \sin{q_1} \, S_{12}( P - q_1 - q_2 - k , q_2 ) }
{ E - \varepsilon(q_1) - \varepsilon(k) - \varepsilon( P - q_1 - q_2 - k ) - \varepsilon(q_2) } 
\nonumber \\
& & - \frac{2V}{N} \sum_{k}  
\frac{ \cos{q_1} \, C_{13}( k , q_2 ) + 
       \sin{q_1} \, S_{13}( k , q_2 ) }
{ E - \varepsilon(q_1) - \varepsilon(k) - \varepsilon( P - q_1 - q_2 - k ) - \varepsilon(q_2) } 
\nonumber \\
& & - \frac{2V}{N} \sum_{k}  
\frac{ \cos{q_1} \, C_{14}( k , P - q_1 - q_2 - k ) + 
       \sin{q_1} \, S_{14}( k , P - q_1 - q_2 - k ) }
{ E - \varepsilon(q_1) - \varepsilon(k) - \varepsilon( P - q_1 - q_2 - k ) - \varepsilon(q_2) } 
\nonumber \\
& & - \frac{2V}{N} \sum_{k}  
\frac{ \cos{k} \, C_{23}( q_1 , q_2 ) + 
       \sin{k} \, S_{23}( q_1 , q_2 ) }
{ E - \varepsilon(q_1) - \varepsilon(k) - \varepsilon( P - q_1 - q_2 - k ) - \varepsilon(q_2) } 
\nonumber \\
& & - \frac{2V}{N} \sum_{k}  
\frac{ \cos{k} \, C_{24}( q_1 , P - q_1 - q_2 - k ) + 
       \sin{k} \, S_{24}( q_1 , P - q_1 - q_2 - k ) }
{ E - \varepsilon(q_1) - \varepsilon(k) - \varepsilon( P - q_1 - q_2 - k ) - \varepsilon(q_2) } 
\nonumber \\
& & - \frac{2V}{N} \sum_{k}  
\frac{ \cos{( P - q_1 - q_2 - k )} \, C_{34}( q_1 , k ) + 
       \sin{( P - q_1 - q_2 - k )} \, S_{34}( q_1 , k ) }
{ E - \varepsilon(q_1) - \varepsilon(k) - \varepsilon( P - q_1 - q_2 - k ) - \varepsilon(q_2) } \: .
\label{4UV2d:eq:sm:ten}    
\end{eqnarray}
\begin{eqnarray}
C_{23}( q_1 , q_2 ) & = &           
      \frac{U}{N} \sum_{k} 
\frac{ \cos{k} \, A_{12}( P - q_1 - q_2 - k , q_2 ) + \cos{k} \, A_{13}( k , q_2 ) +  
       \cos{k} \, A_{14}( k , P - q_1 - q_2 - k ) }
{ E - \varepsilon(q_1) - \varepsilon(k) - \varepsilon( P - q_1 - q_2 - k ) - \varepsilon(q_2) } 
\nonumber \\ 
& & + \frac{U}{N} \sum_{k}  
\frac{ \cos{k} \, A_{23}( q_1 , q_2 ) + \cos{k} \, A_{24}( q_1 , P - q_1 - q_2 - k ) + 
       \cos{k} \, A_{34}( q_1 , k ) }
{ E - \varepsilon(q_1) - \varepsilon(k) - \varepsilon( P - q_1 - q_2 - k ) - \varepsilon(q_2) } 
\nonumber \\
& & - \frac{2V}{N} \sum_{k}  
\frac{ \cos{k}\cos{q_1} \, C_{12}( P - q_1 - q_2 - k , q_2 ) + 
       \cos{k}\sin{q_1} \, S_{12}( P - q_1 - q_2 - k , q_2 ) }
{ E - \varepsilon(q_1) - \varepsilon(k) - \varepsilon( P - q_1 - q_2 - k ) - \varepsilon(q_2) } 
\nonumber \\
& & - \frac{2V}{N} \sum_{k}  
\frac{ \cos{k}\cos{q_1} \, C_{13}( k , q_2 ) + 
       \cos{k}\sin{q_1} \, S_{13}( k , q_2 ) }
{ E - \varepsilon(q_1) - \varepsilon(k) - \varepsilon( P - q_1 - q_2 - k ) - \varepsilon(q_2) } 
\nonumber \\
& & - \frac{2V}{N} \sum_{k}  
\frac{ \cos{k}\cos{q_1} \, C_{14}( k , P - q_1 - q_2 - k ) + 
       \cos{k}\sin{q_1} \, S_{14}( k , P - q_1 - q_2 - k ) }
{ E - \varepsilon(q_1) - \varepsilon(k) - \varepsilon( P - q_1 - q_2 - k ) - \varepsilon(q_2) } 
\nonumber \\
& & - \frac{2V}{N} \sum_{k}  
\frac{ \cos{k}\cos{k} \, C_{23}( q_1 , q_2 ) + 
       \cos{k}\sin{k} \, S_{23}( q_1 , q_2 ) }
{ E - \varepsilon(q_1) - \varepsilon(k) - \varepsilon( P - q_1 - q_2 - k ) - \varepsilon(q_2) } 
\nonumber \\
& & - \frac{2V}{N} \sum_{k}  
\frac{ \cos{k}\cos{k} \, C_{24}( q_1 , P - q_1 - q_2 - k ) + 
       \cos{k}\sin{k} \, S_{24}( q_1 , P - q_1 - q_2 - k ) }
{ E - \varepsilon(q_1) - \varepsilon(k) - \varepsilon( P - q_1 - q_2 - k ) - \varepsilon(q_2) } 
\nonumber \\
& & - \frac{2V}{N} \sum_{k}  
\frac{ \cos{k}\cos{( P - q_1 - q_2 - k )} \, C_{34}( q_1 , k ) + 
       \cos{k}\sin{( P - q_1 - q_2 - k )} \, S_{34}( q_1 , k ) }
{ E - \varepsilon(q_1) - \varepsilon(k) - \varepsilon( P - q_1 - q_2 - k ) - \varepsilon(q_2) } \: .
\label{4UV2d:eq:sm:eleven}    
\end{eqnarray}
\begin{eqnarray}
S_{23}( q_1 , q_2 ) & = &           
      \frac{U}{N} \sum_{k} 
\frac{ \sin{k} \, A_{12}( P - q_1 - q_2 - k , q_2 ) + \sin{k} \, A_{13}( k , q_2 ) +  
       \sin{k} \, A_{14}( k , P - q_1 - q_2 - k ) }
{ E - \varepsilon(q_1) - \varepsilon(k) - \varepsilon( P - q_1 - q_2 - k ) - \varepsilon(q_2) } 
\nonumber \\ 
& & + \frac{U}{N} \sum_{k}  
\frac{ \sin{k} \, A_{23}( q_1 , q_2 ) + \sin{k} \, A_{24}( q_1 , P - q_1 - q_2 - k ) + 
       \sin{k} \, A_{34}( q_1 , k ) }
{ E - \varepsilon(q_1) - \varepsilon(k) - \varepsilon( P - q_1 - q_2 - k ) - \varepsilon(q_2) } 
\nonumber \\
& & - \frac{2V}{N} \sum_{k}  
\frac{ \sin{k}\cos{q_1} \, C_{12}( P - q_1 - q_2 - k , q_2 ) + 
       \sin{k}\sin{q_1} \, S_{12}( P - q_1 - q_2 - k , q_2 ) }
{ E - \varepsilon(q_1) - \varepsilon(k) - \varepsilon( P - q_1 - q_2 - k ) - \varepsilon(q_2) } 
\nonumber \\
& & - \frac{2V}{N} \sum_{k}  
\frac{ \sin{k}\cos{q_1} \, C_{13}( k , q_2 ) + 
       \sin{k}\sin{q_1} \, S_{13}( k , q_2 ) }
{ E - \varepsilon(q_1) - \varepsilon(k) - \varepsilon( P - q_1 - q_2 - k ) - \varepsilon(q_2) } 
\nonumber \\
& & - \frac{2V}{N} \sum_{k}  
\frac{ \sin{k}\cos{q_1} \, C_{14}( k , P - q_1 - q_2 - k ) + 
       \sin{k}\sin{q_1} \, S_{14}( k , P - q_1 - q_2 - k ) }
{ E - \varepsilon(q_1) - \varepsilon(k) - \varepsilon( P - q_1 - q_2 - k ) - \varepsilon(q_2) } 
\nonumber \\
& & - \frac{2V}{N} \sum_{k}  
\frac{ \sin{k}\cos{k} \, C_{23}( q_1 , q_2 ) + 
       \sin{k}\sin{k} \, S_{23}( q_1 , q_2 ) }
{ E - \varepsilon(q_1) - \varepsilon(k) - \varepsilon( P - q_1 - q_2 - k ) - \varepsilon(q_2) } 
\nonumber \\
& & - \frac{2V}{N} \sum_{k}  
\frac{ \sin{k}\cos{k} \, C_{24}( q_1 , P - q_1 - q_2 - k ) + 
       \sin{k}\sin{k} \, S_{24}( q_1 , P - q_1 - q_2 - k ) }
{ E - \varepsilon(q_1) - \varepsilon(k) - \varepsilon( P - q_1 - q_2 - k ) - \varepsilon(q_2) } 
\nonumber \\
& & - \frac{2V}{N} \sum_{k}  
\frac{ \sin{k}\cos{( P - q_1 - q_2 - k )} \, C_{34}( q_1 , k ) + 
       \sin{k}\sin{( P - q_1 - q_2 - k )} \, S_{34}( q_1 , k ) }
{ E - \varepsilon(q_1) - \varepsilon(k) - \varepsilon( P - q_1 - q_2 - k ) - \varepsilon(q_2) } \: .
\label{4UV2d:eq:sm:twelve}    
\end{eqnarray}
\begin{eqnarray}
A_{24}( q_1 , q_2 ) & = &           
      \frac{U}{N} \sum_{k} 
\frac{ A_{12}( q_2 , P - q_1 - q_2 - k ) + A_{13}( k , P - q_1 - q_2 - k ) + A_{14}( k , q_2 ) }
{ E - \varepsilon(q_1) - \varepsilon(k) - \varepsilon(q_2) - \varepsilon( P - q_1 - q_2 - k ) } 
\nonumber \\ 
& & + \frac{U}{N} \sum_{k}  
\frac{ A_{23}( q_1 , P - q_1 - q_2 - k ) + A_{24}( q_1 , q_2 ) + A_{34}( q_1 , k ) }
{ E - \varepsilon(q_1) - \varepsilon(k) - \varepsilon(q_2) - \varepsilon( P - q_1 - q_2 - k ) } 
\nonumber \\
& & - \frac{2V}{N} \sum_{k}  
\frac{ \cos{q_1} \, C_{12}( q_2 , P - q_1 - q_2 - k ) + 
       \sin{q_1} \, S_{12}( q_2 , P - q_1 - q_2 - k ) }
{ E - \varepsilon(q_1) - \varepsilon(k) - \varepsilon(q_2) - \varepsilon( P - q_1 - q_2 - k ) } 
\nonumber \\
& & - \frac{2V}{N} \sum_{k}  
\frac{ \cos{q_1} \, C_{13}( k , P - q_1 - q_2 - k ) + 
       \sin{q_1} \, S_{13}( k , P - q_1 - q_2 - k ) }
{ E - \varepsilon(q_1) - \varepsilon(k) - \varepsilon(q_2) - \varepsilon( P - q_1 - q_2 - k ) } 
\nonumber \\
& & - \frac{2V}{N} \sum_{k}  
\frac{ \cos{q_1} \, C_{14}( k , q_2 ) + 
       \sin{q_1} \, S_{14}( k , q_2 ) }
{ E - \varepsilon(q_1) - \varepsilon(k) - \varepsilon(q_2) - \varepsilon( P - q_1 - q_2 - k ) } 
\nonumber \\
& & - \frac{2V}{N} \sum_{k}  
\frac{ \cos{k} \, C_{23}( q_1 , P - q_1 - q_2 - k ) + 
       \sin{k} \, S_{23}( q_1 , P - q_1 - q_2 - k ) }
{ E - \varepsilon(q_1) - \varepsilon(k) - \varepsilon(q_2) - \varepsilon( P - q_1 - q_2 - k ) } 
\nonumber \\
& & - \frac{2V}{N} \sum_{k}  
\frac{ \cos{k} \, C_{24}( q_1 , q_2 ) + 
       \sin{k} \, S_{24}( q_1 , q_2 ) }
{ E - \varepsilon(q_1) - \varepsilon(k) - \varepsilon(q_2) - \varepsilon( P - q_1 - q_2 - k ) } 
\nonumber \\
& & - \frac{2V}{N} \sum_{k}  
\frac{ \cos{q_2} \, C_{34}( q_1 , k ) + 
       \sin{q_2} \, S_{34}( q_1 , k ) }
{ E - \varepsilon(q_1) - \varepsilon(k) - \varepsilon(q_2) - \varepsilon( P - q_1 - q_2 - k ) }  \: .
\label{4UV2d:eq:sm:thirteen}    
\end{eqnarray}
\begin{eqnarray}
C_{24}( q_1 , q_2 ) & = &           
      \frac{U}{N} \sum_{k} 
\frac{ \cos{k} \, A_{12}( q_2 , P - q_1 - q_2 - k ) + \cos{k} \, A_{13}( k , P - q_1 - q_2 - k ) +  
       \cos{k} \, A_{14}( k , q_2 ) }
{ E - \varepsilon(q_1) - \varepsilon(k) - \varepsilon(q_2) - \varepsilon( P - q_1 - q_2 - k ) } 
\nonumber \\ 
& & + \frac{U}{N} \sum_{k}  
\frac{ \cos{k} \, A_{23}( q_1 , P - q_1 - q_2 - k ) + \cos{k} \, A_{24}( q_1 , q_2 ) + 
       \cos{k} \, A_{34}( q_1 , k ) }
{ E - \varepsilon(q_1) - \varepsilon(k) - \varepsilon(q_2) - \varepsilon( P - q_1 - q_2 - k ) } 
\nonumber \\
& & - \frac{2V}{N} \sum_{k}  
\frac{ \cos{k}\cos{q_1} \, C_{12}( q_2 , P - q_1 - q_2 - k ) + 
       \cos{k}\sin{q_1} \, S_{12}( q_2 , P - q_1 - q_2 - k ) }
{ E - \varepsilon(q_1) - \varepsilon(k) - \varepsilon(q_2) - \varepsilon( P - q_1 - q_2 - k ) } 
\nonumber \\
& & - \frac{2V}{N} \sum_{k}  
\frac{ \cos{k}\cos{q_1} \, C_{13}( k , P - q_1 - q_2 - k ) + 
       \cos{k}\sin{q_1} \, S_{13}( k , P - q_1 - q_2 - k ) }
{ E - \varepsilon(q_1) - \varepsilon(k) - \varepsilon(q_2) - \varepsilon( P - q_1 - q_2 - k ) } 
\nonumber \\
& & - \frac{2V}{N} \sum_{k}  
\frac{ \cos{k}\cos{q_1} \, C_{14}( k , q_2 ) + 
       \cos{k}\sin{q_1} \, S_{14}( k , q_2 ) }
{ E - \varepsilon(q_1) - \varepsilon(k) - \varepsilon(q_2) - \varepsilon( P - q_1 - q_2 - k ) } 
\nonumber \\
& & - \frac{2V}{N} \sum_{k}  
\frac{ \cos{k}\cos{k} \, C_{23}( q_1 , P - q_1 - q_2 - k ) + 
       \cos{k}\sin{k} \, S_{23}( q_1 , P - q_1 - q_2 - k ) }
{ E - \varepsilon(q_1) - \varepsilon(k) - \varepsilon(q_2) - \varepsilon( P - q_1 - q_2 - k ) } 
\nonumber \\
& & - \frac{2V}{N} \sum_{k}  
\frac{ \cos{k}\cos{k} \, C_{24}( q_1 , q_2 ) + 
       \cos{k}\sin{k} \, S_{24}( q_1 , q_2 ) }
{ E - \varepsilon(q_1) - \varepsilon(k) - \varepsilon(q_2) - \varepsilon( P - q_1 - q_2 - k ) } 
\nonumber \\
& & - \frac{2V}{N} \sum_{k}  
\frac{ \cos{k}\cos{q_2} \, C_{34}( q_1 , k ) + 
       \cos{k}\sin{q_2} \, S_{34}( q_1 , k ) }
{ E - \varepsilon(q_1) - \varepsilon(k) - \varepsilon(q_2) - \varepsilon( P - q_1 - q_2 - k ) }  \: .
\label{4UV2d:eq:sm:fourteen}    
\end{eqnarray}
\begin{eqnarray}
S_{24}( q_1 , q_2 ) & = &           
      \frac{U}{N} \sum_{k} 
\frac{ \sin{k} \, A_{12}( q_2 , P - q_1 - q_2 - k ) + \sin{k} \, A_{13}( k , P - q_1 - q_2 - k ) +  
       \sin{k} \, A_{14}( k , q_2 ) }
{ E - \varepsilon(q_1) - \varepsilon(k) - \varepsilon(q_2) - \varepsilon( P - q_1 - q_2 - k ) } 
\nonumber \\ 
& & + \frac{U}{N} \sum_{k}  
\frac{ \sin{k} \, A_{23}( q_1 , P - q_1 - q_2 - k ) + \sin{k} \, A_{24}( q_1 , q_2 ) + 
       \sin{k} \, A_{34}( q_1 , k ) }
{ E - \varepsilon(q_1) - \varepsilon(k) - \varepsilon(q_2) - \varepsilon( P - q_1 - q_2 - k ) } 
\nonumber \\
& & - \frac{2V}{N} \sum_{k}  
\frac{ \sin{k}\cos{q_1} \, C_{12}( q_2 , P - q_1 - q_2 - k ) + 
       \sin{k}\sin{q_1} \, S_{12}( q_2 , P - q_1 - q_2 - k ) }
{ E - \varepsilon(q_1) - \varepsilon(k) - \varepsilon(q_2) - \varepsilon( P - q_1 - q_2 - k ) } 
\nonumber \\
& & - \frac{2V}{N} \sum_{k}  
\frac{ \sin{k}\cos{q_1} \, C_{13}( k , P - q_1 - q_2 - k ) + 
       \sin{k}\sin{q_1} \, S_{13}( k , P - q_1 - q_2 - k ) }
{ E - \varepsilon(q_1) - \varepsilon(k) - \varepsilon(q_2) - \varepsilon( P - q_1 - q_2 - k ) } 
\nonumber \\
& & - \frac{2V}{N} \sum_{k}  
\frac{ \sin{k}\cos{q_1} \, C_{14}( k , q_2 ) + 
       \sin{k}\sin{q_1} \, S_{14}( k , q_2 ) }
{ E - \varepsilon(q_1) - \varepsilon(k) - \varepsilon(q_2) - \varepsilon( P - q_1 - q_2 - k ) } 
\nonumber \\
& & - \frac{2V}{N} \sum_{k}  
\frac{ \sin{k}\cos{k} \, C_{23}( q_1 , P - q_1 - q_2 - k ) + 
       \sin{k}\sin{k} \, S_{23}( q_1 , P - q_1 - q_2 - k ) }
{ E - \varepsilon(q_1) - \varepsilon(k) - \varepsilon(q_2) - \varepsilon( P - q_1 - q_2 - k ) } 
\nonumber \\
& & - \frac{2V}{N} \sum_{k}  
\frac{ \sin{k}\cos{k} \, C_{24}( q_1 , q_2 ) + 
       \sin{k}\sin{k} \, S_{24}( q_1 , q_2 ) }
{ E - \varepsilon(q_1) - \varepsilon(k) - \varepsilon(q_2) - \varepsilon( P - q_1 - q_2 - k ) } 
\nonumber \\
& & - \frac{2V}{N} \sum_{k}  
\frac{ \sin{k}\cos{q_2} \, C_{34}( q_1 , k ) + 
       \sin{k}\sin{q_2} \, S_{34}( q_1 , k ) }
{ E - \varepsilon(q_1) - \varepsilon(k) - \varepsilon(q_2) - \varepsilon( P - q_1 - q_2 - k ) }  \: .
\label{4UV2d:eq:sm:fifteen}    
\end{eqnarray}
\begin{eqnarray}
A_{34}( q_1 , q_2 ) & = &           
      \frac{U}{N} \sum_{k} 
\frac{ A_{12}( k , P - q_1 - q_2 - k ) + A_{13}( q_2 , P - q_1 - q_2 - k ) + A_{14}( q_2 , k ) }
{ E - \varepsilon(q_1) - \varepsilon(q_2) - \varepsilon(k) - \varepsilon( P - q_1 - q_2 - k ) } 
\nonumber \\ 
& & + \frac{U}{N} \sum_{k}  
\frac{ A_{23}( q_1 , P - q_1 - q_2 - k ) + A_{24}( q_1 , k ) + A_{34}( q_1 , q_2 ) }
{ E - \varepsilon(q_1) - \varepsilon(q_2) - \varepsilon(k) - \varepsilon( P - q_1 - q_2 - k ) } 
\nonumber \\
& & - \frac{2V}{N} \sum_{k}  
\frac{ \cos{q_1} \, C_{12}( k , P - q_1 - q_2 - k ) + 
       \sin{q_1} \, S_{12}( k , P - q_1 - q_2 - k ) }
{ E - \varepsilon(q_1) - \varepsilon(q_2) - \varepsilon(k) - \varepsilon( P - q_1 - q_2 - k ) } 
\nonumber \\
& & - \frac{2V}{N} \sum_{k}  
\frac{ \cos{q_1} \, C_{13}( q_2 , P - q_1 - q_2 - k ) + 
       \sin{q_1} \, S_{13}( q_2 , P - q_1 - q_2 - k ) }
{ E - \varepsilon(q_1) - \varepsilon(q_2) - \varepsilon(k) - \varepsilon( P - q_1 - q_2 - k ) } 
\nonumber \\
& & - \frac{2V}{N} \sum_{k}  
\frac{ \cos{q_1} \, C_{14}( q_2 , k ) + 
       \sin{q_1} \, S_{14}( q_2 , k ) }
{ E - \varepsilon(q_1) - \varepsilon(q_2) - \varepsilon(k) - \varepsilon( P - q_1 - q_2 - k ) } 
\nonumber \\
& & - \frac{2V}{N} \sum_{k}  
\frac{ \cos{q_2} \, C_{23}( q_1 , P - q_1 - q_2 - k ) + 
       \sin{q_2} \, S_{23}( q_1 , P - q_1 - q_2 - k ) }
{ E - \varepsilon(q_1) - \varepsilon(q_2) - \varepsilon(k) - \varepsilon( P - q_1 - q_2 - k ) } 
\nonumber \\
& & - \frac{2V}{N} \sum_{k}  
\frac{ \cos{q_2} \, C_{24}( q_1 , k ) + 
       \sin{q_2} \, S_{24}( q_1 , k ) }
{ E - \varepsilon(q_1) - \varepsilon(q_2) - \varepsilon(k) - \varepsilon( P - q_1 - q_2 - k ) } 
\nonumber \\
& & - \frac{2V}{N} \sum_{k}  
\frac{ \cos{k} \, C_{34}( q_1 , q_2 ) + 
       \sin{k} \, S_{34}( q_1 , q_2 ) }
{ E - \varepsilon(q_1) - \varepsilon(q_2) - \varepsilon(k) - \varepsilon( P - q_1 - q_2 - k ) } \: .
\label{4UV2d:eq:sm:sixteen}    
\end{eqnarray}
\begin{eqnarray}
C_{34}( q_1 , q_2 ) & = &           
      \frac{U}{N} \sum_{k} 
\frac{ \cos{k} \, A_{12}( k , P - q_1 - q_2 - k ) + \cos{k} \, A_{13}( q_2 , P - q_1 - q_2 - k ) +  
       \cos{k} \, A_{14}( q_2 , k ) }
{ E - \varepsilon(q_1) - \varepsilon(q_2) - \varepsilon(k) - \varepsilon( P - q_1 - q_2 - k ) } 
\nonumber \\ 
& & + \frac{U}{N} \sum_{k}  
\frac{ \cos{k} \, A_{23}( q_1 , P - q_1 - q_2 - k ) + \cos{k} \, A_{24}( q_1 , k ) + 
       \cos{k} \, A_{34}( q_1 , q_2 ) }
{ E - \varepsilon(q_1) - \varepsilon(q_2) - \varepsilon(k) - \varepsilon( P - q_1 - q_2 - k ) } 
\nonumber \\
& & - \frac{2V}{N} \sum_{k}  
\frac{ \cos{k}\cos{q_1} \, C_{12}( k , P - q_1 - q_2 - k ) + 
       \cos{k}\sin{q_1} \, S_{12}( k , P - q_1 - q_2 - k ) }
{ E - \varepsilon(q_1) - \varepsilon(q_2) - \varepsilon(k) - \varepsilon( P - q_1 - q_2 - k ) } 
\nonumber \\
& & - \frac{2V}{N} \sum_{k}  
\frac{ \cos{k}\cos{q_1} \, C_{13}( q_2 , P - q_1 - q_2 - k ) + 
       \cos{k}\sin{q_1} \, S_{13}( q_2 , P - q_1 - q_2 - k ) }
{ E - \varepsilon(q_1) - \varepsilon(q_2) - \varepsilon(k) - \varepsilon( P - q_1 - q_2 - k ) } 
\nonumber \\
& & - \frac{2V}{N} \sum_{k}  
\frac{ \cos{k}\cos{q_1} \, C_{14}( q_2 , k ) + 
       \cos{k}\sin{q_1} \, S_{14}( q_2 , k ) }
{ E - \varepsilon(q_1) - \varepsilon(q_2) - \varepsilon(k) - \varepsilon( P - q_1 - q_2 - k ) } 
\nonumber \\
& & - \frac{2V}{N} \sum_{k}  
\frac{ \cos{k}\cos{q_2} \, C_{23}( q_1 , P - q_1 - q_2 - k ) + 
       \cos{k}\sin{q_2} \, S_{23}( q_1 , P - q_1 - q_2 - k ) }
{ E - \varepsilon(q_1) - \varepsilon(q_2) - \varepsilon(k) - \varepsilon( P - q_1 - q_2 - k ) } 
\nonumber \\
& & - \frac{2V}{N} \sum_{k}  
\frac{ \cos{k}\cos{q_2} \, C_{24}( q_1 , k ) + 
       \cos{k}\sin{q_2} \, S_{24}( q_1 , k ) }
{ E - \varepsilon(q_1) - \varepsilon(q_2) - \varepsilon(k) - \varepsilon( P - q_1 - q_2 - k ) } 
\nonumber \\
& & - \frac{2V}{N} \sum_{k}  
\frac{ \cos{k}\cos{k} \, C_{34}( q_1 , q_2 ) + 
       \cos{k}\sin{k} \, S_{34}( q_1 , q_2 ) }
{ E - \varepsilon(q_1) - \varepsilon(q_2) - \varepsilon(k) - \varepsilon( P - q_1 - q_2 - k ) } \: .
\label{4UV2d:eq:sm:seventeen}    
\end{eqnarray}
\begin{eqnarray}
S_{34}( q_1 , q_2 ) & = &           
      \frac{U}{N} \sum_{k} 
\frac{ \sin{k} \, A_{12}( k , P - q_1 - q_2 - k ) + \sin{k} \, A_{13}( q_2 , P - q_1 - q_2 - k ) +  
       \sin{k} \, A_{14}( q_2 , k ) }
{ E - \varepsilon(q_1) - \varepsilon(q_2) - \varepsilon(k) - \varepsilon( P - q_1 - q_2 - k ) } 
\nonumber \\ 
& & + \frac{U}{N} \sum_{k}  
\frac{ \sin{k} \, A_{23}( q_1 , P - q_1 - q_2 - k ) + \sin{k} \, A_{24}( q_1 , k ) + 
       \sin{k} \, A_{34}( q_1 , q_2 ) }
{ E - \varepsilon(q_1) - \varepsilon(q_2) - \varepsilon(k) - \varepsilon( P - q_1 - q_2 - k ) } 
\nonumber \\
& & - \frac{2V}{N} \sum_{k}  
\frac{ \sin{k}\cos{q_1} \, C_{12}( k , P - q_1 - q_2 - k ) + 
       \sin{k}\sin{q_1} \, S_{12}( k , P - q_1 - q_2 - k ) }
{ E - \varepsilon(q_1) - \varepsilon(q_2) - \varepsilon(k) - \varepsilon( P - q_1 - q_2 - k ) } 
\nonumber \\
& & - \frac{2V}{N} \sum_{k}  
\frac{ \sin{k}\cos{q_1} \, C_{13}( q_2 , P - q_1 - q_2 - k ) + 
       \sin{k}\sin{q_1} \, S_{13}( q_2 , P - q_1 - q_2 - k ) }
{ E - \varepsilon(q_1) - \varepsilon(q_2) - \varepsilon(k) - \varepsilon( P - q_1 - q_2 - k ) } 
\nonumber \\
& & - \frac{2V}{N} \sum_{k}  
\frac{ \sin{k}\cos{q_1} \, C_{14}( q_2 , k ) + 
       \sin{k}\sin{q_1} \, S_{14}( q_2 , k ) }
{ E - \varepsilon(q_1) - \varepsilon(q_2) - \varepsilon(k) - \varepsilon( P - q_1 - q_2 - k ) } 
\nonumber \\
& & - \frac{2V}{N} \sum_{k}  
\frac{ \sin{k}\cos{q_2} \, C_{23}( q_1 , P - q_1 - q_2 - k ) + 
       \sin{k}\sin{q_2} \, S_{23}( q_1 , P - q_1 - q_2 - k ) }
{ E - \varepsilon(q_1) - \varepsilon(q_2) - \varepsilon(k) - \varepsilon( P - q_1 - q_2 - k ) } 
\nonumber \\
& & - \frac{2V}{N} \sum_{k}  
\frac{ \sin{k}\cos{q_2} \, C_{24}( q_1 , k ) + 
       \sin{k}\sin{q_2} \, S_{24}( q_1 , k ) }
{ E - \varepsilon(q_1) - \varepsilon(q_2) - \varepsilon(k) - \varepsilon( P - q_1 - q_2 - k ) } 
\nonumber \\
& & - \frac{2V}{N} \sum_{k}  
\frac{ \sin{k}\cos{k} \, C_{34}( q_1 , q_2 ) + 
       \sin{k}\sin{k} \, S_{34}( q_1 , q_2 ) }
{ E - \varepsilon(q_1) - \varepsilon(q_2) - \varepsilon(k) - \varepsilon( P - q_1 - q_2 - k ) } \: .
\label{4UV2d:eq:sm:eighteen}    
\end{eqnarray}

\subsection{\label{4UV2d:sec:sm:b}
Four $s = 0$ bosons in 1D
}

Utilizing the definitions, Eqs.~(\ref{4UV2d:eq:twentyfive})-(\ref{4UV2d:eq:twentyseven}), the four-boson wave function is expressed from the Schr\"odinger equation, Eq.~(\ref{4UV2d:eq:twentyone}), as follows
\begin{eqnarray}
\Psi_{s}( q_1 , q_2 , q_3 , q_4 ) & = &           
U \: \frac{ A( q_3 , q_4 ) + A( q_2 , q_4 ) + A( q_2 , q_3 ) + 
            A( q_1 , q_4 ) + A( q_1 , q_3 ) + A( q_1 , q_2 ) }
{ E - \varepsilon(q_1) - \varepsilon(q_2) - \varepsilon(q_3) - \varepsilon(q_4) } 
\nonumber \\ 
& & - \: V \: 
\frac{ ( \cos{q_1} + \cos{q_2} ) \, C( q_3 , q_4 ) + ( \sin{q_1} + \sin{q_2} ) \, S( q_3 , q_4 ) }
{ E - \varepsilon(q_1) - \varepsilon(q_2) - \varepsilon(q_3) - \varepsilon(q_4) }                     
\nonumber \\ 
& & - \: V \: 
\frac{ ( \cos{q_1} + \cos{q_3} ) \, C( q_2 , q_4 ) + ( \sin{q_1} + \sin{q_3} ) \, S( q_2 , q_4 ) }
{ E - \varepsilon(q_1) - \varepsilon(q_2) - \varepsilon(q_3) - \varepsilon(q_4) }    
\nonumber \\ 
& & - \: V \: 
\frac{ ( \cos{q_1} + \cos{q_4} ) \, C( q_2 , q_3 ) + ( \sin{q_1} + \sin{q_4} ) \, S( q_2 , q_3 ) }
{ E - \varepsilon(q_1) - \varepsilon(q_2) - \varepsilon(q_3) - \varepsilon(q_4) }           
\nonumber \\
& & - \: V \: 
\frac{ ( \cos{q_2} + \cos{q_3} ) \, C( q_1 , q_4 ) + ( \sin{q_2} + \sin{q_3} ) \, S( q_1 , q_4 ) }
{ E - \varepsilon(q_1) - \varepsilon(q_2) - \varepsilon(q_3) - \varepsilon(q_4) }                     
\nonumber \\
& & - \: V \: 
\frac{ ( \cos{q_2} + \cos{q_4} ) \, C( q_1 , q_3 ) + ( \sin{q_2} + \sin{q_4} ) \, S( q_1 , q_3 ) }
{ E - \varepsilon(q_1) - \varepsilon(q_2) - \varepsilon(q_3) - \varepsilon(q_4) }           
\nonumber \\
& & - \: V \: 
\frac{ ( \cos{q_3} + \cos{q_4} ) \, C( q_1 , q_2 ) + ( \sin{q_3} + \sin{q_4} ) \, S( q_1 , q_2 ) }
{ E - \varepsilon(q_1) - \varepsilon(q_2) - \varepsilon(q_3) - \varepsilon(q_4) }   \: .                  
\label{4UV2d:eq:sm:nineteen}        
\end{eqnarray}
Substitution of Eq.~(\ref{4UV2d:eq:sm:nineteen}) back into the definitions, Eqs.~(\ref{4UV2d:eq:twentyfive})-(\ref{4UV2d:eq:twentyseven}), leads to three coupled integral equations
\begin{eqnarray}
A( q_1 , q_2 ) & = &           
      \frac{U}{N} \sum_{k} 
\frac{ A( q_2 , P - q_1 - q_2 - k ) + A( q_1 , P - q_1 - q_2 - k ) + A( q_1 , q_2 ) }
{ E - \varepsilon(k) - \varepsilon(q_1) - \varepsilon(q_2) - \varepsilon( P - q_1 - q_2 - k ) } 
\nonumber \\ 
& & + \frac{U}{N} \sum_{k}  
\frac{ A( k , P - q_1 - q_2 - k ) + A( k , q_2 ) + A( k , q_1 ) }
{ E - \varepsilon(k) - \varepsilon(q_1) - \varepsilon(q_2) - \varepsilon( P - q_1 - q_2 - k ) } 
\nonumber \\
& & - \frac{V}{N} \sum_{k}  
\frac{ [ \cos{k} + \cos{q_1} ] \, C( q_2 , P - q_1 - q_2 - k ) + 
       [ \sin{k} + \sin{q_1} ] \, S( q_2 , P - q_1 - q_2 - k ) }
{ E - \varepsilon(k) - \varepsilon(q_1) - \varepsilon(q_2) - \varepsilon( P - q_1 - q_2 - k ) } 
\nonumber \\
& & - \frac{V}{N} \sum_{k}  
\frac{ [ \cos{k} + \cos{q_2} ] \, C( q_1 , P - q_1 - q_2 - k ) + 
       [ \sin{k} + \sin{q_2} ] \, S( q_1 , P - q_1 - q_2 - k ) }
{ E - \varepsilon(k) - \varepsilon(q_1) - \varepsilon(q_2) - \varepsilon( P - q_1 - q_2 - k ) } 
\nonumber \\
& & - \frac{V}{N} \sum_{k}  
\frac{ [ \cos{k} + \cos{( P - q_1 - q_2 - k )} ] \, C( q_1 , q_2 ) + 
       [ \sin{k} + \sin{( P - q_1 - q_2 - k )} ] \, S( q_1 , q_2 ) }
{ E - \varepsilon(k) - \varepsilon(q_1) - \varepsilon(q_2) - \varepsilon( P - q_1 - q_2 - k ) } 
\nonumber \\
& & - \frac{V}{N} \sum_{k}  
\frac{ [ \cos{q_1} + \cos{q_2} ] \, C( k , P - q_1 - q_2 - k ) + 
       [ \sin{q_1} + \sin{q_2} ] \, S( k , P - q_1 - q_2 - k ) }
{ E - \varepsilon(k) - \varepsilon(q_1) - \varepsilon(q_2) - \varepsilon( P - q_1 - q_2 - k ) } 
\nonumber \\
& & - \frac{V}{N} \sum_{k}  
\frac{ [ \cos{q_1} + \cos{( P - q_1 - q_2 - k )} ] \, C( k , q_2 ) + 
       [ \sin{q_1} + \sin{( P - q_1 - q_2 - k )} ] \, S( k , q_2 ) }
{ E - \varepsilon(k) - \varepsilon(q_1) - \varepsilon(q_2) - \varepsilon( P - q_1 - q_2 - k ) } 
\nonumber \\
& & - \frac{V}{N} \sum_{k}  
\frac{ [ \cos{q_2} + \cos{( P - q_1 - q_2 - k )} ] \, C( k , q_1 ) + 
       [ \sin{q_2} + \sin{( P - q_1 - q_2 - k )} ] \, S( k , q_1 ) }
{ E - \varepsilon(k) - \varepsilon(q_1) - \varepsilon(q_2) - \varepsilon( P - q_1 - q_2 - k ) } \: .
\label{4UV2d:eq:sm:twenty}    
\end{eqnarray}
\begin{eqnarray}
C( q_1 , q_2 ) & = &           
      \frac{U}{N} \sum_{k} 
\frac{ \cos{k} \, A( q_2 , P - q_1 - q_2 - k ) + \cos{k} \, A( q_1 , P - q_1 - q_2 - k ) +  
       \cos{k} \, A( q_1 , q_2 ) }
{ E - \varepsilon(k) - \varepsilon(q_1) - \varepsilon(q_2) - \varepsilon( P - q_1 - q_2 - k ) } 
\nonumber \\ 
& & + \frac{U}{N} \sum_{k}  
\frac{ \cos{k} \, A( k , P - q_1 - q_2 - k ) + \cos{k} \, A( k , q_2 ) + 
       \cos{k} \, A( k , q_1 ) }
{ E - \varepsilon(k) - \varepsilon(q_1) - \varepsilon(q_2) - \varepsilon( P - q_1 - q_2 - k ) } 
\nonumber \\
& & - \frac{V}{N} \sum_{k}  
\frac{ \cos{k} \, [ \cos{k} + \cos{q_1} ] \, C( q_2 , P - q_1 - q_2 - k ) + 
       \cos{k} \, [ \sin{k} + \sin{q_1} ] \, S( q_2 , P - q_1 - q_2 - k ) }
{ E - \varepsilon(k) - \varepsilon(q_1) - \varepsilon(q_2) - \varepsilon( P - q_1 - q_2 - k ) } 
\nonumber \\
& & - \frac{V}{N} \sum_{k}  
\frac{ \cos{k} \, [ \cos{k} + \cos{q_2} ] \, C( q_1 , P - q_1 - q_2 - k ) + 
       \cos{k} \, [ \sin{k} + \sin{q_2} ] \, S( q_1 , P - q_1 - q_2 - k ) }
{ E - \varepsilon(k) - \varepsilon(q_1) - \varepsilon(q_2) - \varepsilon( P - q_1 - q_2 - k ) } 
\nonumber \\
& & - \frac{V}{N} \sum_{k}  
\frac{ \cos{k} \, [ \cos{k} + \cos{( P - q_1 - q_2 - k )} ] \, C( q_1 , q_2 ) + 
       \cos{k} \, [ \sin{k} + \sin{( P - q_1 - q_2 - k )} ] \, S( q_1 , q_2 ) }
{ E - \varepsilon(k) - \varepsilon(q_1) - \varepsilon(q_2) - \varepsilon( P - q_1 - q_2 - k ) } 
\nonumber \\
& & - \frac{V}{N} \sum_{k}  
\frac{ \cos{k} \, [ \cos{q_1} + \cos{q_2} ] \, C( k , P - q_1 - q_2 - k ) + 
       \cos{k} \, [ \sin{q_1} + \sin{q_2} ] \, S( k , P - q_1 - q_2 - k ) }
{ E - \varepsilon(k) - \varepsilon(q_1) - \varepsilon(q_2) - \varepsilon( P - q_1 - q_2 - k ) } 
\nonumber \\
& & - \frac{V}{N} \sum_{k}  
\frac{ \cos{k} \, [ \cos{q_1} + \cos{( P - q_1 - q_2 - k )} ] \, C( k , q_2 ) + 
       \cos{k} \, [ \sin{q_1} + \sin{( P - q_1 - q_2 - k )} ] \, S( k , q_2 ) }
{ E - \varepsilon(k) - \varepsilon(q_1) - \varepsilon(q_2) - \varepsilon( P - q_1 - q_2 - k ) } 
\nonumber \\
& & - \frac{V}{N} \sum_{k}  
\frac{ \cos{k} \, [ \cos{q_2} + \cos{( P - q_1 - q_2 - k )} ] \, C( k , q_1 ) + 
       \cos{k} \, [ \sin{q_2} + \sin{( P - q_1 - q_2 - k )} ] \, S( k , q_1 ) }
{ E - \varepsilon(k) - \varepsilon(q_1) - \varepsilon(q_2) - \varepsilon( P - q_1 - q_2 - k ) } \: .
\hspace{0.5cm} 
\label{4UV2d:eq:sm:twentyone}    
\end{eqnarray}
\begin{eqnarray}
S( q_1 , q_2 ) & = &           
      \frac{U}{N} \sum_{k} 
\frac{ \sin{k} \, A( q_2 , P - q_1 - q_2 - k ) + \sin{k} \, A( q_1 , P - q_1 - q_2 - k ) +  
       \sin{k} \, A( q_1 , q_2 ) }
{ E - \varepsilon(k) - \varepsilon(q_1) - \varepsilon(q_2) - \varepsilon( P - q_1 - q_2 - k ) } 
\nonumber \\ 
& & + \frac{U}{N} \sum_{k}  
\frac{ \sin{k} \, A( k , P - q_1 - q_2 - k ) + \sin{k} \, A( k , q_2 ) + 
       \sin{k} \, A( k , q_1 ) }
{ E - \varepsilon(k) - \varepsilon(q_1) - \varepsilon(q_2) - \varepsilon( P - q_1 - q_2 - k ) } 
\nonumber \\
& & - \frac{V}{N} \sum_{k}  
\frac{ \sin{k} \, [ \cos{k} + \cos{q_1} ] \, C( q_2 , P - q_1 - q_2 - k ) + 
       \sin{k} \, [ \sin{k} + \sin{q_1} ] \, S( q_2 , P - q_1 - q_2 - k ) }
{ E - \varepsilon(k) - \varepsilon(q_1) - \varepsilon(q_2) - \varepsilon( P - q_1 - q_2 - k ) } 
\nonumber \\
& & - \frac{V}{N} \sum_{k}  
\frac{ \sin{k} \, [ \cos{k} + \cos{q_2} ] \, C( q_1 , P - q_1 - q_2 - k ) + 
       \sin{k} \, [ \sin{k} + \sin{q_2} ] \, S( q_1 , P - q_1 - q_2 - k ) }
{ E - \varepsilon(k) - \varepsilon(q_1) - \varepsilon(q_2) - \varepsilon( P - q_1 - q_2 - k ) } 
\nonumber \\
& & - \frac{V}{N} \sum_{k}  
\frac{ \sin{k} \, [ \cos{k} + \cos{( P - q_1 - q_2 - k )} ] \, C( q_1 , q_2 ) + 
       \sin{k} \, [ \sin{k} + \sin{( P - q_1 - q_2 - k )} ] \, S( q_1 , q_2 ) }
{ E - \varepsilon(k) - \varepsilon(q_1) - \varepsilon(q_2) - \varepsilon( P - q_1 - q_2 - k ) } 
\nonumber \\
& & - \frac{V}{N} \sum_{k}  
\frac{ \cos{k} \, [ \cos{q_1} + \cos{q_2} ] \, C( k , P - q_1 - q_2 - k ) + 
       \cos{k} \, [ \sin{q_1} + \sin{q_2} ] \, S( k , P - q_1 - q_2 - k ) }
{ E - \varepsilon(k) - \varepsilon(q_1) - \varepsilon(q_2) - \varepsilon( P - q_1 - q_2 - k ) } 
\nonumber \\
& & - \frac{V}{N} \sum_{k}  
\frac{ \sin{k} \, [ \cos{q_1} + \cos{( P - q_1 - q_2 - k )} ] \, C( k , q_2 ) + 
       \sin{k} \, [ \sin{q_1} + \sin{( P - q_1 - q_2 - k )} ] \, S( k , q_2 ) }
{ E - \varepsilon(k) - \varepsilon(q_1) - \varepsilon(q_2) - \varepsilon( P - q_1 - q_2 - k ) } 
\nonumber \\
& & - \frac{V}{N} \sum_{k}  
\frac{ \sin{k} \, [ \cos{q_2} + \cos{( P - q_1 - q_2 - k )} ] \, C( k , q_1 ) + 
       \sin{k} \, [ \sin{q_2} + \sin{( P - q_1 - q_2 - k )} ] \, S( k , q_1 ) }
{ E - \varepsilon(k) - \varepsilon(q_1) - \varepsilon(q_2) - \varepsilon( P - q_1 - q_2 - k ) } \: .
\hspace{0.5cm} 
\label{4UV2d:eq:sm:twentytwo}    
\end{eqnarray}

\subsection{\label{4UV2d:sec:sm:c}
Four $s = 0$ bosons in 2D
}

Based on the form of Eq.~(\ref{4UV2d:eq:twentyone}) after summation over ${\bf b} = \pm {\bf x}, \pm {\bf y}$, we define 5 auxiliary functions:
\begin{eqnarray}
A    ( {\bf q}_1 , {\bf q}_2 ) & = & \frac{1}{N} \sum_{\bf k} 
\hspace{1.1cm} \Psi_{s}( {\bf k} , {\bf q}_1 , {\bf q}_2 , {\bf P} - {\bf q}_1 - {\bf q}_2 - {\bf k} ) \: ,  
\label{4UV2d:eq:sm:twentythree}     \\
C_{x}( {\bf q}_1 , {\bf q}_2 ) & = & \frac{1}{N} \sum_{\bf k} 
\cos{(q_x)}    \Psi_{s}( {\bf k} , {\bf q}_1 , {\bf q}_2 , {\bf P} - {\bf q}_1 - {\bf q}_2 - {\bf k} ) \: ,    
\label{4UV2d:eq:sm:twentyfour}      \\
S_{x}( {\bf q}_1 , {\bf q}_2 ) & = & \frac{1}{N} \sum_{\bf k} 
\sin{(q_x)}    \Psi_{s}( {\bf k} , {\bf q}_1 , {\bf q}_2 , {\bf P} - {\bf q}_1 - {\bf q}_2 - {\bf k} ) \: ,    
\label{4UV2d:eq:sm:twentyfive}      \\
C_{y}( {\bf q}_1 , {\bf q}_2 ) & = & \frac{1}{N} \sum_{\bf k} 
\cos{(q_y)}    \Psi_{s}( {\bf k} , {\bf q}_1 , {\bf q}_2 , {\bf P} - {\bf q}_1 - {\bf q}_2 - {\bf k} ) \: ,    
\label{4UV2d:eq:sm:twentysix}      \\
S_{y}( {\bf q}_1 , {\bf q}_2 ) & = & \frac{1}{N} \sum_{\bf k} 
\sin{(q_y)}    \Psi_{s}( {\bf k} , {\bf q}_1 , {\bf q}_2 , {\bf P} - {\bf q}_1 - {\bf q}_2 - {\bf k} ) \: .    
\label{4UV2d:eq:sm:twentyseven}      
\end{eqnarray}
The four-boson wave function is expressed from the Schr\"odinger equation, Eq.~(\ref{4UV2d:eq:twentyone}): 
\begin{eqnarray}
\Psi_{s}( {\bf q}_1 , {\bf q}_2 , {\bf q}_3 , {\bf q}_4 ) & = & U \: 
\frac{ A( {\bf q}_3 , {\bf q}_4 ) + A( {\bf q}_2 , {\bf q}_4 ) + A( {\bf q}_2 , {\bf q}_3 ) + 
       A( {\bf q}_1 , {\bf q}_4 ) + A( {\bf q}_1 , {\bf q}_3 ) + A( {\bf q}_1 , {\bf q}_2 ) }
{ E - \varepsilon({\bf q}_1) - \varepsilon({\bf q}_2) - \varepsilon({\bf q}_3) - \varepsilon({\bf q}_4) }   
\nonumber \\
& & - V \: 
\frac{ ( \cos{q_{1x}} + \cos{q_{2x}} ) \, C_{x}( {\bf q}_3 , {\bf q}_4 ) + 
       ( \sin{q_{1x}} + \sin{q_{2x}} ) \, S_{x}( {\bf q}_3 , {\bf q}_4 ) }
{ E - \varepsilon({\bf q}_1) - \varepsilon({\bf q}_2) - \varepsilon({\bf q}_3) - \varepsilon({\bf q}_4) }
\nonumber \\
& & - V \: 
\frac{ ( \cos{q_{1y}} + \cos{q_{2y}} ) \, C_{y}( {\bf q}_3 , {\bf q}_4 ) + 
       ( \sin{q_{1y}} + \sin{q_{2y}} ) \, S_{y}( {\bf q}_3 , {\bf q}_4 ) }
{ E - \varepsilon({\bf q}_1) - \varepsilon({\bf q}_2) - \varepsilon({\bf q}_3) - \varepsilon({\bf q}_4) } 
\nonumber \\
& & - V \: 
\frac{ ( \cos{q_{1x}} + \cos{q_{3x}} ) \, C_{x}( {\bf q}_2 , {\bf q}_4 ) + 
       ( \sin{q_{1x}} + \sin{q_{3x}} ) \, S_{x}( {\bf q}_2 , {\bf q}_4 ) }
{ E - \varepsilon({\bf q}_1) - \varepsilon({\bf q}_2) - \varepsilon({\bf q}_3) - \varepsilon({\bf q}_4) }
\nonumber \\
& & - V \: 
\frac{ ( \cos{q_{1y}} + \cos{q_{3y}} ) \, C_{y}( {\bf q}_2 , {\bf q}_4 ) + 
       ( \sin{q_{1y}} + \sin{q_{3y}} ) \, S_{y}( {\bf q}_2 , {\bf q}_4 ) }
{ E - \varepsilon({\bf q}_1) - \varepsilon({\bf q}_2) - \varepsilon({\bf q}_3) - \varepsilon({\bf q}_4) }
\nonumber \\
& & - V \: 
\frac{ ( \cos{q_{1x}} + \cos{q_{4x}} ) \, C_{x}( {\bf q}_2 , {\bf q}_3 ) + 
       ( \sin{q_{1x}} + \sin{q_{4x}} ) \, S_{x}( {\bf q}_2 , {\bf q}_3 ) }
{ E - \varepsilon({\bf q}_1) - \varepsilon({\bf q}_2) - \varepsilon({\bf q}_3) - \varepsilon({\bf q}_4) }
\nonumber \\
& & - V \: 
\frac{ ( \cos{q_{1y}} + \cos{q_{4y}} ) \, C_{y}( {\bf q}_2 , {\bf q}_3 ) + 
       ( \sin{q_{1y}} + \sin{q_{4y}} ) \, S_{y}( {\bf q}_2 , {\bf q}_3 ) }
{ E - \varepsilon({\bf q}_1) - \varepsilon({\bf q}_2) - \varepsilon({\bf q}_3) - \varepsilon({\bf q}_4) }
\nonumber \\
& & - V \: 
\frac{ ( \cos{q_{2x}} + \cos{q_{3x}} ) \, C_{x}( {\bf q}_1 , {\bf q}_4 ) + 
       ( \sin{q_{2x}} + \sin{q_{3x}} ) \, S_{x}( {\bf q}_1 , {\bf q}_4 ) }
{ E - \varepsilon({\bf q}_1) - \varepsilon({\bf q}_2) - \varepsilon({\bf q}_3) - \varepsilon({\bf q}_4) }
\nonumber \\
& & - V \: 
\frac{ ( \cos{q_{2y}} + \cos{q_{3y}} ) \, C_{y}( {\bf q}_1 , {\bf q}_4 ) + 
       ( \sin{q_{2y}} + \sin{q_{3y}} ) \, S_{y}( {\bf q}_1 , {\bf q}_4 ) }
{ E - \varepsilon({\bf q}_1) - \varepsilon({\bf q}_2) - \varepsilon({\bf q}_3) - \varepsilon({\bf q}_4) }
\nonumber \\
& & - V \: 
\frac{ ( \cos{q_{2x}} + \cos{q_{4x}} ) \, C_{x}( {\bf q}_1 , {\bf q}_3 ) + 
       ( \sin{q_{2x}} + \sin{q_{4x}} ) \, S_{x}( {\bf q}_1 , {\bf q}_3 ) }
{ E - \varepsilon({\bf q}_1) - \varepsilon({\bf q}_2) - \varepsilon({\bf q}_3) - \varepsilon({\bf q}_4) }
\nonumber \\
& & - V \: 
\frac{ ( \cos{q_{2y}} + \cos{q_{4y}} ) \, C_{y}( {\bf q}_1 , {\bf q}_3 ) + 
       ( \sin{q_{2y}} + \sin{q_{4y}} ) \, S_{y}( {\bf q}_1 , {\bf q}_3 ) }
{ E - \varepsilon({\bf q}_1) - \varepsilon({\bf q}_2) - \varepsilon({\bf q}_3) - \varepsilon({\bf q}_4) }
\nonumber \\
& & - V \: 
\frac{ ( \cos{q_{3x}} + \cos{q_{4x}} ) \, C_{x}( {\bf q}_1 , {\bf q}_2 ) + 
       ( \sin{q_{3x}} + \sin{q_{4x}} ) \, S_{x}( {\bf q}_1 , {\bf q}_2 ) }
{ E - \varepsilon({\bf q}_1) - \varepsilon({\bf q}_2) - \varepsilon({\bf q}_3) - \varepsilon({\bf q}_4) }
\nonumber \\
& & - V \: 
\frac{ ( \cos{q_{3y}} + \cos{q_{4y}} ) \, C_{y}( {\bf q}_1 , {\bf q}_2 ) + 
       ( \sin{q_{3y}} + \sin{q_{4y}} ) \, S_{y}( {\bf q}_1 , {\bf q}_2 ) }
{ E - \varepsilon({\bf q}_1) - \varepsilon({\bf q}_2) - \varepsilon({\bf q}_3) - \varepsilon({\bf q}_4) } \: .
\label{4UV2d:eq:sm:twentyeight}    
\end{eqnarray}
Substitution of Eq.~(\ref{4UV2d:eq:sm:twentyeight}) in Eq.~(\ref{4UV2d:eq:sm:twentythree}) yields the first of 5 integral equations
\begin{eqnarray}
A( {\bf q}_1 , {\bf q}_2 ) & = & \frac{U}{N} \sum_{\bf k} 
\frac{ f({\bf k}) \, A( {\bf q}_2 , {\bf P} - {\bf q}_1 - {\bf q}_2 - {\bf k} ) + 
       f({\bf k}) \, A( {\bf q}_1 , {\bf P} - {\bf q}_1 - {\bf q}_2 - {\bf k} ) + 
       f({\bf k}) \, A( {\bf q}_1 , {\bf q}_2 ) }
{ E - \varepsilon({\bf k}) - \varepsilon({\bf q}_1) - \varepsilon({\bf q}_2) - 
      \varepsilon({\bf P} - {\bf q}_1 - {\bf q}_2 - {\bf k} ) }   
\nonumber \\
                           &   & + \frac{U}{N} \sum_{\bf k} 
\frac{ f({\bf k}) \, A( {\bf k}   , {\bf P} - {\bf q}_1 - {\bf q}_2 - {\bf k} ) + 
       f({\bf k}) \, A( {\bf k}   , {\bf q}_2 ) + 
       f({\bf k}) \, A( {\bf k}   , {\bf q}_1 ) }
{ E - \varepsilon({\bf k}) - \varepsilon({\bf q}_1) - \varepsilon({\bf q}_2) - 
      \varepsilon({\bf P} - {\bf q}_1 - {\bf q}_2 - {\bf k} ) }                              
\nonumber \\
                           &   & - \frac{V}{N} \sum_{\bf k} 
\frac{ f({\bf k}) \, [ \cos{k_x} + \cos{q_{1x}} ] C_{x}( {\bf q}_2 , {\bf P} - {\bf q}_1 - {\bf q}_2 - {\bf k} ) + 
       f({\bf k}) \, [ \sin{k_x} + \sin{q_{1x}} ] S_{x}( {\bf q}_2 , {\bf P} - {\bf q}_1 - {\bf q}_2 - {\bf k} )}
{ E - \varepsilon({\bf k}) - \varepsilon({\bf q}_1) - \varepsilon({\bf q}_2) - 
      \varepsilon({\bf P} - {\bf q}_1 - {\bf q}_2 - {\bf k} ) }               
\nonumber \\
                           &   & - \frac{V}{N} \sum_{\bf k} 
\frac{ f({\bf k}) \, [ \cos{k_y} + \cos{q_{1y}} ] C_{y}( {\bf q}_2 , {\bf P} - {\bf q}_1 - {\bf q}_2 - {\bf k} ) + 
       f({\bf k}) \, [ \sin{k_y} + \sin{q_{1y}} ] S_{y}( {\bf q}_2 , {\bf P} - {\bf q}_1 - {\bf q}_2 - {\bf k} )}
{ E - \varepsilon({\bf k}) - \varepsilon({\bf q}_1) - \varepsilon({\bf q}_2) - 
      \varepsilon({\bf P} - {\bf q}_1 - {\bf q}_2 - {\bf k} ) }     
\nonumber \\
                           &   & - \frac{V}{N} \sum_{\bf k} 
\frac{ f({\bf k}) \, [ \cos{k_x} + \cos{q_{2x}} ] C_{x}( {\bf q}_1 , {\bf P} - {\bf q}_1 - {\bf q}_2 - {\bf k} ) + 
       f({\bf k}) \, [ \sin{k_x} + \sin{q_{2x}} ] S_{x}( {\bf q}_1 , {\bf P} - {\bf q}_1 - {\bf q}_2 - {\bf k} )}
{ E - \varepsilon({\bf k}) - \varepsilon({\bf q}_1) - \varepsilon({\bf q}_2) - 
      \varepsilon({\bf P} - {\bf q}_1 - {\bf q}_2 - {\bf k} ) }   
\nonumber \\
                           &   & - \frac{V}{N} \sum_{\bf k} 
\frac{ f({\bf k}) \, [ \cos{k_y} + \cos{q_{2y}} ] C_{y}( {\bf q}_1 , {\bf P} - {\bf q}_1 - {\bf q}_2 - {\bf k} ) + 
       f({\bf k}) \, [ \sin{k_y} + \sin{q_{2y}} ] S_{y}( {\bf q}_1 , {\bf P} - {\bf q}_1 - {\bf q}_2 - {\bf k} )}
{ E - \varepsilon({\bf k}) - \varepsilon({\bf q}_1) - \varepsilon({\bf q}_2) - 
      \varepsilon({\bf P} - {\bf q}_1 - {\bf q}_2 - {\bf k} ) }                                                                
\nonumber \\
                           &   & - \frac{V}{N} \sum_{\bf k} 
\frac{ f({\bf k}) \, [ \cos{k_x} + \cos{( P_x - q_{1x} - q_{2x} - k_x )} ] C_{x}( {\bf q}_1 , {\bf q}_2 ) }
{ E - \varepsilon({\bf k}) - \varepsilon({\bf q}_1) - \varepsilon({\bf q}_2) - 
      \varepsilon({\bf P} - {\bf q}_1 - {\bf q}_2 - {\bf k} ) }         
\nonumber \\
                           &   & - \frac{V}{N} \sum_{\bf k} 
\frac{ f({\bf k}) \, [ \sin{k_x} + \sin{( P_x - q_{1x} - q_{2x} - k_x )} ] S_{x}( {\bf q}_1 , {\bf q}_2 )}
{ E - \varepsilon({\bf k}) - \varepsilon({\bf q}_1) - \varepsilon({\bf q}_2) - 
      \varepsilon({\bf P} - {\bf q}_1 - {\bf q}_2 - {\bf k} ) }               
\nonumber \\
                           &   & - \frac{V}{N} \sum_{\bf k} 
\frac{ f({\bf k}) \, [ \cos{k_y} + \cos{( P_y - q_{1y} - q_{2y} - k_y )} ] C_{y}( {\bf q}_1 , {\bf q}_2 ) }
{ E - \varepsilon({\bf k}) - \varepsilon({\bf q}_1) - \varepsilon({\bf q}_2) - 
      \varepsilon({\bf P} - {\bf q}_1 - {\bf q}_2 - {\bf k} ) }         
\nonumber \\
                           &   & - \frac{V}{N} \sum_{\bf k} 
\frac{ f({\bf k}) \, [ \sin{k_y} + \sin{( P_y - q_{1y} - q_{2y} - k_y )} ] S_{y}( {\bf q}_1 , {\bf q}_2 )}
{ E - \varepsilon({\bf k}) - \varepsilon({\bf q}_1) - \varepsilon({\bf q}_2) - 
      \varepsilon({\bf P} - {\bf q}_1 - {\bf q}_2 - {\bf k} ) }                     
\nonumber \\
                           &   & - \frac{V}{N} \sum_{\bf k} 
\frac{ f({\bf k}) \, [ \cos{q_{1x}} + \cos{q_{2x}} ] C_{x}( {\bf k} , {\bf P} - {\bf q}_1 - {\bf q}_2 - {\bf k} ) + 
       f({\bf k}) \, [ \sin{q_{1x}} + \sin{q_{2x}} ] S_{x}( {\bf k} , {\bf P} - {\bf q}_1 - {\bf q}_2 - {\bf k} )}
{ E - \varepsilon({\bf k}) - \varepsilon({\bf q}_1) - \varepsilon({\bf q}_2) - 
      \varepsilon({\bf P} - {\bf q}_1 - {\bf q}_2 - {\bf k} ) }   
\nonumber \\
                           &   & - \frac{V}{N} \sum_{\bf k} 
\frac{ f({\bf k}) \, [ \cos{q_{1y}} + \cos{q_{2y}} ] C_{y}( {\bf k} , {\bf P} - {\bf q}_1 - {\bf q}_2 - {\bf k} ) + 
       f({\bf k}) \, [ \sin{q_{1y}} + \sin{q_{2y}} ] S_{y}( {\bf k} , {\bf P} - {\bf q}_1 - {\bf q}_2 - {\bf k} )}
{ E - \varepsilon({\bf k}) - \varepsilon({\bf q}_1) - \varepsilon({\bf q}_2) - 
      \varepsilon({\bf P} - {\bf q}_1 - {\bf q}_2 - {\bf k} ) }                                          
\nonumber \\
                           &   & - \frac{V}{N} \sum_{\bf k} 
\frac{ f({\bf k}) \, [ \cos{q_{1x}} + \cos{( P_x - q_{1x} - q_{2x} - k_x )} ] C_{x}( {\bf k} , {\bf q}_2 ) }
{ E - \varepsilon({\bf k}) - \varepsilon({\bf q}_1) - \varepsilon({\bf q}_2) - 
      \varepsilon({\bf P} - {\bf q}_1 - {\bf q}_2 - {\bf k} ) }         
\nonumber \\
                           &   & - \frac{V}{N} \sum_{\bf k} 
\frac{ f({\bf k}) \, [ \sin{q_{1x}} + \sin{( P_x - q_{1x} - q_{2x} - k_x )} ] S_{x}( {\bf k} , {\bf q}_2 )}
{ E - \varepsilon({\bf k}) - \varepsilon({\bf q}_1) - \varepsilon({\bf q}_2) - 
      \varepsilon({\bf P} - {\bf q}_1 - {\bf q}_2 - {\bf k} ) }               
\nonumber \\
                           &   & - \frac{V}{N} \sum_{\bf k} 
\frac{ f({\bf k}) \, [ \cos{q_{1y}} + \cos{( P_y - q_{1y} - q_{2y} - k_y )} ] C_{y}( {\bf k} , {\bf q}_2 ) }
{ E - \varepsilon({\bf k}) - \varepsilon({\bf q}_1) - \varepsilon({\bf q}_2) - 
      \varepsilon({\bf P} - {\bf q}_1 - {\bf q}_2 - {\bf k} ) }         
\nonumber \\
                           &   & - \frac{V}{N} \sum_{\bf k} 
\frac{ f({\bf k}) \, [ \sin{q_{1y}} + \sin{( P_y - q_{1y} - q_{2y} - k_y )} ] S_{y}( {\bf k} , {\bf q}_2 )}
{ E - \varepsilon({\bf k}) - \varepsilon({\bf q}_1) - \varepsilon({\bf q}_2) - 
      \varepsilon({\bf P} - {\bf q}_1 - {\bf q}_2 - {\bf k} ) }       
\nonumber \\
                           &   & - \frac{V}{N} \sum_{\bf k} 
\frac{ f({\bf k}) \, [ \cos{q_{2x}} + \cos{( P_x - q_{1x} - q_{2x} - k_x )} ] C_{x}( {\bf k} , {\bf q}_1 ) }
{ E - \varepsilon({\bf k}) - \varepsilon({\bf q}_1) - \varepsilon({\bf q}_2) - 
      \varepsilon({\bf P} - {\bf q}_1 - {\bf q}_2 - {\bf k} ) }         
\nonumber \\
                           &   & - \frac{V}{N} \sum_{\bf k} 
\frac{ f({\bf k}) \, [ \sin{q_{2x}} + \sin{( P_x - q_{1x} - q_{2x} - k_x )} ] S_{x}( {\bf k} , {\bf q}_1 )}
{ E - \varepsilon({\bf k}) - \varepsilon({\bf q}_1) - \varepsilon({\bf q}_2) - 
      \varepsilon({\bf P} - {\bf q}_1 - {\bf q}_2 - {\bf k} ) }               
\nonumber \\
                           &   & - \frac{V}{N} \sum_{\bf k} 
\frac{ f({\bf k}) \, [ \cos{q_{2y}} + \cos{( P_y - q_{1y} - q_{2y} - k_y )} ] C_{y}( {\bf k} , {\bf q}_1 ) }
{ E - \varepsilon({\bf k}) - \varepsilon({\bf q}_1) - \varepsilon({\bf q}_2) - 
      \varepsilon({\bf P} - {\bf q}_1 - {\bf q}_2 - {\bf k} ) }         
\nonumber \\
                           &   & - \frac{V}{N} \sum_{\bf k} 
\frac{ f({\bf k}) \, [ \sin{q_{2y}} + \sin{( P_y - q_{1y} - q_{2y} - k_y )} ] S_{y}( {\bf k} , {\bf q}_1 )}
{ E - \varepsilon({\bf k}) - \varepsilon({\bf q}_1) - \varepsilon({\bf q}_2) - 
      \varepsilon({\bf P} - {\bf q}_1 - {\bf q}_2 - {\bf k} ) }    \: ,   
\label{4UV2d:eq:sm:twentynine}    
\end{eqnarray}
with $f({\bf k}) \equiv 1$. The equations for $C_{x}$, $S_{x}$, $C_{y}$, and $S_{y}$ are also described by Eq.~(\ref{4UV2d:eq:sm:twentynine}) but with $f({\bf k}) = \cos{(k_x)}$, $\sin{(k_x)}$, $\cos{(k_y)}$, and $\sin{(k_y)}$, respectively.

\subsection{\label{4UV2d:sec:sm:d}
Four $s = \frac{1}{2}$ fermions in 1D. $S = 2$
}

Starting with Eq.~(\ref{4UV2d:eq:thirty}), we first sum over ${\bf b} = \pm {\bf x}$ to get a common factor of 2. Then we permute the arguments to bring the wave function to a common form $\Psi_{2}( k , q , q , q + q - k )$, and expand the cosines. Next, utilizing conservation of total momentum, we define 2 auxiliary functions
\begin{eqnarray}
C(q_1,q_2) & = & 
\frac{1}{N} \sum_{k} \cos{(k)} \Psi_{2}( k , q_1 , q_2 , P - q_1 - q_2 - k ) \: ,  
\label{4UV2d:eq:sm:thirty}     \\  
S(q_1,q_2) & = & 
\frac{1}{N} \sum_{k} \sin{(k)} \Psi_{2}( k , q_1 , q_2 , P - q_1 - q_2 - k ) \: .  
\label{4UV2d:eq:sm:thirtyone}   
\end{eqnarray}
Both functions are antisymmetric: $C( q_2 , q_1 ) = - C( q_1 , q_2 )$ and $S( q_2 , q_1 ) = - S( q_1 , q_2 )$. The wave function follows from Eq.~(\ref{4UV2d:eq:thirty})
\begin{eqnarray}
\Psi_{2}( q_1 , q_2 , q_3 , q_4 ) = \hspace{0.2cm}            
& & - \: V \: 
\frac{ ( \cos{q_1} - \cos{q_2} ) \, C( q_3 , q_4 ) + ( \sin{q_1} - \sin{q_2} ) \, S( q_3 , q_4 ) }
{ E - \varepsilon(q_1) - \varepsilon(q_2) - \varepsilon(q_3) - \varepsilon(q_4) }                     
\nonumber \\ 
& & - \: V \: 
\frac{ ( \cos{q_3} - \cos{q_1} ) \, C( q_2 , q_4 ) + ( \sin{q_3} - \sin{q_1} ) \, S( q_2 , q_4 ) }
{ E - \varepsilon(q_1) - \varepsilon(q_2) - \varepsilon(q_3) - \varepsilon(q_4) }    
\nonumber \\ 
& & - \: V \: 
\frac{ ( \cos{q_1} - \cos{q_4} ) \, C( q_2 , q_3 ) + ( \sin{q_1} - \sin{q_4} ) \, S( q_2 , q_3 ) }
{ E - \varepsilon(q_1) - \varepsilon(q_2) - \varepsilon(q_3) - \varepsilon(q_4) }           
\nonumber \\
& & - \: V \: 
\frac{ ( \cos{q_2} - \cos{q_3} ) \, C( q_1 , q_4 ) + ( \sin{q_2} - \sin{q_3} ) \, S( q_1 , q_4 ) }
{ E - \varepsilon(q_1) - \varepsilon(q_2) - \varepsilon(q_3) - \varepsilon(q_4) }                     
\nonumber \\
& & - \: V \: 
\frac{ ( \cos{q_4} - \cos{q_2} ) \, C( q_1 , q_3 ) + ( \sin{q_4} - \sin{q_2} ) \, S( q_1 , q_3 ) }
{ E - \varepsilon(q_1) - \varepsilon(q_2) - \varepsilon(q_3) - \varepsilon(q_4) }           
\nonumber \\
& & - \: V \: 
\frac{ ( \cos{q_3} - \cos{q_4} ) \, C( q_1 , q_2 ) + ( \sin{q_3} - \sin{q_4} ) \, S( q_1 , q_2 ) }
{ E - \varepsilon(q_1) - \varepsilon(q_2) - \varepsilon(q_3) - \varepsilon(q_4) }   \: .                  
\label{4UV2d:eq:sm:thirtytwo}        
\end{eqnarray}
Substitution of Eq.~(\ref{4UV2d:eq:sm:thirtytwo}) in Eqs.~(\ref{4UV2d:eq:sm:thirty}) and (\ref{4UV2d:eq:sm:thirtyone}) yields two coupled integral equations
\begin{eqnarray}
C( q_1 , q_2 ) = \hspace{0.2cm}            
& & - \frac{V}{N} \sum_{k}  
\frac{ \cos{k} \, [ \cos{k} - \cos{q_1} ] \, C( q_2 , P - q_1 - q_2 - k ) + 
       \cos{k} \, [ \sin{k} - \sin{q_1} ] \, S( q_2 , P - q_1 - q_2 - k ) }
{ E - \varepsilon(k) - \varepsilon(q_1) - \varepsilon(q_2) - \varepsilon( P - q_1 - q_2 - k ) } 
\nonumber \\
& & - \frac{V}{N} \sum_{k}  
\frac{ \cos{k} \, [ \cos{q_2} - \cos{k} ] \, C( q_1 , P - q_1 - q_2 - k ) + 
       \cos{k} \, [ \sin{q_2} - \sin{k} ] \, S( q_1 , P - q_1 - q_2 - k ) }
{ E - \varepsilon(k) - \varepsilon(q_1) - \varepsilon(q_2) - \varepsilon( P - q_1 - q_2 - k ) } 
\nonumber \\
& & - \frac{V}{N} \sum_{k}  
\frac{ \cos{k} \, [ \cos{k} - \cos{( P - q_1 - q_2 - k )} ] \, C( q_1 , q_2 ) + 
       \cos{k} \, [ \sin{k} - \sin{( P - q_1 - q_2 - k )} ] \, S( q_1 , q_2 ) }
{ E - \varepsilon(k) - \varepsilon(q_1) - \varepsilon(q_2) - \varepsilon( P - q_1 - q_2 - k ) } 
\nonumber \\
& & - \frac{V}{N} \sum_{k}  
\frac{ \cos{k} \, [ \cos{q_1} - \cos{q_2} ] \, C( k , P - q_1 - q_2 - k ) + 
       \cos{k} \, [ \sin{q_1} - \sin{q_2} ] \, S( k , P - q_1 - q_2 - k ) }
{ E - \varepsilon(k) - \varepsilon(q_1) - \varepsilon(q_2) - \varepsilon( P - q_1 - q_2 - k ) } 
\nonumber \\
& & - \frac{V}{N} \sum_{k}  
\frac{ \cos{k} \, [ \cos{( P - q_1 - q_2 - k )} - \cos{q_1} ] \, C( k , q_2 ) + 
       \cos{k} \, [ \sin{( P - q_1 - q_2 - k )} - \sin{q_1} ] \, S( k , q_2 ) }
{ E - \varepsilon(k) - \varepsilon(q_1) - \varepsilon(q_2) - \varepsilon( P - q_1 - q_2 - k ) } 
\nonumber \\
& & - \frac{V}{N} \sum_{k}  
\frac{ \cos{k} \, [ \cos{q_2} - \cos{( P - q_1 - q_2 - k )} ] \, C( k , q_1 ) + 
       \cos{k} \, [ \sin{q_2} - \sin{( P - q_1 - q_2 - k )} ] \, S( k , q_1 ) }
{ E - \varepsilon(k) - \varepsilon(q_1) - \varepsilon(q_2) - \varepsilon( P - q_1 - q_2 - k ) } \: ,
\hspace{0.5cm} 
\label{4UV2d:eq:sm:thirtythree}    
\end{eqnarray}
\begin{eqnarray}
S( q_1 , q_2 ) = \hspace{0.2cm}            
& & - \frac{V}{N} \sum_{k}  
\frac{ \sin{k} \, [ \cos{k} - \cos{q_1} ] \, C( q_2 , P - q_1 - q_2 - k ) + 
       \sin{k} \, [ \sin{k} - \sin{q_1} ] \, S( q_2 , P - q_1 - q_2 - k ) }
{ E - \varepsilon(k) - \varepsilon(q_1) - \varepsilon(q_2) - \varepsilon( P - q_1 - q_2 - k ) } 
\nonumber \\
& & - \frac{V}{N} \sum_{k}  
\frac{ \sin{k} \, [ \cos{q_2} - \cos{k} ] \, C( q_1 , P - q_1 - q_2 - k ) + 
       \sin{k} \, [ \sin{q_2} - \sin{k} ] \, S( q_1 , P - q_1 - q_2 - k ) }
{ E - \varepsilon(k) - \varepsilon(q_1) - \varepsilon(q_2) - \varepsilon( P - q_1 - q_2 - k ) } 
\nonumber \\
& & - \frac{V}{N} \sum_{k}  
\frac{ \sin{k} \, [ \cos{k} - \cos{( P - q_1 - q_2 - k )} ] \, C( q_1 , q_2 ) + 
       \sin{k} \, [ \sin{k} - \sin{( P - q_1 - q_2 - k )} ] \, S( q_1 , q_2 ) }
{ E - \varepsilon(k) - \varepsilon(q_1) - \varepsilon(q_2) - \varepsilon( P - q_1 - q_2 - k ) } 
\nonumber \\
& & - \frac{V}{N} \sum_{k}  
\frac{ \sin{k} \, [ \cos{q_1} - \cos{q_2} ] \, C( k , P - q_1 - q_2 - k ) + 
       \sin{k} \, [ \sin{q_1} - \sin{q_2} ] \, S( k , P - q_1 - q_2 - k ) }
{ E - \varepsilon(k) - \varepsilon(q_1) - \varepsilon(q_2) - \varepsilon( P - q_1 - q_2 - k ) } 
\nonumber \\
& & - \frac{V}{N} \sum_{k}  
\frac{ \sin{k} \, [ \cos{( P - q_1 - q_2 - k )} - \cos{q_1} ] \, C( k , q_2 ) + 
       \sin{k} \, [ \sin{( P - q_1 - q_2 - k )} - \sin{q_1} ] \, S( k , q_2 ) }
{ E - \varepsilon(k) - \varepsilon(q_1) - \varepsilon(q_2) - \varepsilon( P - q_1 - q_2 - k ) } 
\nonumber \\
& & - \frac{V}{N} \sum_{k}  
\frac{ \sin{k} \, [ \cos{q_2} - \cos{( P - q_1 - q_2 - k )} ] \, C( k , q_1 ) + 
       \sin{k} \, [ \sin{q_2} - \sin{( P - q_1 - q_2 - k )} ] \, S( k , q_1 ) }
{ E - \varepsilon(k) - \varepsilon(q_1) - \varepsilon(q_2) - \varepsilon( P - q_1 - q_2 - k ) } \: .
\hspace{0.5cm} 
\label{4UV2d:eq:sm:thirtyfour}    
\end{eqnarray}

\subsection{\label{4UV2d:sec:sm:e}
Four $s = \frac{1}{2}$ fermions in 2D. $S = 2$
}

Starting with Eq.~(\ref{4UV2d:eq:thirty}), we sum over ${\bf b} = \pm {\bf x}, \pm {\bf y}$, permute the arguments to bring the wave function to a common form $\Psi_{2}( {\bf k} , {\bf q} , {\bf q} , {\bf q} + {\bf q} - {\bf k} )$, and expand the cosines. Utilizing conservation of total momentum, we define 4 auxiliary functions
\begin{eqnarray}
C_{x}( {\bf q}_1 , {\bf q}_2 ) & = & \frac{1}{N} \sum_{\bf k} \cos{(k_x)} 
\Psi_{2}( {\bf k} , {\bf q}_1 , {\bf q}_2 , {\bf P} - {\bf q}_1 - {\bf q}_2 - {\bf k} ) \: ,  
\label{4UV2d:eq:sm:thirtyfive}     \\  
S_{x}( {\bf q}_1 , {\bf q}_2 ) & = & \frac{1}{N} \sum_{\bf k} \sin{(k_x)} 
\Psi_{2}( {\bf k} , {\bf q}_1 , {\bf q}_2 , {\bf P} - {\bf q}_1 - {\bf q}_2 - {\bf k} ) \: ,    
\label{4UV2d:eq:sm:thirtysix}      \\
C_{y}( {\bf q}_1 , {\bf q}_2 ) & = & \frac{1}{N} \sum_{\bf k} \cos{(k_y)} 
\Psi_{2}( {\bf k} , {\bf q}_1 , {\bf q}_2 , {\bf P} - {\bf q}_1 - {\bf q}_2 - {\bf k} ) \: ,  
\label{4UV2d:eq:sm:thirtyseven}     \\  
S_{y}( {\bf q}_1 , {\bf q}_2 ) & = & \frac{1}{N} \sum_{\bf k} \sin{(k_y)} 
\Psi_{2}( {\bf k} , {\bf q}_1 , {\bf q}_2 , {\bf P} - {\bf q}_1 - {\bf q}_2 - {\bf k} ) \: .    
\label{4UV2d:eq:sm:thirtyeight}
\end{eqnarray}
All four functions are antisymmetric. For example, $S_{x}( {\bf q}_2 , {\bf q}_1 ) = - S_{x}( {\bf q}_1 , {\bf q}_2 )$ and so on. The wave function follows from Eq.~(\ref{4UV2d:eq:thirty}) 
\begin{eqnarray}
\Psi_{2}( {\bf q}_1 , {\bf q}_2 , {\bf q}_3 , {\bf q}_4 ) = \hspace{0.2cm}   
& & - V \: 
\frac{ ( \cos{q_{1x}} - \cos{q_{2x}} ) \, C_{x}( {\bf q}_3 , {\bf q}_4 ) + 
       ( \sin{q_{1x}} - \sin{q_{2x}} ) \, S_{x}( {\bf q}_3 , {\bf q}_4 ) }
{ E - \varepsilon({\bf q}_1) - \varepsilon({\bf q}_2) - \varepsilon({\bf q}_3) - \varepsilon({\bf q}_4) }
\nonumber \\
& & - V \: 
\frac{ ( \cos{q_{1y}} - \cos{q_{2y}} ) \, C_{y}( {\bf q}_3 , {\bf q}_4 ) + 
       ( \sin{q_{1y}} - \sin{q_{2y}} ) \, S_{y}( {\bf q}_3 , {\bf q}_4 ) }
{ E - \varepsilon({\bf q}_1) - \varepsilon({\bf q}_2) - \varepsilon({\bf q}_3) - \varepsilon({\bf q}_4) } 
\nonumber \\
& & - V \: 
\frac{ ( \cos{q_{3x}} - \cos{q_{1x}} ) \, C_{x}( {\bf q}_2 , {\bf q}_4 ) + 
       ( \sin{q_{3x}} - \sin{q_{1x}} ) \, S_{x}( {\bf q}_2 , {\bf q}_4 ) }
{ E - \varepsilon({\bf q}_1) - \varepsilon({\bf q}_2) - \varepsilon({\bf q}_3) - \varepsilon({\bf q}_4) }
\nonumber \\
& & - V \: 
\frac{ ( \cos{q_{3y}} - \cos{q_{1y}} ) \, C_{y}( {\bf q}_2 , {\bf q}_4 ) + 
       ( \sin{q_{3y}} - \sin{q_{1y}} ) \, S_{y}( {\bf q}_2 , {\bf q}_4 ) }
{ E - \varepsilon({\bf q}_1) - \varepsilon({\bf q}_2) - \varepsilon({\bf q}_3) - \varepsilon({\bf q}_4) }
\nonumber \\
& & - V \: 
\frac{ ( \cos{q_{1x}} - \cos{q_{4x}} ) \, C_{x}( {\bf q}_2 , {\bf q}_3 ) + 
       ( \sin{q_{1x}} - \sin{q_{4x}} ) \, S_{x}( {\bf q}_2 , {\bf q}_3 ) }
{ E - \varepsilon({\bf q}_1) - \varepsilon({\bf q}_2) - \varepsilon({\bf q}_3) - \varepsilon({\bf q}_4) }
\nonumber \\
& & - V \: 
\frac{ ( \cos{q_{1y}} - \cos{q_{4y}} ) \, C_{y}( {\bf q}_2 , {\bf q}_3 ) + 
       ( \sin{q_{1y}} - \sin{q_{4y}} ) \, S_{y}( {\bf q}_2 , {\bf q}_3 ) }
{ E - \varepsilon({\bf q}_1) - \varepsilon({\bf q}_2) - \varepsilon({\bf q}_3) - \varepsilon({\bf q}_4) }
\nonumber \\
& & - V \: 
\frac{ ( \cos{q_{2x}} - \cos{q_{3x}} ) \, C_{x}( {\bf q}_1 , {\bf q}_4 ) + 
       ( \sin{q_{2x}} - \sin{q_{3x}} ) \, S_{x}( {\bf q}_1 , {\bf q}_4 ) }
{ E - \varepsilon({\bf q}_1) - \varepsilon({\bf q}_2) - \varepsilon({\bf q}_3) - \varepsilon({\bf q}_4) }
\nonumber \\
& & - V \: 
\frac{ ( \cos{q_{2y}} - \cos{q_{3y}} ) \, C_{y}( {\bf q}_1 , {\bf q}_4 ) + 
       ( \sin{q_{2y}} - \sin{q_{3y}} ) \, S_{y}( {\bf q}_1 , {\bf q}_4 ) }
{ E - \varepsilon({\bf q}_1) - \varepsilon({\bf q}_2) - \varepsilon({\bf q}_3) - \varepsilon({\bf q}_4) }
\nonumber \\
& & - V \: 
\frac{ ( \cos{q_{4x}} - \cos{q_{2x}} ) \, C_{x}( {\bf q}_1 , {\bf q}_3 ) + 
       ( \sin{q_{4x}} - \sin{q_{2x}} ) \, S_{x}( {\bf q}_1 , {\bf q}_3 ) }
{ E - \varepsilon({\bf q}_1) - \varepsilon({\bf q}_2) - \varepsilon({\bf q}_3) - \varepsilon({\bf q}_4) }
\nonumber \\
& & - V \: 
\frac{ ( \cos{q_{4y}} - \cos{q_{2y}} ) \, C_{y}( {\bf q}_1 , {\bf q}_3 ) + 
       ( \sin{q_{4y}} - \sin{q_{2y}} ) \, S_{y}( {\bf q}_1 , {\bf q}_3 ) }
{ E - \varepsilon({\bf q}_1) - \varepsilon({\bf q}_2) - \varepsilon({\bf q}_3) - \varepsilon({\bf q}_4) }
\nonumber \\
& & - V \: 
\frac{ ( \cos{q_{3x}} - \cos{q_{4x}} ) \, C_{x}( {\bf q}_1 , {\bf q}_2 ) + 
       ( \sin{q_{3x}} - \sin{q_{4x}} ) \, S_{x}( {\bf q}_1 , {\bf q}_2 ) }
{ E - \varepsilon({\bf q}_1) - \varepsilon({\bf q}_2) - \varepsilon({\bf q}_3) - \varepsilon({\bf q}_4) }
\nonumber \\
& & - V \: 
\frac{ ( \cos{q_{3y}} - \cos{q_{4y}} ) \, C_{y}( {\bf q}_1 , {\bf q}_2 ) + 
       ( \sin{q_{3y}} - \sin{q_{4y}} ) \, S_{y}( {\bf q}_1 , {\bf q}_2 ) }
{ E - \varepsilon({\bf q}_1) - \varepsilon({\bf q}_2) - \varepsilon({\bf q}_3) - \varepsilon({\bf q}_4) } \: .
\label{4UV2d:eq:sm:thirtynine}    
\end{eqnarray}
Substitution of Eq.~(\ref{4UV2d:eq:sm:thirtynine}) in Eq.~(\ref{4UV2d:eq:sm:thirtyfive}) yields the first of 4 integral equations
\begin{eqnarray}
C_{x}( {\bf q}_1 , {\bf q}_2 ) = \hspace{0.2cm} 
                           &   & - \frac{V}{N} \sum_{\bf k} 
\frac{ f({\bf k}) \, [ \cos{k_x} - \cos{q_{1x}} ] C_{x}( {\bf q}_2 , {\bf P} - {\bf q}_1 - {\bf q}_2 - {\bf k} ) + 
       f({\bf k}) \, [ \sin{k_x} - \sin{q_{1x}} ] S_{x}( {\bf q}_2 , {\bf P} - {\bf q}_1 - {\bf q}_2 - {\bf k} )}
{ E - \varepsilon({\bf k}) - \varepsilon({\bf q}_1) - \varepsilon({\bf q}_2) - 
      \varepsilon({\bf P} - {\bf q}_1 - {\bf q}_2 - {\bf k} ) }               
\nonumber \\
                           &   & - \frac{V}{N} \sum_{\bf k} 
\frac{ f({\bf k}) \, [ \cos{k_y} - \cos{q_{1y}} ] C_{y}( {\bf q}_2 , {\bf P} - {\bf q}_1 - {\bf q}_2 - {\bf k} ) + 
       f({\bf k}) \, [ \sin{k_y} - \sin{q_{1y}} ] S_{y}( {\bf q}_2 , {\bf P} - {\bf q}_1 - {\bf q}_2 - {\bf k} )}
{ E - \varepsilon({\bf k}) - \varepsilon({\bf q}_1) - \varepsilon({\bf q}_2) - 
      \varepsilon({\bf P} - {\bf q}_1 - {\bf q}_2 - {\bf k} ) }     
\nonumber \\
                           &   & - \frac{V}{N} \sum_{\bf k} 
\frac{ f({\bf k}) \, [ \cos{q_{2x}} - \cos{k_x} ] C_{x}( {\bf q}_1 , {\bf P} - {\bf q}_1 - {\bf q}_2 - {\bf k} ) + 
       f({\bf k}) \, [ \sin{q_{2x}} - \sin{k_x} ] S_{x}( {\bf q}_1 , {\bf P} - {\bf q}_1 - {\bf q}_2 - {\bf k} )}
{ E - \varepsilon({\bf k}) - \varepsilon({\bf q}_1) - \varepsilon({\bf q}_2) - 
      \varepsilon({\bf P} - {\bf q}_1 - {\bf q}_2 - {\bf k} ) }   
\nonumber \\
                           &   & - \frac{V}{N} \sum_{\bf k} 
\frac{ f({\bf k}) \, [ \cos{q_{2y}} - \cos{k_y} ] C_{y}( {\bf q}_1 , {\bf P} - {\bf q}_1 - {\bf q}_2 - {\bf k} ) + 
       f({\bf k}) \, [ \sin{q_{2y}} - \sin{k_y} ] S_{y}( {\bf q}_1 , {\bf P} - {\bf q}_1 - {\bf q}_2 - {\bf k} )}
{ E - \varepsilon({\bf k}) - \varepsilon({\bf q}_1) - \varepsilon({\bf q}_2) - 
      \varepsilon({\bf P} - {\bf q}_1 - {\bf q}_2 - {\bf k} ) }                                                                
\nonumber \\
                           &   & - \frac{V}{N} \sum_{\bf k} 
\frac{ f({\bf k}) \, [ \cos{k_x} - \cos{( P_x - q_{1x} - q_{2x} - k_x )} ] C_{x}( {\bf q}_1 , {\bf q}_2 ) }
{ E - \varepsilon({\bf k}) - \varepsilon({\bf q}_1) - \varepsilon({\bf q}_2) - 
      \varepsilon({\bf P} - {\bf q}_1 - {\bf q}_2 - {\bf k} ) }         
\nonumber \\
                           &   & - \frac{V}{N} \sum_{\bf k} 
\frac{ f({\bf k}) \, [ \sin{k_x} - \sin{( P_x - q_{1x} - q_{2x} - k_x )} ] S_{x}( {\bf q}_1 , {\bf q}_2 )}
{ E - \varepsilon({\bf k}) - \varepsilon({\bf q}_1) - \varepsilon({\bf q}_2) - 
      \varepsilon({\bf P} - {\bf q}_1 - {\bf q}_2 - {\bf k} ) }               
\nonumber \\
                           &   & - \frac{V}{N} \sum_{\bf k} 
\frac{ f({\bf k}) \, [ \cos{k_y} - \cos{( P_y - q_{1y} - q_{2y} - k_y )} ] C_{y}( {\bf q}_1 , {\bf q}_2 ) }
{ E - \varepsilon({\bf k}) - \varepsilon({\bf q}_1) - \varepsilon({\bf q}_2) - 
      \varepsilon({\bf P} - {\bf q}_1 - {\bf q}_2 - {\bf k} ) }         
\nonumber \\
                           &   & - \frac{V}{N} \sum_{\bf k} 
\frac{ f({\bf k}) \, [ \sin{k_y} - \sin{( P_y - q_{1y} - q_{2y} - k_y )} ] S_{y}( {\bf q}_1 , {\bf q}_2 )}
{ E - \varepsilon({\bf k}) - \varepsilon({\bf q}_1) - \varepsilon({\bf q}_2) - 
      \varepsilon({\bf P} - {\bf q}_1 - {\bf q}_2 - {\bf k} ) }                     
\nonumber \\
                           &   & - \frac{V}{N} \sum_{\bf k} 
\frac{ f({\bf k}) \, [ \cos{q_{1x}} - \cos{q_{2x}} ] C_{x}( {\bf k} , {\bf P} - {\bf q}_1 - {\bf q}_2 - {\bf k} ) + 
       f({\bf k}) \, [ \sin{q_{1x}} - \sin{q_{2x}} ] S_{x}( {\bf k} , {\bf P} - {\bf q}_1 - {\bf q}_2 - {\bf k} )}
{ E - \varepsilon({\bf k}) - \varepsilon({\bf q}_1) - \varepsilon({\bf q}_2) - 
      \varepsilon({\bf P} - {\bf q}_1 - {\bf q}_2 - {\bf k} ) }   
\nonumber \\
                           &   & - \frac{V}{N} \sum_{\bf k} 
\frac{ f({\bf k}) \, [ \cos{q_{1y}} - \cos{q_{2y}} ] C_{y}( {\bf k} , {\bf P} - {\bf q}_1 - {\bf q}_2 - {\bf k} ) + 
       f({\bf k}) \, [ \sin{q_{1y}} - \sin{q_{2y}} ] S_{y}( {\bf k} , {\bf P} - {\bf q}_1 - {\bf q}_2 - {\bf k} )}
{ E - \varepsilon({\bf k}) - \varepsilon({\bf q}_1) - \varepsilon({\bf q}_2) - 
      \varepsilon({\bf P} - {\bf q}_1 - {\bf q}_2 - {\bf k} ) }                                          
\nonumber \\
                           &   & - \frac{V}{N} \sum_{\bf k} 
\frac{ f({\bf k}) \, [ \cos{( P_x - q_{1x} - q_{2x} - k_x )} - \cos{q_{1x}} ] C_{x}( {\bf k} , {\bf q}_2 ) }
{ E - \varepsilon({\bf k}) - \varepsilon({\bf q}_1) - \varepsilon({\bf q}_2) - 
      \varepsilon({\bf P} - {\bf q}_1 - {\bf q}_2 - {\bf k} ) }         
\nonumber \\
                           &   & - \frac{V}{N} \sum_{\bf k} 
\frac{ f({\bf k}) \, [ \sin{( P_x - q_{1x} - q_{2x} - k_x )} - \sin{q_{1x}} ] S_{x}( {\bf k} , {\bf q}_2 )}
{ E - \varepsilon({\bf k}) - \varepsilon({\bf q}_1) - \varepsilon({\bf q}_2) - 
      \varepsilon({\bf P} - {\bf q}_1 - {\bf q}_2 - {\bf k} ) }               
\nonumber \\
                           &   & - \frac{V}{N} \sum_{\bf k} 
\frac{ f({\bf k}) \, [ \cos{( P_y - q_{1y} - q_{2y} - k_y )} - \cos{q_{1y}} ] C_{y}( {\bf k} , {\bf q}_2 ) }
{ E - \varepsilon({\bf k}) - \varepsilon({\bf q}_1) - \varepsilon({\bf q}_2) - 
      \varepsilon({\bf P} - {\bf q}_1 - {\bf q}_2 - {\bf k} ) }         
\nonumber \\
                           &   & - \frac{V}{N} \sum_{\bf k} 
\frac{ f({\bf k}) \, [ \sin{( P_y - q_{1y} - q_{2y} - k_y )} - \sin{q_{1y}} ] S_{y}( {\bf k} , {\bf q}_2 )}
{ E - \varepsilon({\bf k}) - \varepsilon({\bf q}_1) - \varepsilon({\bf q}_2) - 
      \varepsilon({\bf P} - {\bf q}_1 - {\bf q}_2 - {\bf k} ) }       
\nonumber \\
                           &   & - \frac{V}{N} \sum_{\bf k} 
\frac{ f({\bf k}) \, [ \cos{q_{2x}} - \cos{( P_x - q_{1x} - q_{2x} - k_x )} ] C_{x}( {\bf k} , {\bf q}_1 ) }
{ E - \varepsilon({\bf k}) - \varepsilon({\bf q}_1) - \varepsilon({\bf q}_2) - 
      \varepsilon({\bf P} - {\bf q}_1 - {\bf q}_2 - {\bf k} ) }         
\nonumber \\
                           &   & - \frac{V}{N} \sum_{\bf k} 
\frac{ f({\bf k}) \, [ \sin{q_{2x}} - \sin{( P_x - q_{1x} - q_{2x} - k_x )} ] S_{x}( {\bf k} , {\bf q}_1 )}
{ E - \varepsilon({\bf k}) - \varepsilon({\bf q}_1) - \varepsilon({\bf q}_2) - 
      \varepsilon({\bf P} - {\bf q}_1 - {\bf q}_2 - {\bf k} ) }               
\nonumber \\
                           &   & - \frac{V}{N} \sum_{\bf k} 
\frac{ f({\bf k}) \, [ \cos{q_{2y}} - \cos{( P_y - q_{1y} - q_{2y} - k_y )} ] C_{y}( {\bf k} , {\bf q}_1 ) }
{ E - \varepsilon({\bf k}) - \varepsilon({\bf q}_1) - \varepsilon({\bf q}_2) - 
      \varepsilon({\bf P} - {\bf q}_1 - {\bf q}_2 - {\bf k} ) }         
\nonumber \\
                           &   & - \frac{V}{N} \sum_{\bf k} 
\frac{ f({\bf k}) \, [ \sin{q_{2y}} - \sin{( P_y - q_{1y} - q_{2y} - k_y )} ] S_{y}( {\bf k} , {\bf q}_1 )}
{ E - \varepsilon({\bf k}) - \varepsilon({\bf q}_1) - \varepsilon({\bf q}_2) - 
      \varepsilon({\bf P} - {\bf q}_1 - {\bf q}_2 - {\bf k} ) }    \: ,   
\label{4UV2d:eq:sm:forty}    
\end{eqnarray}
with $f({\bf k}) = \cos{(k_x)}$. The equations for $S_{x}$, $C_{y}$, and $S_{y}$ are also described by Eq.~(\ref{4UV2d:eq:sm:forty}) but with $f({\bf k}) = \sin{(k_x)}$, $\cos{(k_y)}$, and $\sin{(k_y)}$, respectively.

\subsection{\label{4UV2d:sec:sm:f}
Four $s = \frac{1}{2}$ fermions in 1D. $S = 1$
}

Summation over ${\bf b} = \pm {\bf x}$ in Eq.~(\ref{4UV2d:eq:thirtytwo}) yields factors $2 \cos{(k-q)}$. The resulting Schr\"odinger equation requires 5 auxiliary functions  
\begin{eqnarray}
A_{14}(q_1,q_2) & = & 
\frac{1}{N} \sum_{k} \hspace{0.8cm} \Psi_{1}( k , q_1 , q_2 ; P - q_1 - q_2 - k ) \: ;  
\hspace{1.0cm}  A_{14}(q_2,q_1) = - A_{14}(q_1,q_2) \: ,
\label{4UV2d:eq:sm:fortyone}     \\  
C_{14}(q_1,q_2) & = & 
\frac{1}{N} \sum_{k} \cos{k} \, \Psi_{1}( k , q_1 , q_2 ; P - q_1 - q_2 - k ) \: ;
\hspace{1.0cm}  C_{14}(q_2,q_1) = - C_{14}(q_1,q_2) \: ,  
\label{4UV2d:eq:sm:fortytwo}     \\  
S_{14}(q_1,q_2) & = & 
\frac{1}{N} \sum_{k} \sin{k} \, \Psi_{1}( k , q_1 , q_2 ; P - q_1 - q_2 - k ) \: ;
\hspace{1.0cm}  S_{14}(q_2,q_1) = - S_{14}(q_1,q_2) \: ,  
\label{4UV2d:eq:sm:fortythree}   \\
C_{13}(q_1,q_2) & = & 
\frac{1}{N} \sum_{k} \cos{k} \, \Psi_{1}( k , q_1 , P - q_1 - q_2 - k ; q_2 ) \: ,  
\label{4UV2d:eq:sm:fortyfour}     \\  
S_{13}(q_1,q_2) & = & 
\frac{1}{N} \sum_{k} \sin{k} \, \Psi_{1}( k , q_1 , P - q_1 - q_2 - k ; q_2 ) \: .  
\label{4UV2d:eq:sm:fortyfive}   
\end{eqnarray}
Note that only the three $\{ 14 \}$ functions are $q_1 \leftrightarrow q_2$ antisymmetric, whereas the two $\{ 13 \}$ functions do not possess any definite symmetry. The wave function is expressed from Eq.~(\ref{4UV2d:eq:thirtytwo}) as follows 
\begin{eqnarray}
\Psi_{1}( q_1 , q_2 , q_3 ; q_4 ) = 
& & \hspace{0.5cm} U \: 
\frac{ A_{14}( q_1 , q_2 ) + A_{14}( q_2 , q_3 ) + A_{14}( q_3 , q_1 ) }
{ E - \varepsilon(q_1) - \varepsilon(q_2) - \varepsilon(q_3) - \varepsilon(q_4) }  
\nonumber \\                 
& & - \: 2 V \: 
\frac{ \cos{q_1} \, C_{14}( q_2 , q_3 ) + \sin{q_1} \, S_{14}( q_2 , q_3 ) }
{ E - \varepsilon(q_1) - \varepsilon(q_2) - \varepsilon(q_3) - \varepsilon(q_4) }                     
\nonumber \\ 
& & - \: 2 V \: 
\frac{ \cos{q_2} \, C_{14}( q_3 , q_1 ) + \sin{q_2} \, S_{14}( q_3 , q_1 ) }
{ E - \varepsilon(q_1) - \varepsilon(q_2) - \varepsilon(q_3) - \varepsilon(q_4) }    
\nonumber \\ 
& & - \: 2 V \: 
\frac{ \cos{q_3} \, C_{14}( q_1 , q_2 ) + \sin{q_3} \, S_{14}( q_1 , q_2 ) }
{ E - \varepsilon(q_1) - \varepsilon(q_2) - \varepsilon(q_3) - \varepsilon(q_4) }                    
\nonumber \\                             
& & \hspace{0.2cm} - \: V \: 
\frac{ ( \cos{q_2} - \cos{q_1} ) \, C_{13}( q_3 , q_4 ) + ( \sin{q_2} - \sin{q_1} ) \, S_{13}( q_3 , q_4 ) }
{ E - \varepsilon(q_1) - \varepsilon(q_2) - \varepsilon(q_3) - \varepsilon(q_4) }                     
\nonumber \\
& & \hspace{0.2cm} - \: V \: 
\frac{ ( \cos{q_1} - \cos{q_3} ) \, C_{13}( q_2 , q_4 ) + ( \sin{q_1} - \sin{q_3} ) \, S_{13}( q_2 , q_4 ) }
{ E - \varepsilon(q_1) - \varepsilon(q_2) - \varepsilon(q_3) - \varepsilon(q_4) }    
\nonumber \\   
& & \hspace{0.2cm} - \: V \: 
\frac{ ( \cos{q_3} - \cos{q_2} ) \, C_{13}( q_1 , q_4 ) + ( \sin{q_3} - \sin{q_2} ) \, S_{13}( q_1 , q_4 ) }
{ E - \varepsilon(q_1) - \varepsilon(q_2) - \varepsilon(q_3) - \varepsilon(q_4) }  \: .  
\label{4UV2d:eq:sm:fortysix}        
\end{eqnarray}
Substitution of Eq.~(\ref{4UV2d:eq:sm:fortysix}) into the definitions, Eqs.~(\ref{4UV2d:eq:sm:fortyone})-(\ref{4UV2d:eq:sm:fortyfive}), yields a system of five integral equations 
\begin{eqnarray}
A_{14}( q_1 , q_2 ) =           
& & \hspace{0.4cm} \frac{U}{N} \sum_{k}  
\frac{ A_{14}( q_1 , q_2 ) - A_{14}( k , q_2 ) + A_{14}( k , q_1 ) }
{ E - \varepsilon(k) - \varepsilon(q_1) - \varepsilon(q_2) - \varepsilon( P - q_1 - q_2 - k ) } 
\nonumber \\
& & - \frac{2V}{N} \sum_{k}  
\frac{ \cos{k}   \, C_{14}( q_1 , q_2 ) 
     - \cos{q_1} \, C_{14}(  k  , q_2 ) 
     + \cos{q_2} \, C_{14}(  k  , q_1 ) }
{ E - \varepsilon(k) - \varepsilon(q_1) - \varepsilon(q_2) - \varepsilon( P - q_1 - q_2 - k ) } 
\nonumber \\
& & - \frac{2V}{N} \sum_{k}  
\frac{ \sin{k}   \, S_{14}( q_1 , q_2 ) 
     - \sin{q_1} \, S_{14}(  k  , q_2 ) 
     + \sin{q_2} \, S_{14}(  k  , q_1 ) }
{ E - \varepsilon(k) - \varepsilon(q_1) - \varepsilon(q_2) - \varepsilon( P - q_1 - q_2 - k ) } 
\nonumber \\
& & \hspace{0.1cm} - \frac{V}{N} \sum_{k}  
\frac{ [ \cos{q_1} - \cos{k} ] \, C_{13}( q_2 , P - q_1 - q_2 - k ) + 
       [ \sin{q_1} - \sin{k} ] \, S_{13}( q_2 , P - q_1 - q_2 - k ) }
{ E - \varepsilon(k) - \varepsilon(q_1) - \varepsilon(q_2) - \varepsilon( P - q_1 - q_2 - k ) } 
\nonumber \\
& & \hspace{0.1cm} - \frac{V}{N} \sum_{k}  
\frac{ [ \cos{k} - \cos{q_2} ] \, C_{13}( q_1 , P - q_1 - q_2 - k ) + 
       [ \sin{k} - \sin{q_2} ] \, S_{13}( q_1 , P - q_1 - q_2 - k ) }
{ E - \varepsilon(k) - \varepsilon(q_1) - \varepsilon(q_2) - \varepsilon( P - q_1 - q_2 - k ) } 
\nonumber \\
& & \hspace{0.1cm} - \frac{V}{N} \sum_{k}  
\frac{ [ \cos{q_2} - \cos{q_1} ] \, C_{13}( k , P - q_1 - q_2 - k ) + 
       [ \sin{q_2} - \sin{q_1} ] \, S_{13}( k , P - q_1 - q_2 - k ) }
{ E - \varepsilon(k) - \varepsilon(q_1) - \varepsilon(q_2) - \varepsilon( P - q_1 - q_2 - k ) } \: ,
\label{4UV2d:eq:sm:fortyseven}    
\end{eqnarray}
\begin{eqnarray}
C_{14}( q_1 , q_2 ) =           
& & \hspace{0.4cm} \frac{U}{N} \sum_{k}  
\frac{ \cos{k} \, A_{14}( q_1 , q_2 ) - \cos{k} \, A_{14}( k , q_2 ) + \cos{k} \, A_{14}( k , q_1 ) }
{ E - \varepsilon(k) - \varepsilon(q_1) - \varepsilon(q_2) - \varepsilon( P - q_1 - q_2 - k ) } 
\nonumber \\
& & - \frac{2V}{N} \sum_{k}  
\frac{ \cos{k} \cos{k}   \, C_{14}( q_1 , q_2 ) 
     - \cos{k} \cos{q_1} \, C_{14}(  k  , q_2 ) 
     + \cos{k} \cos{q_2} \, C_{14}(  k  , q_1 ) }
{ E - \varepsilon(k) - \varepsilon(q_1) - \varepsilon(q_2) - \varepsilon( P - q_1 - q_2 - k ) } 
\nonumber \\
& & - \frac{2V}{N} \sum_{k}  
\frac{ \cos{k} \sin{k}   \, S_{14}( q_1 , q_2 ) 
     - \cos{k} \sin{q_1} \, S_{14}(  k  , q_2 ) 
     + \cos{k} \sin{q_2} \, S_{14}(  k  , q_1 ) }
{ E - \varepsilon(k) - \varepsilon(q_1) - \varepsilon(q_2) - \varepsilon( P - q_1 - q_2 - k ) } 
\nonumber \\
& & \hspace{0.1cm} - \frac{V}{N} \sum_{k}  
\frac{ \cos{k} \, [ \cos{q_1} - \cos{k} ] \, C_{13}( q_2 , P - q_1 - q_2 - k ) + 
       \cos{k} \, [ \sin{q_1} - \sin{k} ] \, S_{13}( q_2 , P - q_1 - q_2 - k ) }
{ E - \varepsilon(k) - \varepsilon(q_1) - \varepsilon(q_2) - \varepsilon( P - q_1 - q_2 - k ) } 
\nonumber \\
& & \hspace{0.1cm} - \frac{V}{N} \sum_{k}  
\frac{ \cos{k} \, [ \cos{k} - \cos{q_2} ] \, C_{13}( q_1 , P - q_1 - q_2 - k ) + 
       \cos{k} \, [ \sin{k} - \sin{q_2} ] \, S_{13}( q_1 , P - q_1 - q_2 - k ) }
{ E - \varepsilon(k) - \varepsilon(q_1) - \varepsilon(q_2) - \varepsilon( P - q_1 - q_2 - k ) } 
\nonumber \\
& & \hspace{0.1cm} - \frac{V}{N} \sum_{k}  
\frac{ \cos{k} \, [ \cos{q_2} - \cos{q_1} ] \, C_{13}( k , P - q_1 - q_2 - k ) + 
       \cos{k} \, [ \sin{q_2} - \sin{q_1} ] \, S_{13}( k , P - q_1 - q_2 - k ) }
{ E - \varepsilon(k) - \varepsilon(q_1) - \varepsilon(q_2) - \varepsilon( P - q_1 - q_2 - k ) } \: ,
\nonumber \\
& & 
\label{4UV2d:eq:sm:fortyeight}    
\end{eqnarray}
\begin{eqnarray}
S_{14}( q_1 , q_2 ) =           
& & \hspace{0.4cm} \frac{U}{N} \sum_{k}  
\frac{ \sin{k} \, A_{14}( q_1 , q_2 ) - \sin{k} \, A_{14}( k , q_2 ) + \sin{k} \, A_{14}( k , q_1 ) }
{ E - \varepsilon(k) - \varepsilon(q_1) - \varepsilon(q_2) - \varepsilon( P - q_1 - q_2 - k ) } 
\nonumber \\
& & - \frac{2V}{N} \sum_{k}  
\frac{ \sin{k} \cos{k}   \, C_{14}( q_1 , q_2 ) 
     - \sin{k} \cos{q_1} \, C_{14}(  k  , q_2 ) 
     + \sin{k} \cos{q_2} \, C_{14}(  k  , q_1 ) }
{ E - \varepsilon(k) - \varepsilon(q_1) - \varepsilon(q_2) - \varepsilon( P - q_1 - q_2 - k ) } 
\nonumber \\
& & - \frac{2V}{N} \sum_{k}  
\frac{ \sin{k} \sin{k}   \, S_{14}( q_1 , q_2 ) 
     - \sin{k} \sin{q_1} \, S_{14}(  k  , q_2 ) 
     + \sin{k} \sin{q_2} \, S_{14}(  k  , q_1 ) }
{ E - \varepsilon(k) - \varepsilon(q_1) - \varepsilon(q_2) - \varepsilon( P - q_1 - q_2 - k ) } 
\nonumber \\
& & \hspace{0.1cm} - \frac{V}{N} \sum_{k}  
\frac{ \sin{k} \, [ \cos{q_1} - \cos{k} ] \, C_{13}( q_2 , P - q_1 - q_2 - k ) + 
       \sin{k} \, [ \sin{q_1} - \sin{k} ] \, S_{13}( q_2 , P - q_1 - q_2 - k ) }
{ E - \varepsilon(k) - \varepsilon(q_1) - \varepsilon(q_2) - \varepsilon( P - q_1 - q_2 - k ) } 
\nonumber \\
& & \hspace{0.1cm} - \frac{V}{N} \sum_{k}  
\frac{ \sin{k} \, [ \cos{k} - \cos{q_2} ] \, C_{13}( q_1 , P - q_1 - q_2 - k ) + 
       \sin{k} \, [ \sin{k} - \sin{q_2} ] \, S_{13}( q_1 , P - q_1 - q_2 - k ) }
{ E - \varepsilon(k) - \varepsilon(q_1) - \varepsilon(q_2) - \varepsilon( P - q_1 - q_2 - k ) } 
\nonumber \\
& & \hspace{0.1cm} - \frac{V}{N} \sum_{k}  
\frac{ \sin{k} \, [ \cos{q_2} - \cos{q_1} ] \, C_{13}( k , P - q_1 - q_2 - k ) + 
       \sin{k} \, [ \sin{q_2} - \sin{q_1} ] \, S_{13}( k , P - q_1 - q_2 - k ) }
{ E - \varepsilon(k) - \varepsilon(q_1) - \varepsilon(q_2) - \varepsilon( P - q_1 - q_2 - k ) } \: ,
\nonumber \\
& & 
\label{4UV2d:eq:sm:fortynine}    
\end{eqnarray}
\begin{eqnarray}
C_{13}( q_1 , q_2 ) =           
& & \hspace{0.4cm} \frac{U}{N} \sum_{k}  
\frac{ \cos{k} \, A_{14}( k   , q_1 ) 
     + \cos{k} \, A_{14}( q_1 , P - q_1 - q_2 - k ) 
     + \cos{k} \, A_{14}( P - q_1 - q_2 - k , k ) }
{ E - \varepsilon(k) - \varepsilon(q_1) - \varepsilon(q_2) - \varepsilon( P - q_1 - q_2 - k ) } 
\nonumber \\
& & - \frac{2V}{N} \sum_{k}  
\frac{ \cos{k} \cos{k} \, C_{14}( q_1 , P - q_1 - q_2 - k ) 
     + \cos{k} \sin{k} \, S_{14}( q_1 , P - q_1 - q_2 - k ) }
{ E - \varepsilon(k) - \varepsilon(q_1) - \varepsilon(q_2) - \varepsilon( P - q_1 - q_2 - k ) } 
\nonumber \\
& & - \frac{2V}{N} \sum_{k}  
\frac{ \cos{k} \cos{q_1} \, C_{14}( P - q_1 - q_2 - k , k ) 
     + \cos{k} \sin{q_1} \, S_{14}( P - q_1 - q_2 - k , k ) }
{ E - \varepsilon(k) - \varepsilon(q_1) - \varepsilon(q_2) - \varepsilon( P - q_1 - q_2 - k ) } 
\nonumber \\
& & - \frac{2V}{N} \sum_{k}  
\frac{ \cos{k} \cos{( P - q_1 - q_2 - k )} \, C_{14}( k , q_1 ) 
     + \cos{k} \sin{( P - q_1 - q_2 - k )} \, S_{14}( k , q_1 ) }
{ E - \varepsilon(k) - \varepsilon(q_1) - \varepsilon(q_2) - \varepsilon( P - q_1 - q_2 - k ) } 
\nonumber \\
& & \hspace{0.1cm} - \frac{V}{N} \sum_{k}  
\frac{ \cos{k} \, [ \cos{q_1} - \cos{k} ] \, C_{13}( P - q_1 - q_2 - k , q_2 ) + 
       \cos{k} \, [ \sin{q_1} - \sin{k} ] \, S_{13}( P - q_1 - q_2 - k , q_2 ) }
{ E - \varepsilon(k) - \varepsilon(q_1) - \varepsilon(q_2) - \varepsilon( P - q_1 - q_2 - k ) } 
\nonumber \\
& & \hspace{0.1cm} - \frac{V}{N} \sum_{k}  
\frac{ \cos{k} \, [ \cos{k} - \cos{( P - q_1 - q_2 - k )} ] \, C_{13}( q_1 , q_2 ) + 
       \cos{k} \, [ \sin{k} - \sin{( P - q_1 - q_2 - k )} ] \, S_{13}( q_1 , q_2 ) }
{ E - \varepsilon(k) - \varepsilon(q_1) - \varepsilon(q_2) - \varepsilon( P - q_1 - q_2 - k ) } 
\nonumber \\
& & \hspace{0.1cm} - \frac{V}{N} \sum_{k}  
\frac{ \cos{k} \, [ \cos{( P - q_1 - q_2 - k )} - \cos{q_1} ] \, C_{13}( k , q_2 ) + 
       \cos{k} \, [ \sin{( P - q_1 - q_2 - k )} - \sin{q_1} ] \, S_{13}( k , q_2 ) }
{ E - \varepsilon(k) - \varepsilon(q_1) - \varepsilon(q_2) - \varepsilon( P - q_1 - q_2 - k ) } ,
\nonumber \\
& & 
\label{4UV2d:eq:sm:fifty}    
\end{eqnarray}
\begin{eqnarray}
S_{13}( q_1 , q_2 ) =           
& & \hspace{0.4cm} \frac{U}{N} \sum_{k}  
\frac{ \sin{k} \, A_{14}( k   , q_1 ) 
     + \sin{k} \, A_{14}( q_1 , P - q_1 - q_2 - k ) 
     + \sin{k} \, A_{14}( P - q_1 - q_2 - k , k ) }
{ E - \varepsilon(k) - \varepsilon(q_1) - \varepsilon(q_2) - \varepsilon( P - q_1 - q_2 - k ) } 
\nonumber \\
& & - \frac{2V}{N} \sum_{k}  
\frac{ \sin{k} \cos{k} \, C_{14}( q_1 , P - q_1 - q_2 - k ) 
     + \sin{k} \sin{k} \, S_{14}( q_1 , P - q_1 - q_2 - k ) }
{ E - \varepsilon(k) - \varepsilon(q_1) - \varepsilon(q_2) - \varepsilon( P - q_1 - q_2 - k ) } 
\nonumber \\
& & - \frac{2V}{N} \sum_{k}  
\frac{ \sin{k} \cos{q_1} \, C_{14}( P - q_1 - q_2 - k , k ) 
     + \sin{k} \sin{q_1} \, S_{14}( P - q_1 - q_2 - k , k ) }
{ E - \varepsilon(k) - \varepsilon(q_1) - \varepsilon(q_2) - \varepsilon( P - q_1 - q_2 - k ) } 
\nonumber \\
& & - \frac{2V}{N} \sum_{k}  
\frac{ \sin{k} \cos{( P - q_1 - q_2 - k )} \, C_{14}( k , q_1 ) 
     + \sin{k} \sin{( P - q_1 - q_2 - k )} \, S_{14}( k , q_1 ) }
{ E - \varepsilon(k) - \varepsilon(q_1) - \varepsilon(q_2) - \varepsilon( P - q_1 - q_2 - k ) } 
\nonumber \\
& & \hspace{0.1cm} - \frac{V}{N} \sum_{k}  
\frac{ \sin{k} \, [ \cos{q_1} - \cos{k} ] \, C_{13}( P - q_1 - q_2 - k , q_2 ) + 
       \sin{k} \, [ \sin{q_1} - \sin{k} ] \, S_{13}( P - q_1 - q_2 - k , q_2 ) }
{ E - \varepsilon(k) - \varepsilon(q_1) - \varepsilon(q_2) - \varepsilon( P - q_1 - q_2 - k ) } 
\nonumber \\
& & \hspace{0.1cm} - \frac{V}{N} \sum_{k}  
\frac{ \sin{k} \, [ \cos{k} - \cos{( P - q_1 - q_2 - k )} ] \, C_{13}( q_1 , q_2 ) + 
       \sin{k} \, [ \sin{k} - \sin{( P - q_1 - q_2 - k )} ] \, S_{13}( q_1 , q_2 ) }
{ E - \varepsilon(k) - \varepsilon(q_1) - \varepsilon(q_2) - \varepsilon( P - q_1 - q_2 - k ) } 
\nonumber \\
& & \hspace{0.1cm} - \frac{V}{N} \sum_{k}  
\frac{ \sin{k} \, [ \cos{( P - q_1 - q_2 - k )} - \cos{q_1} ] \, C_{13}( k , q_2 ) + 
       \sin{k} \, [ \sin{( P - q_1 - q_2 - k )} - \sin{q_1} ] \, S_{13}( k , q_2 ) }
{ E - \varepsilon(k) - \varepsilon(q_1) - \varepsilon(q_2) - \varepsilon( P - q_1 - q_2 - k ) } \, .
\nonumber \\
& & 
\label{4UV2d:eq:sm:fiftyone}    
\end{eqnarray}

\subsection{\label{4UV2d:sec:sm:g}
Four $s = \frac{1}{2}$ fermions in 2D. $S = 1$
}

Summation over ${\bf b} = \pm {\bf x}, \pm {\bf y}$ in Eq.~(\ref{4UV2d:eq:thirtytwo}) yields factors $2 \cos{( k_x - q_x )} + 2 \cos{( k_y - q_y )}$. Utilizing antisymmetry over the first three arguments, transform $\Psi_{1}$ in the $U$ terms and the second $V$ terms to a unified form $\Psi_{1}( {\bf k} , {\bf q} , {\bf q} ; {\bf P} - {\bf q} - {\bf q} - {\bf k} )$, and the first $V$ terms to $\Psi_{1}( {\bf k} , {\bf q} , {\bf P} - {\bf q} - {\bf q} - {\bf k} ; {\bf q} )$, where ${\bf P}$ is the total momentum. Then, we introduce 9 auxiliary functions: 
\begin{eqnarray}
A_{14} ( {\bf q}_1 , {\bf q}_2 ) & = & \frac{1}{N} \sum_{\bf k} 
\hspace{0.9cm} \Psi_{1}( {\bf k} , {\bf q}_1 , {\bf q}_2 ; {\bf P} - {\bf q}_1 - {\bf q}_2 - {\bf k} ) \: ; 
\hspace{0.5cm}    A_{14}( {\bf q}_2 , {\bf q}_1 ) =  - A_{14}( {\bf q}_1 , {\bf q}_2 )  \: , 
\label{4UV2d:eq:sm:fiftytwo}       \\
C_{x14}( {\bf q}_1 , {\bf q}_2 ) & = & \frac{1}{N} \sum_{\bf k} 
\cos{q_x}    \Psi_{1}( {\bf k} , {\bf q}_1 , {\bf q}_2 ; {\bf P} - {\bf q}_1 - {\bf q}_2 - {\bf k} ) \: ;   
\hspace{0.5cm}   C_{x14}( {\bf q}_2 , {\bf q}_1 ) = - C_{x14}( {\bf q}_1 , {\bf q}_2 )  \: , 
\label{4UV2d:eq:sm:fiftythree}     \\
S_{x14}( {\bf q}_1 , {\bf q}_2 ) & = & \frac{1}{N} \sum_{\bf k} 
\sin{q_x}    \Psi_{1}( {\bf k} , {\bf q}_1 , {\bf q}_2 ; {\bf P} - {\bf q}_1 - {\bf q}_2 - {\bf k} ) \: ;
\hspace{0.5cm}   S_{x14}( {\bf q}_2 , {\bf q}_1 ) = - S_{x14}( {\bf q}_1 , {\bf q}_2 )  \: , 
\label{4UV2d:eq:sm:fiftyfour}      \\
C_{y14}( {\bf q}_1 , {\bf q}_2 ) & = & \frac{1}{N} \sum_{\bf k} 
\cos{q_y}    \Psi_{1}( {\bf k} , {\bf q}_1 , {\bf q}_2 ; {\bf P} - {\bf q}_1 - {\bf q}_2 - {\bf k} ) \: ;
\hspace{0.5cm}   C_{y14}( {\bf q}_2 , {\bf q}_1 ) = - C_{y14}( {\bf q}_1 , {\bf q}_2 )  \: ,     
\label{4UV2d:eq:sm:fiftyfive}      \\
S_{y14}( {\bf q}_1 , {\bf q}_2 ) & = & \frac{1}{N} \sum_{\bf k} 
\sin{q_y}    \Psi_{1}( {\bf k} , {\bf q}_1 , {\bf q}_2 ; {\bf P} - {\bf q}_1 - {\bf q}_2 - {\bf k} ) \: ;
\hspace{0.5cm}   S_{y14}( {\bf q}_2 , {\bf q}_1 ) = - S_{y14}( {\bf q}_1 , {\bf q}_2 )  \: ,     
\label{4UV2d:eq:sm:fiftysix}       \\    
C_{x13}( {\bf q}_1 , {\bf q}_2 ) & = & \frac{1}{N} \sum_{\bf k} 
\cos{q_x}    \Psi_{1}( {\bf k} , {\bf q}_1 , {\bf P} - {\bf q}_1 - {\bf q}_2 - {\bf k} ; {\bf q}_2 ) \: ,   
\label{4UV2d:eq:sm:fiftyseven}     \\    
S_{x13}( {\bf q}_1 , {\bf q}_2 ) & = & \frac{1}{N} \sum_{\bf k} 
\sin{q_x}    \Psi_{1}( {\bf k} , {\bf q}_1 , {\bf P} - {\bf q}_1 - {\bf q}_2 - {\bf k} ; {\bf q}_2 ) \: ,   
\label{4UV2d:eq:sm:fiftyeight}     \\    
C_{y13}( {\bf q}_1 , {\bf q}_2 ) & = & \frac{1}{N} \sum_{\bf k} 
\cos{q_y}    \Psi_{1}( {\bf k} , {\bf q}_1 , {\bf P} - {\bf q}_1 - {\bf q}_2 - {\bf k} ; {\bf q}_2 ) \: ,   
\label{4UV2d:eq:sm:fiftynine}      \\    
S_{y13}( {\bf q}_1 , {\bf q}_2 ) & = & \frac{1}{N} \sum_{\bf k} 
\sin{q_y}    \Psi_{1}( {\bf k} , {\bf q}_1 , {\bf P} - {\bf q}_1 - {\bf q}_2 - {\bf k} ; {\bf q}_2 ) \: .   
\label{4UV2d:eq:sm:sixty}
\end{eqnarray}
Utilizing the above definitions and expanding the cosines, the wave function is expressed from Eq.~(\ref{4UV2d:eq:thirtytwo})
\begin{eqnarray}
\Psi_{1}( {\bf q}_1 , {\bf q}_2 , {\bf q}_3 ; {\bf q}_4 ) = \hspace{0.2cm}   
& &  \hspace{0.4cm}  U \:
\frac{ A_{14}( {\bf q}_2 , {\bf q}_3 ) 
     + A_{14}( {\bf q}_3 , {\bf q}_1 ) 
     + A_{14}( {\bf q}_1 , {\bf q}_2 ) }
{ E - \varepsilon({\bf q}_1) - \varepsilon({\bf q}_2) - \varepsilon({\bf q}_3) - \varepsilon({\bf q}_4) }
\nonumber \\
& & - 2 V \: 
\frac{ \cos{q_{1x}} \, C_{x14}( {\bf q}_2 , {\bf q}_3 ) + 
       \sin{q_{1x}} \, S_{x14}( {\bf q}_2 , {\bf q}_3 ) + 
       \cos{q_{1y}} \, C_{y14}( {\bf q}_2 , {\bf q}_3 ) + 
       \sin{q_{1y}} \, S_{y14}( {\bf q}_2 , {\bf q}_3 ) }
{ E - \varepsilon({\bf q}_1) - \varepsilon({\bf q}_2) - \varepsilon({\bf q}_3) - \varepsilon({\bf q}_4) }
\nonumber \\
& & - 2 V \: 
\frac{ \cos{q_{2x}} \, C_{x14}( {\bf q}_3 , {\bf q}_1 ) + 
       \sin{q_{2x}} \, S_{x14}( {\bf q}_3 , {\bf q}_1 ) + 
       \cos{q_{2y}} \, C_{y14}( {\bf q}_3 , {\bf q}_1 ) + 
       \sin{q_{2y}} \, S_{y14}( {\bf q}_3 , {\bf q}_1 ) }
{ E - \varepsilon({\bf q}_1) - \varepsilon({\bf q}_2) - \varepsilon({\bf q}_3) - \varepsilon({\bf q}_4) }
\nonumber \\
& & - 2 V \: 
\frac{ \cos{q_{3x}} \, C_{x14}( {\bf q}_1 , {\bf q}_2 ) + 
       \sin{q_{3x}} \, S_{x14}( {\bf q}_1 , {\bf q}_2 ) + 
       \cos{q_{3y}} \, C_{y14}( {\bf q}_1 , {\bf q}_2 ) + 
       \sin{q_{3y}} \, S_{y14}( {\bf q}_1 , {\bf q}_2 ) }
{ E - \varepsilon({\bf q}_1) - \varepsilon({\bf q}_2) - \varepsilon({\bf q}_3) - \varepsilon({\bf q}_4) }
\nonumber \\
& & \hspace{0.1cm} - V \: 
\frac{ ( \cos{q_{2x}} - \cos{q_{1x}} ) \, C_{x13}( {\bf q}_3 , {\bf q}_4 ) + 
       ( \sin{q_{2x}} - \sin{q_{1x}} ) \, S_{x13}( {\bf q}_3 , {\bf q}_4 ) }
{ E - \varepsilon({\bf q}_1) - \varepsilon({\bf q}_2) - \varepsilon({\bf q}_3) - \varepsilon({\bf q}_4) }
\nonumber \\
& & \hspace{0.1cm} - V \: 
\frac{ ( \cos{q_{2y}} - \cos{q_{1y}} ) \, C_{y13}( {\bf q}_3 , {\bf q}_4 ) + 
       ( \sin{q_{2y}} - \sin{q_{1y}} ) \, S_{y13}( {\bf q}_3 , {\bf q}_4 ) }
{ E - \varepsilon({\bf q}_1) - \varepsilon({\bf q}_2) - \varepsilon({\bf q}_3) - \varepsilon({\bf q}_4) } 
\nonumber \\
& & \hspace{0.1cm} - V \: 
\frac{ ( \cos{q_{1x}} - \cos{q_{3x}} ) \, C_{x13}( {\bf q}_2 , {\bf q}_4 ) + 
       ( \sin{q_{1x}} - \sin{q_{3x}} ) \, S_{x13}( {\bf q}_2 , {\bf q}_4 ) }
{ E - \varepsilon({\bf q}_1) - \varepsilon({\bf q}_2) - \varepsilon({\bf q}_3) - \varepsilon({\bf q}_4) }
\nonumber \\
& & \hspace{0.1cm} - V \: 
\frac{ ( \cos{q_{1y}} - \cos{q_{3y}} ) \, C_{y13}( {\bf q}_2 , {\bf q}_4 ) + 
       ( \sin{q_{1y}} - \sin{q_{3y}} ) \, S_{y13}( {\bf q}_2 , {\bf q}_4 ) }
{ E - \varepsilon({\bf q}_1) - \varepsilon({\bf q}_2) - \varepsilon({\bf q}_3) - \varepsilon({\bf q}_4) }
\nonumber \\
& & \hspace{0.1cm} - V \: 
\frac{ ( \cos{q_{3x}} - \cos{q_{2x}} ) \, C_{x13}( {\bf q}_1 , {\bf q}_4 ) + 
       ( \sin{q_{3x}} - \sin{q_{2x}} ) \, S_{x13}( {\bf q}_1 , {\bf q}_4 ) }
{ E - \varepsilon({\bf q}_1) - \varepsilon({\bf q}_2) - \varepsilon({\bf q}_3) - \varepsilon({\bf q}_4) }
\nonumber \\
& & \hspace{0.1cm} - V \: 
\frac{ ( \cos{q_{3y}} - \cos{q_{2y}} ) \, C_{y13}( {\bf q}_1 , {\bf q}_4 ) + 
       ( \sin{q_{3y}} - \sin{q_{2y}} ) \, S_{y13}( {\bf q}_1 , {\bf q}_4 ) }
{ E - \varepsilon({\bf q}_1) - \varepsilon({\bf q}_2) - \varepsilon({\bf q}_3) - \varepsilon({\bf q}_4) } \: .
\label{4UV2d:eq:sm:sixtyone}    
\end{eqnarray}
Substitution of Eq.~(\ref{4UV2d:eq:sm:sixtyone}) in Eqs.~(\ref{4UV2d:eq:sm:fiftytwo})-(\ref{4UV2d:eq:sm:sixty}) produces nine coupled equations. The first five equations for the $\{ 14 \}$ functions differ only by the form of $f({\bf k})$ under the sum. It is therefore sufficient to present only the first equation for $A_{14}$:
\begin{eqnarray}
A_{14}( {\bf q}_1 , {\bf q}_2 ) = \hspace{0.2cm} 
& & \hspace{0.4cm} \frac{U}{N} \sum_{\bf k} 
\frac{ f({\bf k}) \, A_{14}( {\bf q}_1 , {\bf q}_2 ) + 
       f({\bf k}) \, A_{14}( {\bf q}_2 , {\bf k}   ) + 
       f({\bf k}) \, A_{14}( {\bf k}   , {\bf q}_1 ) }
{ E - \varepsilon({\bf k}) - \varepsilon({\bf q}_1) - \varepsilon({\bf q}_2) - 
      \varepsilon({\bf P} - {\bf q}_1 - {\bf q}_2 - {\bf k} ) }               
\nonumber \\
                           &   & - \frac{2V}{N} \sum_{\bf k} 
\frac{ f({\bf k}) \, \cos{k_x} \, C_{x14}( {\bf q}_1 , {\bf q}_2 ) + 
       f({\bf k}) \, \sin{k_x} \, S_{x14}( {\bf q}_1 , {\bf q}_2 )}
{ E - \varepsilon({\bf k}) - \varepsilon({\bf q}_1) - \varepsilon({\bf q}_2) - 
      \varepsilon({\bf P} - {\bf q}_1 - {\bf q}_2 - {\bf k} ) }               
\nonumber \\
                           &   & - \frac{2V}{N} \sum_{\bf k} 
\frac{ f({\bf k}) \, \cos{k_y} \, C_{y14}( {\bf q}_1 , {\bf q}_2 ) + 
       f({\bf k}) \, \sin{k_y} \, S_{y14}( {\bf q}_1 , {\bf q}_2 )}
{ E - \varepsilon({\bf k}) - \varepsilon({\bf q}_1) - \varepsilon({\bf q}_2) - 
      \varepsilon({\bf P} - {\bf q}_1 - {\bf q}_2 - {\bf k} ) }     
\nonumber \\
                           &   & - \frac{2V}{N} \sum_{\bf k} 
\frac{ f({\bf k}) \, \cos{q_{1x}} \, C_{x14}( {\bf q}_2 , {\bf k} ) + 
       f({\bf k}) \, \sin{q_{1x}} \, S_{x14}( {\bf q}_2 , {\bf k} )}
{ E - \varepsilon({\bf k}) - \varepsilon({\bf q}_1) - \varepsilon({\bf q}_2) - 
      \varepsilon({\bf P} - {\bf q}_1 - {\bf q}_2 - {\bf k} ) }   
\nonumber \\
                           &   & - \frac{2V}{N} \sum_{\bf k} 
\frac{ f({\bf k}) \, \cos{q_{1y}} \, C_{y14}( {\bf q}_2 , {\bf k} ) + 
       f({\bf k}) \, \sin{q_{1y}} \, S_{y14}( {\bf q}_2 , {\bf k} )}
{ E - \varepsilon({\bf k}) - \varepsilon({\bf q}_1) - \varepsilon({\bf q}_2) - 
      \varepsilon({\bf P} - {\bf q}_1 - {\bf q}_2 - {\bf k} ) }                                                                
\nonumber \\
                           &   & - \frac{2V}{N} \sum_{\bf k} 
\frac{ f({\bf k}) \, \cos{q_{2x}} \, C_{x14}( {\bf k} , {\bf q}_1 ) + 
       f({\bf k}) \, \sin{q_{2x}} \, S_{x14}( {\bf k} , {\bf q}_1 )  }
{ E - \varepsilon({\bf k}) - \varepsilon({\bf q}_1) - \varepsilon({\bf q}_2) - 
      \varepsilon({\bf P} - {\bf q}_1 - {\bf q}_2 - {\bf k} ) }         
\nonumber \\
                           &   & - \frac{2V}{N} \sum_{\bf k} 
\frac{ f({\bf k}) \, \cos{q_{2y}} \, C_{y14}( {\bf k} , {\bf q}_1 ) + 
       f({\bf k}) \, \sin{q_{2y}} \, S_{y14}( {\bf k} , {\bf q}_1 )  }
{ E - \varepsilon({\bf k}) - \varepsilon({\bf q}_1) - \varepsilon({\bf q}_2) - 
      \varepsilon({\bf P} - {\bf q}_1 - {\bf q}_2 - {\bf k} ) }              
\nonumber \\
                           &   & \hspace{0.1cm} - \frac{V}{N} \sum_{\bf k} 
\frac{ f({\bf k}) \, ( \cos{q_{1x}} - \cos{k_x} ) \, C_{x13}( {\bf q}_2 , {\bf P} - {\bf q}_1 - {\bf q}_2 - {\bf k} ) }
{ E - \varepsilon({\bf k}) - \varepsilon({\bf q}_1) - \varepsilon({\bf q}_2) - 
      \varepsilon({\bf P} - {\bf q}_1 - {\bf q}_2 - {\bf k} ) }           
\nonumber \\
                           &   & \hspace{0.1cm} - \frac{V}{N} \sum_{\bf k} 
\frac{ f({\bf k}) \, ( \sin{q_{1x}} - \sin{k_x} ) \, S_{x13}( {\bf q}_2 , {\bf P} - {\bf q}_1 - {\bf q}_2 - {\bf k} )  }
{ E - \varepsilon({\bf k}) - \varepsilon({\bf q}_1) - \varepsilon({\bf q}_2) - 
      \varepsilon({\bf P} - {\bf q}_1 - {\bf q}_2 - {\bf k} ) }                   
\nonumber \\
                           &   & \hspace{0.1cm} - \frac{V}{N} \sum_{\bf k} 
\frac{ f({\bf k}) \, ( \cos{q_{1y}} - \cos{k_y} ) \, C_{y13}( {\bf q}_2 , {\bf P} - {\bf q}_1 - {\bf q}_2 - {\bf k} ) }
{ E - \varepsilon({\bf k}) - \varepsilon({\bf q}_1) - \varepsilon({\bf q}_2) - 
      \varepsilon({\bf P} - {\bf q}_1 - {\bf q}_2 - {\bf k} ) }           
\nonumber \\
                           &   & \hspace{0.1cm} - \frac{V}{N} \sum_{\bf k} 
\frac{ f({\bf k}) \, ( \sin{q_{1y}} - \sin{k_y} ) \, S_{y13}( {\bf q}_2 , {\bf P} - {\bf q}_1 - {\bf q}_2 - {\bf k} )  }
{ E - \varepsilon({\bf k}) - \varepsilon({\bf q}_1) - \varepsilon({\bf q}_2) - 
      \varepsilon({\bf P} - {\bf q}_1 - {\bf q}_2 - {\bf k} ) }                    
\nonumber \\
                           &   & \hspace{0.1cm} - \frac{V}{N} \sum_{\bf k} 
\frac{ f({\bf k}) \, ( \cos{k_x} - \cos{q_{2x}} ) \, C_{x13}( {\bf q}_1 , {\bf P} - {\bf q}_1 - {\bf q}_2 - {\bf k} ) }
{ E - \varepsilon({\bf k}) - \varepsilon({\bf q}_1) - \varepsilon({\bf q}_2) - 
      \varepsilon({\bf P} - {\bf q}_1 - {\bf q}_2 - {\bf k} ) }           
\nonumber \\
                           &   & \hspace{0.1cm} - \frac{V}{N} \sum_{\bf k} 
\frac{ f({\bf k}) \, ( \sin{k_x} - \sin{q_{2x}} ) \, S_{x13}( {\bf q}_1 , {\bf P} - {\bf q}_1 - {\bf q}_2 - {\bf k} )  }
{ E - \varepsilon({\bf k}) - \varepsilon({\bf q}_1) - \varepsilon({\bf q}_2) - 
      \varepsilon({\bf P} - {\bf q}_1 - {\bf q}_2 - {\bf k} ) }     
\nonumber \\
                           &   & \hspace{0.1cm} - \frac{V}{N} \sum_{\bf k} 
\frac{ f({\bf k}) \, ( \cos{k_y} - \cos{q_{2y}} ) \, C_{y13}( {\bf q}_1 , {\bf P} - {\bf q}_1 - {\bf q}_2 - {\bf k} ) }
{ E - \varepsilon({\bf k}) - \varepsilon({\bf q}_1) - \varepsilon({\bf q}_2) - 
      \varepsilon({\bf P} - {\bf q}_1 - {\bf q}_2 - {\bf k} ) }           
\nonumber \\
                           &   & \hspace{0.1cm} - \frac{V}{N} \sum_{\bf k} 
\frac{ f({\bf k}) \, ( \sin{k_y} - \sin{q_{2y}} ) \, S_{y13}( {\bf q}_1 , {\bf P} - {\bf q}_1 - {\bf q}_2 - {\bf k} )  }
{ E - \varepsilon({\bf k}) - \varepsilon({\bf q}_1) - \varepsilon({\bf q}_2) - 
      \varepsilon({\bf P} - {\bf q}_1 - {\bf q}_2 - {\bf k} ) }         
\nonumber \\
                           &   & \hspace{0.1cm} - \frac{V}{N} \sum_{\bf k} 
\frac{ f({\bf k}) \, ( \cos{q_{2x}} - \cos{q_{1x}} ) \, C_{x13}( {\bf k} , {\bf P} - {\bf q}_1 - {\bf q}_2 - {\bf k} ) }
{ E - \varepsilon({\bf k}) - \varepsilon({\bf q}_1) - \varepsilon({\bf q}_2) - 
      \varepsilon({\bf P} - {\bf q}_1 - {\bf q}_2 - {\bf k} ) }           
\nonumber \\
                           &   & \hspace{0.1cm} - \frac{V}{N} \sum_{\bf k} 
\frac{ f({\bf k}) \, ( \sin{q_{2x}} - \sin{q_{1x}} ) \, S_{x13}( {\bf k} , {\bf P} - {\bf q}_1 - {\bf q}_2 - {\bf k} ) }
{ E - \varepsilon({\bf k}) - \varepsilon({\bf q}_1) - \varepsilon({\bf q}_2) - 
      \varepsilon({\bf P} - {\bf q}_1 - {\bf q}_2 - {\bf k} ) }         
\nonumber \\
                           &   & \hspace{0.1cm} - \frac{V}{N} \sum_{\bf k} 
\frac{ f({\bf k}) \, ( \cos{q_{2y}} - \cos{q_{1y}} ) \, C_{y13}( {\bf k} , {\bf P} - {\bf q}_1 - {\bf q}_2 - {\bf k} ) }
{ E - \varepsilon({\bf k}) - \varepsilon({\bf q}_1) - \varepsilon({\bf q}_2) - 
      \varepsilon({\bf P} - {\bf q}_1 - {\bf q}_2 - {\bf k} ) }           
\nonumber \\
                           &   & \hspace{0.1cm} - \frac{V}{N} \sum_{\bf k} 
\frac{ f({\bf k}) \, ( \sin{q_{2y}} - \sin{q_{1y}} ) \, S_{y13}( {\bf k} , {\bf P} - {\bf q}_1 - {\bf q}_2 - {\bf k} ) }
{ E - \varepsilon({\bf k}) - \varepsilon({\bf q}_1) - \varepsilon({\bf q}_2) - 
      \varepsilon({\bf P} - {\bf q}_1 - {\bf q}_2 - {\bf k} ) } \: ,         
\label{4UV2d:eq:sm:sixtytwo}    
\end{eqnarray}
where $f({\bf k}) \equiv 1$. The equations for $C_{x14}$, $S_{x14}$, $C_{y14}$, and $S_{y14}$ have the same right-hand-side as Eq.~(\ref{4UV2d:eq:sm:sixtytwo}) but with $f({\bf k}) = \cos{k_x}$, $\sin{k_x}$, $\cos{k_y}$, and $\sin{k_y}$, respectively. 

Likewise, all four equations for the $\{ 13 \}$ functions are similar in the same way. We present only the equation for $C_{x13}$:
\begin{eqnarray}
C_{x13}( {\bf q}_1 , {\bf q}_2 ) = \hspace{0.2cm} 
& & \hspace{0.4cm} \frac{U}{N} \sum_{\bf k} 
\frac{ f({\bf k}) \, A_{14}( {\bf q}_1 , {\bf P} - {\bf q}_1 - {\bf q}_2 - {\bf k} ) + 
       f({\bf k}) \, A_{14}( {\bf P} - {\bf q}_1 - {\bf q}_2 - {\bf k} , {\bf k}   ) + 
       f({\bf k}) \, A_{14}( {\bf k}   , {\bf q}_1 ) }
{ E - \varepsilon({\bf k}) - \varepsilon({\bf q}_1) - \varepsilon({\bf q}_2) - 
      \varepsilon({\bf P} - {\bf q}_1 - {\bf q}_2 - {\bf k} ) }               
\nonumber \\
                           &   & - \frac{2V}{N} \sum_{\bf k} 
\frac{ f({\bf k}) \, \cos{k_x} \, C_{x14}( {\bf q}_1 , {\bf P} - {\bf q}_1 - {\bf q}_2 - {\bf k} ) + 
       f({\bf k}) \, \sin{k_x} \, S_{x14}( {\bf q}_1 , {\bf P} - {\bf q}_1 - {\bf q}_2 - {\bf k} )}
{ E - \varepsilon({\bf k}) - \varepsilon({\bf q}_1) - \varepsilon({\bf q}_2) - 
      \varepsilon({\bf P} - {\bf q}_1 - {\bf q}_2 - {\bf k} ) }               
\nonumber \\
                           &   & - \frac{2V}{N} \sum_{\bf k} 
\frac{ f({\bf k}) \, \cos{k_y} \, C_{y14}( {\bf q}_1 , {\bf P} - {\bf q}_1 - {\bf q}_2 - {\bf k} ) + 
       f({\bf k}) \, \sin{k_y} \, S_{y14}( {\bf q}_1 , {\bf P} - {\bf q}_1 - {\bf q}_2 - {\bf k} )}
{ E - \varepsilon({\bf k}) - \varepsilon({\bf q}_1) - \varepsilon({\bf q}_2) - 
      \varepsilon({\bf P} - {\bf q}_1 - {\bf q}_2 - {\bf k} ) }     
\nonumber \\
                           &   & - \frac{2V}{N} \sum_{\bf k} 
\frac{ f({\bf k}) \, \cos{q_{1x}} \, C_{x14}( {\bf P} - {\bf q}_1 - {\bf q}_2 - {\bf k} , {\bf k} ) + 
       f({\bf k}) \, \sin{q_{1x}} \, S_{x14}( {\bf P} - {\bf q}_1 - {\bf q}_2 - {\bf k} , {\bf k} )}
{ E - \varepsilon({\bf k}) - \varepsilon({\bf q}_1) - \varepsilon({\bf q}_2) - 
      \varepsilon({\bf P} - {\bf q}_1 - {\bf q}_2 - {\bf k} ) }   
\nonumber \\
                           &   & - \frac{2V}{N} \sum_{\bf k} 
\frac{ f({\bf k}) \, \cos{q_{1y}} \, C_{y14}( {\bf P} - {\bf q}_1 - {\bf q}_2 - {\bf k} , {\bf k} ) + 
       f({\bf k}) \, \sin{q_{1y}} \, S_{y14}( {\bf P} - {\bf q}_1 - {\bf q}_2 - {\bf k} , {\bf k} )}
{ E - \varepsilon({\bf k}) - \varepsilon({\bf q}_1) - \varepsilon({\bf q}_2) - 
      \varepsilon({\bf P} - {\bf q}_1 - {\bf q}_2 - {\bf k} ) }                                                                
\nonumber \\
                           &   & - \frac{2V}{N} \sum_{\bf k} 
\frac{ f({\bf k}) \, \cos{ ( P_x - q_{1x} - q_{2x} - k_x ) } \, C_{x14}( {\bf k} , {\bf q}_1 ) + 
       f({\bf k}) \, \sin{ ( P_x - q_{1x} - q_{2x} - k_x ) } \, S_{x14}( {\bf k} , {\bf q}_1 )  }
{ E - \varepsilon({\bf k}) - \varepsilon({\bf q}_1) - \varepsilon({\bf q}_2) - 
      \varepsilon({\bf P} - {\bf q}_1 - {\bf q}_2 - {\bf k} ) }         
\nonumber \\
                           &   & - \frac{2V}{N} \sum_{\bf k} 
\frac{ f({\bf k}) \, \cos{ ( P_y - q_{1y} - q_{2y} - k_y ) } \, C_{y14}( {\bf k} , {\bf q}_1 ) + 
       f({\bf k}) \, \sin{ ( P_y - q_{1y} - q_{2y} - k_y ) } \, S_{y14}( {\bf k} , {\bf q}_1 )  }
{ E - \varepsilon({\bf k}) - \varepsilon({\bf q}_1) - \varepsilon({\bf q}_2) - 
      \varepsilon({\bf P} - {\bf q}_1 - {\bf q}_2 - {\bf k} ) }              
\nonumber \\
                           &   & \hspace{0.1cm} - \frac{V}{N} \sum_{\bf k} 
\frac{ f({\bf k}) \, [ \cos{q_{1x}} - \cos{k_x} ] \, C_{x13}( {\bf P} - {\bf q}_1 - {\bf q}_2 - {\bf k} , {\bf q}_2 ) }
{ E - \varepsilon({\bf k}) - \varepsilon({\bf q}_1) - \varepsilon({\bf q}_2) - 
      \varepsilon({\bf P} - {\bf q}_1 - {\bf q}_2 - {\bf k} ) }           
\nonumber \\
                           &   & \hspace{0.1cm} - \frac{V}{N} \sum_{\bf k} 
\frac{ f({\bf k}) \, [ \sin{q_{1x}} - \sin{k_x} ] \, S_{x13}( {\bf P} - {\bf q}_1 - {\bf q}_2 - {\bf k} , {\bf q}_2 ) }
{ E - \varepsilon({\bf k}) - \varepsilon({\bf q}_1) - \varepsilon({\bf q}_2) - 
      \varepsilon({\bf P} - {\bf q}_1 - {\bf q}_2 - {\bf k} ) }                   
\nonumber \\
                           &   & \hspace{0.1cm} - \frac{V}{N} \sum_{\bf k} 
\frac{ f({\bf k}) \, [ \cos{q_{1y}} - \cos{k_y} ] \, C_{y13}( {\bf P} - {\bf q}_1 - {\bf q}_2 - {\bf k} , {\bf q}_2 ) }
{ E - \varepsilon({\bf k}) - \varepsilon({\bf q}_1) - \varepsilon({\bf q}_2) - 
      \varepsilon({\bf P} - {\bf q}_1 - {\bf q}_2 - {\bf k} ) }           
\nonumber \\
                           &   & \hspace{0.1cm} - \frac{V}{N} \sum_{\bf k} 
\frac{ f({\bf k}) \, [ \sin{q_{1y}} - \sin{k_y} ] \, S_{y13}( {\bf P} - {\bf q}_1 - {\bf q}_2 - {\bf k} , {\bf q}_2 ) }
{ E - \varepsilon({\bf k}) - \varepsilon({\bf q}_1) - \varepsilon({\bf q}_2) - 
      \varepsilon({\bf P} - {\bf q}_1 - {\bf q}_2 - {\bf k} ) }                    
\nonumber \\
                           &   & \hspace{0.1cm} - \frac{V}{N} \sum_{\bf k} 
\frac{ f({\bf k}) \, [ \cos{k_x} - \cos{ ( P_x - q_{1x} - q_{2x} - k_x ) } ] \, C_{x13}( {\bf q}_1 , {\bf q}_2 ) }
{ E - \varepsilon({\bf k}) - \varepsilon({\bf q}_1) - \varepsilon({\bf q}_2) - 
      \varepsilon({\bf P} - {\bf q}_1 - {\bf q}_2 - {\bf k} ) }           
\nonumber \\
                           &   & \hspace{0.1cm} - \frac{V}{N} \sum_{\bf k} 
\frac{ f({\bf k}) \, [ \sin{k_x} - \sin{ ( P_x - q_{1x} - q_{2x} - k_x ) } ] \, S_{x13}( {\bf q}_1 , {\bf q}_2 ) }
{ E - \varepsilon({\bf k}) - \varepsilon({\bf q}_1) - \varepsilon({\bf q}_2) - 
      \varepsilon({\bf P} - {\bf q}_1 - {\bf q}_2 - {\bf k} ) }     
\nonumber \\
                           &   & \hspace{0.1cm} - \frac{V}{N} \sum_{\bf k} 
\frac{ f({\bf k}) \, [ \cos{k_y} - \cos{ ( P_y - q_{1y} - q_{2y} - k_y ) } ] \, C_{y13}( {\bf q}_1 , {\bf q}_2 ) }
{ E - \varepsilon({\bf k}) - \varepsilon({\bf q}_1) - \varepsilon({\bf q}_2) - 
      \varepsilon({\bf P} - {\bf q}_1 - {\bf q}_2 - {\bf k} ) }           
\nonumber \\
                           &   & \hspace{0.1cm} - \frac{V}{N} \sum_{\bf k} 
\frac{ f({\bf k}) \, [ \sin{k_y} - \sin{ ( P_y - q_{1y} - q_{2y} - k_y ) } ] \, S_{y13}( {\bf q}_1 , {\bf q}_2 )  }
{ E - \varepsilon({\bf k}) - \varepsilon({\bf q}_1) - \varepsilon({\bf q}_2) - 
      \varepsilon({\bf P} - {\bf q}_1 - {\bf q}_2 - {\bf k} ) }         
\nonumber \\
                           &   & \hspace{0.1cm} - \frac{V}{N} \sum_{\bf k} 
\frac{ f({\bf k}) \, [ \cos{ ( P_x - q_{1x} - q_{2x} - k_x ) } - \cos{q_{1x}} ] \, C_{x13}( {\bf k} , {\bf q}_2 ) }
{ E - \varepsilon({\bf k}) - \varepsilon({\bf q}_1) - \varepsilon({\bf q}_2) - 
      \varepsilon({\bf P} - {\bf q}_1 - {\bf q}_2 - {\bf k} ) }           
\nonumber \\
                           &   & \hspace{0.1cm} - \frac{V}{N} \sum_{\bf k} 
\frac{ f({\bf k}) \, [ \sin{ ( P_x - q_{1x} - q_{2x} - k_x ) } - \sin{q_{1x}} ] \, S_{x13}( {\bf k} , {\bf q}_2 ) }
{ E - \varepsilon({\bf k}) - \varepsilon({\bf q}_1) - \varepsilon({\bf q}_2) - 
      \varepsilon({\bf P} - {\bf q}_1 - {\bf q}_2 - {\bf k} ) }         
\nonumber \\
                           &   & \hspace{0.1cm} - \frac{V}{N} \sum_{\bf k} 
\frac{ f({\bf k}) \, [ \cos{ ( P_y - q_{1y} - q_{2y} - k_y ) } - \cos{q_{1y}} ] \, C_{y13}( {\bf k} , {\bf q}_2 ) }
{ E - \varepsilon({\bf k}) - \varepsilon({\bf q}_1) - \varepsilon({\bf q}_2) - 
      \varepsilon({\bf P} - {\bf q}_1 - {\bf q}_2 - {\bf k} ) }           
\nonumber \\
                           &   & \hspace{0.1cm} - \frac{V}{N} \sum_{\bf k} 
\frac{ f({\bf k}) \, [ \sin{ ( P_y - q_{1y} - q_{2y} - k_y ) } - \sin{q_{1y}} ] \, S_{y13}( {\bf k} , {\bf q}_2 ) }
{ E - \varepsilon({\bf k}) - \varepsilon({\bf q}_1) - \varepsilon({\bf q}_2) - 
      \varepsilon({\bf P} - {\bf q}_1 - {\bf q}_2 - {\bf k} ) } \: ,         
\label{4UV2d:eq:sm:sixtythree}    
\end{eqnarray}
where $f({\bf k}) = \cos{k_x}$. The equations for $S_{x13}$, $C_{y13}$, and $S_{y13}$ have the same right-hand-side as Eq.~(\ref{4UV2d:eq:sm:sixtythree}) but with $f({\bf k}) = \sin{k_x}$, $\cos{k_y}$, and $\sin{k_y}$, respectively.

\subsection{\label{4UV2d:sec:sm:h}
Four $s = \frac{1}{2}$ fermions in 1D. $S = 0$
}

Summation over ${\bf b} = \pm {\bf x}$ in Eq.~(\ref{4UV2d:eq:thirtyfour}) yields terms like $2 \cos{(k-q)}$, etc. One needs to introduce 7 auxiliary functions: 
\begin{eqnarray}
A_{13}(q_1,q_2) & = & 
\frac{1}{N} \sum_{k} \hspace{0.8cm} \Psi_{0}( k , q_1 ; P - q_1 - q_2 - k , q_2 ) \: ,  
\label{4UV2d:eq:sm:sixtyfour}     \\  
C_{12}(q_1,q_2) & = & 
\frac{1}{N} \sum_{k} \cos{k} \, \Psi_{0}( k , P - q_1 - q_2 - k ; q_1 , q_2 ) \: ;
\hspace{1.0cm}  C_{12}(q_2,q_1) = - C_{12}(q_1,q_2) \: ,  
\label{4UV2d:eq:sm:sixtyfive}     \\  
S_{12}(q_1,q_2) & = & 
\frac{1}{N} \sum_{k} \sin{k} \, \Psi_{0}( k , P - q_1 - q_2 - k ; q_1 , q_2 ) \: ;
\hspace{1.0cm}  S_{12}(q_2,q_1) = - S_{12}(q_1,q_2) \: ,  
\label{4UV2d:eq:sm:sixtysix}      \\
C_{13}(q_1,q_2) & = & 
\frac{1}{N} \sum_{k} \cos{k} \, \Psi_{0}( k , q_1 ; P - q_1 - q_2 - k , q_2 ) \: ,  
\label{4UV2d:eq:sm:sixtyseven}    \\  
S_{13}(q_1,q_2) & = & 
\frac{1}{N} \sum_{k} \sin{k} \, \Psi_{0}( k , q_1 ; P - q_1 - q_2 - k , q_2 ) \: ,  
\label{4UV2d:eq:sm:sixtyeight}    \\
C_{34}(q_1,q_2) & = & 
\frac{1}{N} \sum_{k} \cos{k} \, \Psi_{0}( q_1 , q_2 ; k , P - q_1 - q_2 - k ) \: ;
\hspace{1.0cm}  C_{34}(q_2,q_1) = - C_{34}(q_1,q_2) \: ,  
\label{4UV2d:eq:sm:sixtynine}     \\  
S_{34}(q_1,q_2) & = & 
\frac{1}{N} \sum_{k} \sin{k} \, \Psi_{0}( q_1 , q_2 ; k , P - q_1 - q_2 - k ) \: ;
\hspace{1.0cm}  S_{34}(q_2,q_1) = - S_{34}(q_1,q_2) \: .  
\label{4UV2d:eq:sm:seventy}   
\end{eqnarray}
Expanding the cosines in Eq.~(\ref{4UV2d:eq:thirtyfour}), the wave function is expressed as follows:
\begin{eqnarray}
\Psi_{0}( q_1 , q_2 ; q_3 , q_4 ) = 
& & \hspace{0.5cm} U \: 
\frac{ A_{13}( q_2 , q_4 ) - A_{13}( q_2 , q_3 ) - A_{13}( q_1 , q_4 ) + A_{13}( q_1 , q_3 ) }
{ E - \varepsilon(q_1) - \varepsilon(q_2) - \varepsilon(q_3) - \varepsilon(q_4) }  
\nonumber \\                 
& & \hspace{0.2cm} - \: V \: 
\frac{ ( \cos{q_1} - \cos{q_2} ) \, C_{12}( q_3 , q_4 ) + ( \sin{q_1} - \sin{q_2} ) \, S_{12}( q_3 , q_4 ) }
{ E - \varepsilon(q_1) - \varepsilon(q_2) - \varepsilon(q_3) - \varepsilon(q_4) }                     
\nonumber \\
& & - \: 2 V \: 
\frac{ \cos{q_1} \, C_{13}( q_2 , q_4 ) + \sin{q_1} \, S_{13}( q_2 , q_4 ) 
     - \cos{q_1} \, C_{13}( q_2 , q_3 ) - \sin{q_1} \, S_{13}( q_2 , q_3 ) }
{ E - \varepsilon(q_1) - \varepsilon(q_2) - \varepsilon(q_3) - \varepsilon(q_4) }                     
\nonumber \\ 
& & - \: 2 V \: 
\frac{ - \cos{q_2} \, C_{13}( q_1 , q_4 ) - \sin{q_2} \, S_{13}( q_1 , q_4 ) 
       + \cos{q_2} \, C_{13}( q_1 , q_3 ) + \sin{q_2} \, S_{13}( q_1 , q_3 ) }
{ E - \varepsilon(q_1) - \varepsilon(q_2) - \varepsilon(q_3) - \varepsilon(q_4) }    
\nonumber \\                             
& & \hspace{0.2cm} - \: V \: 
\frac{ ( \cos{q_3} - \cos{q_4} ) \, C_{34}( q_1 , q_2 ) + ( \sin{q_3} - \sin{q_4} ) \, S_{34}( q_1 , q_2 ) }
{ E - \varepsilon(q_1) - \varepsilon(q_2) - \varepsilon(q_3) - \varepsilon(q_4) } \: .  
\label{4UV2d:eq:sm:seventyone}        
\end{eqnarray}
Substitution of wave function, Eq.~(\ref{4UV2d:eq:sm:seventyone}), into definitions, Eqs.~(\ref{4UV2d:eq:sm:sixtyfour})-(\ref{4UV2d:eq:sm:seventy}), produces seven coupled integral equations that split in three groups. The first group includes equations for $C_{12}$ and $S_{12}$. The $C_{12}$ equation reads  
\begin{eqnarray}
C_{12}( q_1 , q_2 ) =           
& & \hspace{0.2cm} \frac{U}{N} \sum_{k}  
\frac{ f(k) \, A_{13}( P - q_1 - q_2 - k , q_2 ) - f(k) \, A_{13}( P - q_1 - q_2 - k , q_1 ) 
     - f(k) \, A_{13}( k , q_2 ) + f(k) \, A_{13}( k , q_1 ) }
{ E - \varepsilon(k) - \varepsilon(q_1) - \varepsilon(q_2) - \varepsilon( P - q_1 - q_2 - k ) } 
\nonumber \\
& & \hspace{0.1cm} - \frac{V}{N} \sum_{k}  
\frac{ f(k) \, [ \cos{k} - \cos{( P - q_1 - q_2 - k )} ] \, C_{12}( q_1 , q_2 ) + 
       f(k) \, [ \sin{k} - \sin{( P - q_1 - q_2 - k )} ] \, S_{12}( q_1 , q_2 ) }
{ E - \varepsilon(k) - \varepsilon(q_1) - \varepsilon(q_2) - \varepsilon( P - q_1 - q_2 - k ) } 
\nonumber \\
& & - \frac{2V}{N} \sum_{k}  
\frac{ f(k) \cos{k} \, C_{13}( P - q_1 - q_2 - k , q_2 ) 
     + f(k) \sin{k} \, S_{13}( P - q_1 - q_2 - k , q_2 ) }
{ E - \varepsilon(k) - \varepsilon(q_1) - \varepsilon(q_2) - \varepsilon( P - q_1 - q_2 - k ) } 
\nonumber \\
& & - \frac{2V}{N} \sum_{k}  
\frac{ - f(k) \cos{k} \, C_{13}( P - q_1 - q_2 - k , q_1 ) 
       - f(k) \sin{k} \, S_{13}( P - q_1 - q_2 - k , q_1 ) }
{ E - \varepsilon(k) - \varepsilon(q_1) - \varepsilon(q_2) - \varepsilon( P - q_1 - q_2 - k ) } 
\nonumber \\
& & - \frac{2V}{N} \sum_{k}  
\frac{ - f(k) \cos{( P - q_1 - q_2 - k )} \, C_{13}( k , q_2 ) 
       - f(k) \sin{( P - q_1 - q_2 - k )} \, S_{13}( k , q_2 ) }
{ E - \varepsilon(k) - \varepsilon(q_1) - \varepsilon(q_2) - \varepsilon( P - q_1 - q_2 - k ) } 
\nonumber \\
& & - \frac{2V}{N} \sum_{k}  
\frac{ f(k) \cos{( P - q_1 - q_2 - k )} \, C_{13}( k , q_1 ) 
     + f(k) \sin{( P - q_1 - q_2 - k )} \, S_{13}( k , q_1 ) }
{ E - \varepsilon(k) - \varepsilon(q_1) - \varepsilon(q_2) - \varepsilon( P - q_1 - q_2 - k ) } 
\nonumber \\
& & \hspace{0.1cm} - \frac{V}{N} \sum_{k}  
\frac{ f(k) \, [ \cos{q_1} - \cos{q_2} ] \, C_{34}( k , P - q_1 - q_2 - k ) + 
       f(k) \, [ \sin{q_1} - \sin{q_2} ] \, S_{34}( k , P - q_1 - q_2 - k ) }
{ E - \varepsilon(k) - \varepsilon(q_1) - \varepsilon(q_2) - \varepsilon( P - q_1 - q_2 - k ) } \: ,
\nonumber \\
& & 
\label{4UV2d:eq:sm:seventytwo}    
\end{eqnarray}
with $f(k) = \cos{k}$. The equation for $S_{12}$ has the same right-hand-side but with $f(k) = \sin{k}$. 

The second group includes equations for $A_{13}$, $C_{13}$, and $S_{13}$. The first one reads:
\begin{eqnarray}
A_{13}( q_1 , q_2 ) =           
& & \hspace{0.2cm} \frac{U}{N} \sum_{k}  
\frac{ f(k) \, A_{13}( q_1 , q_2 ) - f(k) \, A_{13}( q_1 , P - q_1 - q_2 - k ) 
     - f(k) \, A_{13}( k   , q_2 ) + f(k) \, A_{13}( k   , P - q_1 - q_2 - k ) }
{ E - \varepsilon(k) - \varepsilon(q_1) - \varepsilon(q_2) - \varepsilon( P - q_1 - q_2 - k ) } 
\nonumber \\
& & \hspace{0.1cm} - \frac{V}{N} \sum_{k}  
\frac{ f(k) \, [ \cos{k} - \cos{q_1} ] \, C_{12}( P - q_1 - q_2 - k , q_2 ) + 
       f(k) \, [ \sin{k} - \sin{q_1} ] \, S_{12}( P - q_1 - q_2 - k , q_2 ) }
{ E - \varepsilon(k) - \varepsilon(q_1) - \varepsilon(q_2) - \varepsilon( P - q_1 - q_2 - k ) } 
\nonumber \\
& & - \frac{2V}{N} \sum_{k}  
\frac{ f(k) \cos{k} \, C_{13}( q_1 , q_2 ) 
     + f(k) \sin{k} \, S_{13}( q_1 , q_2 ) }
{ E - \varepsilon(k) - \varepsilon(q_1) - \varepsilon(q_2) - \varepsilon( P - q_1 - q_2 - k ) } 
\nonumber \\
& & - \frac{2V}{N} \sum_{k}  
\frac{ - f(k) \cos{k} \, C_{13}( q_1 , P - q_1 - q_2 - k ) 
       - f(k) \sin{k} \, S_{13}( q_1 , P - q_1 - q_2 - k ) }
{ E - \varepsilon(k) - \varepsilon(q_1) - \varepsilon(q_2) - \varepsilon( P - q_1 - q_2 - k ) } 
\nonumber \\
& & - \frac{2V}{N} \sum_{k}  
\frac{ - f(k) \cos{q_1} \, C_{13}( k , q_2 ) 
       - f(k) \sin{q_1} \, S_{13}( k , q_2 ) }
{ E - \varepsilon(k) - \varepsilon(q_1) - \varepsilon(q_2) - \varepsilon( P - q_1 - q_2 - k ) } 
\nonumber \\
& & - \frac{2V}{N} \sum_{k}  
\frac{ f(k) \cos{q_1} \, C_{13}( k , P - q_1 - q_2 - k ) 
     + f(k) \sin{q_1} \, S_{13}( k , P - q_1 - q_2 - k ) }
{ E - \varepsilon(k) - \varepsilon(q_1) - \varepsilon(q_2) - \varepsilon( P - q_1 - q_2 - k ) } 
\nonumber \\
& & \hspace{0.1cm} - \frac{V}{N} \sum_{k}  
\frac{ f(k) \, [ \cos{( P - q_1 - q_2 - k ) } - \cos{q_2} ] \, C_{34}( k , q_1 ) + 
       f(k) \, [ \sin{( P - q_1 - q_2 - k ) } - \sin{q_2} ] \, S_{34}( k , q_1 ) }
{ E - \varepsilon(k) - \varepsilon(q_1) - \varepsilon(q_2) - \varepsilon( P - q_1 - q_2 - k ) } \: ,
\nonumber \\
& & 
\label{4UV2d:eq:sm:seventythree}    
\end{eqnarray}
where $f(k) \equiv 1$. The equations for $C_{13}$ and $S_{13}$ have the same right-hand-side as Eq.~(\ref{4UV2d:eq:sm:seventythree}), but with $f(k) = \cos{k}$ and $f(k) = \sin{k}$, respectively.  

Finally, the third group includes equations for $C_{34}$ and $S_{34}$. The $C_{34}$ equation reads
\begin{eqnarray}
C_{34}( q_1 , q_2 ) =           
& & \hspace{0.2cm} \frac{U}{N} \sum_{k}  
\frac{ f(k) \, A_{13}( q_2 , P - q_1 - q_2 - k ) - f(k) \, A_{13}( q_2 , k ) 
     - f(k) \, A_{13}( q_1 , P - q_1 - q_2 - k ) + f(k) \, A_{13}( q_1 , k ) }
{ E - \varepsilon(k) - \varepsilon(q_1) - \varepsilon(q_2) - \varepsilon( P - q_1 - q_2 - k ) } 
\nonumber \\
& & \hspace{0.1cm} - \frac{V}{N} \sum_{k}  
\frac{ f(k) \, [ \cos{q_1} - \cos{q_2} ] \, C_{12}( k , P - q_1 - q_2 - k ) + 
       f(k) \, [ \sin{q_1} - \sin{q_2} ] \, S_{12}( k , P - q_1 - q_2 - k ) }
{ E - \varepsilon(k) - \varepsilon(q_1) - \varepsilon(q_2) - \varepsilon( P - q_1 - q_2 - k ) } 
\nonumber \\
& & - \frac{2V}{N} \sum_{k}  
\frac{ f(k) \cos{q_1} \, C_{13}( q_2 , P - q_1 - q_2 - k ) 
     + f(k) \sin{q_1} \, S_{13}( q_2 , P - q_1 - q_2 - k ) }
{ E - \varepsilon(k) - \varepsilon(q_1) - \varepsilon(q_2) - \varepsilon( P - q_1 - q_2 - k ) } 
\nonumber \\
& & - \frac{2V}{N} \sum_{k}  
\frac{ - f(k) \cos{q_1} \, C_{13}( q_2 , k ) 
       - f(k) \sin{q_1} \, S_{13}( q_2 , k ) }
{ E - \varepsilon(k) - \varepsilon(q_1) - \varepsilon(q_2) - \varepsilon( P - q_1 - q_2 - k ) } 
\nonumber \\
& & - \frac{2V}{N} \sum_{k}  
\frac{ - f(k) \cos{q_2} \, C_{13}( q_1 , P - q_1 - q_2 - k ) 
       - f(k) \sin{q_2} \, S_{13}( q_1 , P - q_1 - q_2 - k ) }
{ E - \varepsilon(k) - \varepsilon(q_1) - \varepsilon(q_2) - \varepsilon( P - q_1 - q_2 - k ) } 
\nonumber \\
& & - \frac{2V}{N} \sum_{k}  
\frac{ f(k) \cos{q_2} \, C_{13}( q_1 , k ) 
     + f(k) \sin{q_2} \, S_{13}( q_1 , k ) }
{ E - \varepsilon(k) - \varepsilon(q_1) - \varepsilon(q_2) - \varepsilon( P - q_1 - q_2 - k ) } 
\nonumber \\
& & \hspace{0.1cm} - \frac{V}{N} \sum_{k}  
\frac{ f(k) \, [ \cos{k} - \cos{( P - q_1 - q_2 - k ) } ] \, C_{34}( q_1 , q_2 ) + 
       f(k) \, [ \sin{k} - \sin{( P - q_1 - q_2 - k ) } ] \, S_{34}( q_1 , q_2 ) }
{ E - \varepsilon(k) - \varepsilon(q_1) - \varepsilon(q_2) - \varepsilon( P - q_1 - q_2 - k ) } \: ,
\nonumber \\
& & 
\label{4UV2d:eq:sm:seventyfour}    
\end{eqnarray}
where $f(k) = \cos{k}$. The equation for $S_{34}$ has the same right-hand-side but with $f(k) = \sin{k}$.

\subsection{\label{4UV2d:sec:sm:i}
Four $s = \frac{1}{2}$ fermions in 2D. $S = 0$
}

Starting with Eq.~(\ref{4UV2d:eq:thirtyfour}), we first sum over nearest neighbor vectors ${\bf b} = \pm {\bf x}, \pm {\bf y}$ to get factors $2 \cos{(k_x - q_x)} + 2 \cos{(k_y - q_y)}$. Then, we introduce 13 auxiliary functions
\begin{eqnarray}
A_{13} ( {\bf q}_1 , {\bf q}_2 ) & = & \frac{1}{N} \sum_{\bf k} 
\hspace{0.9cm} \Psi_{0}( {\bf k} , {\bf q}_1 ; {\bf P} - {\bf q}_1 - {\bf q}_2 - {\bf k} , {\bf q}_2 ) \: , 
\label{4UV2d:eq:sm:seventyfive}    \\
C_{x12}( {\bf q}_1 , {\bf q}_2 ) & = & \frac{1}{N} \sum_{\bf k} 
\cos{q_x}    \Psi_{0}( {\bf k} , {\bf P} - {\bf q}_1 - {\bf q}_2 - {\bf k} ; {\bf q}_1 , {\bf q}_2 ) \: ;   
\hspace{0.5cm}   C_{x12}( {\bf q}_2 , {\bf q}_1 ) = - C_{x12}( {\bf q}_1 , {\bf q}_2 )  \: , 
\label{4UV2d:eq:sm:seventysix}     \\
S_{x12}( {\bf q}_1 , {\bf q}_2 ) & = & \frac{1}{N} \sum_{\bf k} 
\sin{q_x}    \Psi_{0}( {\bf k} , {\bf P} - {\bf q}_1 - {\bf q}_2 - {\bf k} ; {\bf q}_1 , {\bf q}_2 ) \: ;
\hspace{0.5cm}   S_{x12}( {\bf q}_2 , {\bf q}_1 ) = - S_{x12}( {\bf q}_1 , {\bf q}_2 )  \: , 
\label{4UV2d:eq:sm:seventyseven}   \\
C_{y12}( {\bf q}_1 , {\bf q}_2 ) & = & \frac{1}{N} \sum_{\bf k} 
\cos{q_y}    \Psi_{0}( {\bf k} , {\bf P} - {\bf q}_1 - {\bf q}_2 - {\bf k} ; {\bf q}_1 , {\bf q}_2 ) \: ;
\hspace{0.5cm}   C_{y12}( {\bf q}_2 , {\bf q}_1 ) = - C_{y12}( {\bf q}_1 , {\bf q}_2 )  \: ,     
\label{4UV2d:eq:sm:seventyeight}   \\
S_{y12}( {\bf q}_1 , {\bf q}_2 ) & = & \frac{1}{N} \sum_{\bf k} 
\sin{q_y}    \Psi_{0}( {\bf k} , {\bf P} - {\bf q}_1 - {\bf q}_2 - {\bf k} ; {\bf q}_1 , {\bf q}_2 ) \: ;
\hspace{0.5cm}   S_{y12}( {\bf q}_2 , {\bf q}_1 ) = - S_{y12}( {\bf q}_1 , {\bf q}_2 )  \: ,     
\label{4UV2d:eq:sm:seventynine}    \\    
C_{x13}( {\bf q}_1 , {\bf q}_2 ) & = & \frac{1}{N} \sum_{\bf k} 
\cos{q_x}    \Psi_{0}( {\bf k} , {\bf q}_1 ; {\bf P} - {\bf q}_1 - {\bf q}_2 - {\bf k} , {\bf q}_2 ) \: ,   
\label{4UV2d:eq:sm:eighty}         \\    
S_{x13}( {\bf q}_1 , {\bf q}_2 ) & = & \frac{1}{N} \sum_{\bf k} 
\sin{q_x}    \Psi_{0}( {\bf k} , {\bf q}_1 ; {\bf P} - {\bf q}_1 - {\bf q}_2 - {\bf k} , {\bf q}_2 ) \: ,   
\label{4UV2d:eq:sm:eightyone}      \\    
C_{y13}( {\bf q}_1 , {\bf q}_2 ) & = & \frac{1}{N} \sum_{\bf k} 
\cos{q_y}    \Psi_{0}( {\bf k} , {\bf q}_1 ; {\bf P} - {\bf q}_1 - {\bf q}_2 - {\bf k} , {\bf q}_2 ) \: ,   
\label{4UV2d:eq:sm:eightytwo}      \\    
S_{y13}( {\bf q}_1 , {\bf q}_2 ) & = & \frac{1}{N} \sum_{\bf k} 
\sin{q_y}    \Psi_{0}( {\bf k} , {\bf q}_1 ; {\bf P} - {\bf q}_1 - {\bf q}_2 - {\bf k} , {\bf q}_2 ) \: ,   
\label{4UV2d:eq:sm:eightythree}    \\
C_{x34}( {\bf q}_1 , {\bf q}_2 ) & = & \frac{1}{N} \sum_{\bf k} 
\cos{q_x}    \Psi_{0}( {\bf q}_1 , {\bf q}_2 ; {\bf k} , {\bf P} - {\bf q}_1 - {\bf q}_2 - {\bf k} ) \: ;   
\hspace{0.5cm}   C_{x34}( {\bf q}_2 , {\bf q}_1 ) = - C_{x34}( {\bf q}_1 , {\bf q}_2 )  \: , 
\label{4UV2d:eq:sm:eightyfour}     \\
S_{x34}( {\bf q}_1 , {\bf q}_2 ) & = & \frac{1}{N} \sum_{\bf k} 
\sin{q_x}    \Psi_{0}( {\bf q}_1 , {\bf q}_2 ; {\bf k} , {\bf P} - {\bf q}_1 - {\bf q}_2 - {\bf k} ) \: ;   
\hspace{0.5cm}   S_{x34}( {\bf q}_2 , {\bf q}_1 ) = - S_{x34}( {\bf q}_1 , {\bf q}_2 )  \: , 
\label{4UV2d:eq:sm:eightyfive}     \\
C_{y34}( {\bf q}_1 , {\bf q}_2 ) & = & \frac{1}{N} \sum_{\bf k} 
\cos{q_y}    \Psi_{0}( {\bf q}_1 , {\bf q}_2 ; {\bf k} , {\bf P} - {\bf q}_1 - {\bf q}_2 - {\bf k} ) \: ;   
\hspace{0.5cm}   C_{y34}( {\bf q}_2 , {\bf q}_1 ) = - C_{y34}( {\bf q}_1 , {\bf q}_2 )  \: , 
\label{4UV2d:eq:sm:eightysix}      \\
S_{y34}( {\bf q}_1 , {\bf q}_2 ) & = & \frac{1}{N} \sum_{\bf k} 
\sin{q_y}    \Psi_{0}( {\bf q}_1 , {\bf q}_2 ; {\bf k} , {\bf P} - {\bf q}_1 - {\bf q}_2 - {\bf k} ) \: ;   
\hspace{0.5cm}   S_{y34}( {\bf q}_2 , {\bf q}_1 ) = - S_{y34}( {\bf q}_1 , {\bf q}_2 )  \: . 
\label{4UV2d:eq:sm:eightyseven}     
\end{eqnarray}
With these definitions, the wave function is expressed from Eq.~(\ref{4UV2d:eq:thirtyfour}) as follows
\begin{eqnarray}
\Psi_{0}( {\bf q}_1 , {\bf q}_2 ; {\bf q}_3 , {\bf q}_4 ) = \hspace{0.2cm}   
& &  \hspace{0.4cm}  U \:
\frac{ A_{13}( {\bf q}_2 , {\bf q}_4 ) - A_{13}( {\bf q}_2 , {\bf q}_3 ) 
     - A_{13}( {\bf q}_1 , {\bf q}_4 ) + A_{13}( {\bf q}_1 , {\bf q}_3 ) }
{ E - \varepsilon({\bf q}_1) - \varepsilon({\bf q}_2) - \varepsilon({\bf q}_3) - \varepsilon({\bf q}_4) }
\nonumber \\
& & \hspace{0.1cm} - V \: 
\frac{ ( \cos{q_{1x}} - \cos{q_{2x}} ) \, C_{x12}( {\bf q}_3 , {\bf q}_4 ) + 
       ( \sin{q_{1x}} - \sin{q_{2x}} ) \, S_{x12}( {\bf q}_3 , {\bf q}_4 ) }
{ E - \varepsilon({\bf q}_1) - \varepsilon({\bf q}_2) - \varepsilon({\bf q}_3) - \varepsilon({\bf q}_4) }
\nonumber \\
& & \hspace{0.1cm} - V \: 
\frac{ ( \cos{q_{1y}} - \cos{q_{2y}} ) \, C_{y12}( {\bf q}_3 , {\bf q}_4 ) + 
       ( \sin{q_{1y}} - \sin{q_{2y}} ) \, S_{y12}( {\bf q}_3 , {\bf q}_4 ) }
{ E - \varepsilon({\bf q}_1) - \varepsilon({\bf q}_2) - \varepsilon({\bf q}_3) - \varepsilon({\bf q}_4) }
\nonumber \\
& & - 2 V \: 
\frac{ \cos{q_{1x}} \, C_{x13}( {\bf q}_2 , {\bf q}_4 ) + 
       \sin{q_{1x}} \, S_{x13}( {\bf q}_2 , {\bf q}_4 ) + 
       \cos{q_{1y}} \, C_{y13}( {\bf q}_2 , {\bf q}_4 ) + 
       \sin{q_{1y}} \, S_{y13}( {\bf q}_2 , {\bf q}_4 ) }
{ E - \varepsilon({\bf q}_1) - \varepsilon({\bf q}_2) - \varepsilon({\bf q}_3) - \varepsilon({\bf q}_4) }
\nonumber \\
& & - 2 V \: 
\frac{ - \cos{q_{1x}} \, C_{x13}( {\bf q}_2 , {\bf q}_3 )  
       - \sin{q_{1x}} \, S_{x13}( {\bf q}_2 , {\bf q}_3 )  
       - \cos{q_{1y}} \, C_{y13}( {\bf q}_2 , {\bf q}_3 )  
       - \sin{q_{1y}} \, S_{y13}( {\bf q}_2 , {\bf q}_3 ) }
{ E - \varepsilon({\bf q}_1) - \varepsilon({\bf q}_2) - \varepsilon({\bf q}_3) - \varepsilon({\bf q}_4) }
\nonumber \\
& & - 2 V \: 
\frac{ - \cos{q_{2x}} \, C_{x13}( {\bf q}_1 , {\bf q}_4 )  
       - \sin{q_{2x}} \, S_{x13}( {\bf q}_1 , {\bf q}_4 )  
       - \cos{q_{2y}} \, C_{y13}( {\bf q}_1 , {\bf q}_4 )  
       - \sin{q_{2y}} \, S_{y13}( {\bf q}_1 , {\bf q}_4 ) }
{ E - \varepsilon({\bf q}_1) - \varepsilon({\bf q}_2) - \varepsilon({\bf q}_3) - \varepsilon({\bf q}_4) }
\nonumber \\
& & - 2 V \: 
\frac{ \cos{q_{2x}} \, C_{x13}( {\bf q}_1 , {\bf q}_3 ) + 
       \sin{q_{2x}} \, S_{x13}( {\bf q}_1 , {\bf q}_3 ) + 
       \cos{q_{2y}} \, C_{y13}( {\bf q}_1 , {\bf q}_3 ) + 
       \sin{q_{2y}} \, S_{y13}( {\bf q}_1 , {\bf q}_3 ) }
{ E - \varepsilon({\bf q}_1) - \varepsilon({\bf q}_2) - \varepsilon({\bf q}_3) - \varepsilon({\bf q}_4) }
\nonumber \\
& & \hspace{0.1cm} - V \: 
\frac{ ( \cos{q_{3x}} - \cos{q_{4x}} ) \, C_{x34}( {\bf q}_1 , {\bf q}_2 ) + 
       ( \sin{q_{3x}} - \sin{q_{4x}} ) \, S_{x34}( {\bf q}_1 , {\bf q}_2 ) }
{ E - \varepsilon({\bf q}_1) - \varepsilon({\bf q}_2) - \varepsilon({\bf q}_3) - \varepsilon({\bf q}_4) }
\nonumber \\
& & \hspace{0.1cm} - V \: 
\frac{ ( \cos{q_{3y}} - \cos{q_{4y}} ) \, C_{y34}( {\bf q}_1 , {\bf q}_2 ) + 
       ( \sin{q_{3y}} - \sin{q_{4y}} ) \, S_{y34}( {\bf q}_1 , {\bf q}_2 ) }
{ E - \varepsilon({\bf q}_1) - \varepsilon({\bf q}_2) - \varepsilon({\bf q}_3) - \varepsilon({\bf q}_4) } \: .
\label{4UV2d:eq:sm:eightyeight}    
\end{eqnarray}
Substitution of Eq.~(\ref{4UV2d:eq:sm:eightyeight}) in Eqs.~(\ref{4UV2d:eq:sm:seventyfive})-(\ref{4UV2d:eq:sm:eightyseven}) yields a system of 13 coupled integral equations that split in three groups. The first group includes the equations for $C_{x12}$, $S_{x12}$, $C_{y12}$, and $S_{y12}$. The equation for $C_{x12}$ reads 
\begin{eqnarray}
C_{x12}( {\bf q}_1 , {\bf q}_2 ) = \hspace{0.2cm} 
& & \hspace{0.4cm} \frac{U}{N} \sum_{\bf k} 
\frac{ f({\bf k}) \, A_{13}( {\bf P} - {\bf q}_1 - {\bf q}_2 - {\bf k} , {\bf q}_2 ) - 
       f({\bf k}) \, A_{13}( {\bf P} - {\bf q}_1 - {\bf q}_2 - {\bf k} , {\bf q}_1 ) }
{ E - \varepsilon({\bf k}) - \varepsilon({\bf q}_1) - \varepsilon({\bf q}_2) - 
      \varepsilon({\bf P} - {\bf q}_1 - {\bf q}_2 - {\bf k} ) }               
\nonumber \\
& & \hspace{0.1cm} + \frac{U}{N} \sum_{\bf k} 
\frac{ - f({\bf k}) \, A_{13}( {\bf k} , {\bf q}_2 ) + 
         f({\bf k}) \, A_{13}( {\bf k} , {\bf q}_1 ) }
{ E - \varepsilon({\bf k}) - \varepsilon({\bf q}_1) - \varepsilon({\bf q}_2) - 
      \varepsilon({\bf P} - {\bf q}_1 - {\bf q}_2 - {\bf k} ) }               
\nonumber \\
                           &   & \hspace{0.1cm} - \frac{V}{N} \sum_{\bf k} 
\frac{ f({\bf k}) \, [ \cos{k_x} - \cos{( P_x - q_{1x} - q_{2x} - k_x )} ] \, 
C_{x12}( {\bf q}_1 , {\bf q}_2 ) }
{ E - \varepsilon({\bf k}) - \varepsilon({\bf q}_1) - \varepsilon({\bf q}_2) - 
      \varepsilon({\bf P} - {\bf q}_1 - {\bf q}_2 - {\bf k} ) }               
\nonumber \\
                           &   & \hspace{0.1cm} - \frac{V}{N} \sum_{\bf k} 
\frac{ f({\bf k}) \, [ \sin{k_x} - \sin{( P_x - q_{1x} - q_{2x} - k_x )} ] \, 
S_{x12}( {\bf q}_1 , {\bf q}_2 ) }
{ E - \varepsilon({\bf k}) - \varepsilon({\bf q}_1) - \varepsilon({\bf q}_2) - 
      \varepsilon({\bf P} - {\bf q}_1 - {\bf q}_2 - {\bf k} ) }     
\nonumber \\
                           &   & \hspace{0.1cm} - \frac{V}{N} \sum_{\bf k} 
\frac{ f({\bf k}) \, [ \cos{k_y} - \cos{( P_y - q_{1y} - q_{2y} - k_y )} ] \, 
C_{y12}( {\bf q}_1 , {\bf q}_2 ) }
{ E - \varepsilon({\bf k}) - \varepsilon({\bf q}_1) - \varepsilon({\bf q}_2) - 
      \varepsilon({\bf P} - {\bf q}_1 - {\bf q}_2 - {\bf k} ) }               
\nonumber \\
                           &   & \hspace{0.1cm} - \frac{V}{N} \sum_{\bf k} 
\frac{ f({\bf k}) \, [ \sin{k_y} - \sin{( P_y - q_{1y} - q_{2y} - k_y )} ] \, 
S_{y12}( {\bf q}_1 , {\bf q}_2 ) }
{ E - \varepsilon({\bf k}) - \varepsilon({\bf q}_1) - \varepsilon({\bf q}_2) - 
      \varepsilon({\bf P} - {\bf q}_1 - {\bf q}_2 - {\bf k} ) }     
\nonumber \\
                           &   & - \frac{2V}{N} \sum_{\bf k} 
\frac{ f({\bf k}) \, \cos{k_{x}} \, C_{x13}( {\bf P} - {\bf q}_1 - {\bf q}_2 - {\bf k} , {\bf q}_2 ) + 
       f({\bf k}) \, \sin{k_{x}} \, S_{x13}( {\bf P} - {\bf q}_1 - {\bf q}_2 - {\bf k} , {\bf q}_2 )}
{ E - \varepsilon({\bf k}) - \varepsilon({\bf q}_1) - \varepsilon({\bf q}_2) - 
      \varepsilon({\bf P} - {\bf q}_1 - {\bf q}_2 - {\bf k} ) }   
\nonumber \\
                           &   & - \frac{2V}{N} \sum_{\bf k} 
\frac{ f({\bf k}) \, \cos{k_{y}} \, C_{y13}( {\bf P} - {\bf q}_1 - {\bf q}_2 - {\bf k} , {\bf q}_2 ) + 
       f({\bf k}) \, \sin{k_{y}} \, S_{y13}( {\bf P} - {\bf q}_1 - {\bf q}_2 - {\bf k} , {\bf q}_2 )}
{ E - \varepsilon({\bf k}) - \varepsilon({\bf q}_1) - \varepsilon({\bf q}_2) - 
      \varepsilon({\bf P} - {\bf q}_1 - {\bf q}_2 - {\bf k} ) }                                                                
\nonumber \\
                           &   & - \frac{2V}{N} \sum_{\bf k} 
\frac{ - f({\bf k}) \, \cos{k_x} \, C_{x13}( {\bf P} - {\bf q}_1 - {\bf q}_2 - {\bf k} , {\bf q}_1 )  
       - f({\bf k}) \, \sin{k_x} \, S_{x13}( {\bf P} - {\bf q}_1 - {\bf q}_2 - {\bf k} , {\bf q}_1 ) }
{ E - \varepsilon({\bf k}) - \varepsilon({\bf q}_1) - \varepsilon({\bf q}_2) - 
      \varepsilon({\bf P} - {\bf q}_1 - {\bf q}_2 - {\bf k} ) }         
\nonumber \\
                           &   & - \frac{2V}{N} \sum_{\bf k} 
\frac{ - f({\bf k}) \, \cos{k_y} \, C_{y13}( {\bf P} - {\bf q}_1 - {\bf q}_2 - {\bf k} , {\bf q}_1 )  
       - f({\bf k}) \, \sin{k_y} \, S_{y13}( {\bf P} - {\bf q}_1 - {\bf q}_2 - {\bf k} , {\bf q}_1 ) }
{ E - \varepsilon({\bf k}) - \varepsilon({\bf q}_1) - \varepsilon({\bf q}_2) - 
      \varepsilon({\bf P} - {\bf q}_1 - {\bf q}_2 - {\bf k} ) }         
\nonumber \\
                           &   & - \frac{2V}{N} \sum_{\bf k} 
\frac{ - f({\bf k}) \, \cos{ ( P_x - q_{1x} - q_{2x} - k_x ) } \, C_{x13}( {\bf k} , {\bf q}_2 ) 
       - f({\bf k}) \, \sin{ ( P_x - q_{1x} - q_{2x} - k_x ) } \, S_{x13}( {\bf k} , {\bf q}_2 ) }
{ E - \varepsilon({\bf k}) - \varepsilon({\bf q}_1) - \varepsilon({\bf q}_2) - 
      \varepsilon({\bf P} - {\bf q}_1 - {\bf q}_2 - {\bf k} ) }           
\nonumber \\
                           &   & - \frac{2V}{N} \sum_{\bf k} 
\frac{ - f({\bf k}) \, \cos{ ( P_y - q_{1y} - q_{2y} - k_y ) } \, C_{y13}( {\bf k} , {\bf q}_2 ) 
       - f({\bf k}) \, \sin{ ( P_y - q_{1y} - q_{2y} - k_y ) } \, S_{y13}( {\bf k} , {\bf q}_2 ) }
{ E - \varepsilon({\bf k}) - \varepsilon({\bf q}_1) - \varepsilon({\bf q}_2) - 
      \varepsilon({\bf P} - {\bf q}_1 - {\bf q}_2 - {\bf k} ) }           
\nonumber \\
                           &   & - \frac{2V}{N} \sum_{\bf k} 
\frac{   f({\bf k}) \, \cos{ ( P_x - q_{1x} - q_{2x} - k_x ) } \, C_{x13}( {\bf k} , {\bf q}_1 ) 
       + f({\bf k}) \, \sin{ ( P_x - q_{1x} - q_{2x} - k_x ) } \, S_{x13}( {\bf k} , {\bf q}_1 ) }
{ E - \varepsilon({\bf k}) - \varepsilon({\bf q}_1) - \varepsilon({\bf q}_2) - 
      \varepsilon({\bf P} - {\bf q}_1 - {\bf q}_2 - {\bf k} ) }           
\nonumber \\
                           &   & - \frac{2V}{N} \sum_{\bf k} 
\frac{   f({\bf k}) \, \cos{ ( P_y - q_{1y} - q_{2y} - k_y ) } \, C_{y13}( {\bf k} , {\bf q}_1 ) 
       + f({\bf k}) \, \sin{ ( P_y - q_{1y} - q_{2y} - k_y ) } \, S_{y13}( {\bf k} , {\bf q}_1 ) }
{ E - \varepsilon({\bf k}) - \varepsilon({\bf q}_1) - \varepsilon({\bf q}_2) - 
      \varepsilon({\bf P} - {\bf q}_1 - {\bf q}_2 - {\bf k} ) }           
\nonumber \\
                           &   & \hspace{0.1cm} - \frac{V}{N} \sum_{\bf k} 
\frac{ f({\bf k}) \, ( \cos{q_{1x}} - \cos{q_{2x}} ) \, 
C_{x34}( {\bf k} , {\bf P} - {\bf q}_1 - {\bf q}_2 - {\bf k} ) }
{ E - \varepsilon({\bf k}) - \varepsilon({\bf q}_1) - \varepsilon({\bf q}_2) - 
      \varepsilon({\bf P} - {\bf q}_1 - {\bf q}_2 - {\bf k} ) }           
\nonumber \\
                           &   & \hspace{0.1cm} - \frac{V}{N} \sum_{\bf k} 
\frac{ f({\bf k}) \, ( \sin{q_{1x}} - \sin{q_{2x}} ) \, 
S_{x34}( {\bf k} , {\bf P} - {\bf q}_1 - {\bf q}_2 - {\bf k} ) }
{ E - \varepsilon({\bf k}) - \varepsilon({\bf q}_1) - \varepsilon({\bf q}_2) - 
      \varepsilon({\bf P} - {\bf q}_1 - {\bf q}_2 - {\bf k} ) }     
\nonumber \\
                           &   & \hspace{0.1cm} - \frac{V}{N} \sum_{\bf k} 
\frac{ f({\bf k}) \, ( \cos{q_{1y}} - \cos{q_{2y}} ) \, 
C_{y34}( {\bf k} , {\bf P} - {\bf q}_1 - {\bf q}_2 - {\bf k} ) }
{ E - \varepsilon({\bf k}) - \varepsilon({\bf q}_1) - \varepsilon({\bf q}_2) - 
      \varepsilon({\bf P} - {\bf q}_1 - {\bf q}_2 - {\bf k} ) }           
\nonumber \\
                           &   & \hspace{0.1cm} - \frac{V}{N} \sum_{\bf k} 
\frac{ f({\bf k}) \, ( \sin{q_{1y}} - \sin{q_{2y}} ) \, 
S_{y34}( {\bf k} , {\bf P} - {\bf q}_1 - {\bf q}_2 - {\bf k} ) }
{ E - \varepsilon({\bf k}) - \varepsilon({\bf q}_1) - \varepsilon({\bf q}_2) - 
      \varepsilon({\bf P} - {\bf q}_1 - {\bf q}_2 - {\bf k} ) }   \: ,   
\label{4UV2d:eq:sm:eightynine}    
\end{eqnarray}
where $f({\bf k}) = \cos{k_x}$. The equations for $S_{x12}$, $C_{y12}$, and $S_{y12}$ have the same right-hand-side as Eq.~(\ref{4UV2d:eq:sm:eightynine}) but with $f({\bf k}) = \sin{k_x}$, $f({\bf k}) = \cos{k_y}$, and $f({\bf k}) = \sin{k_y}$, respectively. 

The second group includes the equations for $A_{13}$, $C_{x13}$, $S_{x13}$, $C_{y13}$, and $S_{y13}$. The equation for $A_{13}$ reads 
\begin{eqnarray}
A_{13}( {\bf q}_1 , {\bf q}_2 ) = \hspace{0.2cm} 
& & \hspace{0.4cm} \frac{U}{N} \sum_{\bf k} 
\frac{ f({\bf k}) \, A_{13}( {\bf q}_1 , {\bf q}_2 ) - 
       f({\bf k}) \, A_{13}( {\bf q}_1 , {\bf P} - {\bf q}_1 - {\bf q}_2 - {\bf k} ) }
{ E - \varepsilon({\bf k}) - \varepsilon({\bf q}_1) - \varepsilon({\bf q}_2) - 
      \varepsilon({\bf P} - {\bf q}_1 - {\bf q}_2 - {\bf k} ) }               
\nonumber \\
& & \hspace{0.1cm} + \frac{U}{N} \sum_{\bf k} 
\frac{ - f({\bf k}) \, A_{13}( {\bf k} , {\bf q}_2 ) + 
         f({\bf k}) \, A_{13}( {\bf k} , {\bf P} - {\bf q}_1 - {\bf q}_2 - {\bf k} ) }
{ E - \varepsilon({\bf k}) - \varepsilon({\bf q}_1) - \varepsilon({\bf q}_2) - 
      \varepsilon({\bf P} - {\bf q}_1 - {\bf q}_2 - {\bf k} ) }               
\nonumber \\
                           &   & \hspace{0.1cm} - \frac{V}{N} \sum_{\bf k} 
\frac{ f({\bf k}) \, ( \cos{k_x} - \cos{q_{1x}} ) \, 
C_{x12}( {\bf P} - {\bf q}_1 - {\bf q}_2 - {\bf k} , {\bf q}_2 ) }
{ E - \varepsilon({\bf k}) - \varepsilon({\bf q}_1) - \varepsilon({\bf q}_2) - 
      \varepsilon({\bf P} - {\bf q}_1 - {\bf q}_2 - {\bf k} ) }               
\nonumber \\
                           &   & \hspace{0.1cm} - \frac{V}{N} \sum_{\bf k} 
\frac{ f({\bf k}) \, ( \sin{k_x} - \sin{q_{1x}} ) \, 
S_{x12}( {\bf P} - {\bf q}_1 - {\bf q}_2 - {\bf k} , {\bf q}_2 ) }
{ E - \varepsilon({\bf k}) - \varepsilon({\bf q}_1) - \varepsilon({\bf q}_2) - 
      \varepsilon({\bf P} - {\bf q}_1 - {\bf q}_2 - {\bf k} ) }     
\nonumber \\
                           &   & \hspace{0.1cm} - \frac{V}{N} \sum_{\bf k} 
\frac{ f({\bf k}) \, ( \cos{k_y} - \cos{q_{1y}} ) \, 
C_{y12}( {\bf P} - {\bf q}_1 - {\bf q}_2 - {\bf k} , {\bf q}_2 ) }
{ E - \varepsilon({\bf k}) - \varepsilon({\bf q}_1) - \varepsilon({\bf q}_2) - 
      \varepsilon({\bf P} - {\bf q}_1 - {\bf q}_2 - {\bf k} ) }               
\nonumber \\
                           &   & \hspace{0.1cm} - \frac{V}{N} \sum_{\bf k} 
\frac{ f({\bf k}) \, ( \sin{k_y} - \sin{q_{1y}} ) \, 
S_{y12}( {\bf P} - {\bf q}_1 - {\bf q}_2 - {\bf k} , {\bf q}_2 ) }
{ E - \varepsilon({\bf k}) - \varepsilon({\bf q}_1) - \varepsilon({\bf q}_2) - 
      \varepsilon({\bf P} - {\bf q}_1 - {\bf q}_2 - {\bf k} ) }     
\nonumber \\
                           &   & - \frac{2V}{N} \sum_{\bf k} 
\frac{ f({\bf k}) \, \cos{k_{x}} \, C_{x13}( {\bf q}_1 , {\bf q}_2 ) + 
       f({\bf k}) \, \sin{k_{x}} \, S_{x13}( {\bf q}_1 , {\bf q}_2 ) }
{ E - \varepsilon({\bf k}) - \varepsilon({\bf q}_1) - \varepsilon({\bf q}_2) - 
      \varepsilon({\bf P} - {\bf q}_1 - {\bf q}_2 - {\bf k} ) }   
\nonumber \\
                           &   & - \frac{2V}{N} \sum_{\bf k} 
\frac{ f({\bf k}) \, \cos{k_{y}} \, C_{y13}( {\bf q}_1 , {\bf q}_2 ) + 
       f({\bf k}) \, \sin{k_{y}} \, S_{y13}( {\bf q}_1 , {\bf q}_2 )}
{ E - \varepsilon({\bf k}) - \varepsilon({\bf q}_1) - \varepsilon({\bf q}_2) - 
      \varepsilon({\bf P} - {\bf q}_1 - {\bf q}_2 - {\bf k} ) }                                                                
\nonumber \\
                           &   & - \frac{2V}{N} \sum_{\bf k} 
\frac{ - f({\bf k}) \, \cos{k_x} \, C_{x13}( {\bf q}_1, {\bf P} - {\bf q}_1 - {\bf q}_2 - {\bf k} )  
       - f({\bf k}) \, \sin{k_x} \, S_{x13}( {\bf q}_1, {\bf P} - {\bf q}_1 - {\bf q}_2 - {\bf k} ) }
{ E - \varepsilon({\bf k}) - \varepsilon({\bf q}_1) - \varepsilon({\bf q}_2) - 
      \varepsilon({\bf P} - {\bf q}_1 - {\bf q}_2 - {\bf k} ) }         
\nonumber \\
                           &   & - \frac{2V}{N} \sum_{\bf k} 
\frac{ - f({\bf k}) \, \cos{k_y} \, C_{y13}( {\bf q}_1 , {\bf P} - {\bf q}_1 - {\bf q}_2 - {\bf k} )  
       - f({\bf k}) \, \sin{k_y} \, S_{y13}( {\bf q}_1 , {\bf P} - {\bf q}_1 - {\bf q}_2 - {\bf k} ) }
{ E - \varepsilon({\bf k}) - \varepsilon({\bf q}_1) - \varepsilon({\bf q}_2) - 
      \varepsilon({\bf P} - {\bf q}_1 - {\bf q}_2 - {\bf k} ) }         
\nonumber \\
                           &   & - \frac{2V}{N} \sum_{\bf k} 
\frac{ - f({\bf k}) \, \cos{q_{1x}} \, C_{x13}( {\bf k} , {\bf q}_2 ) 
       - f({\bf k}) \, \sin{q_{1x}} \, S_{x13}( {\bf k} , {\bf q}_2 ) }
{ E - \varepsilon({\bf k}) - \varepsilon({\bf q}_1) - \varepsilon({\bf q}_2) - 
      \varepsilon({\bf P} - {\bf q}_1 - {\bf q}_2 - {\bf k} ) }           
\nonumber \\
                           &   & - \frac{2V}{N} \sum_{\bf k} 
\frac{ - f({\bf k}) \, \cos{q_{1y}} \, C_{y13}( {\bf k} , {\bf q}_2 ) 
       - f({\bf k}) \, \sin{q_{1y}} \, S_{y13}( {\bf k} , {\bf q}_2 ) }
{ E - \varepsilon({\bf k}) - \varepsilon({\bf q}_1) - \varepsilon({\bf q}_2) - 
      \varepsilon({\bf P} - {\bf q}_1 - {\bf q}_2 - {\bf k} ) }           
\nonumber \\
                           &   & - \frac{2V}{N} \sum_{\bf k} 
\frac{   f({\bf k}) \, \cos{q_{1x}} \, C_{x13}( {\bf k} , {\bf P} - {\bf q}_1 - {\bf q}_2 - {\bf k} ) 
       + f({\bf k}) \, \sin{q_{1x}} \, S_{x13}( {\bf k} , {\bf P} - {\bf q}_1 - {\bf q}_2 - {\bf k} ) }
{ E - \varepsilon({\bf k}) - \varepsilon({\bf q}_1) - \varepsilon({\bf q}_2) - 
      \varepsilon({\bf P} - {\bf q}_1 - {\bf q}_2 - {\bf k} ) }           
\nonumber \\
                           &   & - \frac{2V}{N} \sum_{\bf k} 
\frac{   f({\bf k}) \, \cos{q_{1y}} \, C_{y13}( {\bf k} , {\bf P} - {\bf q}_1 - {\bf q}_2 - {\bf k} ) 
       + f({\bf k}) \, \sin{q_{1y}} \, S_{y13}( {\bf k} , {\bf P} - {\bf q}_1 - {\bf q}_2 - {\bf k} ) }
{ E - \varepsilon({\bf k}) - \varepsilon({\bf q}_1) - \varepsilon({\bf q}_2) - 
      \varepsilon({\bf P} - {\bf q}_1 - {\bf q}_2 - {\bf k} ) }           
\nonumber \\
                           &   & \hspace{0.1cm} - \frac{V}{N} \sum_{\bf k} 
\frac{ f({\bf k}) \, [ \cos{( P_x - q_{1x} - q_{2x} - k_x )} - \cos{q_{2x}} ] \, 
C_{x34}( {\bf k} , {\bf q}_1 ) }
{ E - \varepsilon({\bf k}) - \varepsilon({\bf q}_1) - \varepsilon({\bf q}_2) - 
      \varepsilon({\bf P} - {\bf q}_1 - {\bf q}_2 - {\bf k} ) }           
\nonumber \\
                           &   & \hspace{0.1cm} - \frac{V}{N} \sum_{\bf k} 
\frac{ f({\bf k}) \, [ \sin{( P_x - q_{1x} - q_{2x} - k_x )} - \sin{q_{2x}} ] \, 
S_{x34}( {\bf k} , {\bf q}_1 ) }
{ E - \varepsilon({\bf k}) - \varepsilon({\bf q}_1) - \varepsilon({\bf q}_2) - 
      \varepsilon({\bf P} - {\bf q}_1 - {\bf q}_2 - {\bf k} ) }     
\nonumber \\
                           &   & \hspace{0.1cm} - \frac{V}{N} \sum_{\bf k} 
\frac{ f({\bf k}) \, [ \cos{( P_y - q_{1y} - q_{2y} - k_y )} - \cos{q_{2y}} ] \, 
C_{y34}( {\bf k} , {\bf q}_1 ) }
{ E - \varepsilon({\bf k}) - \varepsilon({\bf q}_1) - \varepsilon({\bf q}_2) - 
      \varepsilon({\bf P} - {\bf q}_1 - {\bf q}_2 - {\bf k} ) }           
\nonumber \\
                           &   & \hspace{0.1cm} - \frac{V}{N} \sum_{\bf k} 
\frac{ f({\bf k}) \, [ \sin{( P_y - q_{1y} - q_{2y} - k_y )} - \sin{q_{2y}} ] \, 
S_{y34}( {\bf k} , {\bf q}_1 ) }
{ E - \varepsilon({\bf k}) - \varepsilon({\bf q}_1) - \varepsilon({\bf q}_2) - 
      \varepsilon({\bf P} - {\bf q}_1 - {\bf q}_2 - {\bf k} ) }   \: ,   
\label{4UV2d:eq:sm:ninety}    
\end{eqnarray}
where $f({\bf k}) \equiv 1$. The equations for $C_{x13}$, $S_{x13}$, $C_{y13}$, and $S_{y13}$ have the same right-hand-side as Eq.~(\ref{4UV2d:eq:sm:ninety}) but with $f({\bf k}) = \cos{k_x}$, $f({\bf k}) = \sin{k_x}$, $f({\bf k}) = \cos{k_y}$, and $f({\bf k}) = \sin{k_y}$, respectively. 

The third group includes the equations for $C_{x34}$, $S_{x34}$, $C_{y34}$, and $S_{y34}$. The equation for $C_{x34}$ reads 
\begin{eqnarray}
C_{x34}( {\bf q}_1 , {\bf q}_2 ) = \hspace{0.2cm} 
& & \hspace{0.4cm} \frac{U}{N} \sum_{\bf k} 
\frac{ f({\bf k}) \, A_{13}( {\bf q}_2 , {\bf P} - {\bf q}_1 - {\bf q}_2 - {\bf k} ) - 
       f({\bf k}) \, A_{13}( {\bf q}_2 , {\bf k} ) }
{ E - \varepsilon({\bf k}) - \varepsilon({\bf q}_1) - \varepsilon({\bf q}_2) - 
      \varepsilon({\bf P} - {\bf q}_1 - {\bf q}_2 - {\bf k} ) }               
\nonumber \\
& & \hspace{0.1cm} + \frac{U}{N} \sum_{\bf k} 
\frac{ - f({\bf k}) \, A_{13}( {\bf q}_1 , {\bf P} - {\bf q}_1 - {\bf q}_2 - {\bf k} ) + 
         f({\bf k}) \, A_{13}( {\bf q}_1 , {\bf k} ) }
{ E - \varepsilon({\bf k}) - \varepsilon({\bf q}_1) - \varepsilon({\bf q}_2) - 
      \varepsilon({\bf P} - {\bf q}_1 - {\bf q}_2 - {\bf k} ) }               
\nonumber \\
                           &   & \hspace{0.1cm} - \frac{V}{N} \sum_{\bf k} 
\frac{ f({\bf k}) \, ( \cos{q_{1x}} - \cos{q_{2x}} ) \, 
C_{x12}( {\bf k} , {\bf P} - {\bf q}_1 - {\bf q}_2 - {\bf k} ) }
{ E - \varepsilon({\bf k}) - \varepsilon({\bf q}_1) - \varepsilon({\bf q}_2) - 
      \varepsilon({\bf P} - {\bf q}_1 - {\bf q}_2 - {\bf k} ) }               
\nonumber \\
                           &   & \hspace{0.1cm} - \frac{V}{N} \sum_{\bf k} 
\frac{ f({\bf k}) \, ( \sin{q_{1x}} - \sin{q_{2x}} ) \, 
S_{x12}( {\bf k} , {\bf P} - {\bf q}_1 - {\bf q}_2 - {\bf k} ) }
{ E - \varepsilon({\bf k}) - \varepsilon({\bf q}_1) - \varepsilon({\bf q}_2) - 
      \varepsilon({\bf P} - {\bf q}_1 - {\bf q}_2 - {\bf k} ) }     
\nonumber \\
                           &   & \hspace{0.1cm} - \frac{V}{N} \sum_{\bf k} 
\frac{ f({\bf k}) \, ( \cos{q_{1y}} - \cos{q_{2y}} ) \, 
C_{y12}( {\bf k} , {\bf P} - {\bf q}_1 - {\bf q}_2 - {\bf k} ) }
{ E - \varepsilon({\bf k}) - \varepsilon({\bf q}_1) - \varepsilon({\bf q}_2) - 
      \varepsilon({\bf P} - {\bf q}_1 - {\bf q}_2 - {\bf k} ) }               
\nonumber \\
                           &   & \hspace{0.1cm} - \frac{V}{N} \sum_{\bf k} 
\frac{ f({\bf k}) \, ( \sin{q_{1y}} - \sin{q_{2y}} ) \, 
S_{y12}( {\bf k} , {\bf P} - {\bf q}_1 - {\bf q}_2 - {\bf k} ) }
{ E - \varepsilon({\bf k}) - \varepsilon({\bf q}_1) - \varepsilon({\bf q}_2) - 
      \varepsilon({\bf P} - {\bf q}_1 - {\bf q}_2 - {\bf k} ) }     
\nonumber \\
                           &   & - \frac{2V}{N} \sum_{\bf k} 
\frac{ f({\bf k}) \, \cos{q_{1x}} \, C_{x13}( {\bf q}_2 , {\bf P} - {\bf q}_1 - {\bf q}_2 - {\bf k} ) + 
       f({\bf k}) \, \sin{q_{1x}} \, S_{x13}( {\bf q}_2 , {\bf P} - {\bf q}_1 - {\bf q}_2 - {\bf k} ) }
{ E - \varepsilon({\bf k}) - \varepsilon({\bf q}_1) - \varepsilon({\bf q}_2) - 
      \varepsilon({\bf P} - {\bf q}_1 - {\bf q}_2 - {\bf k} ) }   
\nonumber \\
                           &   & - \frac{2V}{N} \sum_{\bf k} 
\frac{ f({\bf k}) \, \cos{q_{1y}} \, C_{y13}( {\bf q}_2 , {\bf P} - {\bf q}_1 - {\bf q}_2 - {\bf k} ) + 
       f({\bf k}) \, \sin{q_{1y}} \, S_{y13}( {\bf q}_2 , {\bf P} - {\bf q}_1 - {\bf q}_2 - {\bf k} )}
{ E - \varepsilon({\bf k}) - \varepsilon({\bf q}_1) - \varepsilon({\bf q}_2) - 
      \varepsilon({\bf P} - {\bf q}_1 - {\bf q}_2 - {\bf k} ) }                                                                
\nonumber \\
                           &   & - \frac{2V}{N} \sum_{\bf k} 
\frac{ - f({\bf k}) \, \cos{q_{1x}} \, C_{x13}( {\bf q}_2, {\bf k} )  
       - f({\bf k}) \, \sin{q_{1x}} \, S_{x13}( {\bf q}_2, {\bf k} ) }
{ E - \varepsilon({\bf k}) - \varepsilon({\bf q}_1) - \varepsilon({\bf q}_2) - 
      \varepsilon({\bf P} - {\bf q}_1 - {\bf q}_2 - {\bf k} ) }         
\nonumber \\
                           &   & - \frac{2V}{N} \sum_{\bf k} 
\frac{ - f({\bf k}) \, \cos{q_{1y}} \, C_{y13}( {\bf q}_2 , {\bf k} )  
       - f({\bf k}) \, \sin{q_{1y}} \, S_{y13}( {\bf q}_2 , {\bf k} ) }
{ E - \varepsilon({\bf k}) - \varepsilon({\bf q}_1) - \varepsilon({\bf q}_2) - 
      \varepsilon({\bf P} - {\bf q}_1 - {\bf q}_2 - {\bf k} ) }         
\nonumber \\
                           &   & - \frac{2V}{N} \sum_{\bf k} 
\frac{ - f({\bf k}) \, \cos{q_{2x}} \, C_{x13}( {\bf q}_1 , {\bf P} - {\bf q}_1 - {\bf q}_2 - {\bf k} ) 
       - f({\bf k}) \, \sin{q_{2x}} \, S_{x13}( {\bf q}_1 , {\bf P} - {\bf q}_1 - {\bf q}_2 - {\bf k} ) }
{ E - \varepsilon({\bf k}) - \varepsilon({\bf q}_1) - \varepsilon({\bf q}_2) - 
      \varepsilon({\bf P} - {\bf q}_1 - {\bf q}_2 - {\bf k} ) }           
\nonumber \\
                           &   & - \frac{2V}{N} \sum_{\bf k} 
\frac{ - f({\bf k}) \, \cos{q_{2y}} \, C_{y13}( {\bf q}_1 , {\bf P} - {\bf q}_1 - {\bf q}_2 - {\bf k} ) 
       - f({\bf k}) \, \sin{q_{2y}} \, S_{y13}( {\bf q}_1 , {\bf P} - {\bf q}_1 - {\bf q}_2 - {\bf k} ) }
{ E - \varepsilon({\bf k}) - \varepsilon({\bf q}_1) - \varepsilon({\bf q}_2) - 
      \varepsilon({\bf P} - {\bf q}_1 - {\bf q}_2 - {\bf k} ) }           
\nonumber \\
                           &   & - \frac{2V}{N} \sum_{\bf k} 
\frac{   f({\bf k}) \, \cos{q_{2x}} \, C_{x13}( {\bf q}_1 , {\bf k} ) 
       + f({\bf k}) \, \sin{q_{2x}} \, S_{x13}( {\bf q}_1 , {\bf k} ) }
{ E - \varepsilon({\bf k}) - \varepsilon({\bf q}_1) - \varepsilon({\bf q}_2) - 
      \varepsilon({\bf P} - {\bf q}_1 - {\bf q}_2 - {\bf k} ) }           
\nonumber \\
                           &   & - \frac{2V}{N} \sum_{\bf k} 
\frac{   f({\bf k}) \, \cos{q_{2y}} \, C_{y13}( {\bf q}_1 , {\bf k} ) 
       + f({\bf k}) \, \sin{q_{2y}} \, S_{y13}( {\bf q}_1 , {\bf k} ) }
{ E - \varepsilon({\bf k}) - \varepsilon({\bf q}_1) - \varepsilon({\bf q}_2) - 
      \varepsilon({\bf P} - {\bf q}_1 - {\bf q}_2 - {\bf k} ) }           
\nonumber \\
                           &   & \hspace{0.1cm} - \frac{V}{N} \sum_{\bf k} 
\frac{ f({\bf k}) \, [ \cos{k_{x}} - \cos{( P_x - q_{1x} - q_{2x} - k_x )} ] \, 
C_{x34}( {\bf q}_1 , {\bf q}_2 ) }
{ E - \varepsilon({\bf k}) - \varepsilon({\bf q}_1) - \varepsilon({\bf q}_2) - 
      \varepsilon({\bf P} - {\bf q}_1 - {\bf q}_2 - {\bf k} ) }           
\nonumber \\
                           &   & \hspace{0.1cm} - \frac{V}{N} \sum_{\bf k} 
\frac{ f({\bf k}) \, [ \sin{k_{x}} - \sin{( P_x - q_{1x} - q_{2x} - k_x )} ] \, 
S_{x34}( {\bf q}_1 , {\bf q}_2 ) }
{ E - \varepsilon({\bf k}) - \varepsilon({\bf q}_1) - \varepsilon({\bf q}_2) - 
      \varepsilon({\bf P} - {\bf q}_1 - {\bf q}_2 - {\bf k} ) }     
\nonumber \\
                           &   & \hspace{0.1cm} - \frac{V}{N} \sum_{\bf k} 
\frac{ f({\bf k}) \, [ \cos{k_{y}} - \cos{( P_y - q_{1y} - q_{2y} - k_y )} ] \, 
C_{y34}( {\bf q}_1 , {\bf q}_2 ) }
{ E - \varepsilon({\bf k}) - \varepsilon({\bf q}_1) - \varepsilon({\bf q}_2) - 
      \varepsilon({\bf P} - {\bf q}_1 - {\bf q}_2 - {\bf k} ) }           
\nonumber \\
                           &   & \hspace{0.1cm} - \frac{V}{N} \sum_{\bf k} 
\frac{ f({\bf k}) \, [ \sin{k_{y}} - \sin{( P_y - q_{1y} - q_{2y} - k_y )} ] \, 
S_{y34}( {\bf q}_1 , {\bf q}_2 ) }
{ E - \varepsilon({\bf k}) - \varepsilon({\bf q}_1) - \varepsilon({\bf q}_2) - 
      \varepsilon({\bf P} - {\bf q}_1 - {\bf q}_2 - {\bf k} ) }   \: ,   
\label{4UV2d:eq:sm:ninetyone}    
\end{eqnarray}
where $f({\bf k}) = \cos{k_x}$. The equations for $S_{x34}$, $C_{y34}$, and $S_{y34}$ have the same right-hand-side as Eq.~(\ref{4UV2d:eq:sm:ninety}) but with $f({\bf k}) = \sin{k_x}$, $f({\bf k}) = \cos{k_y}$, and $f({\bf k}) = \sin{k_y}$, respectively.

\subsection{\label{4UV2d:sec:sm:j}
Quad threshold values. Bosons in 1D
}

Table~\ref{4UV2d:tab:four} contains numerical values of quad threshold $V_{4}$ in the 1D Bose-$UV$ model. The $U = 10 \, t$ and $U = 15 \,t$ values are plotted in Fig.~\ref{4UV2d:fig:five} of the main text. The $N = \infty$ values are plotted in Fig.~\ref{4UV2d:fig:six} of the main text. 
\begin{table*}[ht]
\renewcommand{\tabcolsep}{0.2cm}
\renewcommand{\arraystretch}{1.3}
\begin{tabular}{|c||c|c|c|c|c||c|}
\hline\hline
 $U/t$  &   $N = 12$     &  $N = 16$      &  $N = 24$      &  $N = 40$      &  $N = 64$      & $N = \infty$   \\  \hline
\hline 
   2    &  $0.67978759$  &  $0.67612518$  &  $0.67270458$  &  $0.67014501$  &  $0.66877532$  &  $0.66658022$  \\  \hline
   4    &  $1.02817626$  &  $1.01974822$  &  $1.01222705$  &  $1.00683837$  &  $1.00405731$  &  $0.99976860$  \\  \hline
   6    &  $1.23405810$  &  $1.22362104$  &  $1.21446240$  &  $1.20800341$  &  $1.20471282$  &  $1.19973379$  \\  \hline
   8    &  $1.36847607$  &  $1.35763258$  &  $1.34816455$  &  $1.34151708$  &  $1.33814361$  &  $1.33306700$  \\  \hline
  10    &  $1.46284765$  &  $1.45226239$  &  $1.44303076$  &  $1.43655322$  &  $1.43326507$  &  $1.42832728$  \\  \hline
  15    &  $1.60892578$  &  $1.59973220$  &  $1.59168310$  &  $1.58600927$  &  $1.58311676$  &  $1.57875036$  \\  \hline
  20    &  $1.69251171$  &  $1.68464892$  &  $1.67773486$  &  $1.67283544$  &  $1.67032440$  &  $1.66651501$  \\  \hline
  30    &  $1.78456640$  &  $1.77859973$  &  $1.77331787$  &  $1.76954504$  &  $1.76759368$  &  $1.76461773$  \\  \hline
  40    &  $1.83419140$  &  $1.82942474$  &  $1.82518408$  &  $1.82213818$  &  $1.82055267$  &  $1.81812714$  \\  \hline
  50    &  $1.86522705$  &  $1.86127285$  &  $1.85773974$  &  $1.85519128$  &  $1.85385955$  &  $1.85181436$  \\  \hline
\hline 
\end{tabular}
\caption{
Quad threshold $V_{4}$ in the 1D Bose-$UV$ model as a function of $U$ and linear chain size $N$.   
} 
\label{4UV2d:tab:four}
\end{table*}

\subsection{\label{4UV2d:sec:sm:k}
Quad threshold values. Fermions in 1D
}

Table~\ref{4UV2d:tab:five} contains numerical values of quad threshold $V_{4}$ in the 1D Fermi-$UV$ model. The $N = \infty$ values are plotted in Fig.~\ref{4UV2d:fig:eight} of the main text. 
\begin{table*}[ht]
\renewcommand{\tabcolsep}{0.2cm}
\renewcommand{\arraystretch}{1.3}
\begin{tabular}{|c||c|c|c|c|c|c|c||c|}
\hline\hline
 $U/t$  &  $N = 24$      &  $N = 32$      &  $N = 40$      &  $N = 48$      &  $N = 56$      &  $N = 60$      &  $N = 64$      & $N = \infty$ \\  \hline
\hline 
   0    &  $0.88211547$  &  $0.85984001$  &  $0.85564994$  &  $0.85507316$  &  $$            &                &  $$            &  $0.852189$  \\  \hline 
   2    &  $1.21607065$  &  $1.16893796$  &  $1.14527320$  &  $1.13458999$  &  $$            &                &  $$            &  $1.081174$  \\  \hline
   4    &  $1.43205537$  &  $1.38314580$  &  $1.35440101$  &  $1.33662063$  &  $$            &                &  $$            &  $1.247719$  \\  \hline
   6    &  $1.56603403$  &  $1.51955604$  &  $1.49149818$  &  $1.47287612$  &  $$            &                &  $$            &  $1.379766$  \\  \hline
   8    &  $1.65504789$  &  $1.61142616$  &  $1.58500526$  &  $1.56714515$  &  $$            &                &  $$            &  $1.477845$  \\  \hline
  10    &  $1.71805805$  &  $1.67699165$  &  $1.65216560$  &  $1.63531539$  &  $$            &                &  $$            &  $1.551064$  \\  \hline
  15    &  $1.81611190$  &  $1.77985420$  &  $1.75806541$  &  $1.74332412$  &  $$            &                &  $$            &  $1.669618$  \\  \hline
  20    &  $$            &  $1.83922004$  &  $1.81942634$  &  $1.80609588$  &  $$            &                &  $$            &  $1.739443$  \\  \hline
  25    &  $$            &  $1.87543426$  &  $1.85933685$  &  $1.84697967$  &  $$            &                &  $1.83121109$  &  $1.783905$  \\  \hline
  30    &  $$            &  $$            &  $1.88731109$  &  $1.87567047$  &  $$            &                &  $1.86086181$  &  $1.816436$  \\  \hline
  40    &  $$            &  $$            &  $$            &  $1.91247665$  &  $1.90551257$  &                &  $1.89968902$  &  $1.861326$  \\  \hline
  50    &  $$            &  $$            &  $$            &  $$            &  $$            &  $1.92647302$  &  $1.92392529$  &  $1.885709$  \\  \hline
\hline 
\end{tabular}
\caption{
Quad threshold $V_{4}$ in the 1D Fermi-$UV$ model as a function of $U$ and linear chain size $N$.   
} 
\label{4UV2d:tab:five}
\end{table*}

\subsection{\label{4UV2d:sec:sm:l}
Quad threshold values. Bosons in 2D
}

Table~\ref{4UV2d:tab:six} contains numerical values of quad threshold $V_{4}$ in the 2D Bose-$UV$ model at ${\bf P} = (0,0)$. The decay channel is $[ 4 ] \rightarrow [ 22 ]$. A parabolic fit $V_{4}(N_x) = V_{4\infty} + c \cdot N^{-2}_{x}$ was utilized to estimate $N_x = \infty$ limit values. The latter are plotted in Fig.~\ref{4UV2d:fig:eleven} of the main text. 
\begin{table*}[ht]
\renewcommand{\tabcolsep}{0.2cm}
\renewcommand{\arraystretch}{1.3}
\begin{tabular}{|c||c|c|c||c|}
\hline\hline
 $U/t$  &  $N_x = N_y = 6$   &  $N_x = N_y = 8$   &  $N_x = N_y = 10$   &  $N_x = N_y = \infty$   \\  \hline
\hline 
   0    &  $0.0$             &  $0.0$             &  $0.0$              &  $0.0$                  \\  \hline
   1    &  $0.22246818$      &  $0.22235984$      &  $0.22231025$       &  $0.22222209$           \\  \hline
   2    &  $0.40120290$      &  $0.40067121$      &  $0.40042762$       &  $0.39999457$           \\  \hline
   3    &  $0.54807853$      &  $0.54691200$      &  $0.54638175$       &  $0.54543908$           \\  \hline
   4    &  $0.67085990$      &  $0.66898689$      &  $$                 &  $0.66657873$           \\  \hline
   5    &  $0.77494392$      &  $0.77238273$      &  $$                 &  $0.76908977$           \\  \hline
   7    &  $0.94161033$      &  $0.93787956$      &  $$                 &  $0.93308285$           \\  \hline
  10    &  $1.12189064$      &  $1.11701011$      &  $$                 &  $1.11073514$           \\  \hline
  15    &  $1.31682319$      &  $1.31115760$      &  $$                 &  $1.30387327$           \\  \hline
  20    &  $1.44122962$      &  $1.43547554$      &  $$                 &  $1.42807743$           \\  \hline
  25    &  $1.52740440$      &  $1.52183418$      &  $$                 &  $1.51467246$           \\  \hline
  30    &  $1.59059104$      &  $1.58530158$      &  $$                 &  $1.57850084$           \\  \hline
  40    &  $1.67701606$      &  $1.67233543$      &  $$                 &  $1.66631747$           \\  \hline
  50    &  $1.73333091$      &  $1.72918968$      &  $$                 &  $1.72386524$           \\  \hline
 100    &  $1.85756568$      &  $1.85503349$      &  $$                 &  $1.85177781$           \\  \hline
\hline 
\end{tabular}
\caption{
Quad threshold $V_{4}$ in the 2D Bose-$UV$ model as a function of $U$ and linear lattice size $N_x$. The decay channel is $[ 4 ] \rightarrow [ 22 ]$.  
} 
\label{4UV2d:tab:six}
\end{table*}

Table~\ref{4UV2d:tab:seven} contains numerical values of {\em trion} threshold $V_{3}$ in the 2D Bose-$UV$ model at ${\bf P} = (0,0)$. The decay channel is $[ 3 ] \rightarrow [ 21 ]$. A parabolic fit $V_{3}(N_x) = V_{3\infty} + c \cdot N^{-2}_{x}$ was utilized to estimate the $N_x = \infty$ limit values. Some of $N_{3\infty}$ are plotted in Fig.~\ref{4UV2d:fig:eleven} of the main text. The trion energies were computed by solving the trion integral equations developed earlier.~\cite{SupplMat2014}

\begin{table*}[ht]
\renewcommand{\tabcolsep}{0.2cm}
\renewcommand{\arraystretch}{1.3}
\begin{tabular}{|c||c|c|c||c|}
\hline\hline
 $U/t$  &  $N_x = N_y = 8$   &  $N_x = N_y = 10$   &  $N_x = N_y = 12$   &  $N_x = N_y = \infty$   \\  \hline
\hline 
   0    &  $0.0$             &  $0.0$              &  $0.0$              &  $0.0$                  \\  \hline
   1    &  $0.22223014$      &  $0.22222785$       &  $0.22222633$       &  $0.22222286$           \\  \hline
   2    &  $0.40003776$      &  $0.40002403$       &  $0.40001716$       &  $0.40000156$           \\  \hline
   3    &  $0.54553489$      &  $0.54550666$       &  $0.54549140$       &  $0.54545673$           \\  \hline
   4    &  $0.66679344$      &  $0.66674842$       &  $0.66672401$       &  $0.66666852$           \\  \hline
   5    &  $0.76940269$      &  $0.76934089$       &  $0.76930809$       &  $0.76923353$           \\  \hline
   6    &  $0.85735435$      &  $0.85727882$       &  $0.85723762$       &  $0.85714399$           \\  \hline
   7    &  $0.93357963$      &  $0.93349113$       &  $0.93344383$       &  $0.93333632$           \\  \hline
   8    &  $1.00027553$      &  $1.00017650$       &  $1.00012279$       &  $1.00000072$           \\  \hline
   9    &  $1.05912284$      &  $1.05901527$       &  $1.05895729$       &  $1.05882551$           \\  \hline
  10    &  $1.11142921$      &  $1.11131477$       &  $1.11125297$       &  $1.11111252$           \\  \hline
  13    &  $1.23844947$      &  $1.23832206$       &  $1.23825340$       &  $1.23809734$           \\  \hline
  15    &  $1.30471382$      &  $1.30458259$       &  $1.30451087$       &  $1.30434788$           \\  \hline
  18    &  $1.38498725$      &  $1.38485374$       &  $1.38478126$       &  $1.38461653$           \\  \hline
  20    &  $1.42894248$      &  $1.42880897$       &  $1.42873649$       &  $1.42857177$           \\  \hline
  23    &  $1.48423576$      &  $1.48410453$       &  $1.48403358$       &  $1.48387232$           \\  \hline
  25    &  $1.51551017$      &  $1.51538124$       &  $1.51531181$       &  $1.51515402$           \\  \hline
  30    &  $1.57931175$      &  $1.57920188$       &  $1.57915077$       &  $1.57903459$           \\  \hline
  35    &  $1.62838478$      &  $1.62835578$       &  $1.62840080$       &  $1.62850310$           \\  \hline
  40    &  $1.66722831$      &  $1.66726570$       &  $1.66738548$       &  $1.66765771$           \\  \hline
  45    &  $1.69874076$      &  $1.69883079$       &  $1.69901161$       &  $1.69942255$           \\  \hline
  50    &  $1.72481727$      &  $1.72495155$       &  $1.72518272$       &  $1.72570811$           \\  \hline
\hline 
\end{tabular}
\caption{
{\em Trion} threshold $V_{3}$ in the 2D Bose-$UV$ model as a function of $U$ and linear lattice size $N_x$. The decay channel is $[ 3 ] \rightarrow [ 21 ]$.  
} 
\label{4UV2d:tab:seven}
\end{table*}

\subsection{\label{4UV2d:sec:sm:m}
Quad threshold values. Fermions in 2D
}

Table~\ref{4UV2d:tab:eight} lists quad threshold values for total spin $S = 2$. Table~\ref{4UV2d:tab:nine} lists quad threshold values for total spin $S = 1$. Table~\ref{4UV2d:tab:ten} lists quad threshold values for total spin $S = 0$.

\begin{table*}[ht]
\renewcommand{\tabcolsep}{0.2cm}
\renewcommand{\arraystretch}{1.3}
\begin{tabular}{|c||c|c|c||c|}
\hline\hline
 $U/t$  &  $N_x = N_y = 6$   &  $N_x = N_y = 8$   &  $N_x = N_y = 10$   &  $N_x = N_y = \infty$   \\  \hline
\hline 
   0    &  $4.23495798$      &  $4.23253755$      &  $4.23239784$       &  $4.23214946$           \\  \hline
   1    &  $4.16412010$      &  $4.16136894$      &  $4.16119422$       &  $4.16088360$           \\  \hline
   2    &  $4.10662269$      &  $4.10356407$      &  $4.10335433$       &  $4.10298145$           \\  \hline
   3    &  $4.05961408$      &  $4.05627527$      &  $4.05602828$       &  $4.05558918$           \\  \hline
   4    &  $4.02085123$      &  $4.01725950$      &  $4.01697723$       &  $4.01647541$           \\  \hline
   5    &  $3.98859672$      &  $3.98477821$      &  $3.98446159$       &  $3.98389871$           \\  \hline
   7    &  $3.93856315$      &  $3.93436164$      &  $3.93398246$       &  $3.93330836$           \\  \hline
  10    &  $3.88740577$      &  $3.88277244$      &  $3.88231315$       &  $3.88149663$           \\  \hline
  15    &  $3.83654565$      &  $3.83144216$      &  $3.83088340$       &  $3.82989004$           \\  \hline
  20    &  $3.80696897$      &  $3.80157423$      &  $3.80094757$       &  $3.79983350$           \\  \hline
  25    &  $3.78795127$      &  $3.78236503$      &  $3.78168914$       &  $3.78048755$           \\  \hline
  30    &  $3.77481956$      &  $3.76909980$      &  $3.76838714$       &  $3.76712018$           \\  \hline
  40    &  $3.75785264$      &  $3.75212840$      &  $3.75136688$       &  $3.75001306$           \\  \hline
  50    &  $3.73931236$      &  $3.74182090$      &  $3.74102752$       &  $3.73961706$           \\  \hline
 100    &  $3.70311670$      &  $3.72124185$      &  $3.72038215$       &  $3.71885379$           \\  \hline
\hline 
\end{tabular}
\caption{
$S = 2$ quad threshold $V_{4}$ in the 2D Fermi-$UV$ model as a function of $U$ and linear lattice size $N_x$. The decay channel is $[ 4 ] \rightarrow [ 22 ]$.  
} 
\label{4UV2d:tab:eight}
\end{table*}
\begin{table*}[ht]
\renewcommand{\tabcolsep}{0.2cm}
\renewcommand{\arraystretch}{1.3}
\begin{tabular}{|c||c|c|c||c|}
\hline\hline
 $U/t$  &  $N_x = N_y = 6$   &  $N_x = N_y = 8$   &  $N_x = N_y = 10$   &  $N_x = N_y = \infty$   \\  \hline
\hline 
   0    &  $2.79217109$      &  $2.71363258$      &  $2.69079933$       &  $2.65020688$           \\  \hline
   1    &  $2.91020317$      &  $2.82616157$      &  $2.80076408$       &  $2.75561298$           \\  \hline
   2    &  $3.00757026$      &  $2.91925239$      &  $2.89169731$       &  $2.84271050$           \\  \hline
   3    &  $3.08517265$      &  $2.99404563$      &  $2.96484260$       &  $2.91292610$           \\  \hline
   4    &  $3.14600944$      &  $3.05341911$      &  $3.02306556$       &  $2.96910369$           \\  \hline
   5    &  $3.19359321$      &  $3.10057182$      &  $3.06948738$       &  $3.01422615$           \\  \hline
   7    &  $3.26096611$      &  $3.16894111$      &  $3.13727073$       &  $3.08096783$           \\  \hline
  10    &  $3.32159080$      &  $3.23281059$      &  $3.20139427$       &  $3.14554303$           \\  \hline
  15    &  $3.37492027$      &  $3.29161109$      &  $3.26146736$       &  $3.20787850$           \\  \hline
  20    &  $3.40389671$      &  $3.32482872$      &  $3.29597206$       &  $3.24467133$           \\  \hline
  25    &  $3.42226600$      &  $3.34637641$      &  $3.31859626$       &  $3.26920932$           \\  \hline
  30    &  $3.43504753$      &  $3.36158409$      &  $3.33467597$       &  $3.28683931$           \\  \hline
  40    &  $3.45179634$      &  $3.38175163$      &  $3.35613441$       &  $3.31059268$           \\  \hline
  50    &  $3.46236000$      &  $3.39458885$      &  $3.36986579$       &  $3.32591368$           \\  \hline
 100    &  $3.48501167$      &  $3.42233924$      &  $3.39970359$       &  $3.35946243$           \\  \hline
\hline 
\end{tabular}
\caption{
$S = 1$ quad threshold $V_{4}$ in the 2D Fermi-$UV$ model as a function of $U$ and linear lattice size $N_x$. The decay channel is $[ 4 ] \rightarrow [ 22 ]$.  
} 
\label{4UV2d:tab:nine}
\end{table*}
\begin{table*}[ht]
\renewcommand{\tabcolsep}{0.2cm}
\renewcommand{\arraystretch}{1.3}
\begin{tabular}{|c||c|c|c||c|}
\hline\hline
 $U/t$  &  $N_x = N_y = 6$   &  $N_x = N_y = 8$   &  $N_x = N_y = 10$   &  $N_x = N_y = \infty$   \\  \hline
\hline 
   0    &  $2.66593208$      &  $2.61358757$      &  $2.58745155$       &  $2.54404696$           \\  \hline
   1    &  $2.81226234$      &  $2.75910224$      &  $2.73014411$       &  $2.68564234$           \\  \hline
   2    &  $2.91970024$      &  $2.86858024$      &  $2.83836708$       &  $2.79516071$           \\  \hline
   3    &  $3.00006752$      &  $2.95212364$      &  $2.92164649$       &  $2.88075139$           \\  \hline
   4    &  $3.06182060$      &  $3.01722526$      &  $2.98702659$       &  $2.94866527$           \\  \hline
   5    &  $3.11056785$      &  $3.06908912$      &  $3.03943290$       &  $3.00347923$           \\  \hline
   7    &  $3.18252897$      &  $3.14620018$      &  $3.11788673$       &  $3.08598931$           \\  \hline
  10    &  $3.25325965$      &  $3.22232856$      &  $3.19594940$       &  $3.16839521$           \\  \hline
  15    &  $3.32317291$      &  $3.29758251$      &  $3.27366851$       &  $3.23115473$           \\  \hline
  20    &  $3.36497001$      &  $3.34246360$      &  $3.32025880$       &  $3.28078360$           \\  \hline
  25    &  $3.39286384$      &  $3.37234098$      &  $3.35136372$       &  $3.31407081$           \\  \hline
  30    &  $3.41283416$      &  $3.39368852$      &  $3.37362678$       &  $3.33796146$           \\  \hline
  40    &  $3.43955657$      &  $3.42219071$      &  $3.40339570$       &  $3.36998234$           \\  \hline
  50    &  $3.45663726$      &  $3.44036781$      &  $3.42240440$       &  $3.39046944$           \\  \hline
 100    &  $3.49350006$      &  $3.47947845$      &  $3.46335113$       &  $3.43468033$           \\  \hline
\hline 
\end{tabular}
\caption{
$S = 0$ quad threshold $V_{4}$ in the 2D Fermi-$UV$ model as a function of $U$ and linear lattice size $N_x$. The decay channel is $[ 4 ] \rightarrow [ 22 ]$.  
} 
\label{4UV2d:tab:ten}
\end{table*}

\end{widetext}

\end{document}